\documentclass{aa}  
\usepackage[english] {babel}
\usepackage{astron}
\usepackage[dvips]{graphicx}
\usepackage{ae}
\usepackage{color}
\usepackage{amsfonts}
\usepackage{tabularx}
\usepackage{natbib}
\usepackage{amssymb,amsmath}
\usepackage{listliketab}
\usepackage[toc,page]{appendix}
\usepackage{txfonts}
\usepackage{placeins}
\usepackage{float}
\usepackage{afterpage}
\usepackage{longtable}
\usepackage{supertabular}
\usepackage{url}
\usepackage{rotating}
\usepackage{wasysym}
\usepackage{adjustbox,lipsum}
\usepackage{pdflscape}
\usepackage{supertabular}
\begin{document}
\bibliographystyle{aa} 

   \title{Dozens of compact and high velocity-dispersion, early-type galaxies in Sloan Digital Sky Survey}

   \author{Christoph Saulder
          \inst{1,4}
          \and
          Remco C.~E. van den Bosch
          \inst{2}
          \and
          Steffen Mieske
          \inst{3}              
          }

   \institute{
   European Southern Observatory,
   Karl-Schwarzschild-Stra\ss e 2, 85748 Garching bei M\"unchen, Germany\\
    \email{christoph.saulder@equinoxomega.net}\\
    \and   
  Max-Planck-Institut f\"ur Astronomie, K\"onigstuhl 17, 69117 Heidelberg, Germany\\    
    \and   
   European Southern Observatory,
   Alonso de C\'{o}rdova 3107, Vitacura, Casilla 19001, Santiago, Chile\\
    \and
   Department of Astrophysics, University of Vienna,
   T\"urkenschanzstra\ss e 17, 1180 Vienna, Austria\\
             }
   \date{Received December 5, 2014 ; accepted March 17, 2015}

\abstract
{Passive galaxies at high redshift are much smaller than equally massive early types today. If this size evolution is caused by stochastic merging processes, then a small fraction of the compact galaxies should persist until today. Up to now it has not been possible to systematically identify the existence of such objects in SDSS.}
{We aim at finding potential survivors of these compact galaxies in SDSS, as targets for more detailed follow-up observations. }
{From the virial theorem, it is expected that for a given mass, compact galaxies have stellar velocity dispersion higher than the mean owing to their smaller sizes. Therefore velocity dispersion, coupled with size (or mass), is an appropriate method of selecting relics, independent of the stellar population properties. Based on these considerations, we designed a set of criteria the use the distribution of early-type galaxies from SDSS on the $\textrm{log}_{10}(R_{0})$-$\textrm{log}_{10}(\sigma_{0})$ plane to find the most extreme objects on it. We thus selected compact massive galaxy candidates by restricting them to high velocity dispersions $\sigma_{0} >$323.2 km s$^{-1}$ and small sizes $R_{0}<$2.18 kpc.}
{We find 76 galaxies at 0.05 < z < 0.2, which have properties that are similar to the typical quiescent galaxies at high redshift. We discuss how these galaxies relate to average present-day early-type galaxies. We study how well these galaxies fit on well-known local universe relations of early-type galaxies, such as the fundamental plane, the red sequence, or mass-size relations. As expected from the selection criteria, the candidates are located in an extreme corner of the mass-size plane. However, they do not extend as deeply into the so-called zone of exclusion as some of the red nuggets found at high redshift, since they are a factor 2-3 less massive on a given intrinsic scale size. Several of our candidates are close to the size resolution limit of SDSS, but are not so small that they are classified as point sources. We find that our candidates are systematically offset on a scaling relation compared to the average early-type galaxies, but still within the general range of other early-type galaxies. Furthermore, our candidates are similar to the mass-size range expected for passive evolution of the red nuggets from their high redshift to the present.}
{The 76 selected candidates form an appropriate set of objects for further follow-up observations. They do not constitute a separate population of peculiar galaxies, but form the extreme tail of a continuous distribution of early-type galaxies. We argue that selecting a high-velocity dispersion is the best way to find analogues of compact high redshift galaxies in the local universe.} 

   \keywords{ galaxies: elliptical and lenticular, cD --
    surveys --
     galaxies: fundamental parameters --
      galaxies: peculiar 
               }

   \maketitle

\section{Introduction}
%\subsection{Compact and massive early-type galaxies}
Compact massive early-type galaxies are common at high redshifts ($z>1$) \citep{Trujillo:2006,vanDokkum:2008}. At those epochs, an average $10^{11}$~M$_{\astrosun}$ passive galaxy has a size of 1~kpc, whereas today's early-type galaxies are three to ten times larger \citep{Taylor:2010}, indicating that galaxies undergo a significant amount of size evolution \citep{vanderWel:2014b}. If a small fraction of those early galaxies, which are also called red nuggets, evolve completely passively, without any mergers, then some of them must have remained compact until today. Those objects would be pristine relics, which would allow direct insight into how these objects formed long ago. It is therefore interesting to find out if any of those systems still remain today. 

Several studies have been done to find such objects using the SDSS photometry with varying success. \citet{Taylor:2010} found no analogues of the $z\sim2$ early-type galaxies. Similarly, \citet{Trujillo:2009} found few objects, but those turned out to be young with ages of 2 Gyr. \citet{Damjanov:2009} found nine objects -- some of which were old -- indicating  that some of the relics must exist. \citet{Damjanov:2014} found several object in the BOSS survey and measured a space density of $10^{-6}$ galaxies Mpc$^{-3}$, which is consistent with expectations from semi-analytic models \citep{Quilis:13}. In contrast to this, there are also claims of detecting high number densities of these compact massive early-type galaxies in cluster environments \citep{Valentinuzzi:2010}, and even in the field \citep{Poggianti:2013}, that are in tension with the upper limits of the model predictions \citep{Damjanov:2014}. The results from \citet{Valentinuzzi:2010} have been debated, and various inconsistencies with other works are pointed out in \citet{Taylor:2010}.

The virial theorem (and its observational projection, the fundamental plane) predicts that these small galaxies must have high-velocity dispersions. This has been directly confirmed with deep spectroscopy of a handful of these objects \citep{vanDokkum:2009, vdSande:2013, Toft:2012}.  This makes the dispersion a very good discriminator for finding these very dense objects in the local universe because measuring dispersion is much easier at low redshift. The other advantage is that such a selection is independent of uncertainties in photometric stellar masses. The stellar velocity dispersion was used in \citet{vdBosch:2012} as a discriminator and found six compact objects -- including NGC 1277 -- in the HETMGS survey \citep{vdBosch:2015}. These objects appear to be consistent with being relics, given their size, mass, and velocity dispersion \citep{vdBosch:2012}. In particular, NGC 1277 has a high dynamical mass-to-light ratio \citep{Emsellem:2013,Yildirim:2014}. Subsequently, \citet{Trujillo:2014} revealed that its photometry is similar to the nuggets and the stellar population has a uniformly old age. The galaxies of \citet{vdBosch:2012} are lenticular, which agrees well with most compact massive high redshift galaxies being disc-dominated \citep{vdWel:2011,Chevance:2012}.  

This begs the question of whether relics of such compact high-z nuggets can be found as high-dispersion galaxies in SDSS. The starting point of our investigation is [BHF2008] 19, which is the galaxy with ID number 19 on the list of the highest dispersion galaxies in SDSS by \citet{Bernardi:2008}. It is a very compact and massive early-type galaxy with a size of $R_e$=2.17 kpc and mass of $10^{11}$ M$_{\astrosun}$. We refer to this object as b19 in this paper. It was investigated in great detail in \citet{Lasker:2013} and it was found that b19 has a high stellar mass-to-ratio of $\sim 7$ $M_{\astrosun}/L_{\astrosun,\textrm{i}}$ and probably a bottom-heavy initial mass function\footnote{A large black hole could not be ruled out by the observations, but even if this system has a large black hole, then the dynamical and stellar population models still work better with a bottom heavy initial mass function.}. The object is located at a redshift $z=0.1166$ and is considered to be one of the most compact galaxies for its given mass in the local universe. 

%\subsection{Goal of this paper}
%\label{goal_intro}
In this paper, we performed a systematic search in SDSS to find objects similar to b19, so as to have a broader basis for future investigations of compact, high-dispersion, massive early-type galaxies. The other object, NGC 1277, could not be used for this, because it is not in the main SDSS survey\footnote{NGC 1277 was observed by SDSS as part of a auxiliary Perseus survey.}. 

While this paper is primarily a sample selection for follow-up observations, we also touch on the following important questions. Is b19 the most extreme (in the sense of mass and compactness) early-type galaxy in the local universe? Do objects like b19 just form the compact-massive tail of the general distribution of elliptical galaxies, or are they outliers known scaling relations for early-type galaxies? In which aspects do b19-like objects differ from other present-day, early-type galaxies, and are they related to red nuggets from the early universe? 

We used selection criteria based on size and central velocity dispersion to find potential red nuggets in the local universe. This is different from what was done by other authors, such as \citet{Trujillo:2009} and \citet{Taylor:2010}, who used selection criteria based on size and stellar mass. It is difficult to measure stellar masses without additional follow-up on the SDSS, and the uncertainties are high with at least 0.1 dex statistical and 0.2 dex systematic error for the stellar masses of \citet{Blanton:2007}, which were used by \citet{Trujillo:2009}: about 0.1 dex for the stellar masses used by \citet{Taylor:2010}, which were based on method of \citet{Kauffmann:2003} and \citet{Salim:2007}, and 0.15 dex for the stellar masses of \citet{Mendel:2014}, which we used in this paper. To avoid these uncertainties in our sample definition, we selected with more directly measured quantities, such as the physical radii and the central velocity dispersion. The galaxies of \citet{vdBosch:2012} and b19 are usually assumed to be relics of the red nuggets and if this is the case, there might be more galaxies with similar properties, and some of them may have the same origins. Our method is additionally motivated by the fact that at least a subgroup of red nuggets possess high central velocity dispersions \citep{Newman:2010,Bezanson:2013} and that the stellar-to-dynamical mass-to-light ratio decreases slightly over time \citep{vdSande:2013}. We set out to find them and provide a new selection method that is capable of discovering possible remnants of compact massive red galaxies from the early universe that would have been missed in previous investigations that used different selection criteria. Our goal is to define a sample to be used for follow-up observations to determine whether there is a systematic variation in the initial mass function, such as the bottom-heavy initial mass function of b19 \citep{Lasker:2013}, and if they host over-massive central black hole such as the one in NGC 1277 \citep{vdBosch:2012}. Furthermore, follow-up observation will also be required to clean the sample from high central velocity dispersion galaxies that are superpositions of two or more galaxies, which is a known issue with this kind of galaxies in SDSS \citep{Bernardi:2008}. 

%\subsection{Structure of this paper}
In Section \ref{basic_sample}, we describe the basic sample used for this investigation. After it is calibrated as explained in Section \ref{method}, we discuss the selection of our candidates in Section \ref{sec_candsel} using various cuts, which are defined there. We investigate the global properties of our candidates and their relation to the basic sample in Section \ref{sec_results}. We discuss our candidates and their relation to other samples of potential red nugget galaxies in Section \ref{sec_discussion}. In Section \ref{summary}, we provide a summary and some concluding remarks on our work. We supplement our paper with three appendices that provide updated fundamental-plane coefficients in Appendix \ref{newFPparameter}, additional tables of other samples and their cross-matches with our candidates in Appendix \ref{taylor_cross}, and an alternative candidate sample using Sersic profiles instead of the de Vaucouleurs profile in Appendix \ref{sersic_cand}.

Throughout this paper, we assume a $\Lambda$-CDM cosmology with the following parameters: relative dark energy density $\Omega_{\Lambda}=0.7$, relative matter density $\Omega_{M}=0.3$, and a Hubble parameter $H_{0}=70$ km s$^{-1}$ Mpc$^{-1}$.

\section{Basic sample}
\label{basic_sample}
As the baseline sample of our search for b19 analogues, we made broad use of the Sloan Digital Sky Surveys (SDSS) and especially of its tenth \citep{SDSS_DR10} and seventh \citep{SDSS_DR7} data releases (DR10 and DR7). Furthermore, we used GalaxyZoo \citep{GalaxyZoo,GalaxyZoo_data} for our galaxy classifications, the refits of SDSS DR7 using Sersic profiles done by \citet{Simard:2011}, and the stellar masses from \citet{Mendel:2014}, which is itself based on the previous work of \citet{Simard:2011}. For comparison, we also used the list of 63 compact massive galaxies from \citet{Taylor:2010}, which is based on SDSS DR7 as well as a list of 29 compact massive galaxies from \citet{Trujillo:2009}, which is based on the NYU Value-Added Galaxy Catalog \citep{Blanton:2005} and covers a sub-sample of SDSS. 

\begin{table}
\begin{center}
\begin{tabular}{cc}
parameter & condition \\ \hline
 \emph{SpecObj.z} &> 0 \\
 \emph{SpecObj.z} &< 0.5 \\
 \emph{SpecObj.zWarning} &= 0 \\
 \emph{SpecObj.veldisp} &> 100\\
\emph{SpecObj.snMedian} &> 10\\
\emph{SpecObj.class} &='GALAXY'\\
(\emph{P.flags\_r} \& 0x40000) &= 0
\end{tabular}
\end{center}
\caption{Selection criteria given in the language of the SDSS CAS-job queries.}
\label{criteria}
\end{table}

We selected all galaxies from the SDSS database that fulfil the following criteria, which are summarized in Table \ref{criteria}: spectroscopic data has to be available and redshifts have to be reliably obtained (\emph{zWarning}-flag set to zero). We required that the measured redshifts lie between 0 and 0.5. By using spectroscopic data from SDSS, we implicitly introduced the selection criteria of SDSS spectroscopy on our data, which are a minimum apparent magnitude in the r band of 17.77 mag \citep{SDSS_spectarget} and a saturation limit, which corresponds to a maximum magnitude of either 13 mag in the u band, 14 mag in the g, r, or i band, or 12 mag in the z band. The central velocity dispersion has to be higher than 100 km s$^{-1}$, the signal-to-noise for spectra has to be better than 10, and the automatic spectral classification has to confirm that the object is a galaxy. To ensure reliable photometric measurements, we required that there are no saturated objects in our sample (\emph{P.flags\_r} not set to \emph{SATURATED}). As a direct consequence of these requirements, we required that there must be spectroscopic data for every galaxy in our sample. We imposed the target limit for galaxy spectroscopy of SDSS on our sample, which is a minimum Petrosian magnitude in the r band of 17.77 mag \citep{SDSS_spectarget} and saturation limit of 13 mag in the u band, 14 mag in the g, r, or i band, or 12 mag in the z band.

With these criteria, we found 393 033 galaxies in SDSS DR10. For these galaxies, we downloaded the SDSS DR10 object ID, the galactic and equatorial coordinates, the redshift, the central velocity dispersion, and the following photometric quantities for the g, r, i, and z filters each: the axis-ratios, the de Vaucouleurs radii, the de Vaucouleurs model magnitudes, the galactic extinction, the likelihoods for a de Vaucouleurs profile and for an exponential profile, and the probability of its being an early-type galaxy based on GalaxyZoo.

We used the Sersic fit radii and magnitudes, as well as the Sersic indices from the catalogue by \citet{Simard:2011}. This catalogue is based on SDSS DR7, and it only provides the SDSS DR7 ID, which differs from the SDSS DR10 object IDs, to identify the galaxies in the catalogue, but no coordinates to do a direct cross-match. We could revert to SDSS DR7, but we prefer to take advantage of the updated photometry of SDSS DR10 \citep{SDSS_DR10}. We solved this problem by using a complete set of all SDSS DR7 galaxies with object IDs, equatorial coordinates, and redshifts to create a bridge between our data and the catalogue, which allows for direct cross-identification between them. 

We used the stellar masses of SDSS galaxies based on the dusty models of the catalogue by \citet{Mendel:2014}. Their estimates for stellar masses were derived using a stellar population synthesis based on the code of \citet{Conroy:2009} with spectral energy distributions based on the SDSS broadband photometry. We cross-matched this catalogue with the one of \citet{Simard:2011} and our DR10 sample. Since the \citet{Mendel:2014} catalogue has stricter redshift limits than our SDSS DR10 sample, the measured redshifts of the combined sample have to lie between 0.005 and 0.4\footnote{The reduced upper redshift limit is no concern to us, since we do not expect to detect any intrinsically small galaxies at redshifts higher than 0.2 anyway.} now. This also removed all galaxies that might be blended with a nearby star \citep{Mendel:2014}. We used the SDSS DR7 Object ID (to cross-match the catalogue with the other samples) and the logarithm of the stellar masses derived from the Sersic profiles and the composite profiles of the \citet{Mendel:2014} catalogue. After all these cross-matching, we ended up with a sample of 370 159 galaxies.

Additional constraints were applied to the data after the calculation of several parameters from the observed values. Galaxies with a velocity dispersion of higher than 420 km s$^{-1}$ were removed from the sample, because these values would be outside the trusted margin of SDSS algorithm for measuring the central velocity dispersion. We checked that dropping this upper dispersion limit would only contribute galaxies with unreasonably high central velocity dispersions. Furthermore, we eliminated all galaxies with an absolute magnitude either brighter than -25 mag or fainter than -15 mag in any of the used filters. A handful of galaxies with physical radii of more than $10^{2.5}$ kpc were also removed to avoid contamination from incorrectly measured radii. 

Furthermore, the selected galaxies must have been identified as an elliptical galaxy with a probability greater than 0.5 based on GalaxyZoo. We carefully investigated the effect of different values of the criterion on our sample and on the candidates we want to find (see Section \ref{sec_candsel}). Values higher than 0.5 will remove too many promising candidates from our sample, while for values below 0.5, the candidates in our sample will be heavily contaminated by galaxies that are superimposed on another galaxy in the line of sight or near neighbours, galaxies close to a very bright foreground star, and star-burst galaxies. We required that the likelihood for a de Vaucouleurs profile is greater than the likelihood for an exponential profile in every filter except the u band, owing to known problems\footnote{See: \url{https://www.sdss3.org/dr10/imaging/caveats.php#usky}.} with this filter. This criterion was necessary because we used parameters, such as the radii and magnitudes obtained by de Vaucouleurs fits in this paper. If the likelihood for another profile is indeed higher, it would result in poorly derived values for our parameters. 

In light of our comparison with other samples of compact massive galaxies in the local universe (see Section \ref{sec_results} for details), we found that the vast majority of these galaxies in the literature are within our basic sample and therefore best described by a de Vaucouleurs profile. Since we were searching for potential survivors of the red nuggets, we limited our sample to red sequence galaxies. We did this by removing all galaxies bluer than the lower 3-$\sigma$ limit of the red sequence fit performed in \citet{Saulder:2013}. After this filtering, there were 233 833 galaxies between a redshift of 0.005 and 0.4 left (about $59.5\%$ of the first selection and $63\%$ of the cross-matched sample). Those form the basic sample that were used for the further analysis in this paper. 
\section{Method}
\label{method}
One has to carefully calibrate the parameters obtained from the SDSS database and the refits done by \citet{Simard:2011} before using them to classify and characterize the galaxies. The following quantities are calculated for all sources in every band used and for every set of fit parameters. 

One starts off by considering the galactic extinction by using the Schlegel maps \citep{Schlegelmaps}:
\begin{equation}
m_{\textrm{extcor}}=m_{\textrm{sdss}} - A_\textrm{Schlegel},
\label{extinctioncorrection}
\end{equation}
where $m_{\textrm{extcor}}$ denotes the extinction correct magnitude, $m_{\textrm{sdss}}$ the observed apparent magnitude, and $A_\textrm{Schlegel}$ the extinction according to Schlegel maps.

The K-correction used in this paper,
\begin{equation}
K(z_{obs},m_{f_{1}}-m_{f_{2}})=\sum\limits_{i,j} B_{ij} z_{obs}^{i} (m_{f_{1}}-m_{f_{2}})^{j}
\label{Kcorrection}
\end{equation}
follows the model of \citet{Chilingarian:2010}, but with updated coefficients $B_{ij}$ from \citet{Saulder:2013}. It requires the extinction-corrected magnitudes, $m_{f_{1}}$ and $m_{f_{2}}$, of two different filters, $f_{1}$ and $f_{2}$, and the observed redshift $z_{obs}$. 

In the next step, one obtains the fully corrected rest-frame magnitude $m_{\textrm{app}}$ by considering the K correction $K(z_{obs},m_{f_{1}}-m_{f_{2}})$:
\begin{equation}
m_{\textrm{cor}} = m_{\textrm{extcor}} - K(z_{obs},m_{f_{1}}-m_{f_{2}}).
\label{apperantmag}
\end{equation}
The redshift $z$ is corrected for our motion relative to the cosmic microwave background (CMB). 

The measured model semi-major from the SDSS data $a_{\textrm{sdss}}$ has to be renormalized to account for the different ellipticities of the galaxies in the following way:
\begin{equation}
r_{\textrm{circ}} = a_{\textrm{sdss}} \sqrt{q_{b/a}} .
\label{rcor}
\end{equation}
We follow \citet{Bernardi:2003c} and get a comparable quantity for all types of elliptical galaxies, the circularized radius $r_{\textrm{circ}}$, with the help of the minor semi-axis to the major semi-axis ratio $q_{b/a}$. 

Because of the fixed fibre size of SDSS, an additional correction on the measured central velocity dispersion $\sigma_{\textrm{sdss}}$ is required and we take advantage of the work of \citet{Jorgensen:1995} and \citet{Wegner:1999} to use 
\begin{equation}
\sigma_{0}=\sigma_{\textrm{sdss}} \cdot \left( \frac{a_{\textrm{fiber}}}{r_{\textrm{circ}}/8} . \right)^{0.04},
\label{sigmacor}
\end{equation}
where $\sigma_{0}$ denotes the corrected central velocity dispersion and $a_{\textrm{fiber}}$ stands for the radius of the SDSS fibres, which is 1.5 arcseconds for the galaxies in our sample. Here, $\sigma_{0}$ is typically about 10$\%$ higher than the measured value $\sigma_{\textrm{sdss}}$ \citep{Saulder:2013}.

For the following calculations, one requires the luminosity distance $D_{L}$, which is given by
\begin{equation}
D_{L}(z)=\frac{c \cdot z}{H_{0}} \left(1+\left( \frac{z \cdot (1-q_{0})}{\sqrt{1+2 q_{0} \cdot z}+1+q_{0} \cdot z} \right)\right)
\label{lumdist}
\end{equation}
with $H_{0}$ being the present day Hubble parameter and $q_{0}=\frac{\Omega_{M}}{2}-\Omega_{\Lambda}$ the current declaration parameter, which depends on the cosmological parameters $\Omega_{M}$ and $\Omega_{\Lambda}$. 

With the luminosity distance at hand, the angular diameter distance is given by
\begin{equation}
D_{A}(z)=D_{L}(z) \cdot (1+z)^{-2} .
\label{angdist}
\end{equation}

\noindent The physical radius $R_{0}$ of the galaxy is obtained using simple trigonometry: 
\begin{equation}
R_{0}=D_{A}(z) \cdot \textrm{tan}\left(r_{\textrm{circ}}\right) .
\label{realradius}
\end{equation}

\noindent The measured surface brightness $\mu_{0}$ is defined in the following way:
\begin{equation}
\mu_{0}=m_{\textrm{cor}} + 2.5\cdot \textrm{log}_{10}\left( 2\pi \cdot r_{\textrm{circ}}^{2} \right) - 10\cdot \textrm{log}_{10} \left( 1 + z \right)  +Q \cdot z 
\label{surfacebrightness}
\end{equation}
with the term $-10\cdot \textrm{log}_{10} \left( 1 + z \right)$ correcting for cosmological dimming of surface brightnesses. Since we only intend to use the surface brightness, we include a parameter that corrects for the secular evolution evolution of early-type galaxies, when applying or calculating the fundamental plane (see Appendix \ref{newFPparameter} and \citet{Saulder:2013}). The evolution parameter $Q=1.07$ mag per $z$ was derived in \citet{Saulder:2013} for early-type galaxies. 

Another quantity that is required for our investigations is the absolute magnitude $M_{\textrm{abs}}$, which is calculated using the distance module:
\begin{equation}
m_{\textrm{cor}}-M_{\textrm{abs}}=5 \cdot \textrm{log}_{10}(D_{L}/\textrm{pc})-5.
\label{distmod}
\end{equation}

The dynamical mass is given by
\begin{equation}
M_{\textrm{dyn}}=\frac{\beta(n) \sigma_{0}^{2} \cdot R_{0}}{G}
\label{dynmass}
\end{equation}
with $G$ being the gravitational constant. The function $\beta(n_{\textrm{S}})$ is defined by:
\begin{equation}
\beta(n) = 8.87 - 0.831 \cdot n_{\textrm{S}} + 0.0241 \cdot n_{\textrm{S}}^{2}
\label{betafunction}
\end{equation}
according to \citet{Cappellari:2006}, based on results from \citet{Bertin:2002}. It depends on the Sersic-index $n_{\textrm{S}}$, if a Sersic profile was used to obtain the effective radius. For de Vaucouleurs profiles, which are Sersic profiles with Sersic-indices $n_{\textrm{S}}=4$, one would expect a $\beta$ of 5.953; however, it has been found by observations \citep{Cappellari:2006} that a $\beta$ of five works better. \citet{Cappellari:2006} argue that this deviation is due to differences between the idealised simulation Equation \ref{betafunction} is based on and real observational data. \citet{Belli:2014} find that the equations works well for spherical systems, but has problems if discs are present. We therefore decided to use the $\beta(n_{\textrm{S}})$ from Equation \ref{betafunction}, when using a Sersic profile, but we adopt a $\beta$ of 5, when using a de Vaucouleurs profile in our analysis.

With all the equations and definitions given in this section, we now proceed to the selection and analysis of galaxies similar to b19. 

 % \begin{table*}
%   \afterpage{
   \onecolumn
 \begin{landscape}
 \centering
  \bottomcaption{List of the basic parameters of our candidate galaxies. First column: internal IDs, which are used to identify the galaxies. The numbering is essentially random and only based on the order the galaxies were drawn from the basic sample. The galaxy b19 has the internal ID 2. Second column: object ID used by SDSS DR10. Third and fourth column: equatorial coordinates of the galaxies. Fifth column: redshift $z$, already corrected for our motion relative to the CMB. Sixth, seventh, and eighth columns: observed uncorrected refitted SDSS parameters in the following order: observed apparent magnitude $m_{\textrm{sdss}}$, angular semi-major axis $a_{\textrm{sdss}}$, central velocity dispersion $\sigma_{\textrm{sdss}}$. Ninth column: axis ratio $q_{b/a}$. Tenth column: GalaxyZoo probability $\mathcal{L}_{\textrm{ETG}}$ of the galaxy being classified as an early-type.}
 \tablehead{Internal ID & SDSS DR10 ID & ra & dec & $z$ & $m_{\textrm{sdss},r}$ &$a_{\textrm{sdss}}$ & $\sigma_{\textrm{sdss}}$ & $q_{b/a}$ & $\mathcal{L}_{\textrm{ETG}}$\\ 
  &  & [$^{\circ}$]&  [$^{\circ}$]&   & [mag] & [arcsec] & [km/s] & & \\ \hline }
 \begin{supertabular}[p]{cccccccccc}
           1 &1237648721255596242&  236.8072  &   -0.1422  &    0.1138  &   17.18    $\pm$    0.00    &    1.17    $\pm$    0.02    &         315$\pm$          14&    0.67    &    0.88    \\
           2 &1237648703523520846&  229.4240  &   -0.7049  &    0.1166  &   17.04    $\pm$    0.00    &    1.25    $\pm$    0.02    &         336$\pm$          12&    0.67    &    0.81    \\
           3 &1237651191892607189&  125.5691  &   48.2553  &    0.1276  &   17.56    $\pm$    0.01    &    1.39    $\pm$    0.02    &         351$\pm$          14&    0.46    &    0.75    \\
           4 &1237651753466462236&  164.0158  &    1.9983  &    0.1153  &   17.72    $\pm$    0.01    &    0.95    $\pm$    0.03    &         297$\pm$          22&    0.66    &    0.82    \\
           5 &1237652934037536913&  327.3491  &   -8.6752  &    0.1014  &   17.39    $\pm$    0.01    &    0.81    $\pm$    0.03    &         320$\pm$          16&    0.88    &    0.70    \\
           6 &1237652900773298301&   58.0541  &   -5.8611  &    0.1137  &   17.25    $\pm$    0.01    &    1.03    $\pm$    0.02    &         306$\pm$          14&    0.38    &    0.66    \\
           7 &1237652629102067836&    8.1716  &  -10.6661  &    0.1557  &   17.60    $\pm$    0.01    &    1.00    $\pm$    0.03    &         355$\pm$          18&    0.63    &    1.00    \\
           8 &1237651252589363420&  247.9117  &   46.2683  &    0.1321  &   17.67    $\pm$    0.01    &    0.78    $\pm$    0.01    &         311$\pm$          14&    0.30    &    0.76    \\
           9 &1237655502424769160&  256.4241  &   33.4779  &    0.1022  &   17.33    $\pm$    0.01    &    1.51    $\pm$    0.02    &         326$\pm$          16&    0.52    &    0.77    \\
          10 &1237651539246186637&  167.7205  &   66.7862  &    0.1362  &   17.57    $\pm$    0.01    &    1.09    $\pm$    0.02    &         350$\pm$          14&    0.29    &    0.59    \\
          11 &1237651735773708418&  218.3124  &    1.5053  &    0.1096  &   17.33    $\pm$    0.01    &    0.96    $\pm$    0.02    &         291$\pm$          16&    0.68    &    0.74    \\
          12 &1237659329240236080&  243.4534  &   41.1059  &    0.1381  &   17.66    $\pm$    0.01    &    0.85    $\pm$    0.03    &         290$\pm$          17&    0.56    &    0.78    \\
          13 &1237666339727671425&   20.8205  &    0.2955  &    0.0928  &   17.10    $\pm$    0.00    &    1.11    $\pm$    0.02    &         296$\pm$          11&    0.73    &    0.88    \\
          14 &1237651714798125236&  248.3287  &   47.1274  &    0.1229  &   17.63    $\pm$    0.01    &    0.60    $\pm$    0.02    &         335$\pm$          12&    0.85    &    0.66    \\
          15 &1237658206124507259&  193.5474  &   50.8170  &    0.1209  &   17.22    $\pm$    0.00    &    1.17    $\pm$    0.01    &         341$\pm$          16&    0.41    &    0.80    \\
          16 &1237652944786424004&    1.1323  &   16.0719  &    0.1144  &   17.58    $\pm$    0.01    &    0.91    $\pm$    0.01    &         291$\pm$          15&    0.29    &    0.55    \\
          17 &1237662267540570526&  235.5841  &    4.7666  &    0.1105  &   17.15    $\pm$    0.01    &    1.20    $\pm$    0.02    &         302$\pm$          10&    0.56    &    0.77    \\
          18 &1237652948530102500&   10.3768  &   -9.2352  &    0.0538  &   15.24    $\pm$    0.00    &    3.09    $\pm$    0.02    &         310$\pm$           5&    0.33    &    0.53    \\
          19 &1237656241159995854&  331.7753  &   12.0459  &    0.1607  &   17.91    $\pm$    0.01    &    1.00    $\pm$    0.03    &         306$\pm$          16&    0.57    &    0.89    \\
          20 &1237656243317113067&  354.1646  &   15.8222  &    0.1179  &   17.56    $\pm$    0.01    &    1.16    $\pm$    0.02    &         290$\pm$          16&    0.41    &    0.73    \\
          21 &1237655474503024820&  245.6049  &   44.7856  &    0.0716  &   15.84    $\pm$    0.00    &    1.86    $\pm$    0.02    &         333$\pm$           8&    0.61    &    0.81    \\
          22 &1237657596224209238&  123.8014  &   38.6793  &    0.1259  &   17.16    $\pm$    0.00    &    1.26    $\pm$    0.02    &         333$\pm$          13&    0.48    &    0.89    \\
          23 &1237662264318034136&  217.8880  &    8.9225  &    0.1108  &   17.09    $\pm$    0.01    &    1.46    $\pm$    0.02    &         384$\pm$          15&    0.53    &    0.72    \\
          24 &1237665569297203655&  254.5120  &   41.8378  &    0.0375  &   15.37    $\pm$    0.00    &    1.62    $\pm$    0.01    &         303$\pm$           7&    0.48    &    0.64    \\
          25 &1237654605857751221&  148.8860  &    4.3722  &    0.0937  &   16.41    $\pm$    0.00    &    1.48    $\pm$    0.01    &         352$\pm$           9&    0.39    &    0.52    \\
          26 &1237655465916170402&  184.8400  &   63.5358  &    0.1039  &   17.46    $\pm$    0.01    &    0.76    $\pm$    0.01    &         292$\pm$          14&    0.39    &    0.52    \\
          27 &1237657628456190055&  187.6884  &   51.7060  &    0.1517  &   17.61    $\pm$    0.01    &    0.90    $\pm$    0.01    &         307$\pm$          14&    0.43    &    0.62    \\
          28 &1237660025032081578&  340.4373  &   -0.8113  &    0.1293  &   17.68    $\pm$    0.01    &    0.63    $\pm$    0.02    &         373$\pm$          22&    0.86    &    0.77    \\
          29 &1237661064411349290&  138.3286  &    8.1161  &    0.0934  &   16.82    $\pm$    0.00    &    1.41    $\pm$    0.01    &         295$\pm$           9&    0.28    &    0.61    \\
          30 &1237661849849430137&  156.3195  &   40.3153  &    0.0682  &   16.57    $\pm$    0.00    &    1.53    $\pm$    0.02    &         317$\pm$          10&    0.78    &    0.58    \\
          31 &1237663277928022281&    0.6027  &    0.5352  &    0.0784  &   17.56    $\pm$    0.01    &    0.64    $\pm$    0.02    &         331$\pm$          17&    0.68    &    0.77    \\
          32 &1237661383314702588&  160.1959  &   39.9311  &    0.1394  &   17.77    $\pm$    0.01    &    0.86    $\pm$    0.02    &         324$\pm$          15&    0.36    &    0.69    \\
          33 &1237662697568796852&  226.2857  &   30.1184  &    0.1450  &   17.24    $\pm$    0.01    &    0.86    $\pm$    0.02    &         314$\pm$           9&    0.73    &    0.71    \\
          34 &1237661812272857187&  180.2528  &   12.2175  &    0.1295  &   17.73    $\pm$    0.01    &    1.00    $\pm$    0.03    &         291$\pm$          17&    0.52    &    0.81    \\
          35 &1237665532252520624&  223.1388  &   22.5927  &    0.1551  &   17.66    $\pm$    0.01    &    1.32    $\pm$    0.02    &         318$\pm$          16&    0.34    &    0.54    \\
          36 &1237662224087974057&  238.7278  &   25.4691  &    0.1556  &   17.84    $\pm$    0.01    &    1.18    $\pm$    0.02    &         308$\pm$          17&    0.45    &    0.76    \\
          37 &1237664130483618005&  166.7737  &   13.3182  &    0.1188  &   17.10    $\pm$    0.00    &    1.40    $\pm$    0.02    &         328$\pm$          13&    0.49    &    0.76    \\
          38 &1237664669510074510&  158.0224  &   37.4689  &    0.1043  &   16.60    $\pm$    0.00    &    0.99    $\pm$    0.01    &         385$\pm$          12&    0.63    &    0.71    \\
          39 &1237665549429899544&  223.0734  &   22.4871  &    0.1165  &   17.39    $\pm$    0.01    &    0.78    $\pm$    0.01    &         335$\pm$          13&    0.29    &    0.62    \\
          40 &1237667209978380503&  149.1117  &   23.9641  &    0.1193  &   17.33    $\pm$    0.01    &    0.95    $\pm$    0.02    &         356$\pm$          25&    0.81    &    0.68    \\
          41 &1237663278461944053&  353.8668  &    1.0467  &    0.0827  &   16.35    $\pm$    0.00    &    1.71    $\pm$    0.02    &         320$\pm$           9&    0.55    &    0.80    \\
          42 &1237662340012638220&  239.5694  &   27.2367  &    0.0896  &   17.01    $\pm$    0.00    &    0.81    $\pm$    0.01    &         296$\pm$          12&    0.55    &    0.75    \\
          43 &1237664667887140986&  128.6548  &   24.3250  &    0.0705  &   16.42    $\pm$    0.00    &    1.22    $\pm$    0.02    &         296$\pm$           9&    0.71    &    0.77    \\
          44 &1237664093432119636&  121.7265  &   20.7624  &    0.1247  &   17.64    $\pm$    0.01    &    0.93    $\pm$    0.02    &         299$\pm$          14&    0.44    &    0.66    \\
          45 &1237661850400260193&  199.4989  &   43.6141  &    0.1140  &   17.76    $\pm$    0.01    &    0.64    $\pm$    0.02    &         287$\pm$          16&    0.59    &    0.62    \\
          46 &1237664852035174654&  219.1545  &   31.3943  &    0.0850  &   16.14    $\pm$    0.00    &    1.46    $\pm$    0.01    &         331$\pm$           9&    0.36    &    0.77    \\
          47 &1237667429035540562&  178.7061  &   26.4323  &    0.1108  &   17.05    $\pm$    0.00    &    1.12    $\pm$    0.01    &         316$\pm$          12&    0.62    &    0.67    \\
          48 &1237673808655221213&  121.7151  &   19.4664  &    0.1242  &   18.05    $\pm$    0.01    &    0.69    $\pm$    0.03    &         294$\pm$          19&    0.62    &    0.73    \\
          49 &1237664854715727968&  210.0376  &   35.9503  &    0.1494  &   17.76    $\pm$    0.01    &    1.01    $\pm$    0.03    &         317$\pm$          14&    0.66    &    0.65    \\
          50 &1237665535469486145&  243.3042  &   17.8080  &    0.0374  &   14.72    $\pm$    0.00    &    3.44    $\pm$    0.01    &         316$\pm$           7&    0.44    &    0.68    \\
          51 &1237663478723969457&  338.0784  &   -0.4059  &    0.0865  &   17.04    $\pm$    0.00    &    1.23    $\pm$    0.02    &         327$\pm$          17&    0.54    &    0.80    \\
          52 &1237665440978698364&  194.2722  &   28.9814  &    0.0686  &   15.45    $\pm$    0.00    &    2.19    $\pm$    0.01    &         340$\pm$           8&    0.57    &    0.78    \\
          53 &1237667910055100586&  181.7985  &   23.8744  &    0.0775  &   16.55    $\pm$    0.00    &    1.21    $\pm$    0.02    &         328$\pm$          11&    0.77    &    0.86    \\
          54 &1237667734526492801&  227.3075  &   16.4333  &    0.1159  &   17.56    $\pm$    0.01    &    1.02    $\pm$    0.02    &         310$\pm$          17&    0.49    &    0.61    \\
          55 &1237662619725005006&  240.2092  &   29.2028  &    0.0913  &   16.65    $\pm$    0.00    &    1.40    $\pm$    0.01    &         327$\pm$          11&    0.62    &    0.84    \\
          56 &1237664869745230095&  128.9418  &   34.2085  &    0.1978  &   18.43    $\pm$    0.01    &    0.68    $\pm$    0.04    &         316$\pm$          24&    0.77    &    0.65    \\
          57 &1237665429169242591&  209.7906  &   27.9501  &    0.0811  &   17.16    $\pm$    0.00    &    0.63    $\pm$    0.01    &         287$\pm$          10&    0.46    &    0.62    \\
          58 &1237665440975224988&  185.1490  &   29.2998  &    0.0908  &   16.49    $\pm$    0.00    &    1.45    $\pm$    0.02    &         332$\pm$          13&    0.74    &    0.85    \\
          59 &1237668299281662070&  194.2881  &   20.8064  &    0.0868  &   16.76    $\pm$    0.00    &    1.17    $\pm$    0.01    &         307$\pm$           9&    0.51    &    0.56    \\
          60 &1237668349753950509&  232.0499  &   12.1307  &    0.1225  &   17.63    $\pm$    0.01    &    1.16    $\pm$    0.02    &         311$\pm$          14&    0.38    &    0.81    \\
          61 &1237668271372501042&  227.9714  &   14.2653  &    0.1221  &   17.70    $\pm$    0.01    &    0.92    $\pm$    0.02    &         291$\pm$          16&    0.54    &    0.75    \\
          62 &1237648721758978188&  160.3022  &    0.2285  &    0.1300  &   17.60    $\pm$    0.01    &    1.24    $\pm$    0.02    &         305$\pm$          15&    0.37    &    0.64    \\
          63 &1237664671640715458&  191.2284  &   36.1838  &    0.0877  &   17.64    $\pm$    0.01    &    0.47    $\pm$    0.01    &         293$\pm$          15&    0.65    &    0.68    \\
          64 &1237667735062708393&  225.9192  &   17.2367  &    0.1505  &   17.73    $\pm$    0.01    &    1.07    $\pm$    0.03    &         309$\pm$          17&    0.56    &    0.74    \\
          65 &1237662335717015837&  236.8248  &   33.1773  &    0.1265  &   17.74    $\pm$    0.01    &    0.84    $\pm$    0.03    &         296$\pm$          16&    0.64    &    0.78    \\
          66 &1237668310021440087&  245.6255  &    9.3970  &    0.2018  &   17.70    $\pm$    0.01    &    0.85    $\pm$    0.02    &         302$\pm$          13&    0.51    &    0.79    \\
          67 &1237661358617067696&  181.3091  &   48.4216  &    0.0648  &   15.89    $\pm$    0.00    &    2.01    $\pm$    0.02    &         311$\pm$           8&    0.54    &    0.71    \\
          68 &1237668298203070641&  182.4650  &   20.0535  &    0.1116  &   17.58    $\pm$    0.01    &    1.02    $\pm$    0.02    &         293$\pm$          12&    0.56    &    0.72    \\
          69 &1237662336794820961&  245.8542  &   28.0910  &    0.1233  &   17.01    $\pm$    0.00    &    1.01    $\pm$    0.02    &         306$\pm$          11&    0.73    &    0.65    \\
          70 &1237667917032980629&  189.9670  &   21.1529  &    0.1085  &   16.78    $\pm$    0.00    &    1.43    $\pm$    0.01    &         321$\pm$           9&    0.51    &    0.74    \\
          71 &1237662224614490342&  214.0046  &   35.9910  &    0.1271  &   17.52    $\pm$    0.01    &    1.12    $\pm$    0.02    &         300$\pm$          14&    0.54    &    0.78    \\
          72 &1237661950244945934&  162.5130  &   11.8190  &    0.0812  &   16.56    $\pm$    0.00    &    1.75    $\pm$    0.02    &         340$\pm$          11&    0.46    &    0.88    \\
          73 &1237668333640810655&  225.5537  &   14.6343  &    0.0697  &   16.62    $\pm$    0.00    &    0.95    $\pm$    0.01    &         351$\pm$          14&    0.58    &    0.52    \\
          74 &1237662236410577091&  226.1287  &    6.6601  &    0.1439  &   17.76    $\pm$    0.01    &    0.82    $\pm$    0.03    &         316$\pm$          17&    0.78    &    0.80    \\
          75 &1237662302971691136&  214.9301  &   49.2366  &    0.0260  &   14.54    $\pm$    0.00    &    3.59    $\pm$    0.02    &         378$\pm$           2&    0.93    &    0.51    \\
          76 &1237667917030555837&  184.0304  &   21.1393  &    0.1278  &   17.04    $\pm$    0.00    &    1.24    $\pm$    0.02    &         389$\pm$          16&    0.53    &    0.79    \\
 \end{supertabular}
 \label{list_candiates_basics_dV}
%  \twocolumn
% \end{landscape}}
 
 % \afterpage{\begin{landscape}
% \onecolumn
 %\centering
   \bottomcaption{List of the derived parameters based on the de Vaucouleurs fits from SDSS for our candidate galaxies. First column: internal IDs of our galaxies. Second column: scale radius $R_{\textrm{r}}$ of the galaxies measured in the SDSS r band (in kpc). Third column: corrected central velocity dispersion $\sigma_{0}$ (in km/s). Fourth column: surface brightness $\mu_{r}$ measured in the SDSS r band (in mag/arcsec$^{2}$). Fifth column: absolute magnitude in r band $M_{\textrm{r}}$. Sixth column: g-r colour $(M_{\textrm{g}}-M_{\textrm{r}})$ (in mag). Seventh column: logarithm of the dynamical mass $M_{\textrm{dyn}}$ (in solar masses). Eighth column: logarithm of the stellar mass $M_{\textrm{*}}$ (in solar masses). Ninth column: dynamical mass-to-light ratio $\vernal_{\textrm{dyn}}$ (in solar units $M_{\astrosun}/L_{\astrosun ,\textrm{r}}$). Tenth column: stellar mass-to-light ratio $\vernal_{*}$ (in solar units $M_{\astrosun}/L_{\astrosun ,\textrm{r}}$).}
 \tablehead{internal ID & $R_{\textrm{r}}$ & $\sigma_{0}$ & $\mu_{r}$ & $M_{\textrm{r}}$ & $(M_{\textrm{g}}-M_{\textrm{r}})$ & $M_{\textrm{z}}$ & log$_{10}$($M_{\textrm{dyn}}$) &  log$_{10}$($M_{*}$) & $\vernal_{\textrm{dyn}}$ & $\vernal_{*}$ \\
  & [kpc] & [km s$^{-1}$] & [$\frac{\textrm{mag}}{\textrm{arcsec}^{2}}$]  & [mag] & [mag] & [mag] & [log$_{10}(M_{\astrosun})$] & [log$_{10}(M_{\astrosun})$] & [$M_{\astrosun}/L_{\astrosun,\textrm{r}}$]& [$M_{\astrosun}/L_{\astrosun,\textrm{r}}$] \\ \hline }
 \begin{supertabular}{ccccccccccc}
           1 &    2.00    $\pm$    0.04    &         348$\pm$          16&   18.38    $\pm$    0.04    &  -21.81    $\pm$    0.01    &    0.83    $\pm$    0.01    &  -22.46    $\pm$    0.01    &   11.45    $\pm$    0.02    &   11.04    $\pm$    0.15    &    6.95    $\pm$    0.37    &    2.71    $\pm$    1.12    \\
           2 &    2.17    $\pm$    0.05    &         371$\pm$          13&   18.43    $\pm$    0.05    &  -21.95    $\pm$    0.01    &    0.80    $\pm$    0.01    &  -22.61    $\pm$    0.01    &   11.54    $\pm$    0.02    &   11.15    $\pm$    0.15    &    7.55    $\pm$    0.35    &    3.06    $\pm$    1.26    \\
           3 &    2.17    $\pm$    0.05    &         389$\pm$          16&   18.78    $\pm$    0.05    &  -21.60    $\pm$    0.01    &    0.77    $\pm$    0.02    &  -22.31    $\pm$    0.01    &   11.58    $\pm$    0.02    &   10.98    $\pm$    0.15    &   11.39    $\pm$    0.61    &    2.85    $\pm$    1.18    \\
           4 &    1.62    $\pm$    0.06    &         332$\pm$          25&   18.60    $\pm$    0.07    &  -21.15    $\pm$    0.01    &    0.77    $\pm$    0.02    &  -21.83    $\pm$    0.02    &   11.32    $\pm$    0.03    &   10.77    $\pm$    0.15    &    9.41    $\pm$    0.83    &    2.64    $\pm$    1.09    \\
           5 &    1.44    $\pm$    0.05    &         358$\pm$          18&   18.29    $\pm$    0.08    &  -21.18    $\pm$    0.01    &    0.62    $\pm$    0.01    &  -22.00    $\pm$    0.01    &   11.33    $\pm$    0.03    &   10.83    $\pm$    0.15    &    9.44    $\pm$    0.61    &    2.99    $\pm$    1.23    \\
           6 &    1.32    $\pm$    0.04    &         344$\pm$          15&   17.50    $\pm$    0.06    &  -21.79    $\pm$    0.01    &    0.77    $\pm$    0.01    &  -22.46    $\pm$    0.01    &   11.26    $\pm$    0.02    &   10.95    $\pm$    0.15    &    4.54    $\pm$    0.26    &    2.23    $\pm$    0.92    \\
           7 &    2.18    $\pm$    0.09    &         395$\pm$          20&   18.41    $\pm$    0.09    &  -22.01    $\pm$    0.01    &    0.78    $\pm$    0.02    &  -22.80    $\pm$    0.02    &   11.60    $\pm$    0.03    &   11.16    $\pm$    0.15    &    8.13    $\pm$    0.55    &    2.97    $\pm$    1.23    \\
           8 &    1.02    $\pm$    0.03    &         355$\pm$          16&   17.27    $\pm$    0.06    &  -21.48    $\pm$    0.01    &    0.83    $\pm$    0.01    &  -22.19    $\pm$    0.01    &   11.18    $\pm$    0.02    &   11.00    $\pm$    0.15    &    5.01    $\pm$    0.28    &    3.32    $\pm$    1.37    \\
           9 &    2.05    $\pm$    0.05    &         358$\pm$          17&   19.02    $\pm$    0.05    &  -21.22    $\pm$    0.01    &    0.84    $\pm$    0.02    &  -21.89    $\pm$    0.01    &   11.49    $\pm$    0.02    &   10.85    $\pm$    0.15    &   13.02    $\pm$    0.77    &    3.00    $\pm$    1.24    \\
          10 &    1.42    $\pm$    0.05    &         395$\pm$          16&   17.81    $\pm$    0.08    &  -21.66    $\pm$    0.01    &    0.83    $\pm$    0.01    &  -22.39    $\pm$    0.02    &   11.41    $\pm$    0.02    &   10.99    $\pm$    0.15    &    7.28    $\pm$    0.43    &    2.75    $\pm$    1.14    \\
          11 &    1.58    $\pm$    0.04    &         324$\pm$          18&   18.26    $\pm$    0.06    &  -21.42    $\pm$    0.01    &    0.73    $\pm$    0.02    &  -22.06    $\pm$    0.01    &   11.29    $\pm$    0.03    &   10.88    $\pm$    0.15    &    6.80    $\pm$    0.45    &    2.65    $\pm$    1.09    \\
          12 &    1.59    $\pm$    0.07    &         327$\pm$          19&   18.10    $\pm$    0.09    &  -21.62    $\pm$    0.01    &    1.10    $\pm$    0.02    &  -22.16    $\pm$    0.02    &   11.29    $\pm$    0.03    &   11.01    $\pm$    0.15    &    5.78    $\pm$    0.44    &    2.97    $\pm$    1.23    \\
          13 &    1.66    $\pm$    0.03    &         327$\pm$          13&   18.54    $\pm$    0.04    &  -21.22    $\pm$    0.01    &    0.76    $\pm$    0.01    &  -21.87    $\pm$    0.01    &   11.31    $\pm$    0.02    &   10.78    $\pm$    0.15    &    8.76    $\pm$    0.41    &    2.58    $\pm$    1.07    \\
          14 &    1.24    $\pm$    0.03    &         379$\pm$          14&   17.88    $\pm$    0.06    &  -21.29    $\pm$    0.01    &    0.59    $\pm$    0.01    &  -21.91    $\pm$    0.01    &   11.31    $\pm$    0.02    &   10.66    $\pm$    0.15    &    8.23    $\pm$    0.39    &    1.82    $\pm$    0.75    \\
          15 &    1.65    $\pm$    0.03    &         381$\pm$          18&   18.09    $\pm$    0.04    &  -21.70    $\pm$    0.01    &    0.83    $\pm$    0.01    &  -22.39    $\pm$    0.01    &   11.45    $\pm$    0.02    &   11.00    $\pm$    0.15    &    7.60    $\pm$    0.41    &    2.74    $\pm$    1.13    \\
          16 &    1.03    $\pm$    0.02    &         331$\pm$          17&   17.46    $\pm$    0.05    &  -21.29    $\pm$    0.01    &    0.81    $\pm$    0.02    &  -22.01    $\pm$    0.01    &   11.12    $\pm$    0.02    &   10.85    $\pm$    0.15    &    5.21    $\pm$    0.31    &    2.80    $\pm$    1.16    \\
          17 &    1.83    $\pm$    0.04    &         335$\pm$          11&   18.33    $\pm$    0.05    &  -21.66    $\pm$    0.01    &    0.83    $\pm$    0.01    &  -22.33    $\pm$    0.01    &   11.38    $\pm$    0.02    &   10.95    $\pm$    0.15    &    6.72    $\pm$    0.30    &    2.51    $\pm$    1.04    \\
          18 &    1.85    $\pm$    0.02    &         334$\pm$           6&   18.15    $\pm$    0.02    &  -21.82    $\pm$    0.01    &    0.82    $\pm$    0.01    &  -22.51    $\pm$    0.01    &   11.38    $\pm$    0.01    &   11.04    $\pm$    0.15    &    5.89    $\pm$    0.14    &    2.69    $\pm$    1.11    \\
          19 &    2.13    $\pm$    0.07    &         342$\pm$          18&   18.44    $\pm$    0.07    &  -21.94    $\pm$    0.01    &    0.85    $\pm$    0.02    &  -22.69    $\pm$    0.02    &   11.46    $\pm$    0.03    &   11.15    $\pm$    0.15    &    6.35    $\pm$    0.42    &    3.07    $\pm$    1.27    \\
          20 &    1.60    $\pm$    0.04    &         324$\pm$          18&   18.26    $\pm$    0.05    &  -21.45    $\pm$    0.01    &    0.80    $\pm$    0.02    &  -22.18    $\pm$    0.01    &   11.29    $\pm$    0.03    &   10.88    $\pm$    0.15    &    6.69    $\pm$    0.43    &    2.64    $\pm$    1.09    \\
          21 &    2.00    $\pm$    0.02    &         362$\pm$           9&   18.34    $\pm$    0.02    &  -21.81    $\pm$    0.01    &    0.81    $\pm$    0.01    &  -22.46    $\pm$    0.01    &   11.48    $\pm$    0.01    &   11.02    $\pm$    0.15    &    7.54    $\pm$    0.24    &    2.57    $\pm$    1.06    \\
          22 &    2.00    $\pm$    0.04    &         370$\pm$          14&   18.27    $\pm$    0.05    &  -21.94    $\pm$    0.01    &    0.82    $\pm$    0.01    &  -22.63    $\pm$    0.01    &   11.50    $\pm$    0.02    &   11.14    $\pm$    0.15    &    6.99    $\pm$    0.34    &    3.01    $\pm$    1.24    \\
          23 &    2.17    $\pm$    0.04    &         423$\pm$          17&   18.70    $\pm$    0.04    &  -21.67    $\pm$    0.01    &    0.82    $\pm$    0.02    &  -22.42    $\pm$    0.03    &   11.66    $\pm$    0.02    &   11.02    $\pm$    0.15    &   12.70    $\pm$    0.64    &    2.96    $\pm$    1.22    \\
          24 &    0.84    $\pm$    0.01    &         333$\pm$           7&   17.39    $\pm$    0.02    &  -20.83    $\pm$    0.01    &    0.82    $\pm$    0.01    &  -21.59    $\pm$    0.01    &   11.03    $\pm$    0.01    &   10.60    $\pm$    0.15    &    6.53    $\pm$    0.20    &    2.42    $\pm$    1.00    \\
          25 &    1.62    $\pm$    0.02    &         390$\pm$          10&   17.72    $\pm$    0.03    &  -22.00    $\pm$    0.01    &    0.82    $\pm$    0.01    &  -22.73    $\pm$    0.01    &   11.46    $\pm$    0.01    &   11.11    $\pm$    0.15    &    5.94    $\pm$    0.19    &    2.68    $\pm$    1.11    \\
          26 &    0.92    $\pm$    0.02    &         333$\pm$          16&   17.37    $\pm$    0.05    &  -21.12    $\pm$    0.01    &    0.82    $\pm$    0.01    &  -21.86    $\pm$    0.01    &   11.07    $\pm$    0.02    &   10.77    $\pm$    0.15    &    5.49    $\pm$    0.31    &    2.74    $\pm$    1.13    \\
          27 &    1.58    $\pm$    0.04    &         346$\pm$          16&   17.81    $\pm$    0.05    &  -21.91    $\pm$    0.01    &    0.83    $\pm$    0.01    &  -22.60    $\pm$    0.01    &   11.34    $\pm$    0.02    &   11.06    $\pm$    0.15    &    4.96    $\pm$    0.27    &    2.58    $\pm$    1.06    \\
          28 &    1.36    $\pm$    0.05    &         421$\pm$          25&   17.83    $\pm$    0.07    &  -21.53    $\pm$    0.01    &    0.77    $\pm$    0.02    &  -22.23    $\pm$    0.02    &   11.45    $\pm$    0.03    &   10.93    $\pm$    0.15    &    8.88    $\pm$    0.65    &    2.73    $\pm$    1.13    \\
          29 &    1.29    $\pm$    0.02    &         329$\pm$          10&   17.65    $\pm$    0.03    &  -21.57    $\pm$    0.01    &    0.81    $\pm$    0.01    &  -22.29    $\pm$    0.01    &   11.21    $\pm$    0.01    &   10.85    $\pm$    0.15    &    4.99    $\pm$    0.19    &    2.15    $\pm$    0.89    \\
          30 &    1.76    $\pm$    0.03    &         346$\pm$          11&   18.91    $\pm$    0.04    &  -20.96    $\pm$    0.01    &    0.66    $\pm$    0.01    &  -21.57    $\pm$    0.01    &   11.39    $\pm$    0.02    &   10.57    $\pm$    0.15    &   13.21    $\pm$    0.53    &    1.99    $\pm$    0.82    \\
          31 &    0.78    $\pm$    0.02    &         375$\pm$          19&   17.76    $\pm$    0.06    &  -20.35    $\pm$    0.02    &    0.74    $\pm$    0.02    &  -21.00    $\pm$    0.02    &   11.10    $\pm$    0.02    &   10.41    $\pm$    0.15    &   11.99    $\pm$    0.82    &    2.40    $\pm$    0.99    \\
          32 &    1.29    $\pm$    0.05    &         368$\pm$          17&   17.73    $\pm$    0.08    &  -21.54    $\pm$    0.01    &    0.82    $\pm$    0.02    &  -22.27    $\pm$    0.01    &   11.31    $\pm$    0.02    &   10.96    $\pm$    0.15    &    6.41    $\pm$    0.40    &    2.90    $\pm$    1.20    \\
          33 &    1.89    $\pm$    0.06    &         351$\pm$          10&   17.95    $\pm$    0.07    &  -22.16    $\pm$    0.01    &    0.79    $\pm$    0.01    &  -22.71    $\pm$    0.01    &   11.43    $\pm$    0.02    &   11.15    $\pm$    0.15    &    4.87    $\pm$    0.22    &    2.51    $\pm$    1.04    \\
          34 &    1.68    $\pm$    0.06    &         326$\pm$          19&   18.41    $\pm$    0.08    &  -21.43    $\pm$    0.01    &    0.83    $\pm$    0.02    &  -22.18    $\pm$    0.02    &   11.32    $\pm$    0.03    &   10.94    $\pm$    0.15    &    7.28    $\pm$    0.53    &    3.05    $\pm$    1.26    \\
          35 &    2.09    $\pm$    0.07    &         355$\pm$          18&   18.36    $\pm$    0.07    &  -21.97    $\pm$    0.01    &    0.86    $\pm$    0.02    &  -22.75    $\pm$    0.01    &   11.49    $\pm$    0.02    &   11.18    $\pm$    0.15    &    6.51    $\pm$    0.41    &    3.22    $\pm$    1.33    \\
          36 &    2.16    $\pm$    0.06    &         343$\pm$          19&   18.53    $\pm$    0.06    &  -21.88    $\pm$    0.01    &    0.82    $\pm$    0.02    &  -22.54    $\pm$    0.02    &   11.47    $\pm$    0.03    &   11.09    $\pm$    0.15    &    6.86    $\pm$    0.47    &    2.88    $\pm$    1.19    \\
          37 &    2.13    $\pm$    0.04    &         363$\pm$          15&   18.57    $\pm$    0.04    &  -21.77    $\pm$    0.01    &    0.77    $\pm$    0.01    &  -22.39    $\pm$    0.01    &   11.51    $\pm$    0.02    &   10.96    $\pm$    0.15    &    8.32    $\pm$    0.39    &    2.33    $\pm$    0.96    \\
          38 &    1.51    $\pm$    0.02    &         429$\pm$          13&   17.61    $\pm$    0.03    &  -21.97    $\pm$    0.01    &    0.78    $\pm$    0.01    &  -22.62    $\pm$    0.01    &   11.51    $\pm$    0.01    &   11.06    $\pm$    0.15    &    6.88    $\pm$    0.27    &    2.44    $\pm$    1.01    \\
          39 &    0.89    $\pm$    0.02    &         383$\pm$          15&   16.94    $\pm$    0.05    &  -21.50    $\pm$    0.01    &    0.80    $\pm$    0.01    &  -22.20    $\pm$    0.01    &   11.18    $\pm$    0.02    &   10.87    $\pm$    0.15    &    4.96    $\pm$    0.25    &    2.45    $\pm$    1.01    \\
          40 &    1.87    $\pm$    0.05    &         395$\pm$          27&   18.48    $\pm$    0.06    &  -21.57    $\pm$    0.01    &    0.66    $\pm$    0.01    &  -22.23    $\pm$    0.01    &   11.53    $\pm$    0.03    &   10.80    $\pm$    0.15    &   10.44    $\pm$    0.79    &    1.94    $\pm$    0.80    \\
          41 &    1.97    $\pm$    0.02    &         350$\pm$          10&   18.38    $\pm$    0.03    &  -21.75    $\pm$    0.01    &    0.81    $\pm$    0.01    &  -22.50    $\pm$    0.01    &   11.45    $\pm$    0.01    &   11.09    $\pm$    0.15    &    7.33    $\pm$    0.25    &    3.21    $\pm$    1.33    \\
          42 &    1.01    $\pm$    0.02    &         334$\pm$          13&   17.41    $\pm$    0.05    &  -21.27    $\pm$    0.01    &    0.84    $\pm$    0.01    &  -22.04    $\pm$    0.01    &   11.12    $\pm$    0.02    &   10.88    $\pm$    0.15    &    5.28    $\pm$    0.26    &    3.05    $\pm$    1.26    \\
          43 &    1.39    $\pm$    0.02    &         327$\pm$          10&   18.09    $\pm$    0.03    &  -21.27    $\pm$    0.01    &    0.79    $\pm$    0.01    &  -21.91    $\pm$    0.01    &   11.24    $\pm$    0.01    &   10.84    $\pm$    0.15    &    7.01    $\pm$    0.27    &    2.79    $\pm$    1.15    \\
          44 &    1.40    $\pm$    0.03    &         336$\pm$          15&   18.01    $\pm$    0.05    &  -21.42    $\pm$    0.01    &    0.78    $\pm$    0.01    &  -22.17    $\pm$    0.01    &   11.26    $\pm$    0.02    &   10.91    $\pm$    0.15    &    6.49    $\pm$    0.36    &    2.87    $\pm$    1.19    \\
          45 &    1.03    $\pm$    0.04    &         326$\pm$          18&   17.72    $\pm$    0.08    &  -21.03    $\pm$    0.01    &    0.77    $\pm$    0.02    &  -21.78    $\pm$    0.01    &   11.10    $\pm$    0.03    &   10.71    $\pm$    0.15    &    6.42    $\pm$    0.46    &    2.61    $\pm$    1.08    \\
          46 &    1.39    $\pm$    0.01    &         368$\pm$          11&   17.45    $\pm$    0.02    &  -21.93    $\pm$    0.01    &    0.80    $\pm$    0.01    &  -22.63    $\pm$    0.01    &   11.34    $\pm$    0.01    &   11.04    $\pm$    0.15    &    4.86    $\pm$    0.17    &    2.43    $\pm$    1.00    \\
          47 &    1.80    $\pm$    0.03    &         350$\pm$          14&   18.28    $\pm$    0.04    &  -21.68    $\pm$    0.01    &    0.76    $\pm$    0.01    &  -22.24    $\pm$    0.01    &   11.41    $\pm$    0.02    &   10.90    $\pm$    0.15    &    7.12    $\pm$    0.33    &    2.21    $\pm$    0.91    \\
          48 &    1.22    $\pm$    0.06    &         333$\pm$          21&   18.11    $\pm$    0.11    &  -21.02    $\pm$    0.01    &    0.86    $\pm$    0.02    &  -21.57    $\pm$    0.02    &   11.20    $\pm$    0.03    &   10.70    $\pm$    0.15    &    8.01    $\pm$    0.68    &    2.54    $\pm$    1.05    \\
          49 &    2.17    $\pm$    0.07    &         352$\pm$          16&   18.70    $\pm$    0.08    &  -21.71    $\pm$    0.01    &    0.85    $\pm$    0.02    &  -22.39    $\pm$    0.01    &   11.50    $\pm$    0.02    &   11.05    $\pm$    0.15    &    8.50    $\pm$    0.51    &    3.04    $\pm$    1.26    \\
          50 &    1.69    $\pm$    0.01    &         338$\pm$           7&   18.18    $\pm$    0.01    &  -21.56    $\pm$    0.01    &    0.84    $\pm$    0.01    &  -22.22    $\pm$    0.01    &   11.35    $\pm$    0.01    &   10.91    $\pm$    0.15    &    6.96    $\pm$    0.21    &    2.55    $\pm$    1.05    \\
          51 &    1.48    $\pm$    0.03    &         363$\pm$          18&   18.32    $\pm$    0.05    &  -21.19    $\pm$    0.01    &    0.77    $\pm$    0.02    &  -21.86    $\pm$    0.02    &   11.35    $\pm$    0.02    &   10.74    $\pm$    0.15    &    9.88    $\pm$    0.62    &    2.43    $\pm$    1.00    \\
          52 &    2.17    $\pm$    0.02    &         368$\pm$           9&   18.22    $\pm$    0.02    &  -22.10    $\pm$    0.01    &    0.84    $\pm$    0.01    &  -22.80    $\pm$    0.01    &   11.53    $\pm$    0.01    &   11.10    $\pm$    0.15    &    6.43    $\pm$    0.19    &    2.37    $\pm$    0.98    \\
          53 &    1.57    $\pm$    0.03    &         361$\pm$          12&   18.30    $\pm$    0.03    &  -21.33    $\pm$    0.01    &    0.76    $\pm$    0.01    &  -21.91    $\pm$    0.01    &   11.38    $\pm$    0.02    &   10.85    $\pm$    0.15    &    9.11    $\pm$    0.39    &    2.70    $\pm$    1.12    \\
          54 &    1.52    $\pm$    0.05    &         347$\pm$          19&   18.31    $\pm$    0.07    &  -21.30    $\pm$    0.01    &    0.82    $\pm$    0.02    &  -21.97    $\pm$    0.02    &   11.33    $\pm$    0.03    &   10.86    $\pm$    0.15    &    8.44    $\pm$    0.58    &    2.87    $\pm$    1.19    \\
          55 &    1.89    $\pm$    0.03    &         360$\pm$          12&   18.39    $\pm$    0.03    &  -21.66    $\pm$    0.01    &    0.79    $\pm$    0.01    &  -22.30    $\pm$    0.01    &   11.45    $\pm$    0.02    &   10.92    $\pm$    0.15    &    8.07    $\pm$    0.33    &    2.34    $\pm$    0.97    \\
          56 &    2.00    $\pm$    0.14    &         356$\pm$          27&   18.38    $\pm$    0.15    &  -21.90    $\pm$    0.01    &    1.04    $\pm$    0.03    &  -22.49    $\pm$    0.02    &   11.47    $\pm$    0.04    &   11.28    $\pm$    0.15    &    6.67    $\pm$    0.71    &    4.27    $\pm$    1.77    \\
          57 &    0.66    $\pm$    0.01    &         328$\pm$          12&   16.93    $\pm$    0.04    &  -20.81    $\pm$    0.01    &    0.79    $\pm$    0.01    &  -21.47    $\pm$    0.01    &   10.92    $\pm$    0.02    &   10.58    $\pm$    0.15    &    5.09    $\pm$    0.23    &    2.36    $\pm$    0.98    \\
          58 &    2.12    $\pm$    0.03    &         364$\pm$          15&   18.55    $\pm$    0.03    &  -21.75    $\pm$    0.01    &    0.77    $\pm$    0.01    &  -22.39    $\pm$    0.01    &   11.51    $\pm$    0.02    &   10.97    $\pm$    0.15    &    8.51    $\pm$    0.40    &    2.45    $\pm$    1.01    \\
          59 &    1.37    $\pm$    0.02    &         342$\pm$          11&   17.92    $\pm$    0.04    &  -21.41    $\pm$    0.01    &    0.80    $\pm$    0.01    &  -22.15    $\pm$    0.01    &   11.27    $\pm$    0.01    &   10.88    $\pm$    0.15    &    6.59    $\pm$    0.26    &    2.67    $\pm$    1.10    \\
          60 &    1.59    $\pm$    0.04    &         348$\pm$          16&   18.29    $\pm$    0.05    &  -21.42    $\pm$    0.01    &    0.79    $\pm$    0.02    &  -22.20    $\pm$    0.02    &   11.35    $\pm$    0.02    &   10.96    $\pm$    0.15    &    7.95    $\pm$    0.47    &    3.27    $\pm$    1.35    \\
          61 &    1.50    $\pm$    0.04    &         326$\pm$          18&   18.27    $\pm$    0.05    &  -21.31    $\pm$    0.01    &    0.78    $\pm$    0.02    &  -21.96    $\pm$    0.01    &   11.27    $\pm$    0.03    &   10.86    $\pm$    0.15    &    7.25    $\pm$    0.47    &    2.80    $\pm$    1.16    \\
          62 &    1.77    $\pm$    0.05    &         341$\pm$          16&   18.32    $\pm$    0.06    &  -21.63    $\pm$    0.01    &    0.84    $\pm$    0.02    &  -22.36    $\pm$    0.01    &   11.38    $\pm$    0.02    &   11.00    $\pm$    0.15    &    6.97    $\pm$    0.41    &    2.90    $\pm$    1.20    \\
          63 &    0.62    $\pm$    0.01    &         337$\pm$          17&   17.12    $\pm$    0.05    &  -20.51    $\pm$    0.01    &    0.80    $\pm$    0.02    &  -21.14    $\pm$    0.01    &   10.91    $\pm$    0.02    &   10.45    $\pm$    0.15    &    6.71    $\pm$    0.40    &    2.32    $\pm$    0.96    \\
          64 &    2.14    $\pm$    0.08    &         344$\pm$          19&   18.58    $\pm$    0.08    &  -21.80    $\pm$    0.01    &    0.81    $\pm$    0.02    &  -22.46    $\pm$    0.01    &   11.47    $\pm$    0.03    &   11.06    $\pm$    0.15    &    7.33    $\pm$    0.52    &    2.87    $\pm$    1.18    \\
          65 &    1.54    $\pm$    0.06    &         332$\pm$          18&   18.30    $\pm$    0.08    &  -21.34    $\pm$    0.01    &    0.79    $\pm$    0.02    &  -22.07    $\pm$    0.02    &   11.29    $\pm$    0.03    &   10.80    $\pm$    0.15    &    7.46    $\pm$    0.54    &    2.38    $\pm$    0.99    \\
          66 &    2.06    $\pm$    0.08    &         340$\pm$          15&   17.66    $\pm$    0.09    &  -22.69    $\pm$    0.01    &    0.80    $\pm$    0.02    &  -23.36    $\pm$    0.01    &   11.44    $\pm$    0.03    &   11.27    $\pm$    0.15    &    3.03    $\pm$    0.19    &    2.05    $\pm$    0.84    \\
          67 &    1.83    $\pm$    0.02    &         338$\pm$           9&   18.39    $\pm$    0.03    &  -21.57    $\pm$    0.01    &    0.78    $\pm$    0.01    &  -22.25    $\pm$    0.01    &   11.39    $\pm$    0.01    &   10.89    $\pm$    0.15    &    7.52    $\pm$    0.26    &    2.38    $\pm$    0.98    \\
          68 &    1.56    $\pm$    0.05    &         328$\pm$          13&   18.46    $\pm$    0.07    &  -21.19    $\pm$    0.01    &    0.75    $\pm$    0.01    &  -21.88    $\pm$    0.01    &   11.29    $\pm$    0.02    &   10.79    $\pm$    0.15    &    8.45    $\pm$    0.46    &    2.69    $\pm$    1.11    \\
          69 &    1.94    $\pm$    0.04    &         339$\pm$          12&   18.06    $\pm$    0.05    &  -22.07    $\pm$    0.01    &    0.62    $\pm$    0.01    &  -22.66    $\pm$    0.01    &   11.41    $\pm$    0.02    &   10.96    $\pm$    0.15    &    5.02    $\pm$    0.22    &    1.77    $\pm$    0.73    \\
          70 &    2.03    $\pm$    0.03    &         355$\pm$          10&   18.26    $\pm$    0.03    &  -21.95    $\pm$    0.01    &    0.80    $\pm$    0.01    &  -22.70    $\pm$    0.01    &   11.47    $\pm$    0.01    &   11.17    $\pm$    0.15    &    6.40    $\pm$    0.23    &    3.23    $\pm$    1.33    \\
          71 &    1.90    $\pm$    0.04    &         334$\pm$          15&   18.57    $\pm$    0.05    &  -21.52    $\pm$    0.01    &    0.80    $\pm$    0.02    &  -22.12    $\pm$    0.01    &   11.39    $\pm$    0.02    &   10.91    $\pm$    0.15    &    7.93    $\pm$    0.43    &    2.61    $\pm$    1.08    \\
          72 &    1.83    $\pm$    0.03    &         373$\pm$          12&   18.53    $\pm$    0.03    &  -21.43    $\pm$    0.01    &    0.77    $\pm$    0.01    &  -22.13    $\pm$    0.01    &   11.47    $\pm$    0.01    &   10.87    $\pm$    0.15    &   10.33    $\pm$    0.41    &    2.62    $\pm$    1.08    \\
          73 &    0.97    $\pm$    0.01    &         393$\pm$          15&   17.56    $\pm$    0.03    &  -21.01    $\pm$    0.01    &    0.76    $\pm$    0.01    &  -21.64    $\pm$    0.01    &   11.24    $\pm$    0.02    &   10.61    $\pm$    0.15    &    8.94    $\pm$    0.42    &    2.08    $\pm$    0.86    \\
          74 &    1.85    $\pm$    0.07    &         353$\pm$          19&   18.41    $\pm$    0.09    &  -21.65    $\pm$    0.01    &    0.80    $\pm$    0.02    &  -22.36    $\pm$    0.01    &   11.43    $\pm$    0.03    &   10.99    $\pm$    0.15    &    7.70    $\pm$    0.55    &    2.80    $\pm$    1.15    \\
          75 &    1.81    $\pm$    0.01    &         397$\pm$           3&   19.07    $\pm$    0.02    &  -20.81    $\pm$    0.04    &    0.73    $\pm$    0.06    &  -21.39    $\pm$    0.04    &   11.52    $\pm$    0.00    &   10.59    $\pm$    0.15    &   20.56    $\pm$    2.13    &    2.42    $\pm$    1.03    \\
          76 &    2.08    $\pm$    0.04    &         432$\pm$          18&   18.22    $\pm$    0.04    &  -22.07    $\pm$    0.01    &    0.79    $\pm$    0.01    &  -22.77    $\pm$    0.01    &   11.65    $\pm$    0.02    &   11.10    $\pm$    0.15    &    8.76    $\pm$    0.44    &    2.47    $\pm$    1.02    \\
 \end{supertabular}
 \label{list_candiates_dV}
  \twocolumn
 \end{landscape}%}

\section{Candidate selection}
\label{sec_candsel}
\begin{figure}[ht]
\begin{center}
\includegraphics[width=0.45\textwidth]{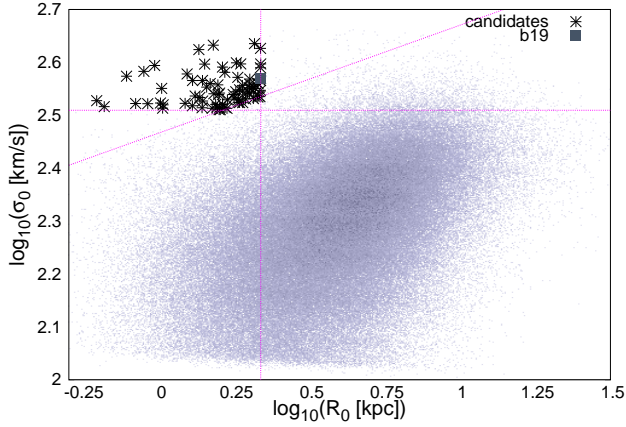}\\ 
\caption{Selection criteria for our compact and massive galaxy candidates indicated by the dashed magenta lines. The black stars represent the 75 new candidates for galaxies with similar properties in de Vaucouleurs fit parameters as b19, while b19 itself is represented by a grey filled square in the plot.}
\label{R0_sigma_V}
\end{center}
\end{figure}

\begin{figure*}[ht]
\begin{center}
\includegraphics[width=0.12\textwidth]{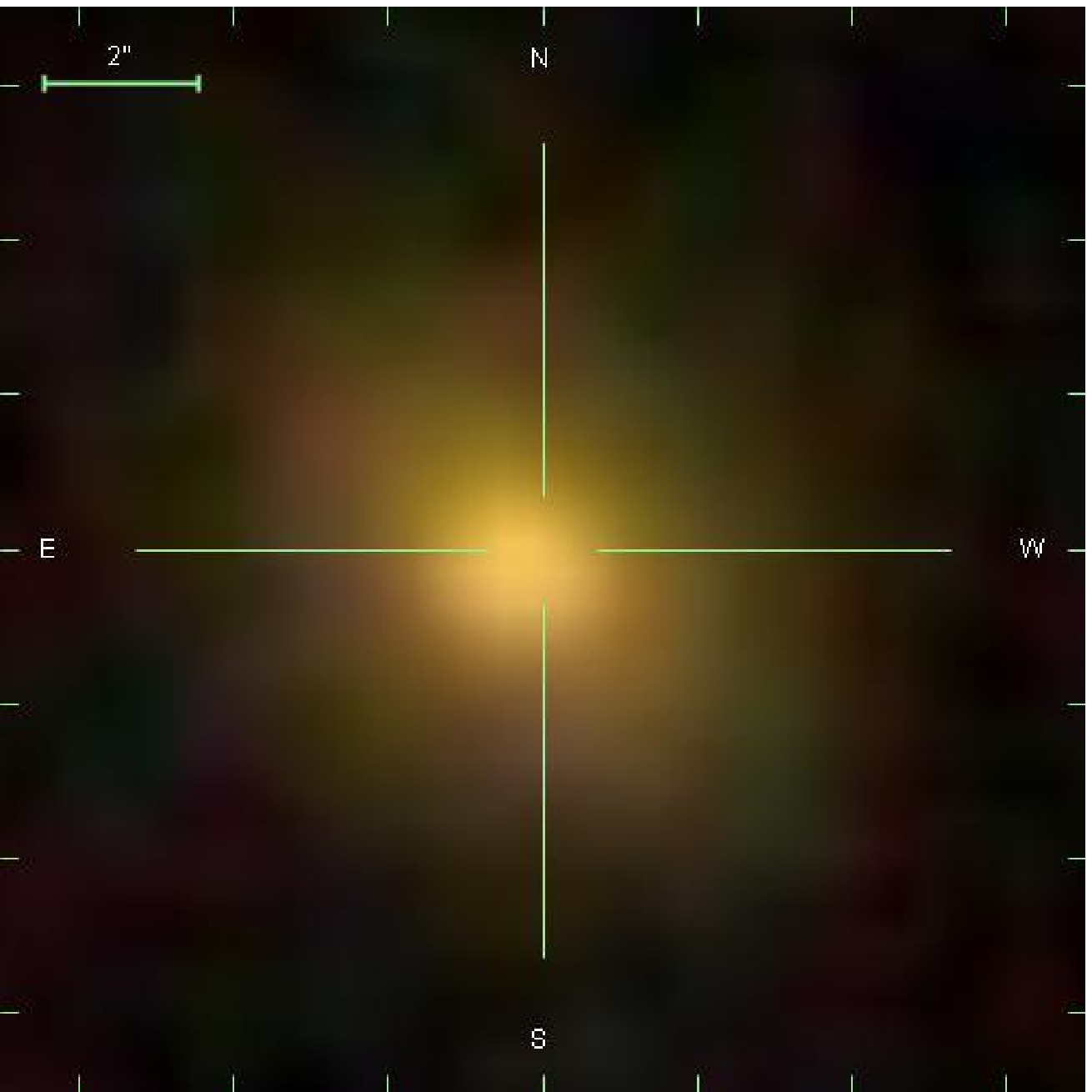}
\includegraphics[width=0.12\textwidth]{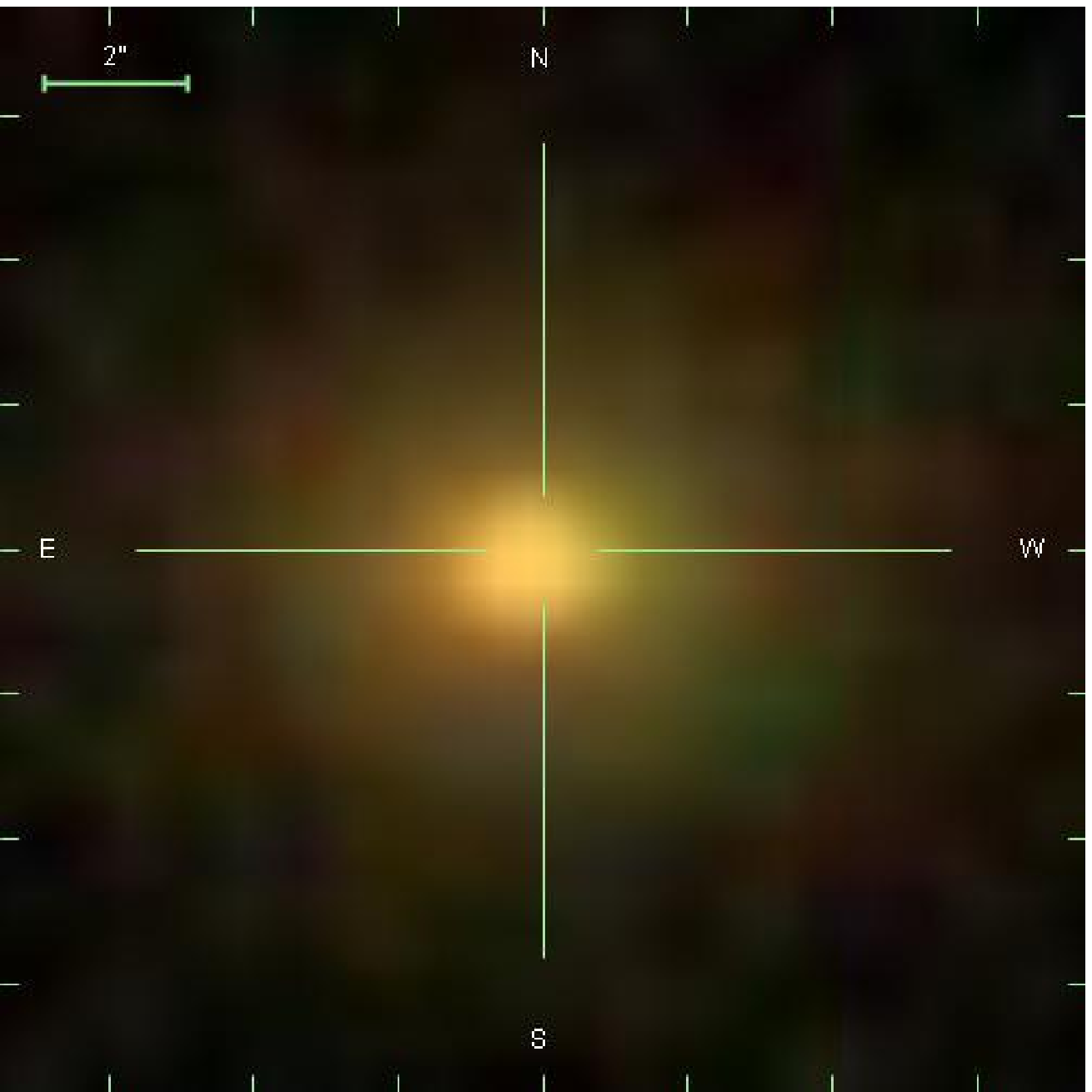}
\includegraphics[width=0.12\textwidth]{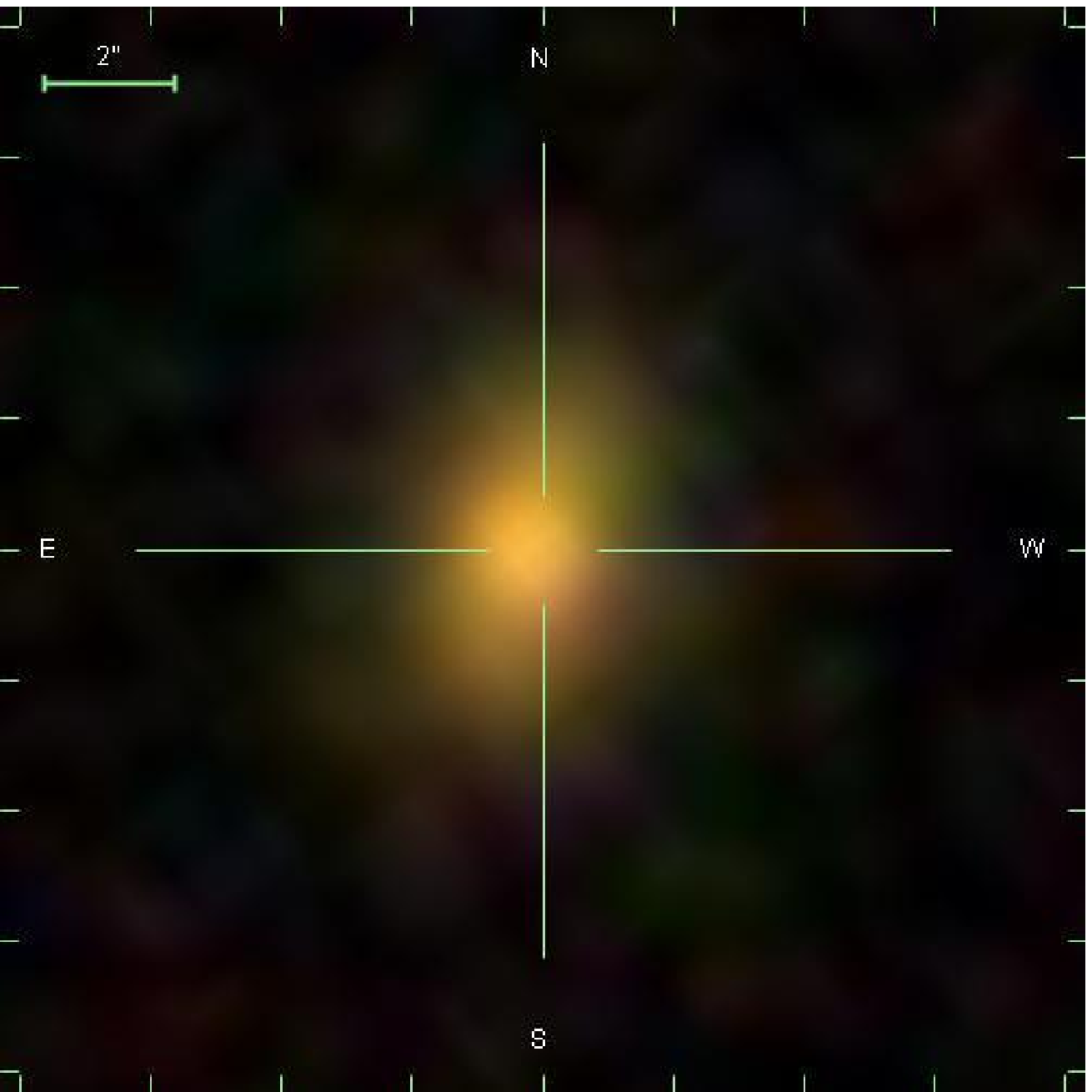}
\includegraphics[width=0.12\textwidth]{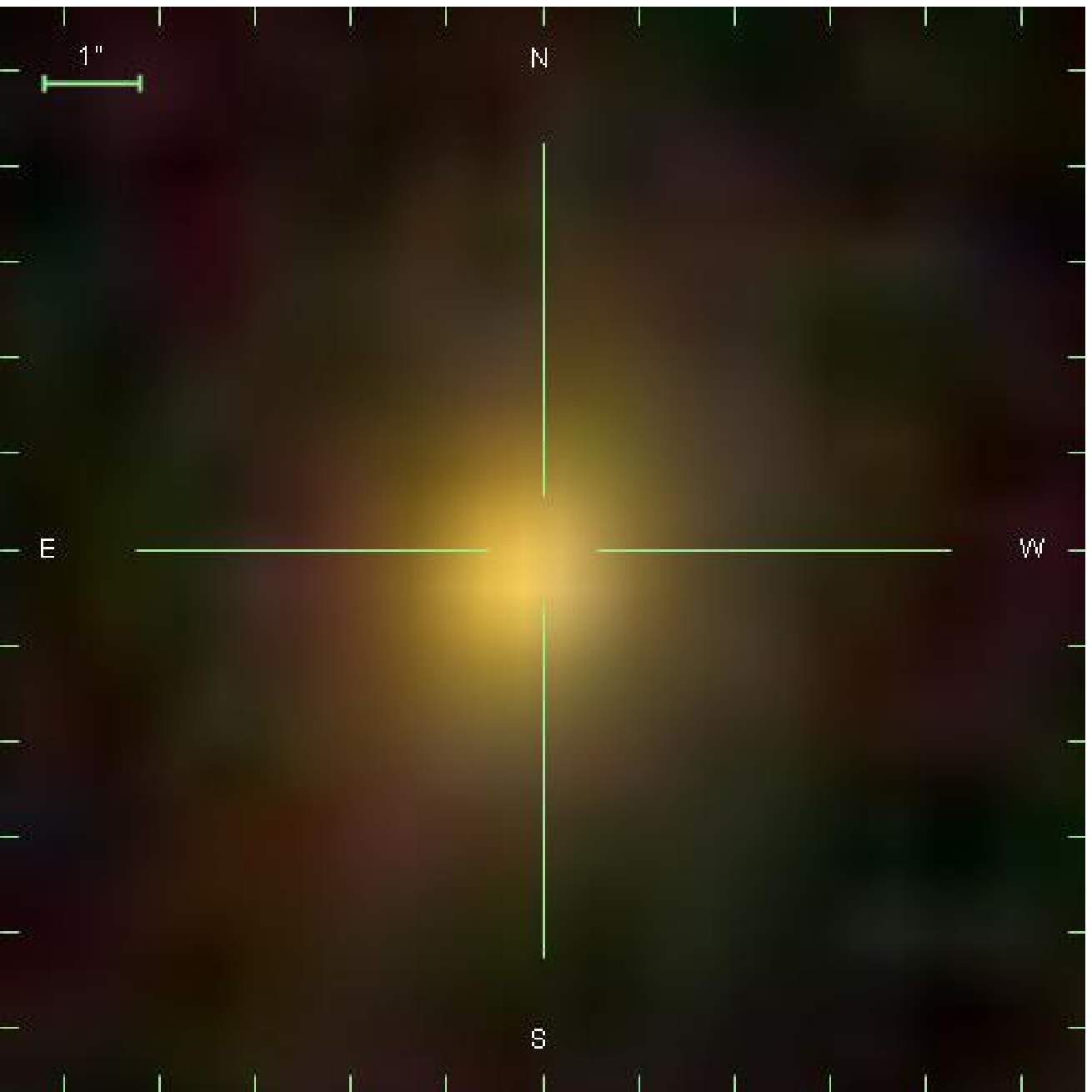}
\includegraphics[width=0.12\textwidth]{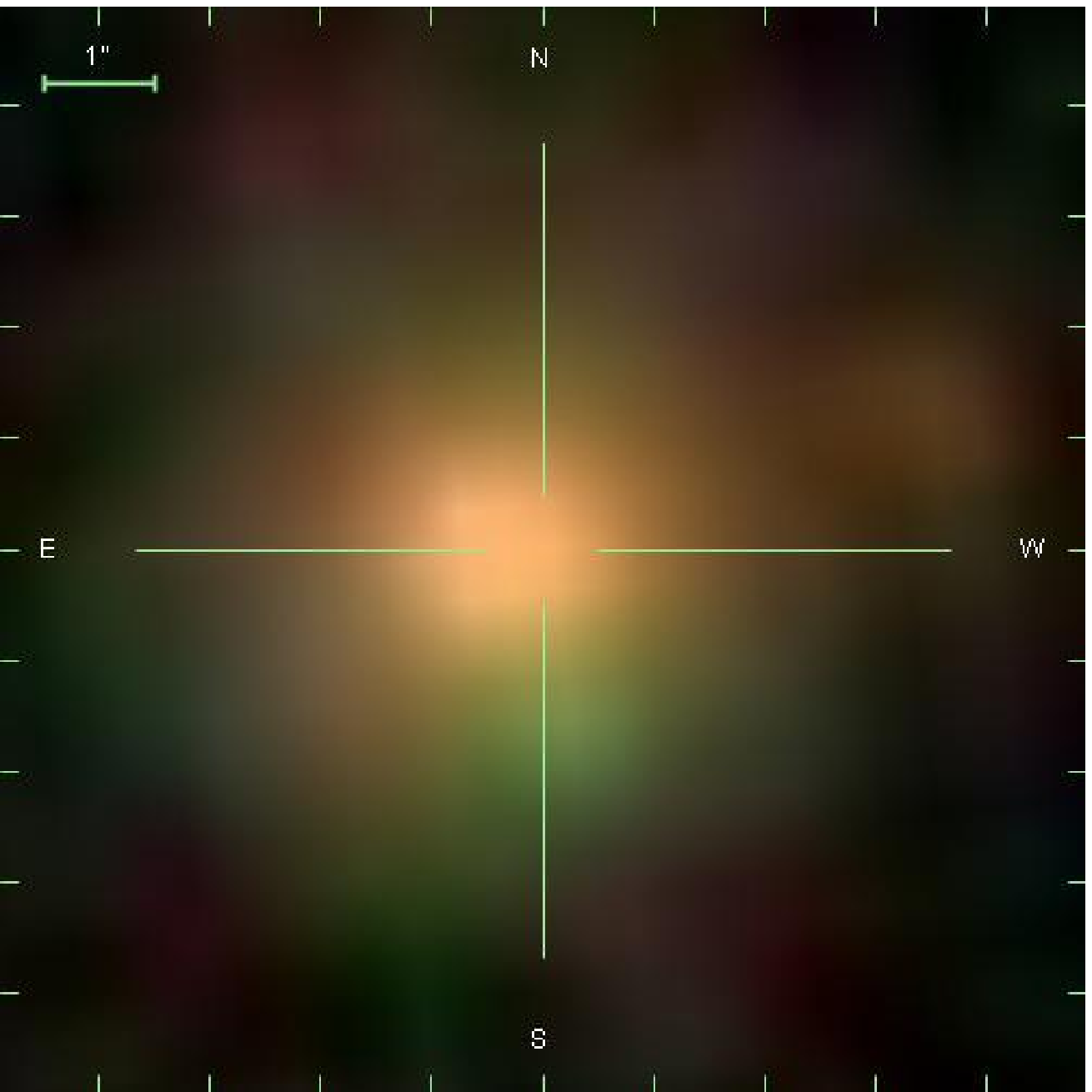}
\includegraphics[width=0.12\textwidth]{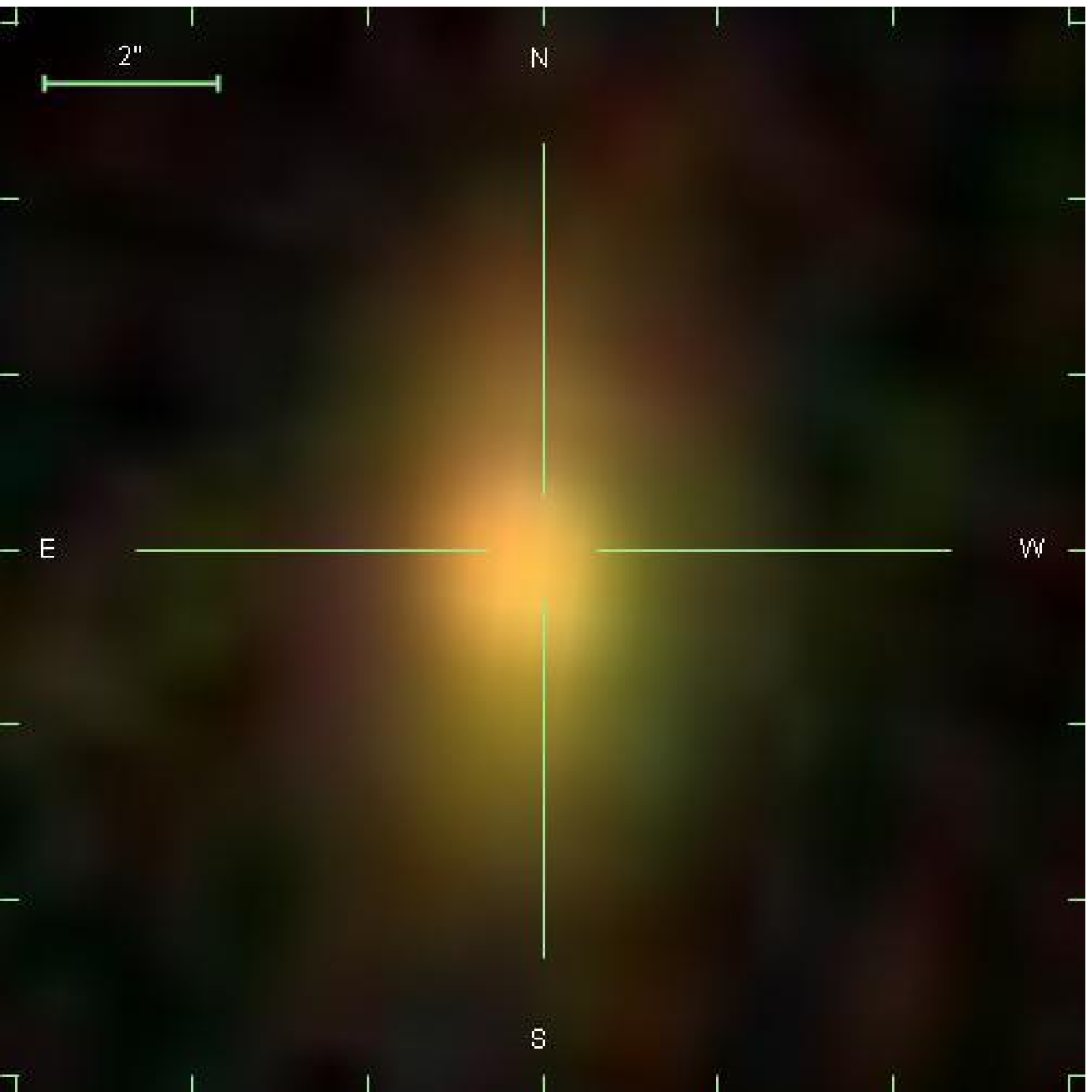}
\includegraphics[width=0.12\textwidth]{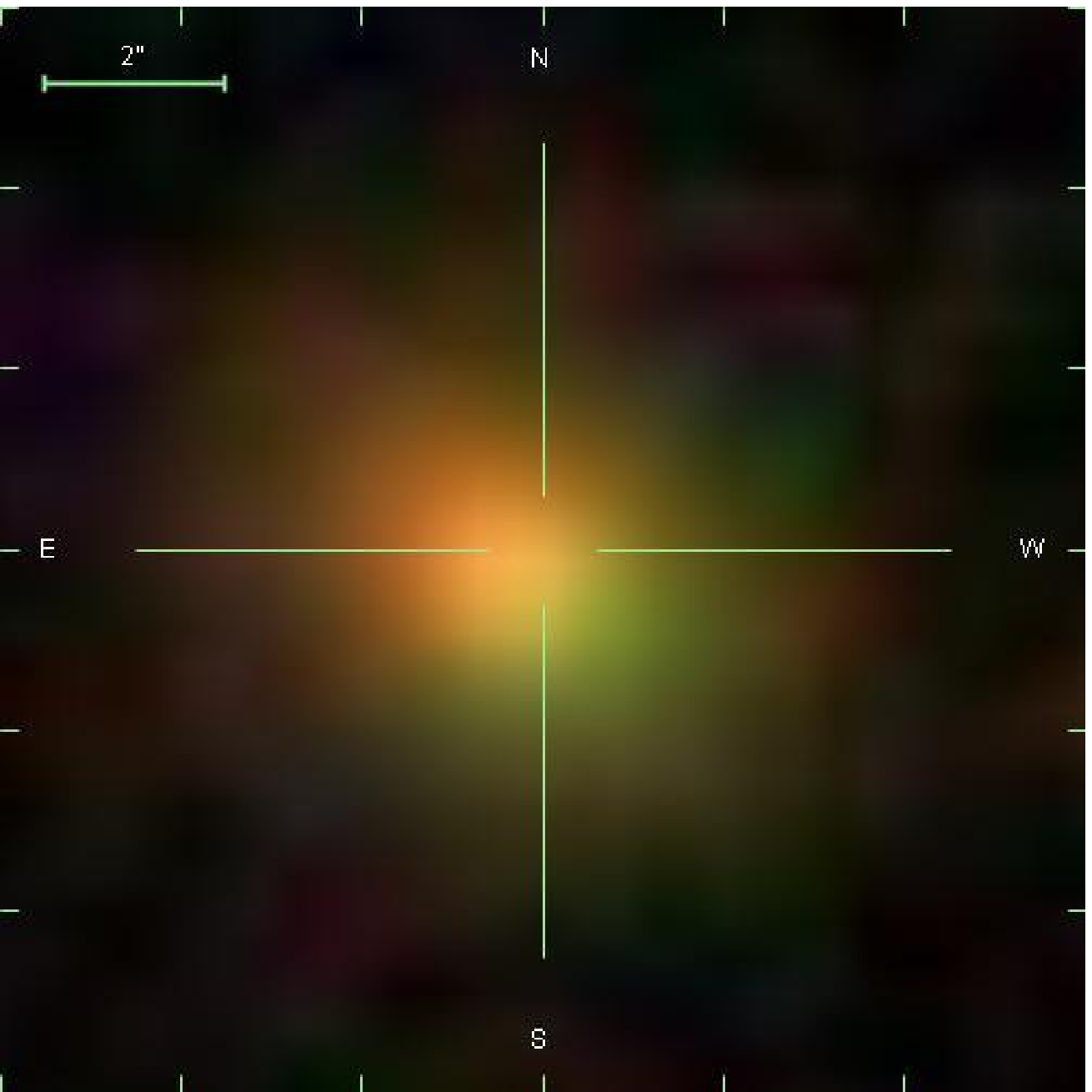}
\includegraphics[width=0.12\textwidth]{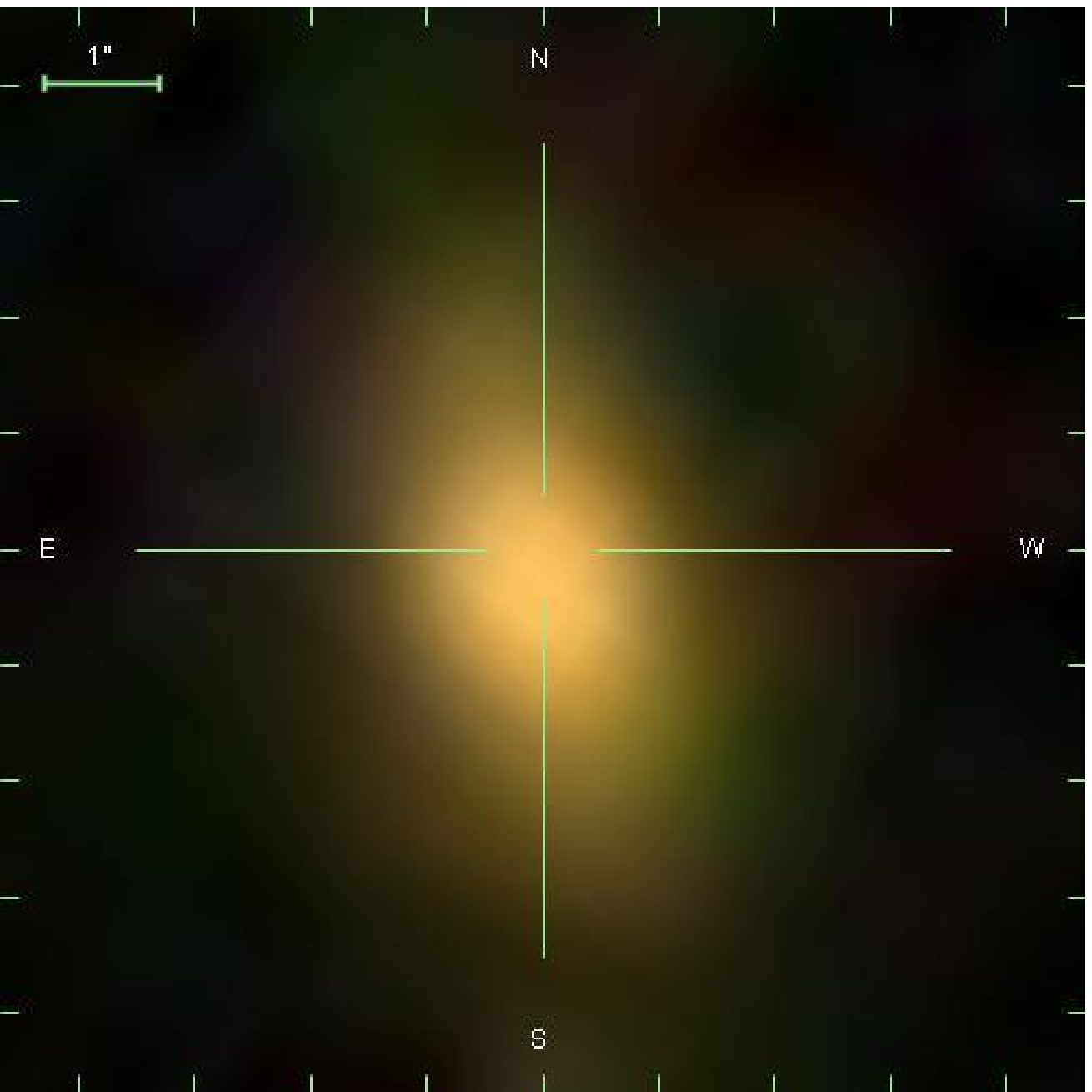}\\
\includegraphics[width=0.12\textwidth]{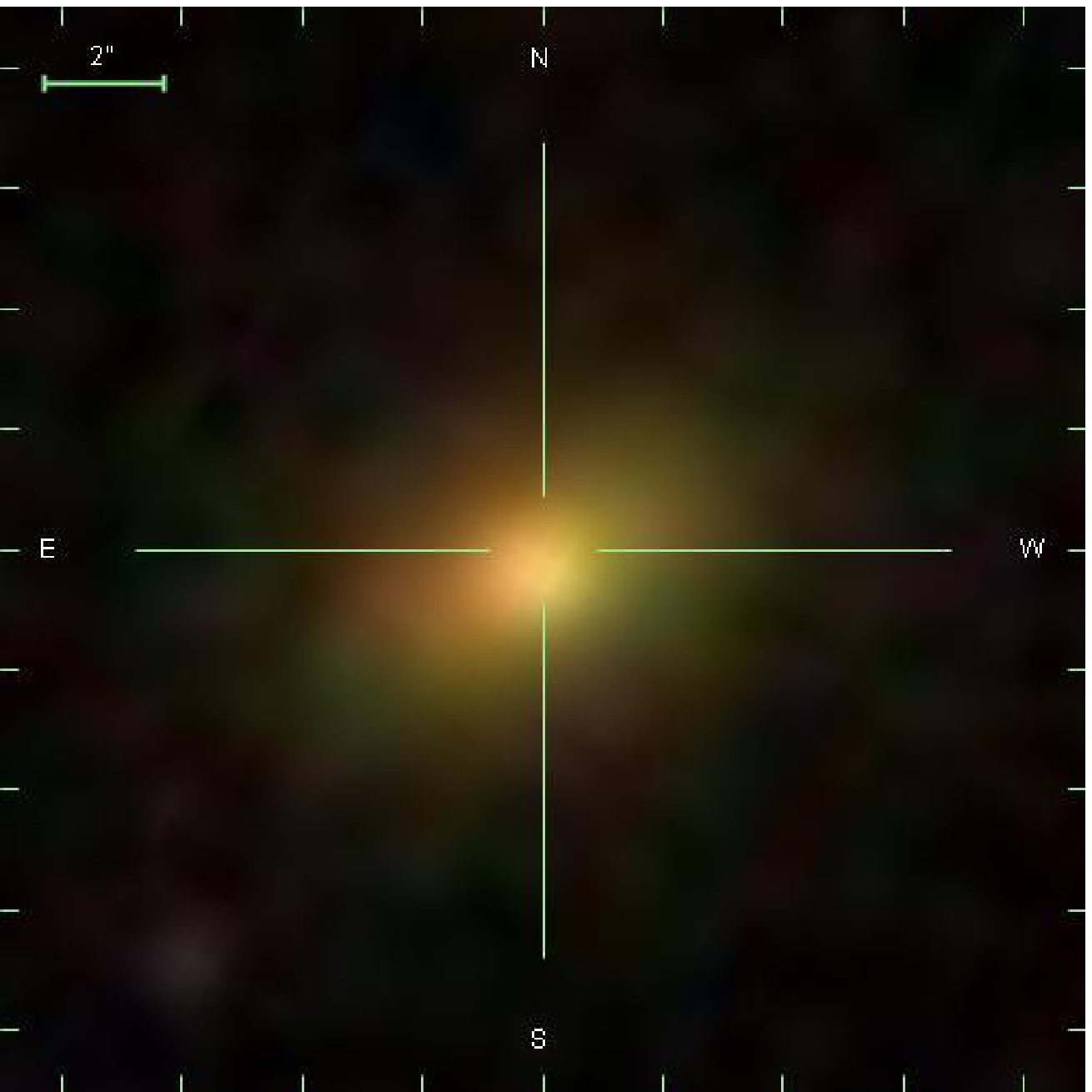}
\includegraphics[width=0.12\textwidth]{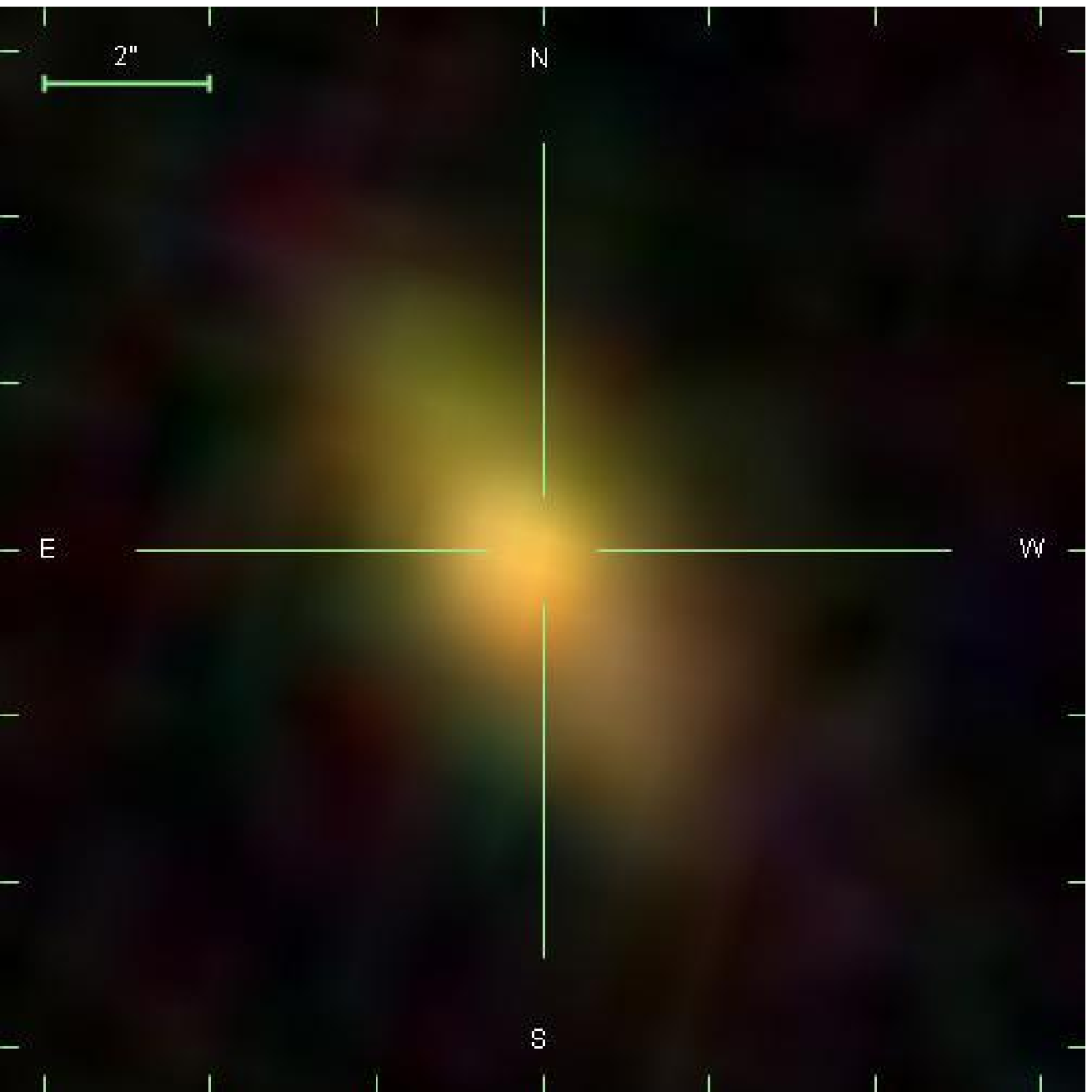}
\includegraphics[width=0.12\textwidth]{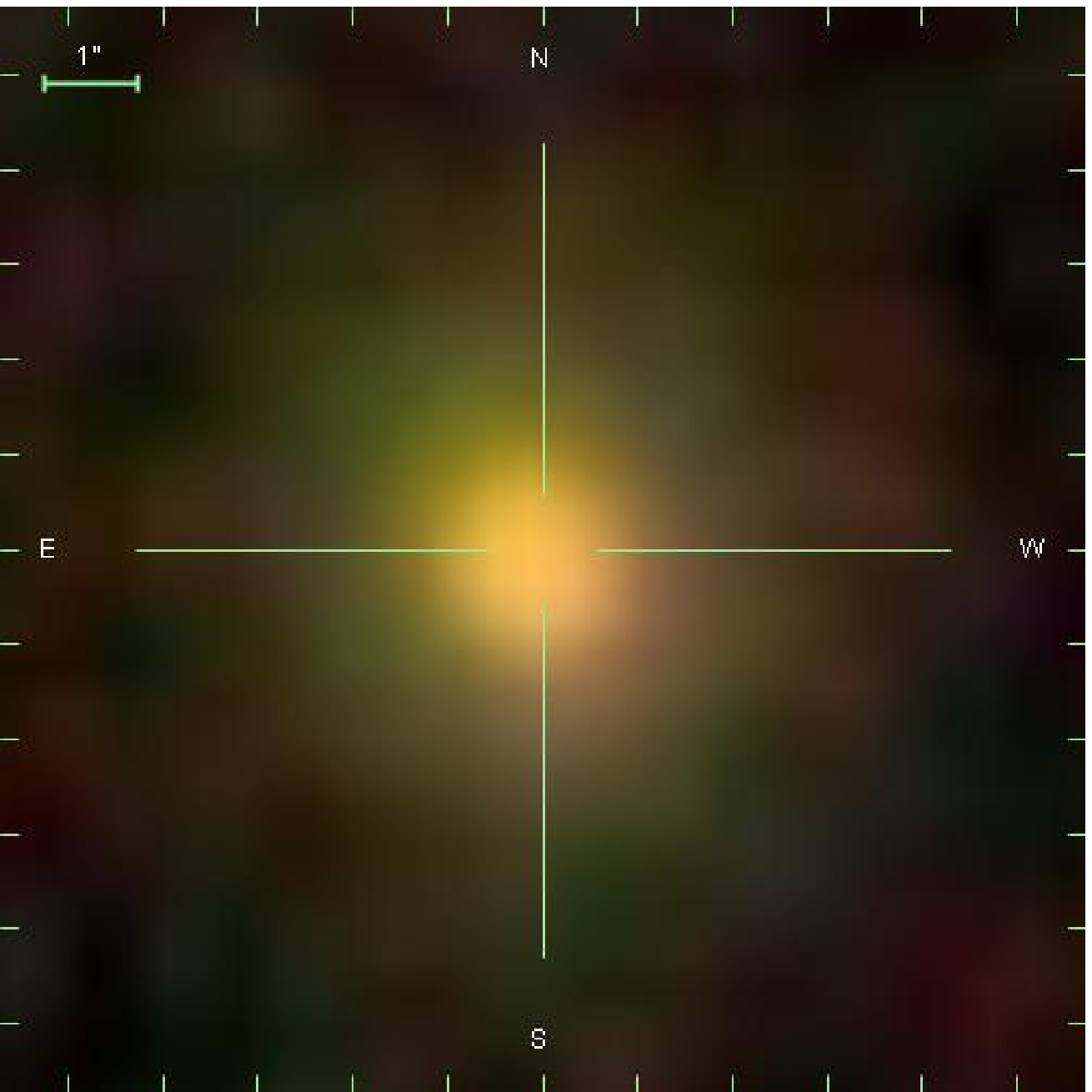}
\includegraphics[width=0.12\textwidth]{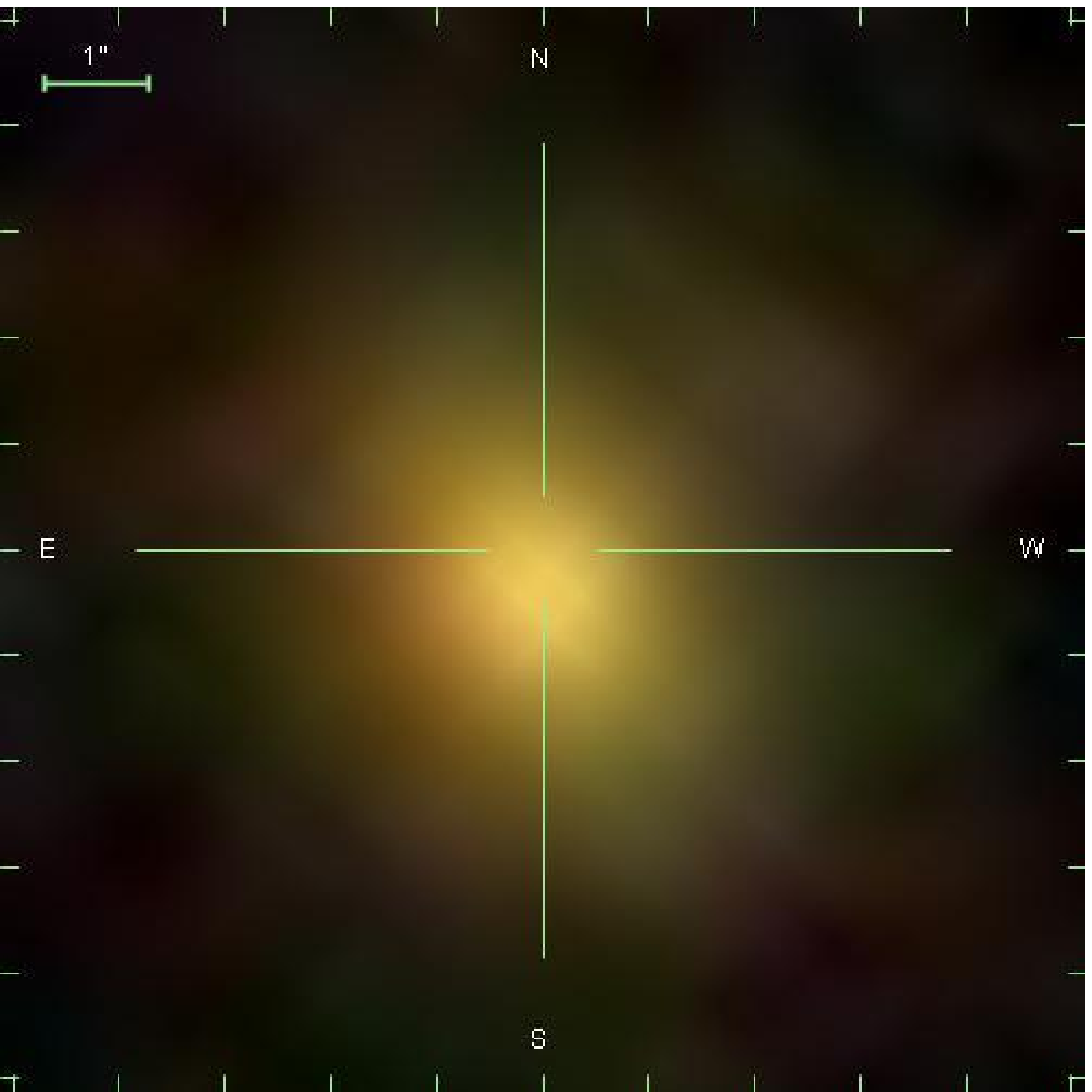}
\includegraphics[width=0.12\textwidth]{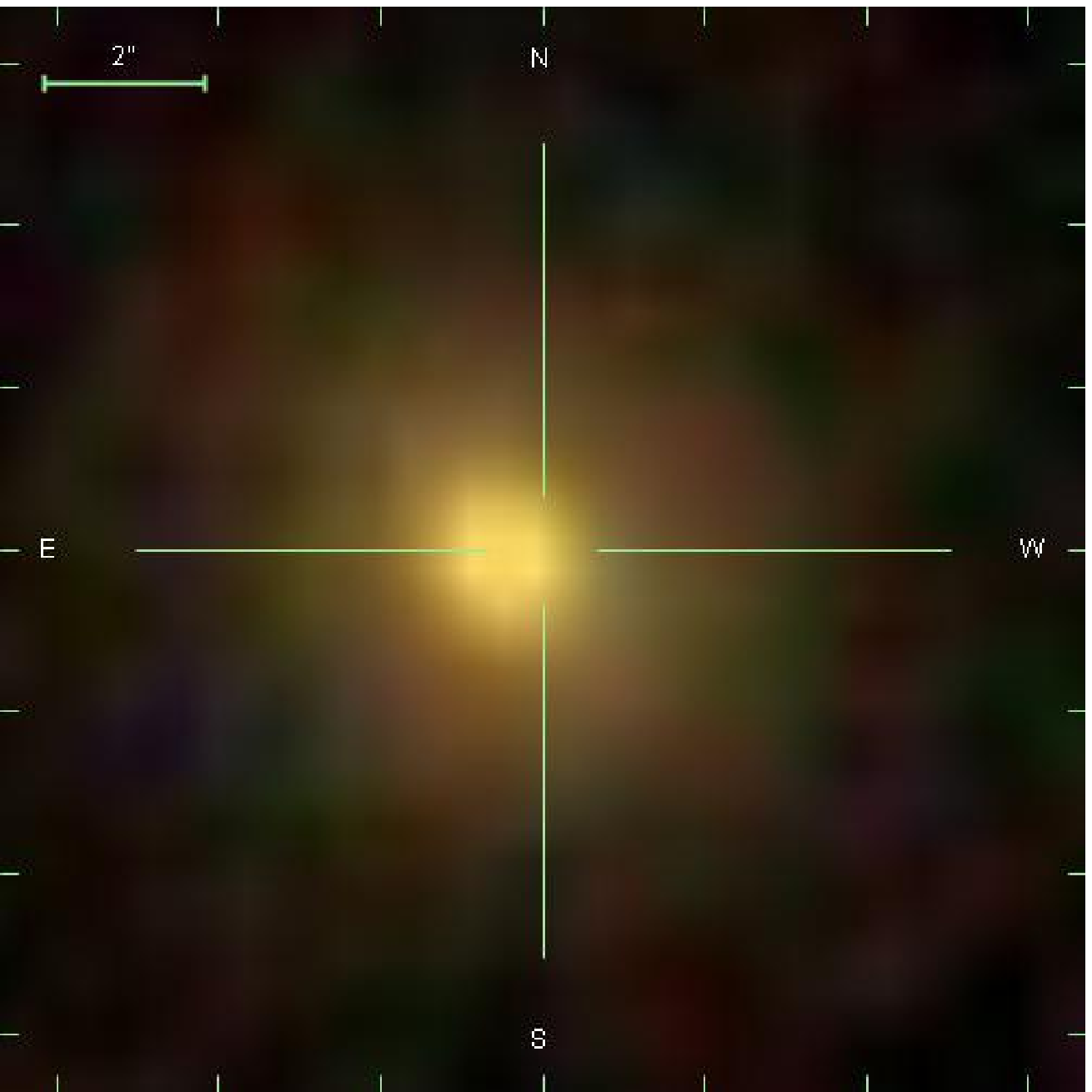}
\includegraphics[width=0.12\textwidth]{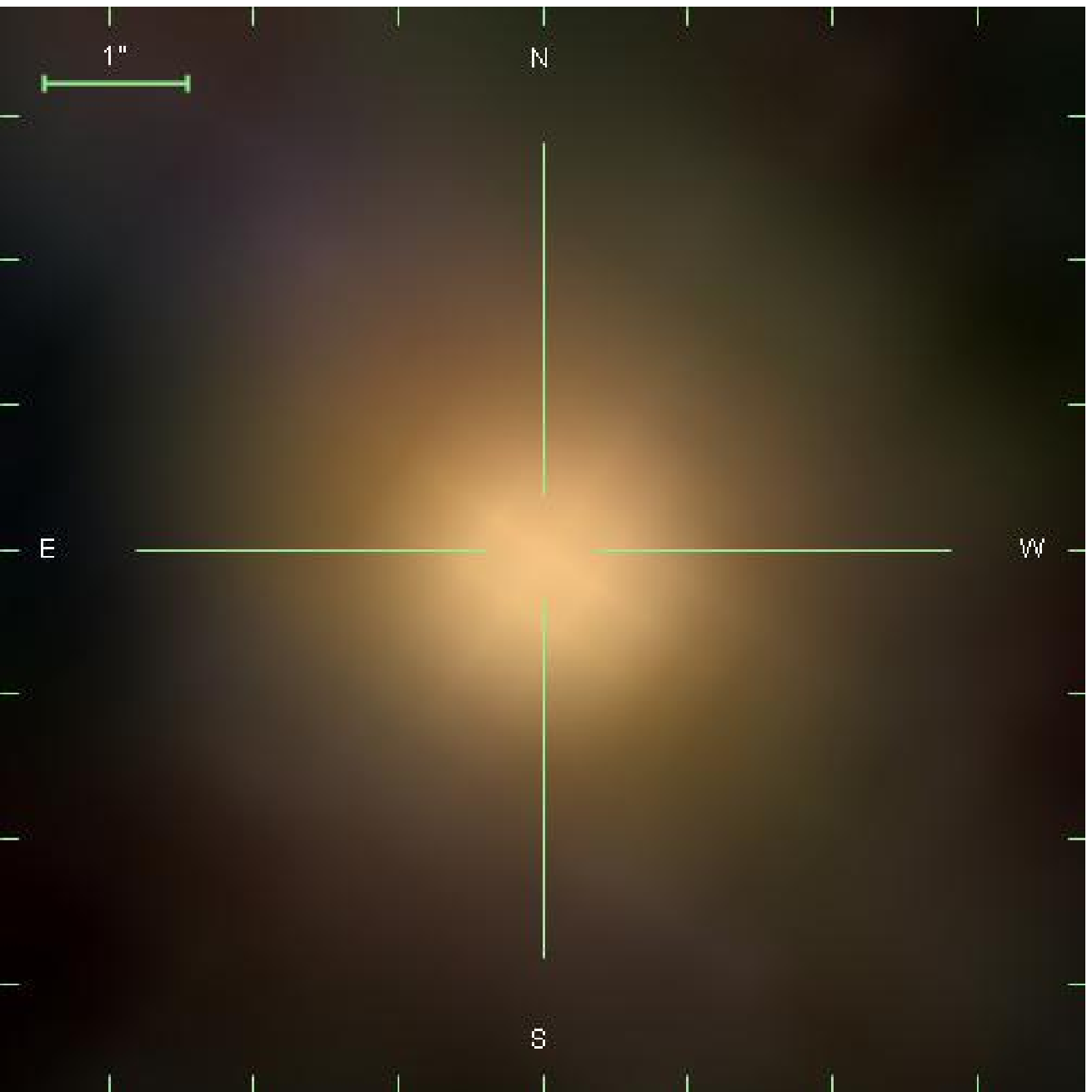}
\includegraphics[width=0.12\textwidth]{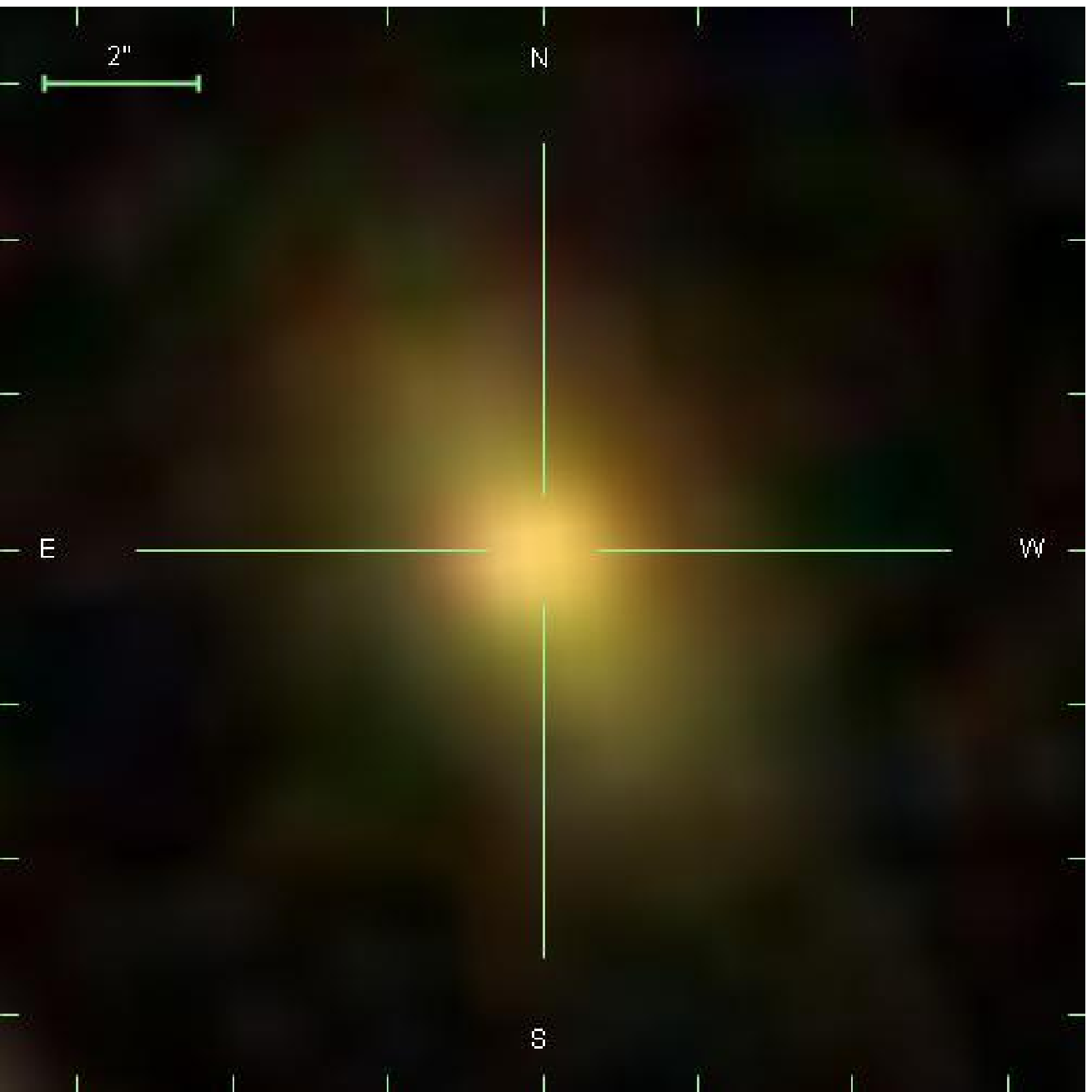}
\includegraphics[width=0.12\textwidth]{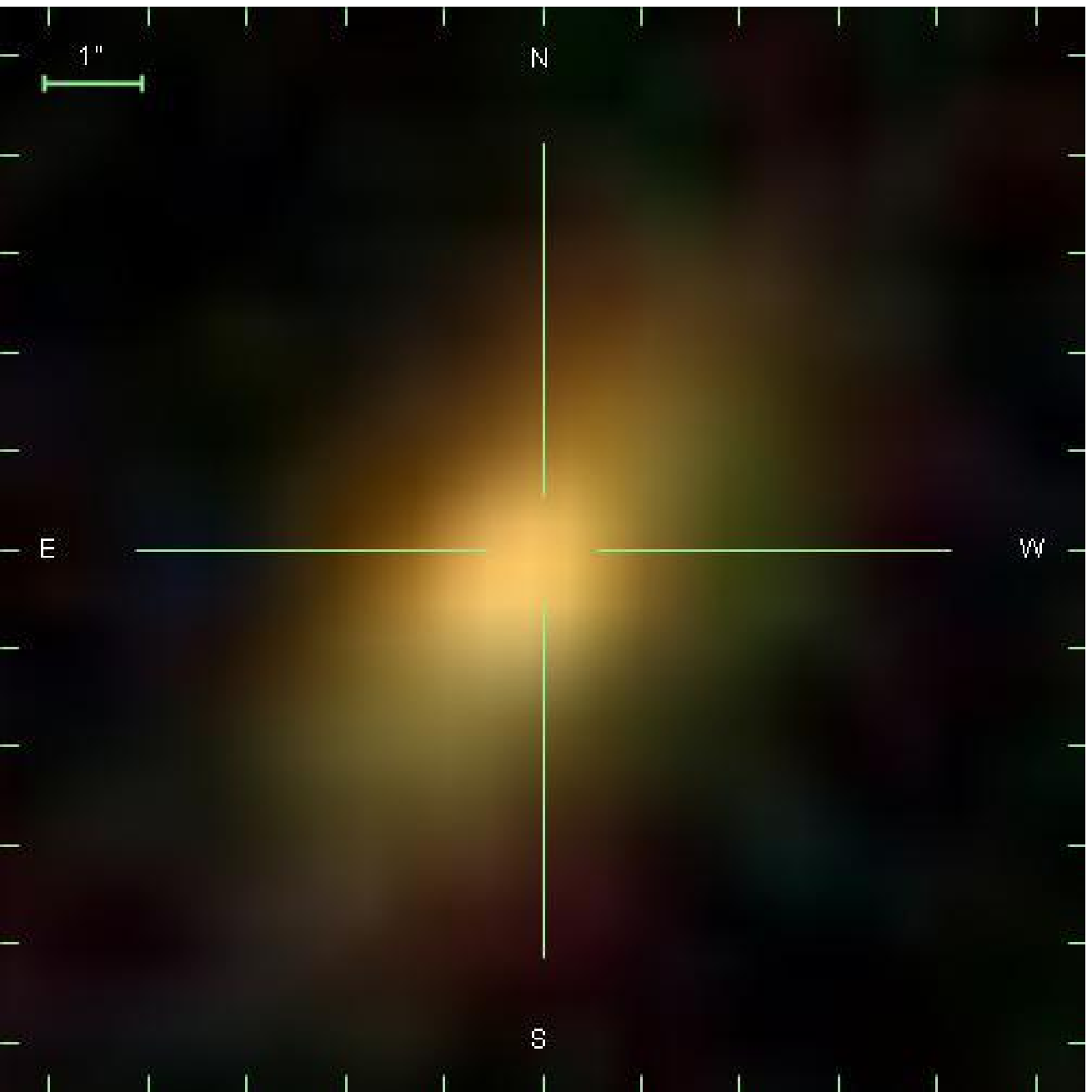}\\
\includegraphics[width=0.12\textwidth]{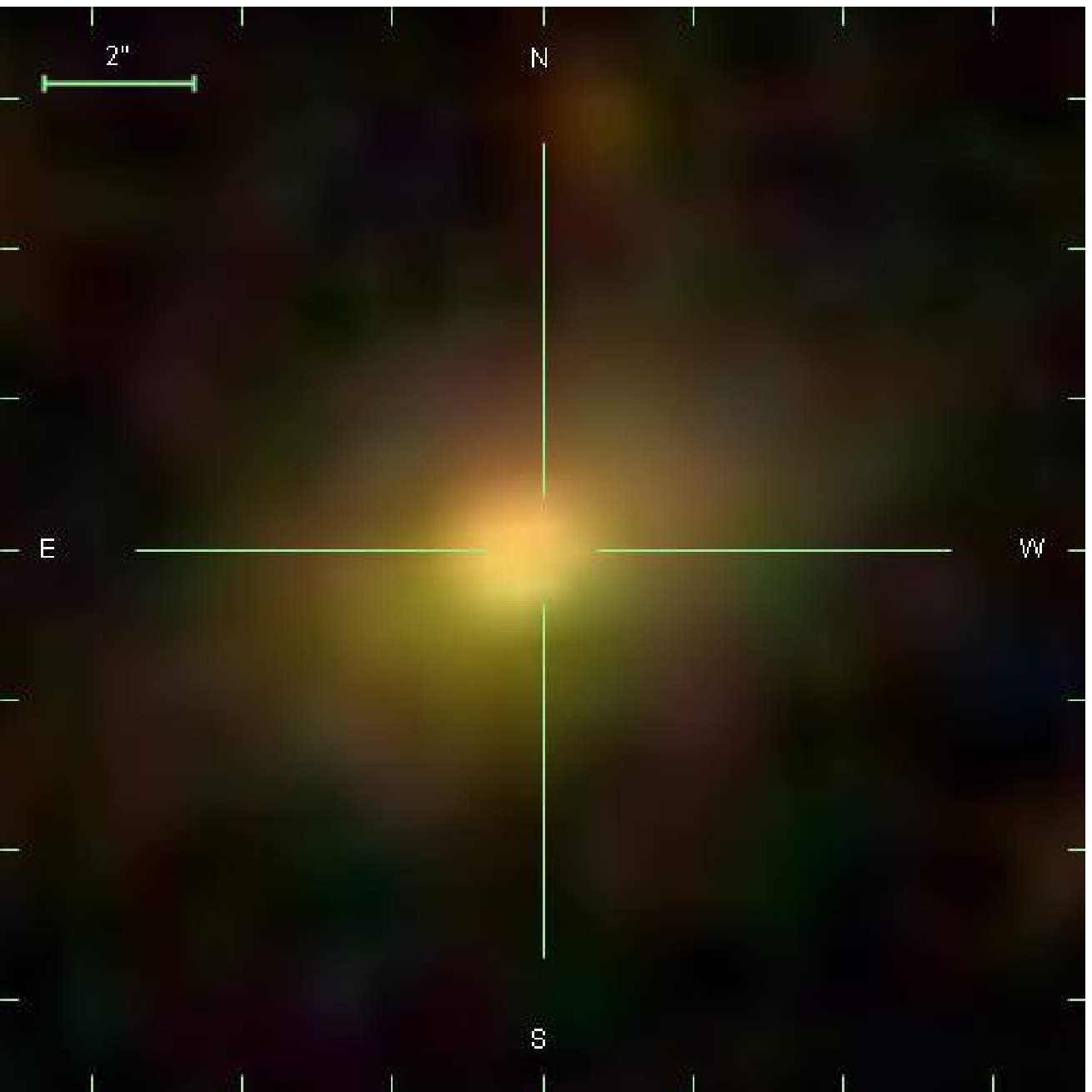}
\includegraphics[width=0.12\textwidth]{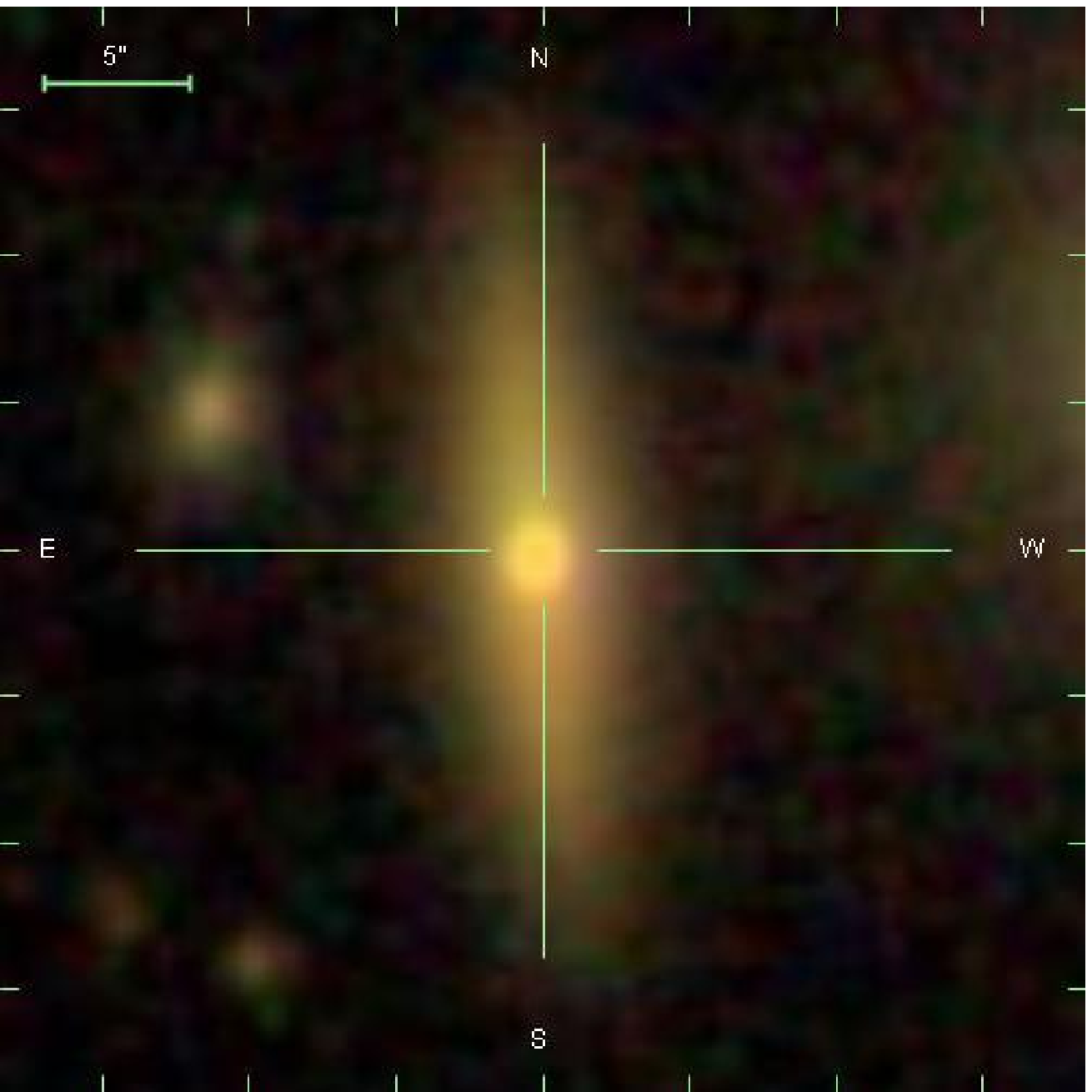}
\includegraphics[width=0.12\textwidth]{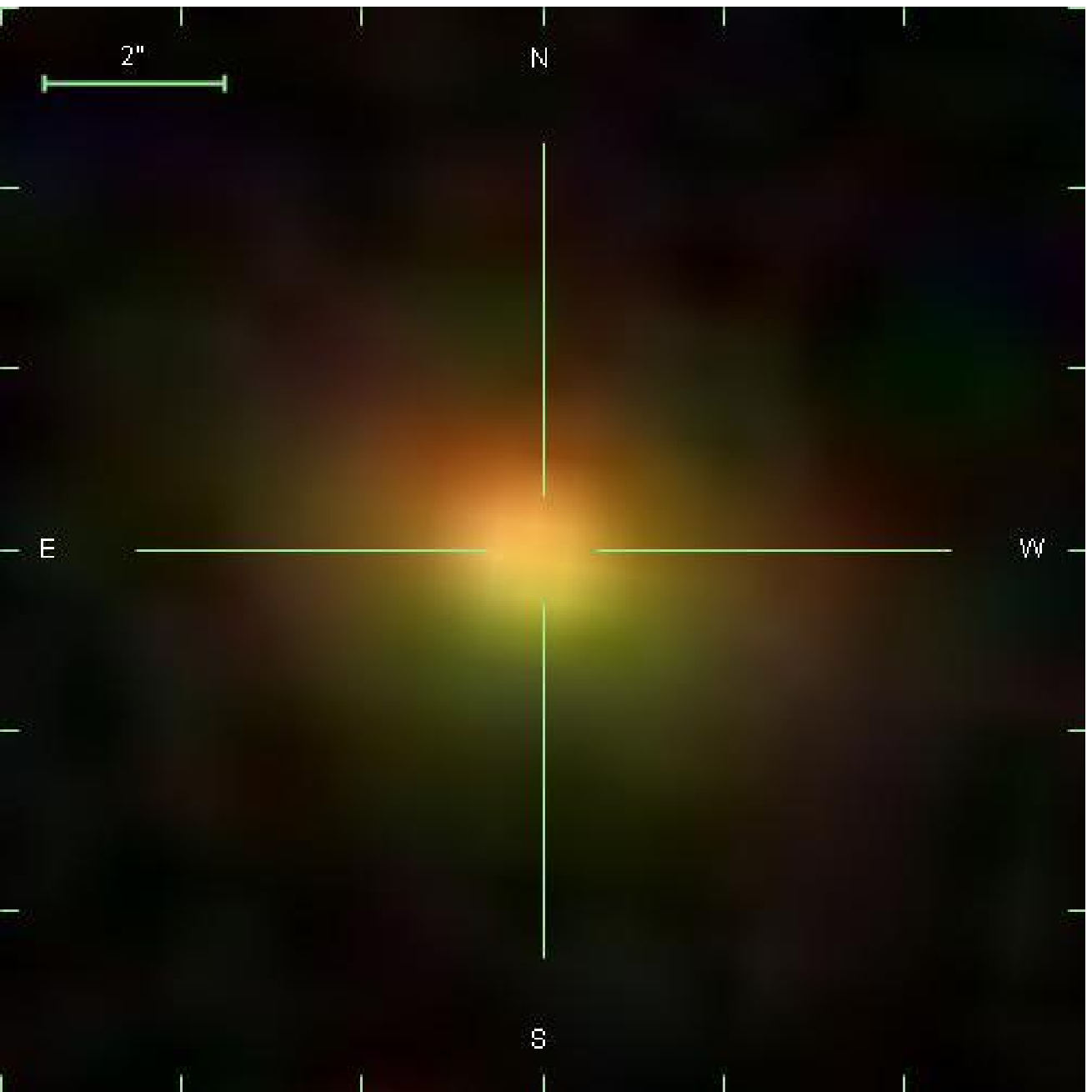}
\includegraphics[width=0.12\textwidth]{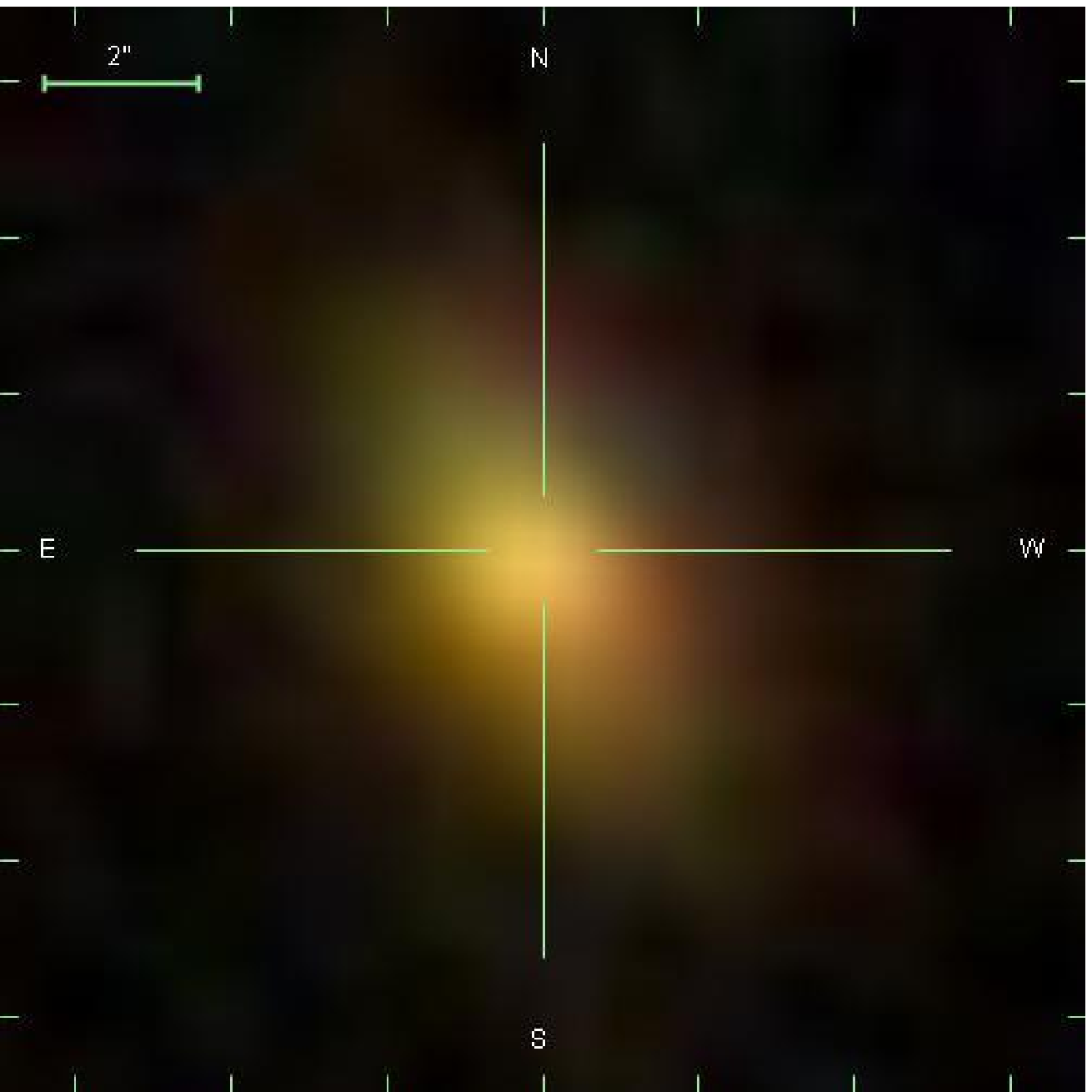}
\includegraphics[width=0.12\textwidth]{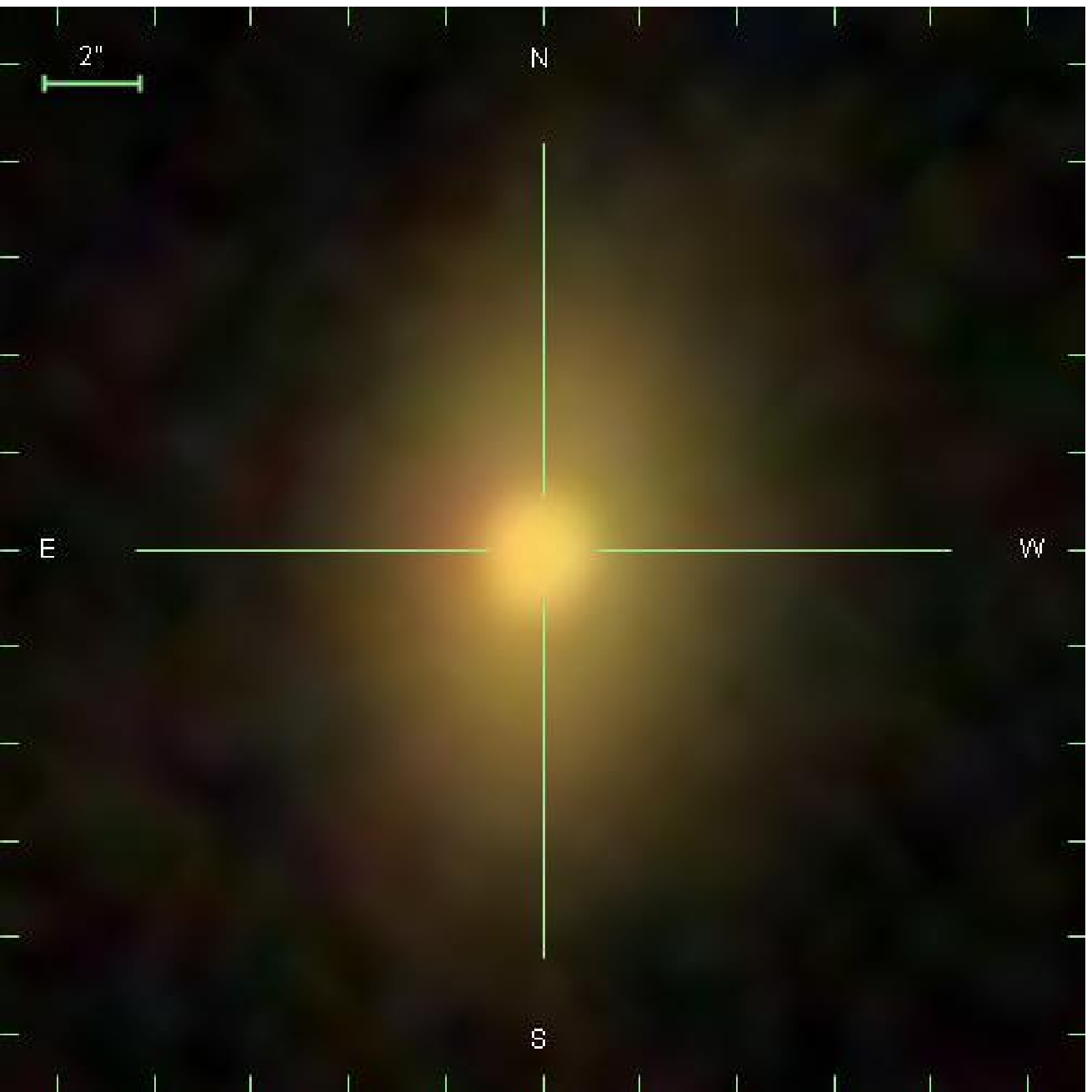}
\includegraphics[width=0.12\textwidth]{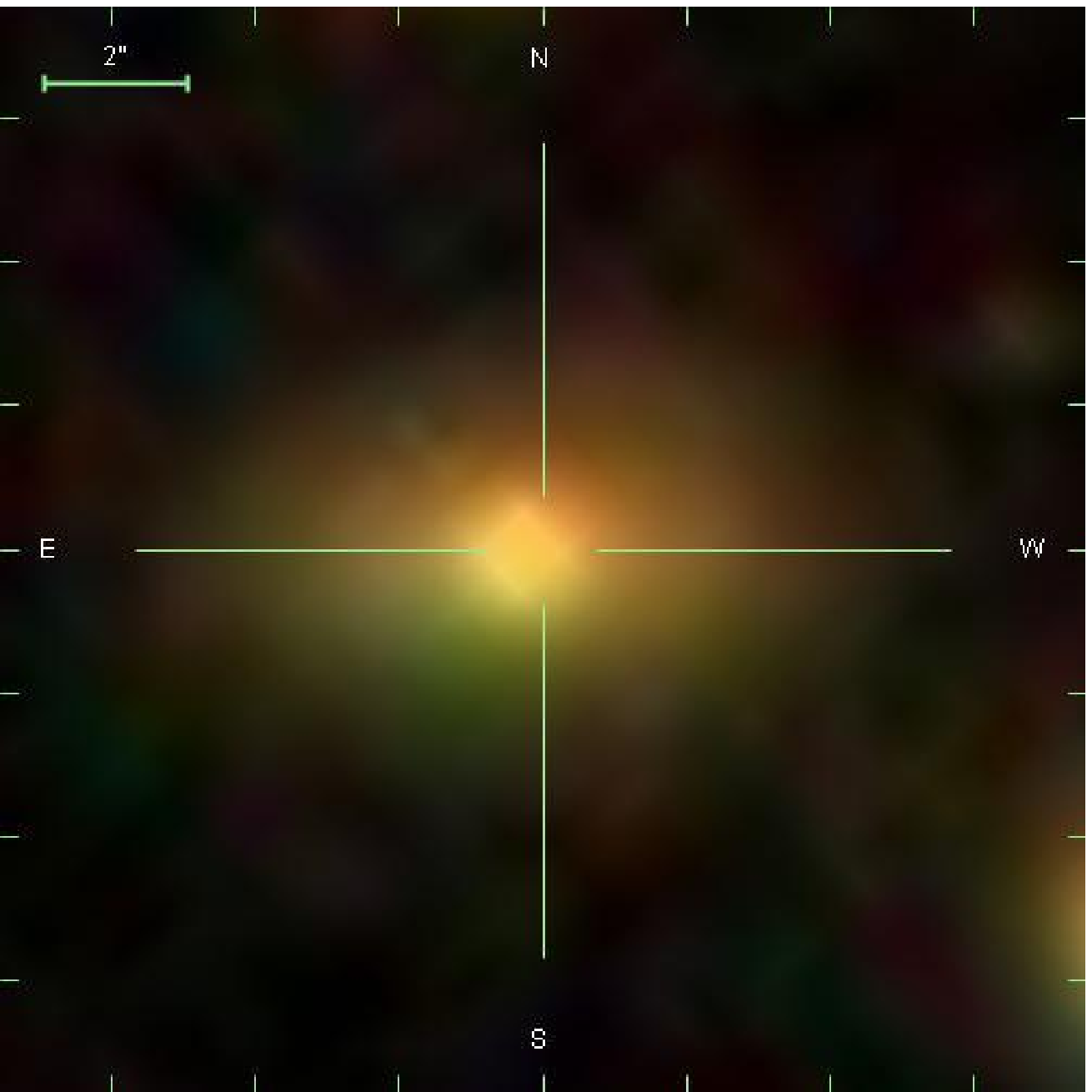}
\includegraphics[width=0.12\textwidth]{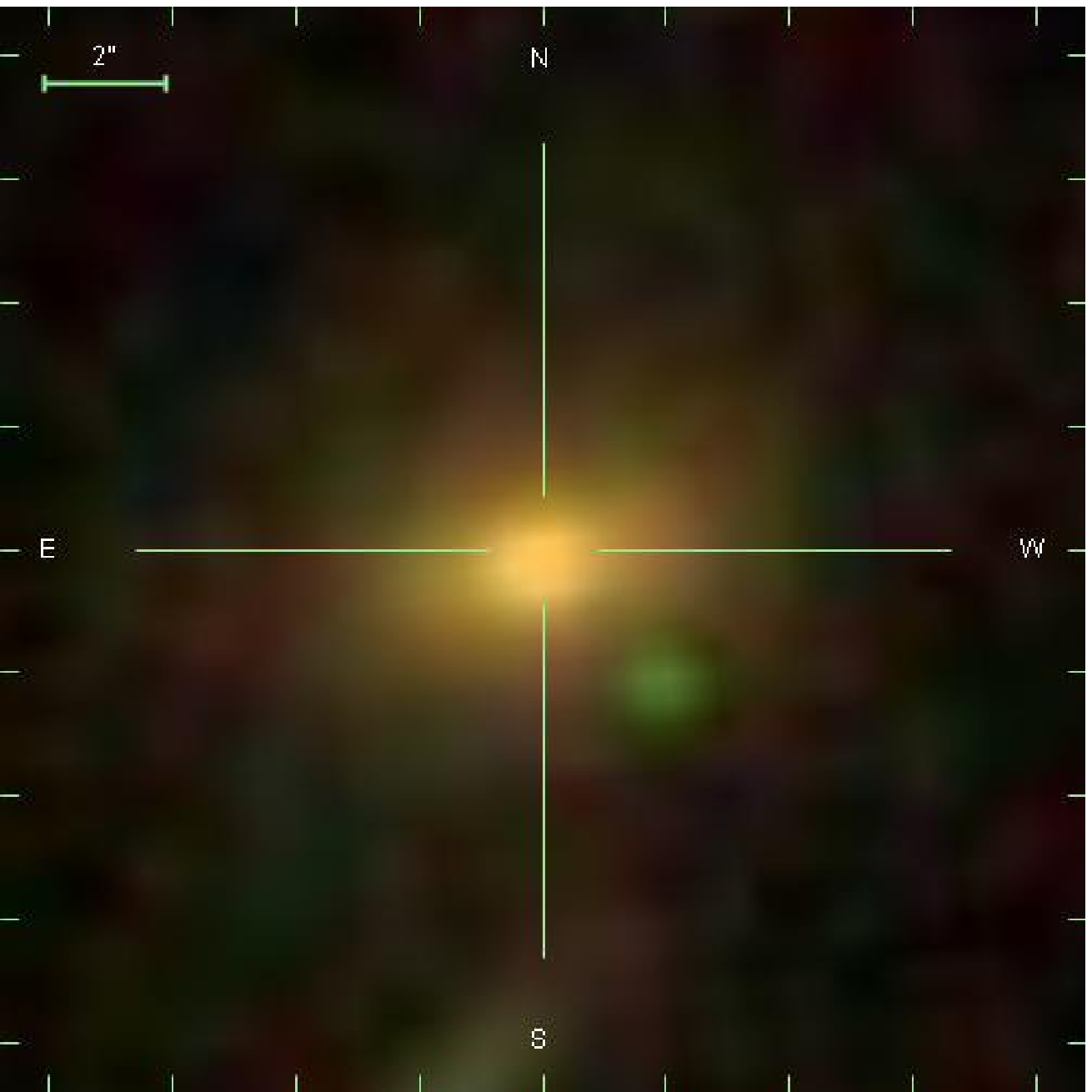}
\includegraphics[width=0.12\textwidth]{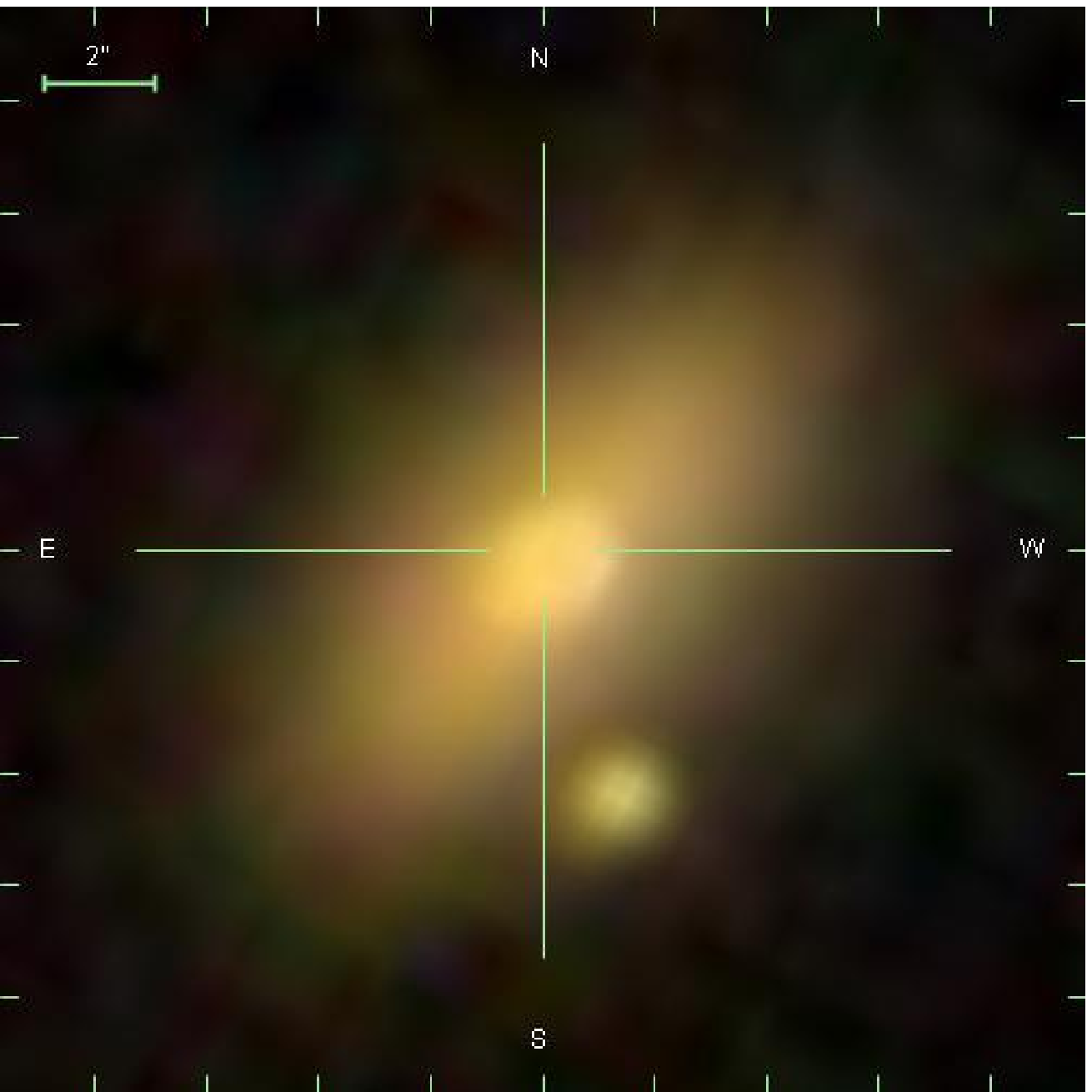}\\
\includegraphics[width=0.12\textwidth]{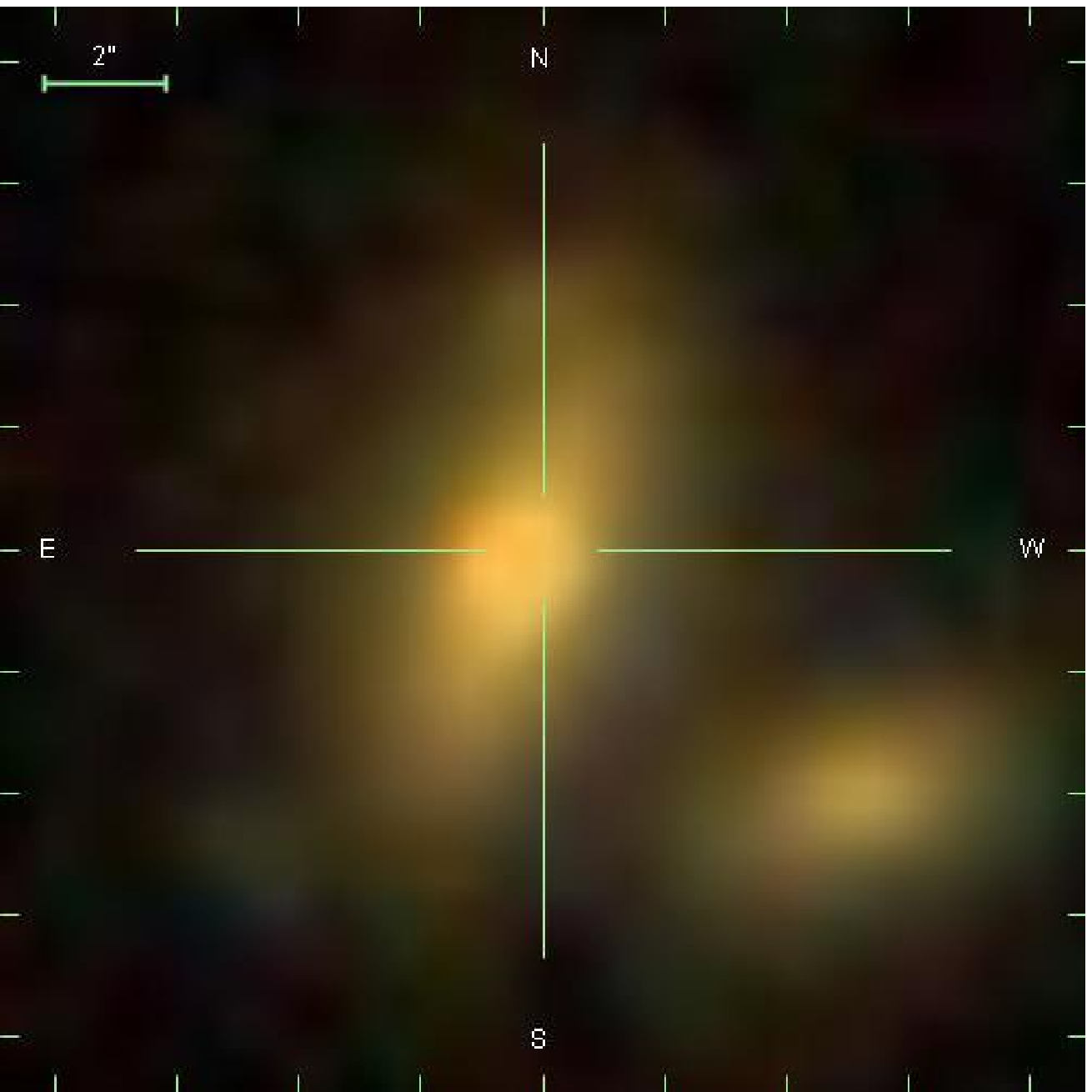}
\includegraphics[width=0.12\textwidth]{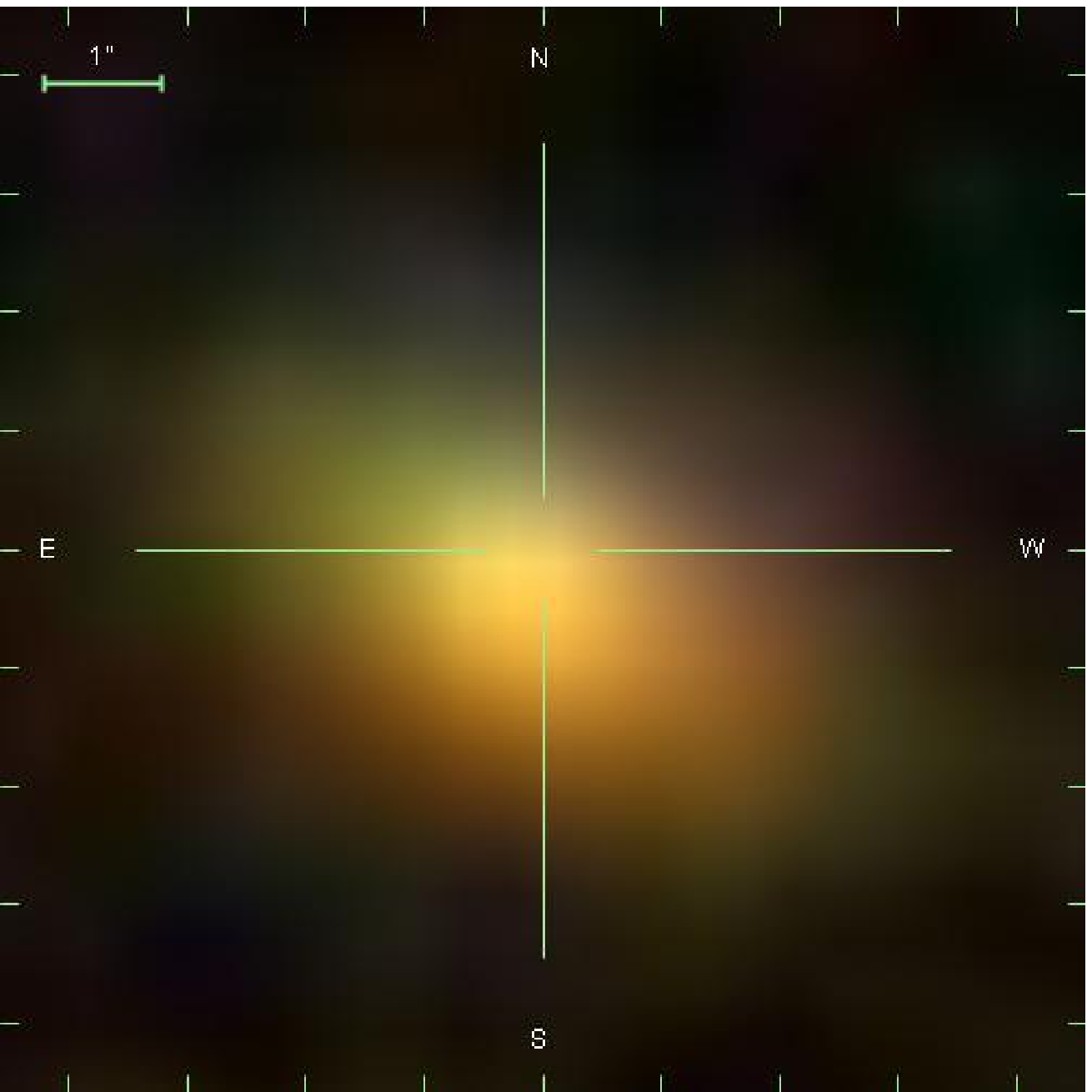}
\includegraphics[width=0.12\textwidth]{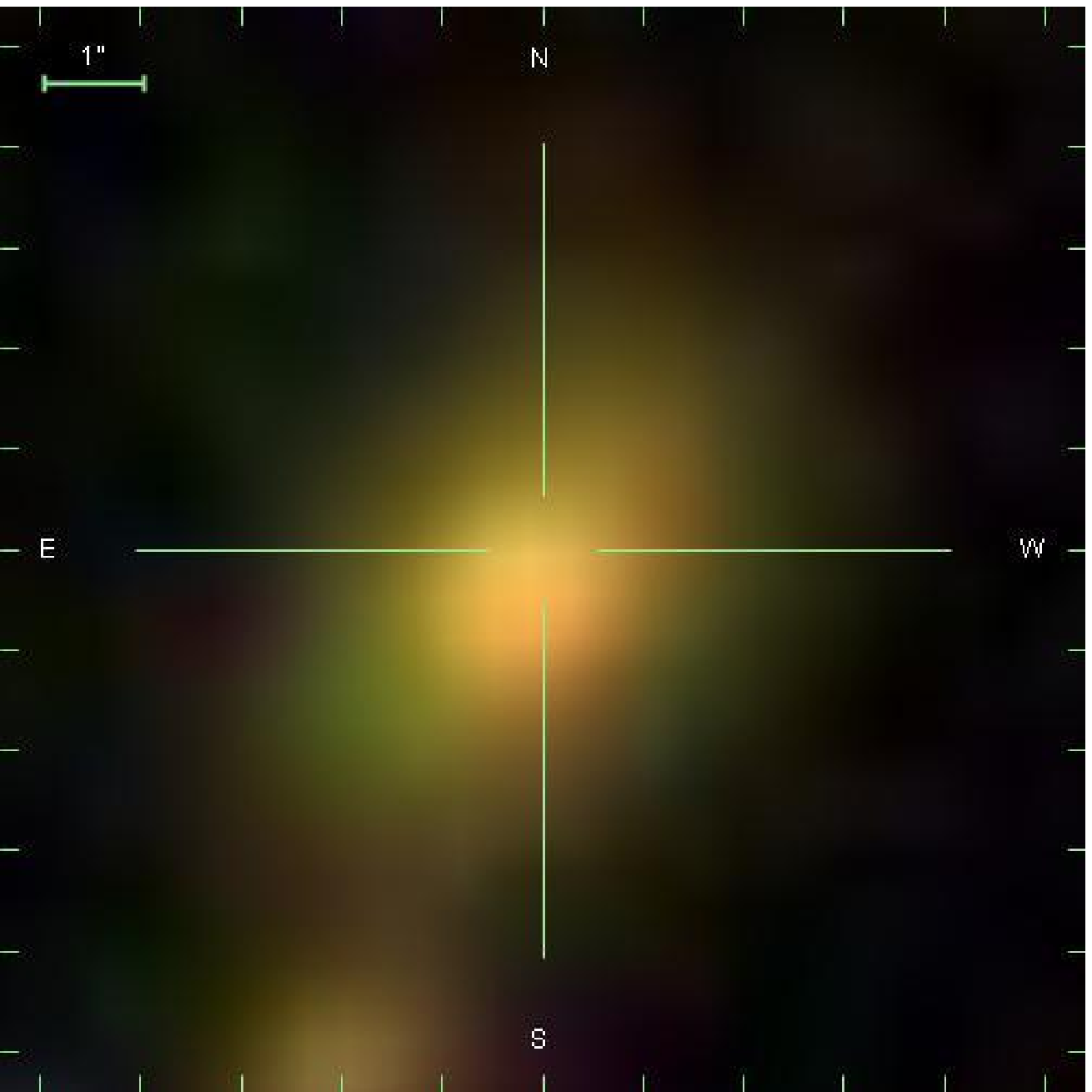}
\includegraphics[width=0.12\textwidth]{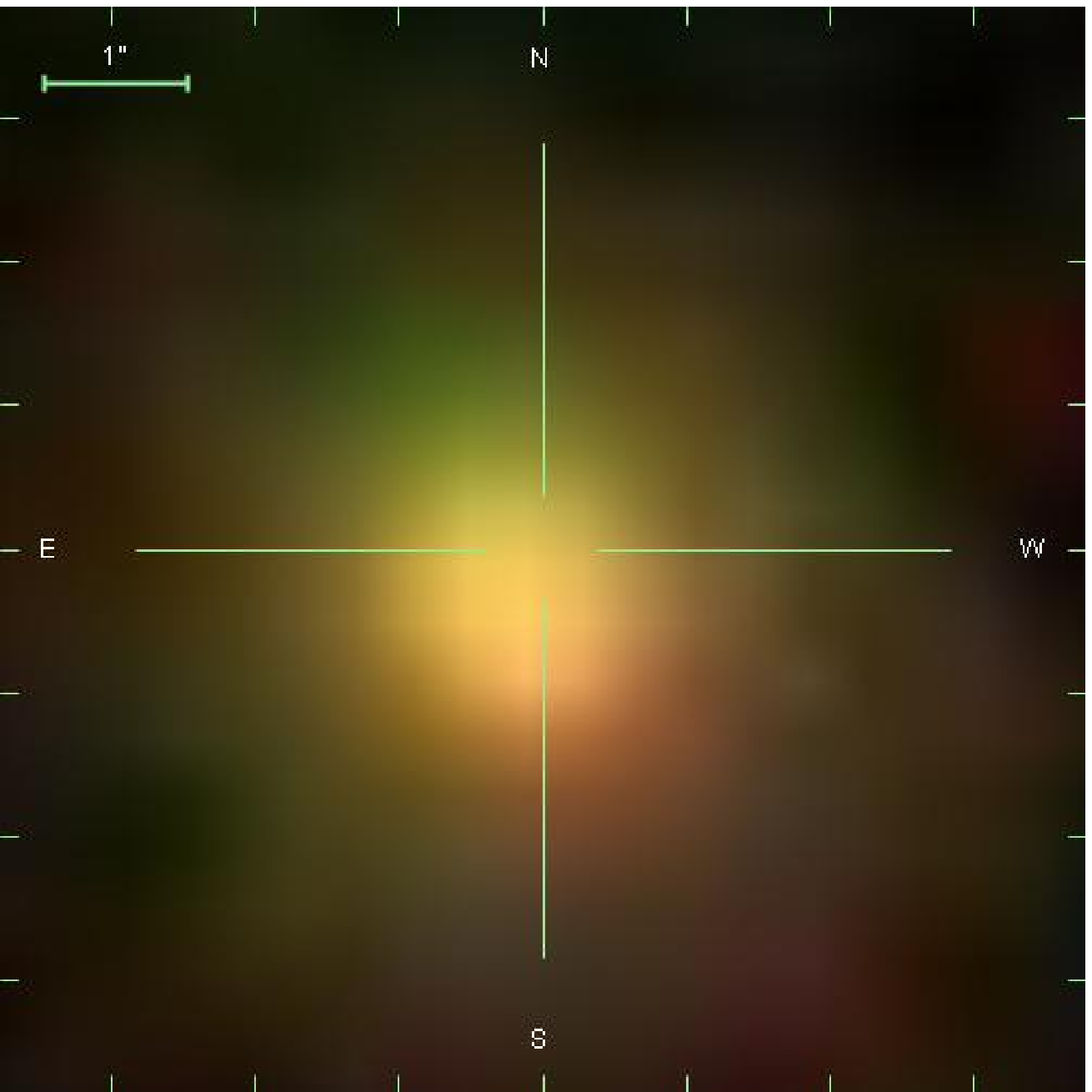}
\includegraphics[width=0.12\textwidth]{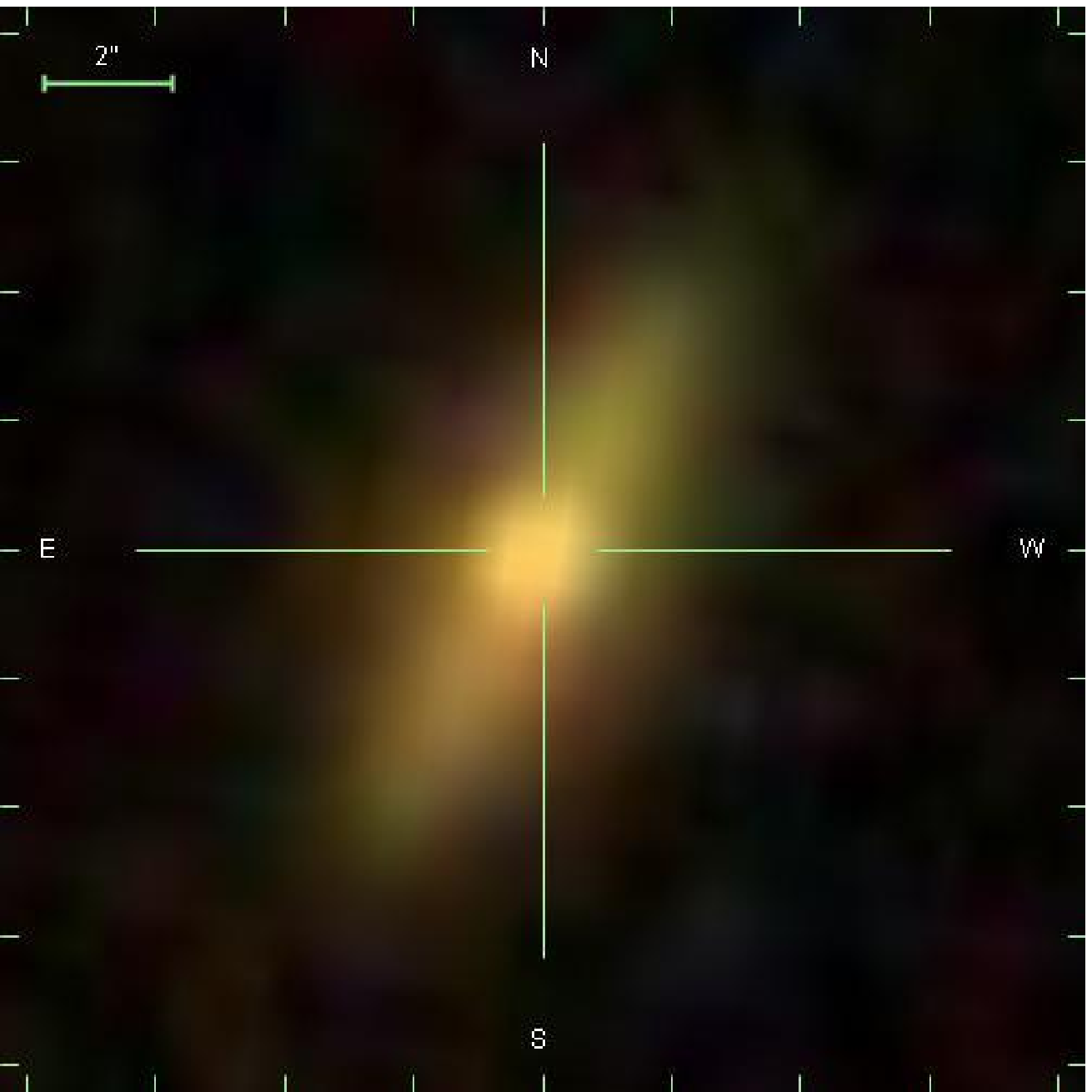}
\includegraphics[width=0.12\textwidth]{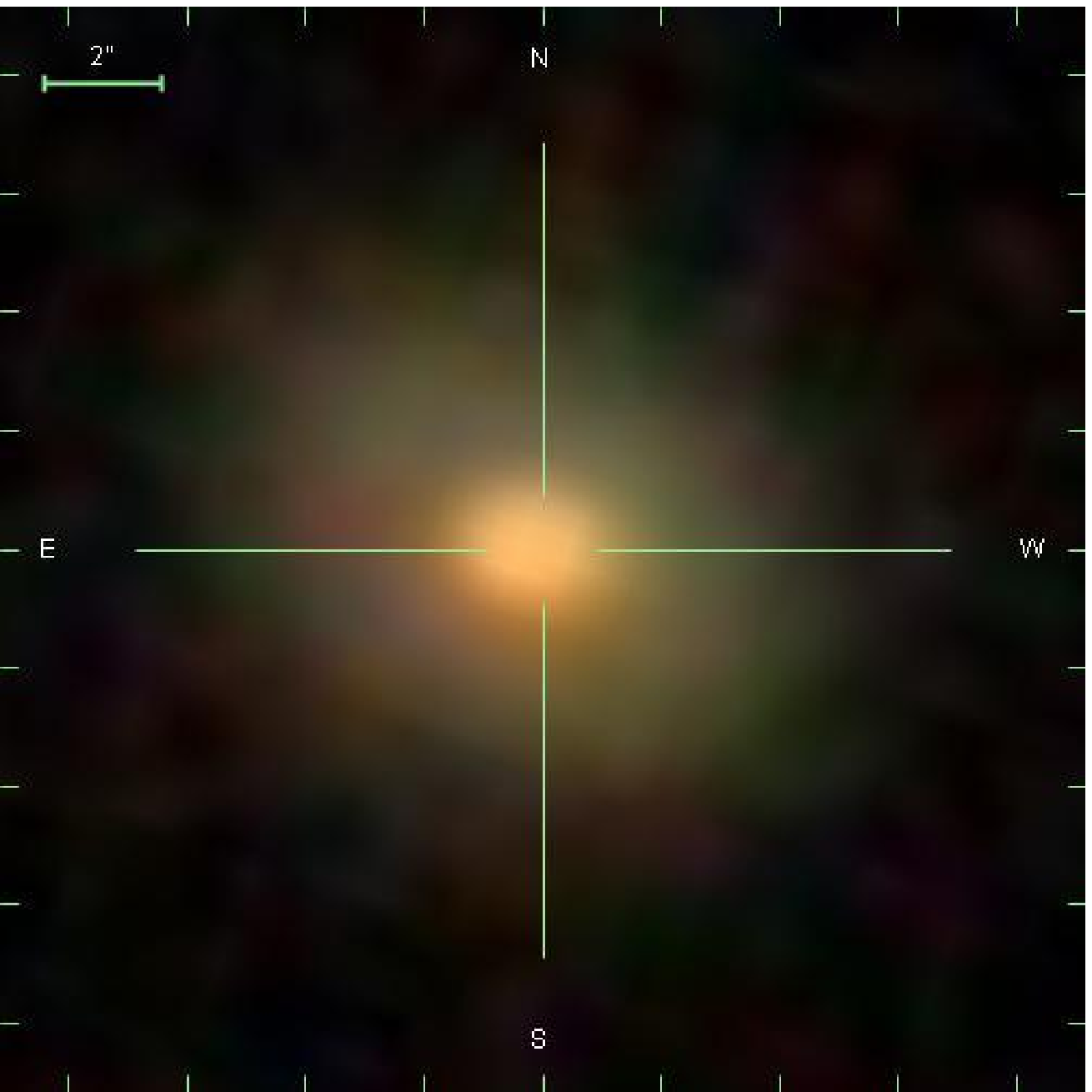}
\includegraphics[width=0.12\textwidth]{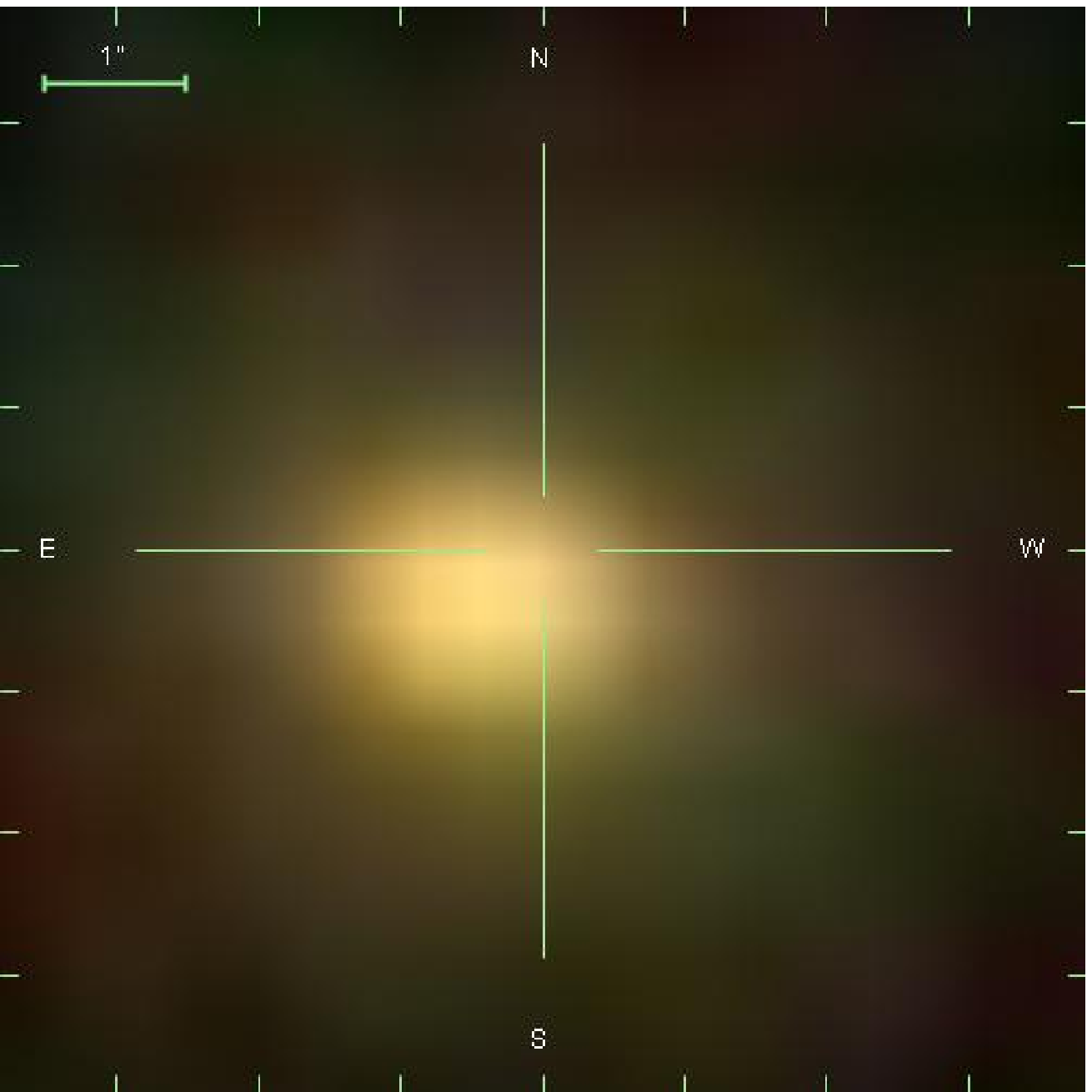}
\includegraphics[width=0.12\textwidth]{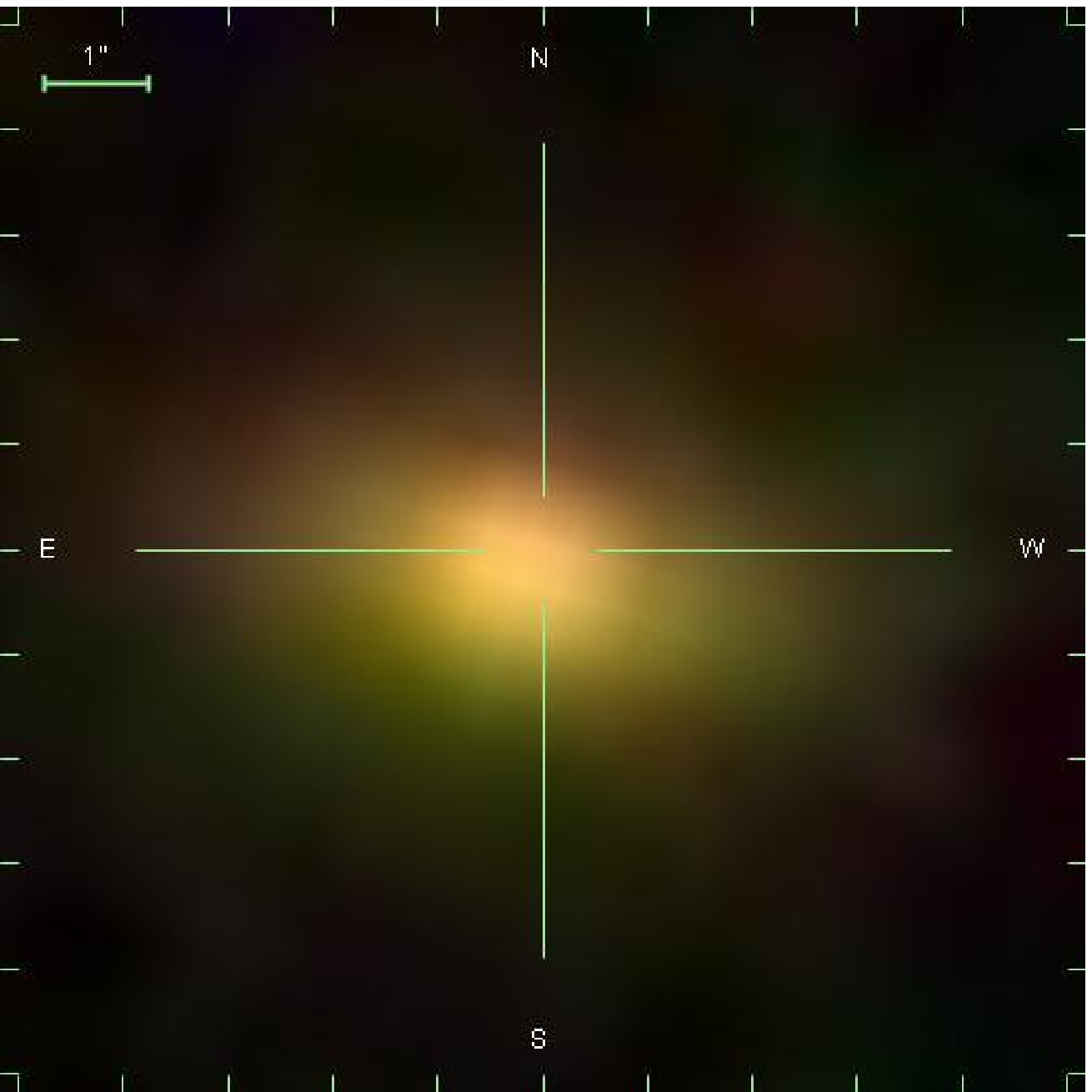}\\
\includegraphics[width=0.12\textwidth]{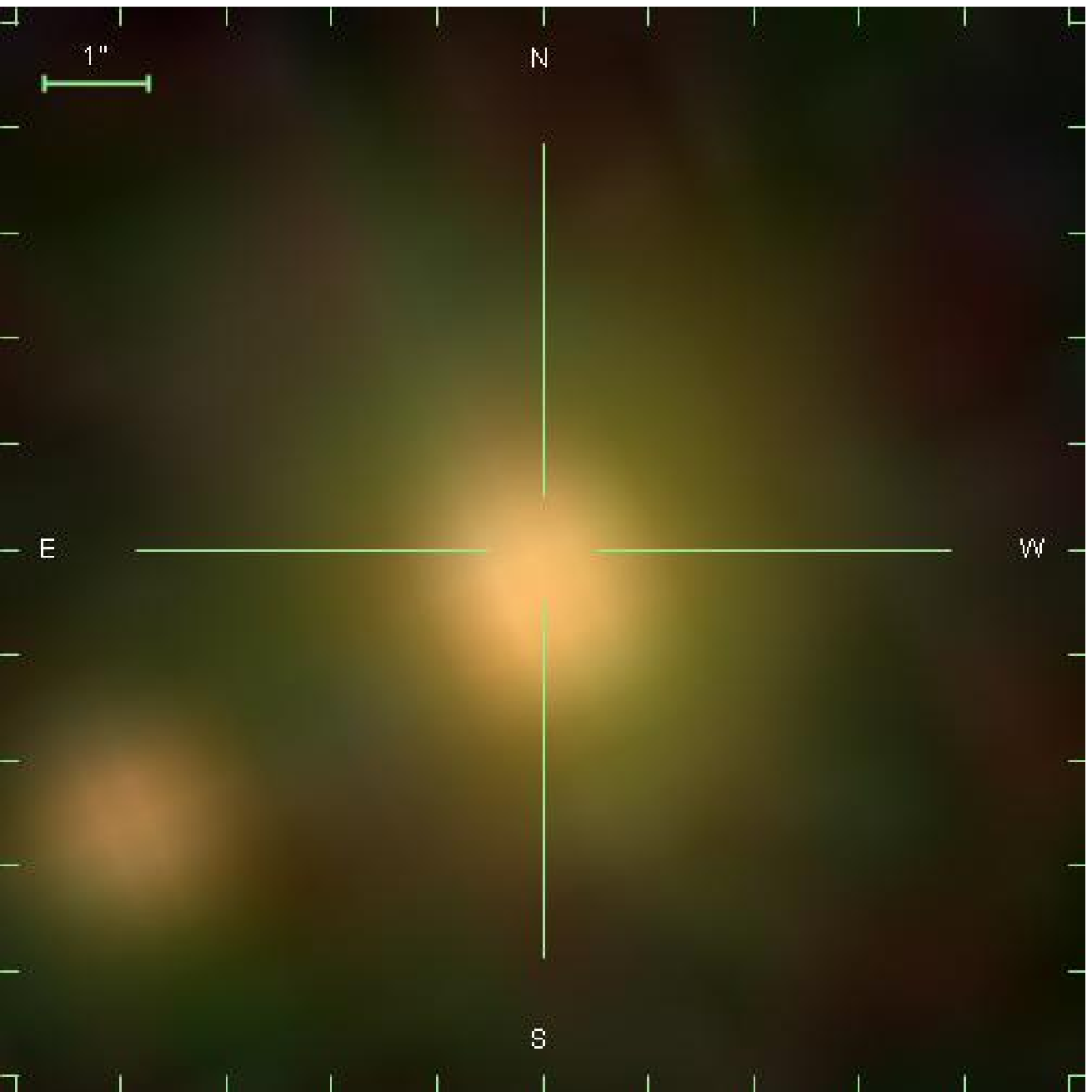}
\includegraphics[width=0.12\textwidth]{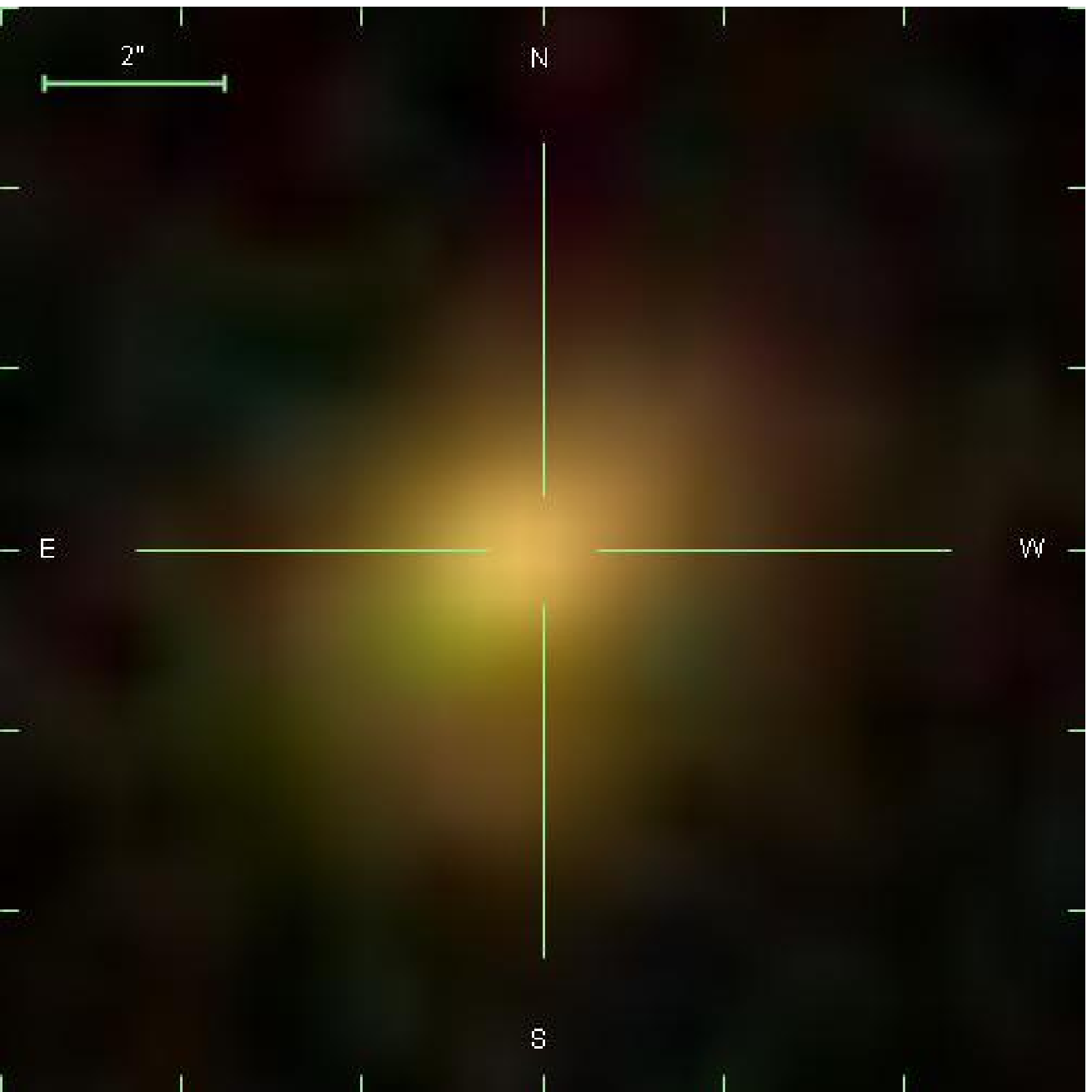}
\includegraphics[width=0.12\textwidth]{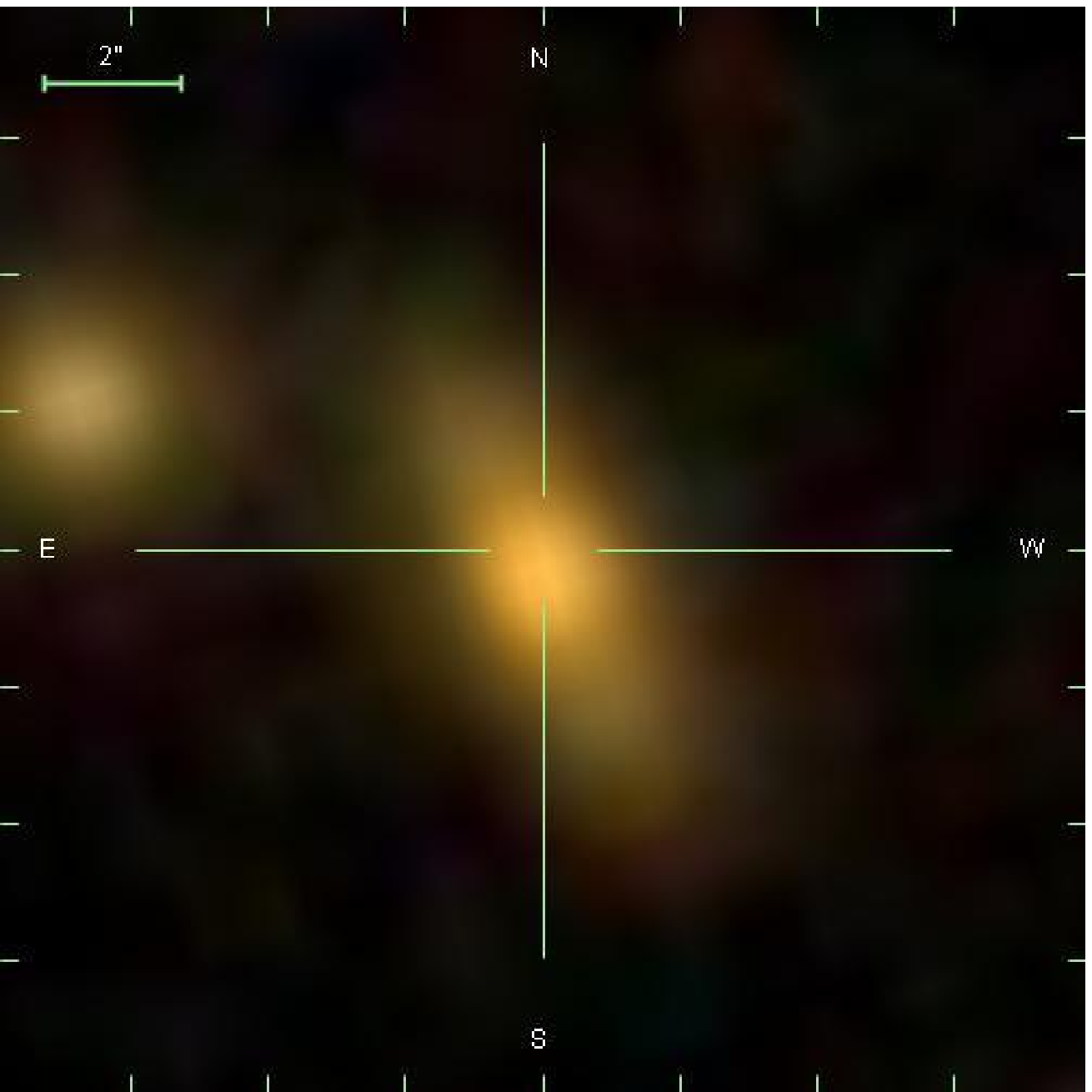}
\includegraphics[width=0.12\textwidth]{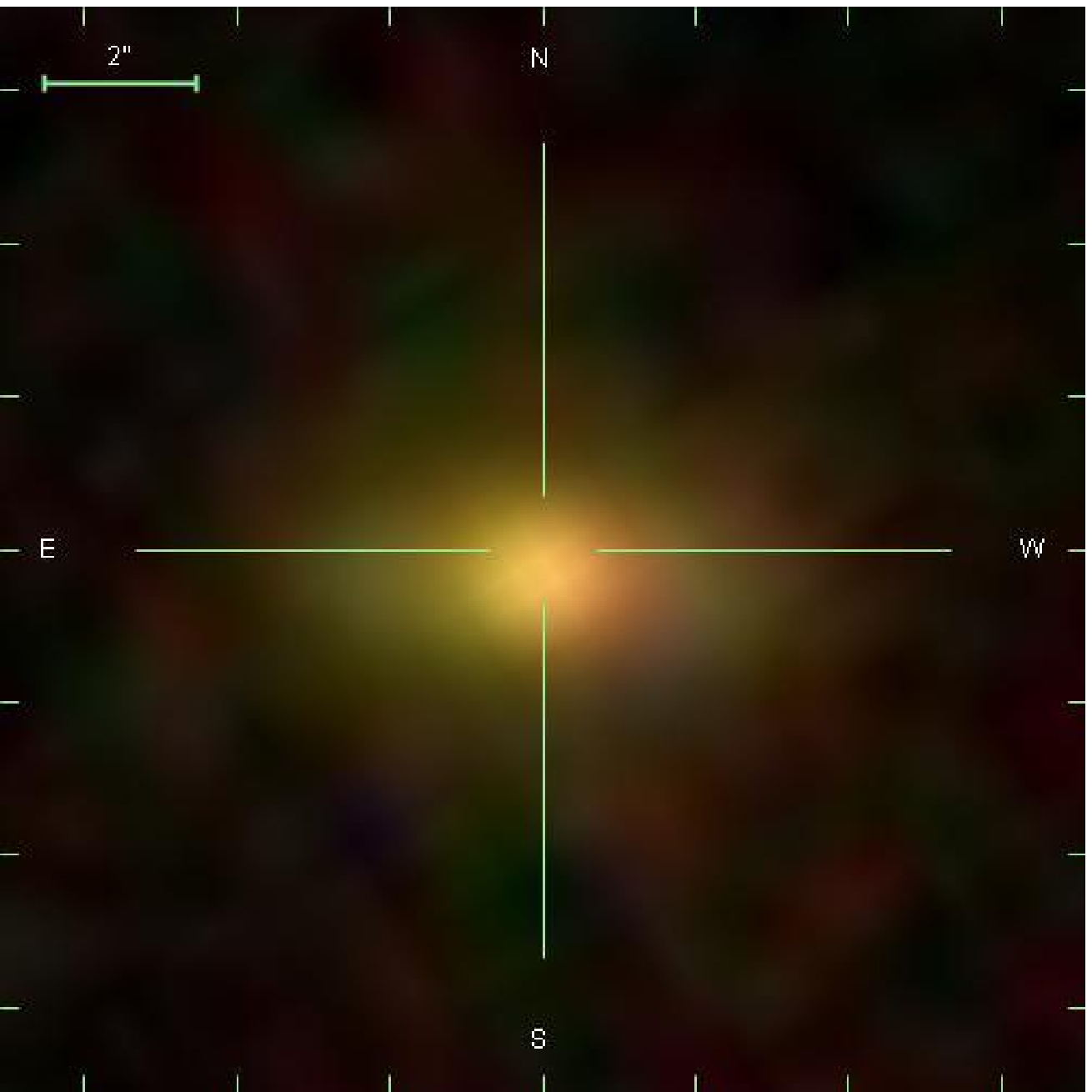}
\includegraphics[width=0.12\textwidth]{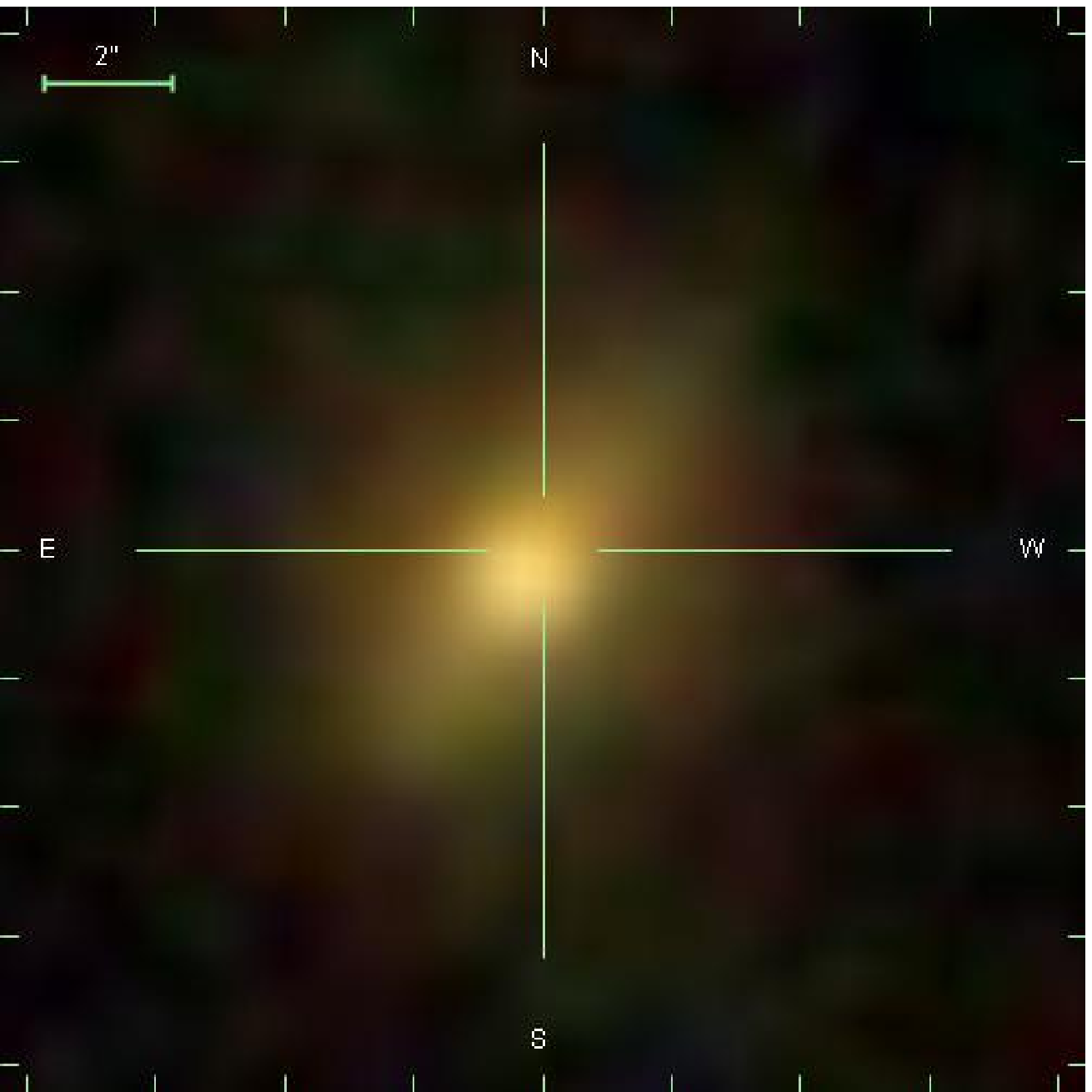}
\includegraphics[width=0.12\textwidth]{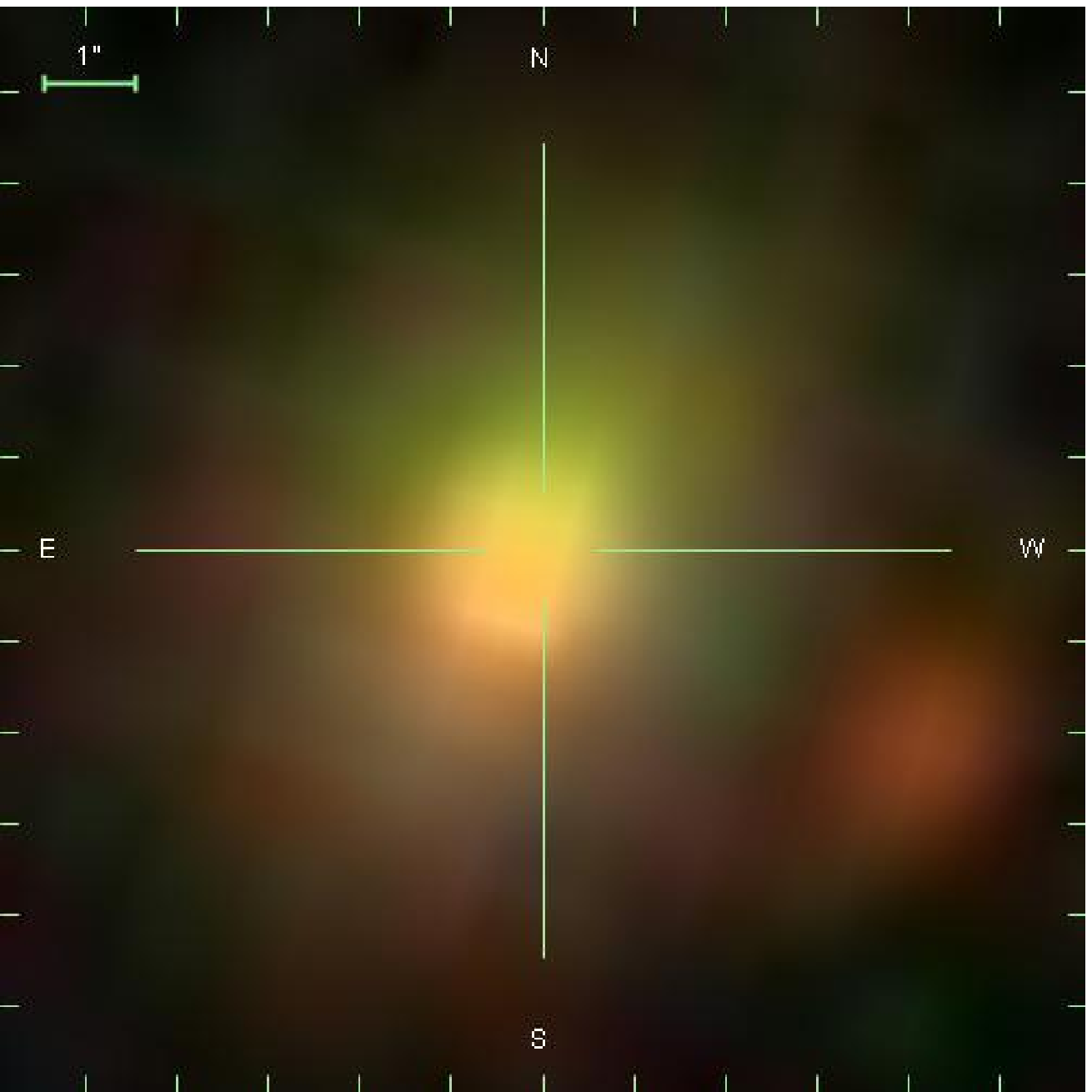}
\includegraphics[width=0.12\textwidth]{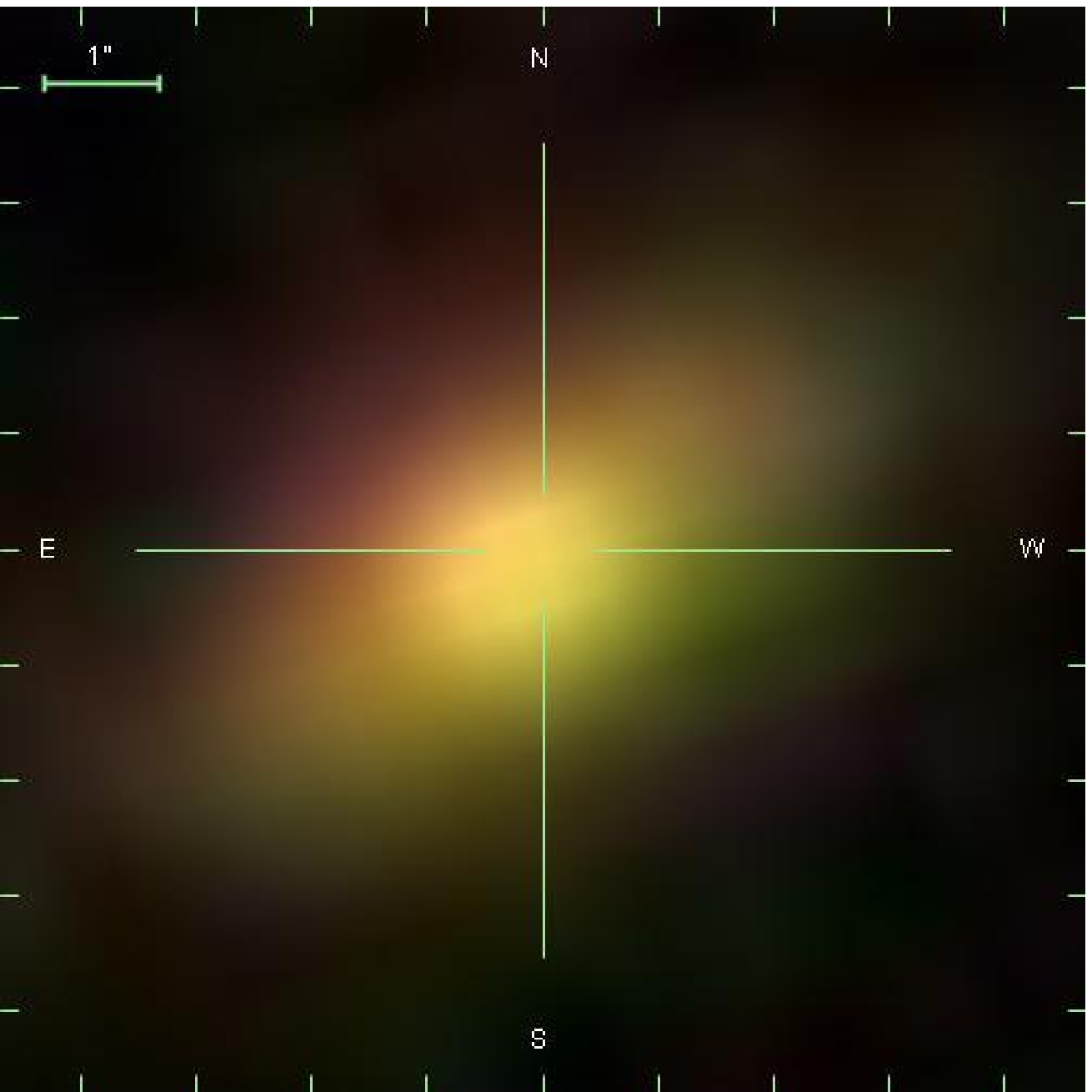}
\includegraphics[width=0.12\textwidth]{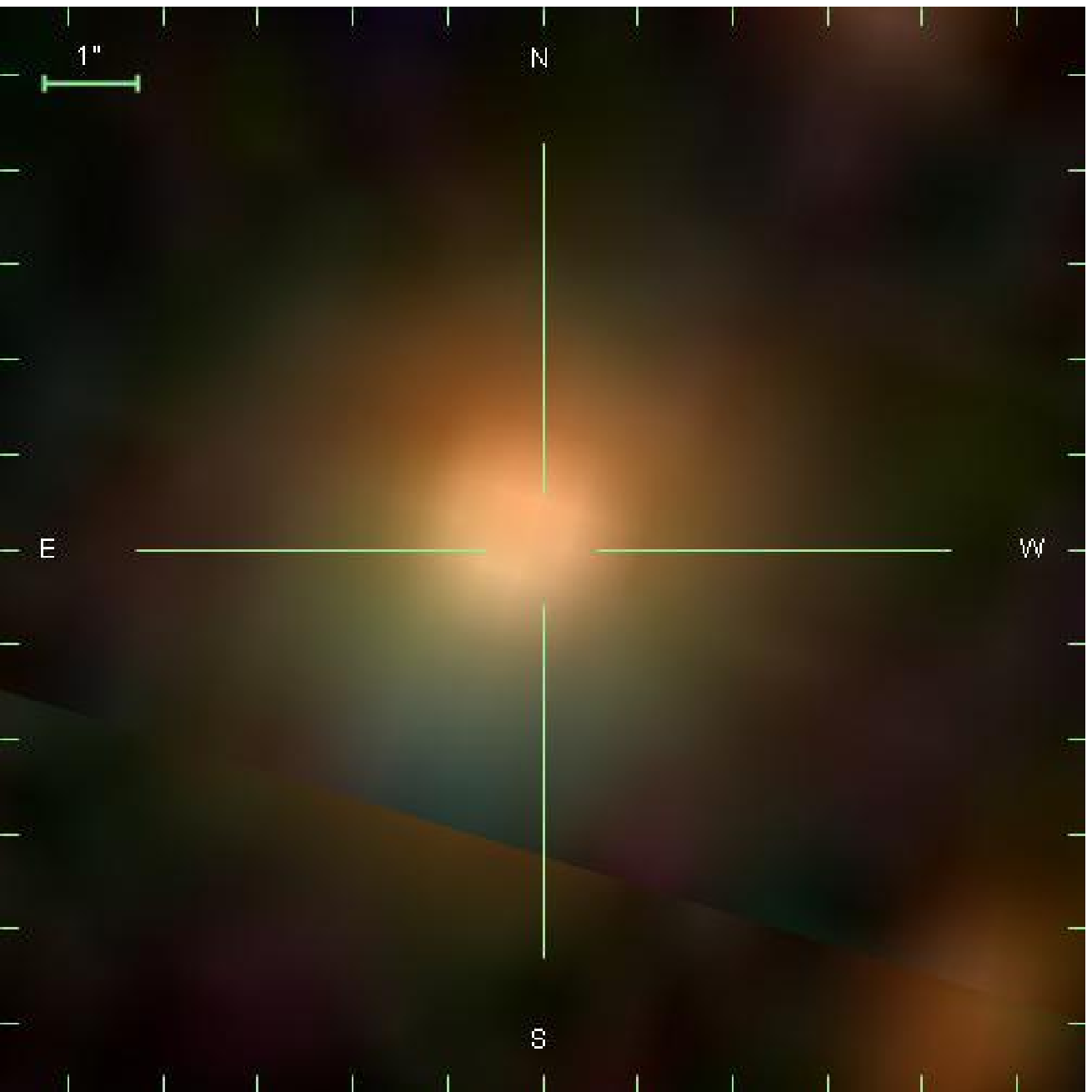}\\
\includegraphics[width=0.12\textwidth]{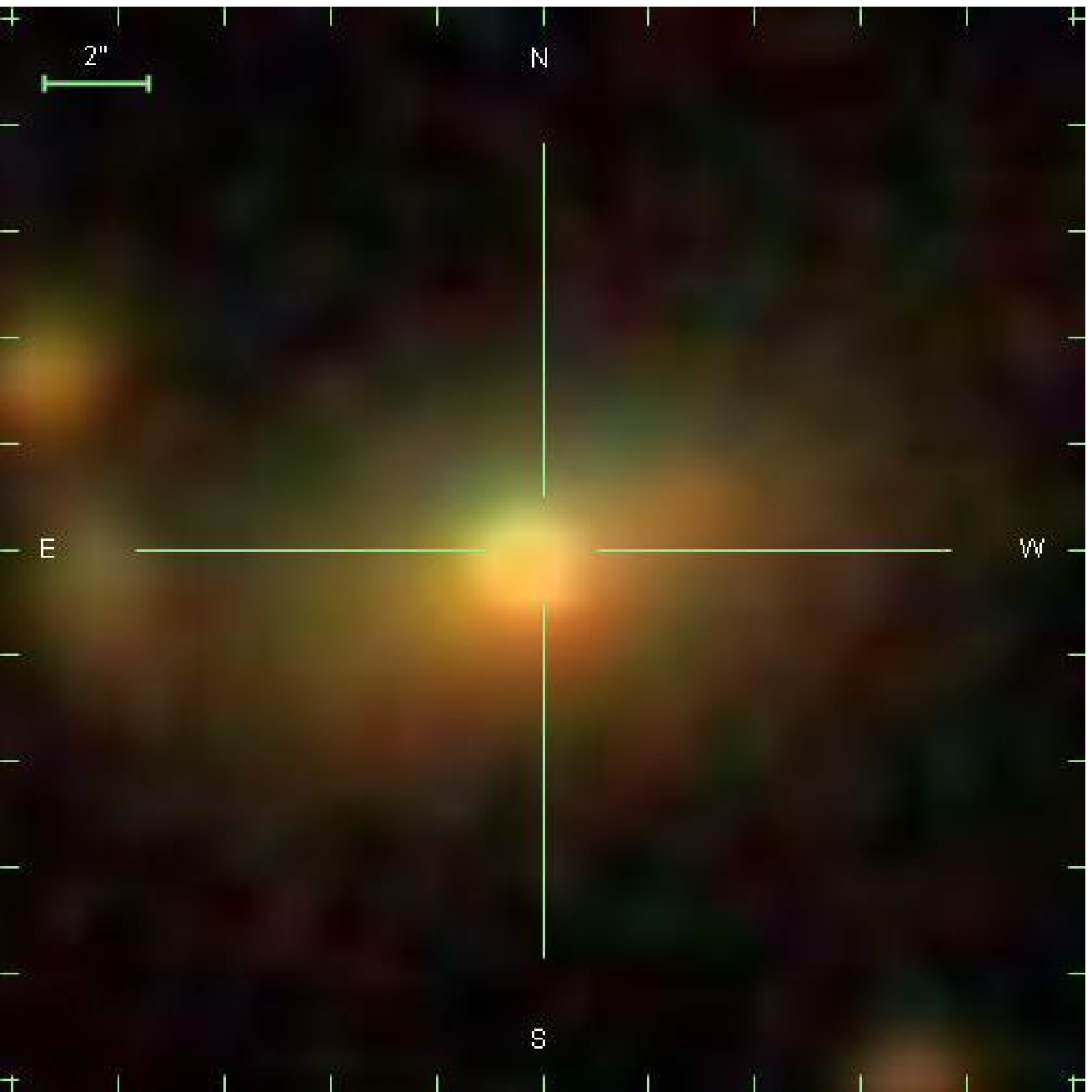}
\includegraphics[width=0.12\textwidth]{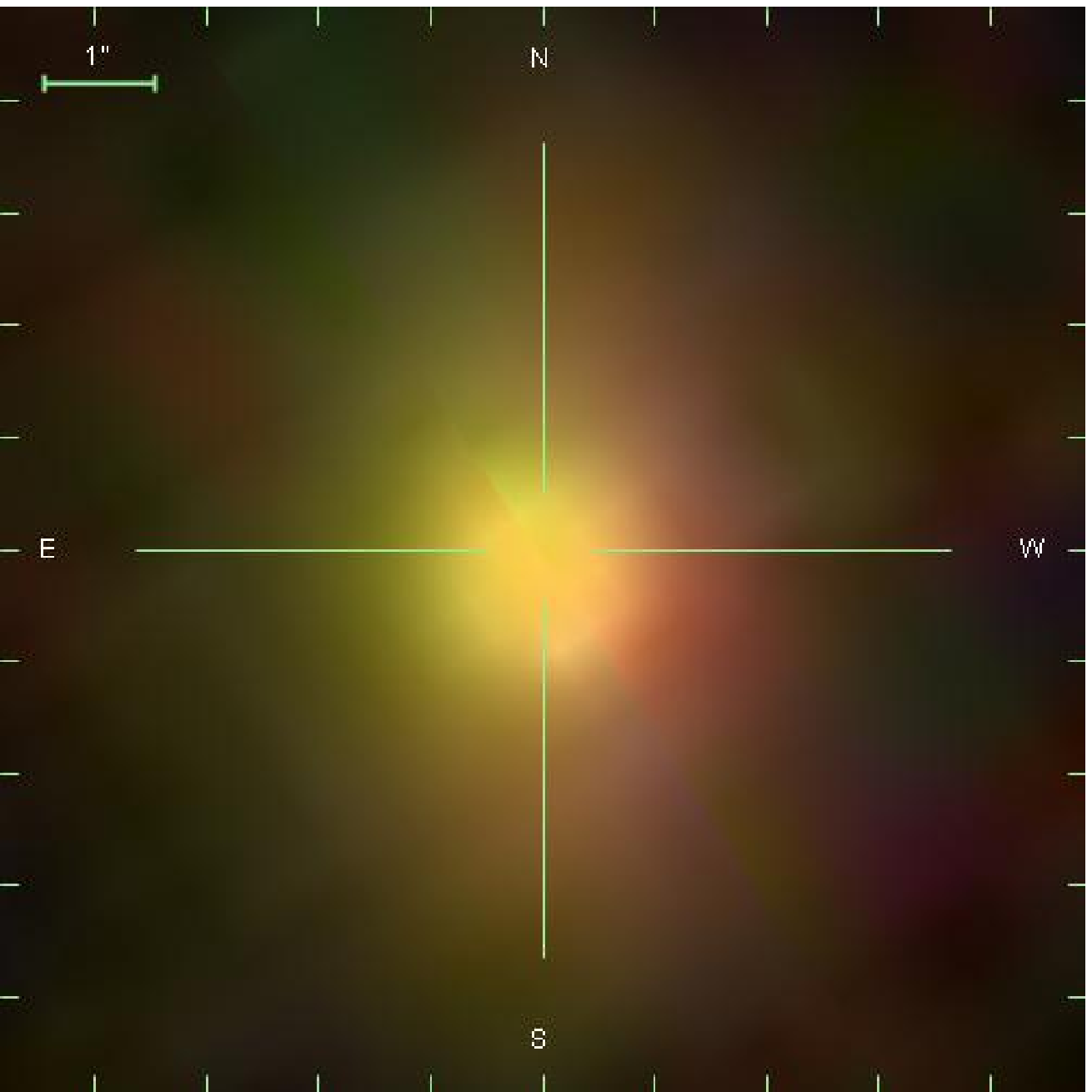}
\includegraphics[width=0.12\textwidth]{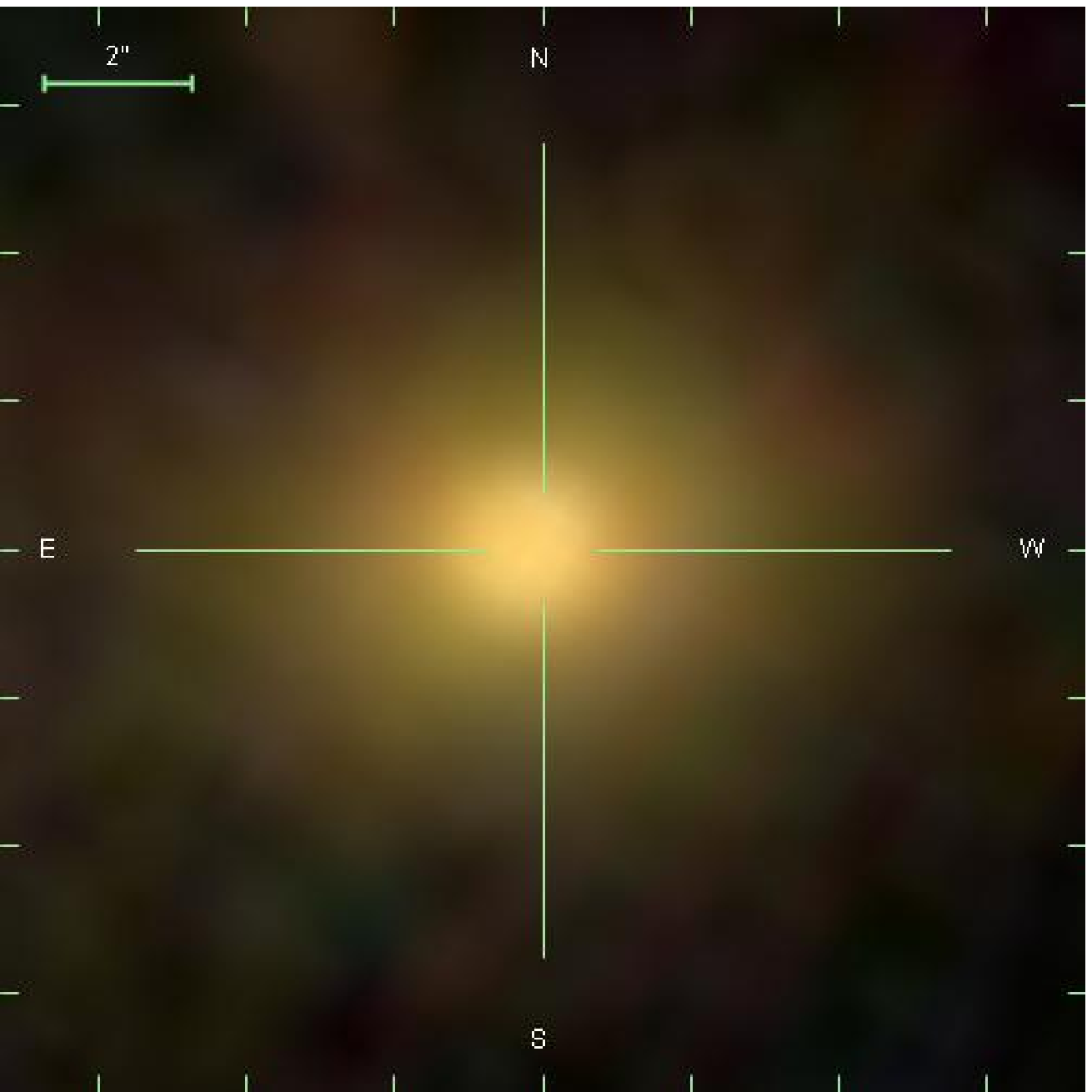}
\includegraphics[width=0.12\textwidth]{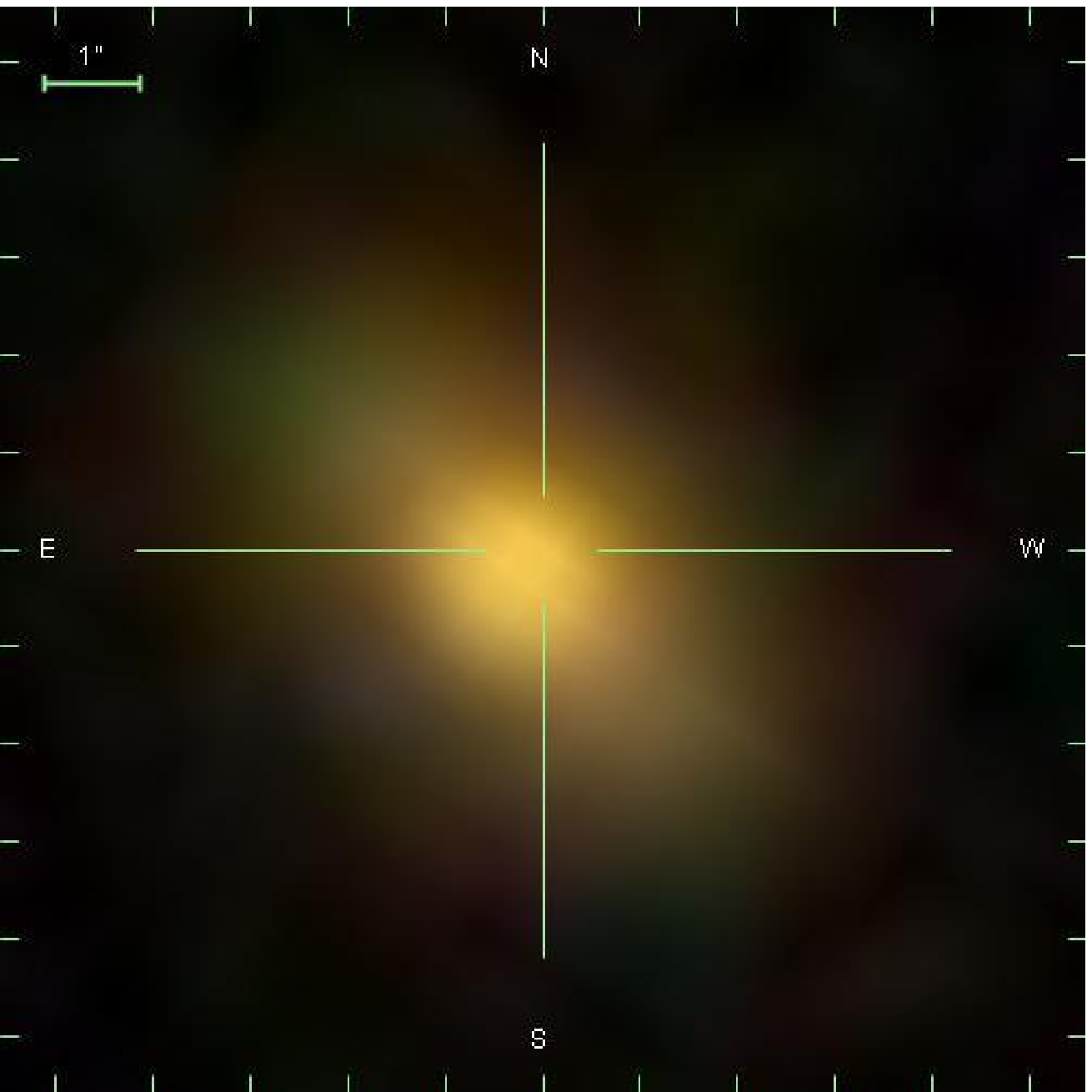}
\includegraphics[width=0.12\textwidth]{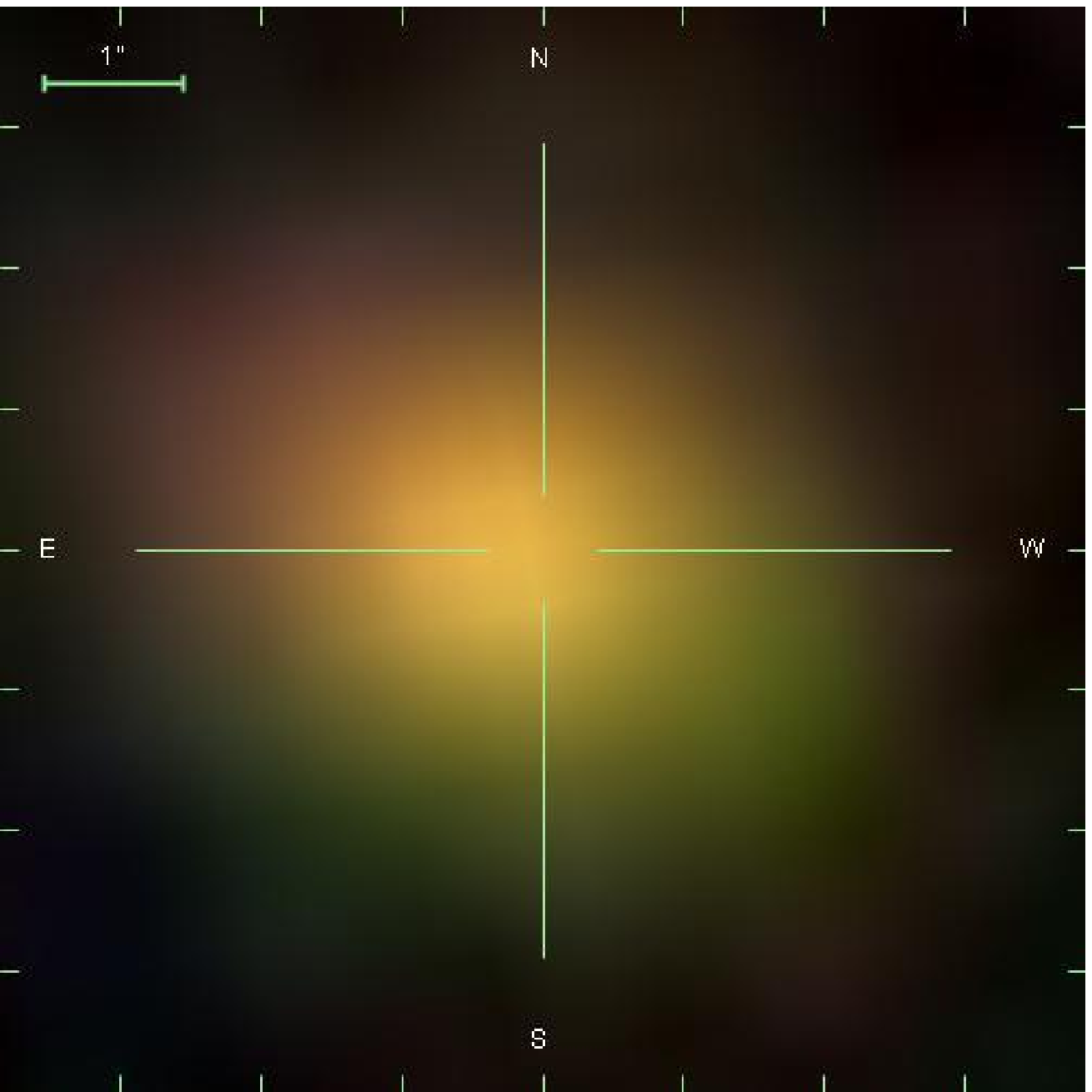}
\includegraphics[width=0.12\textwidth]{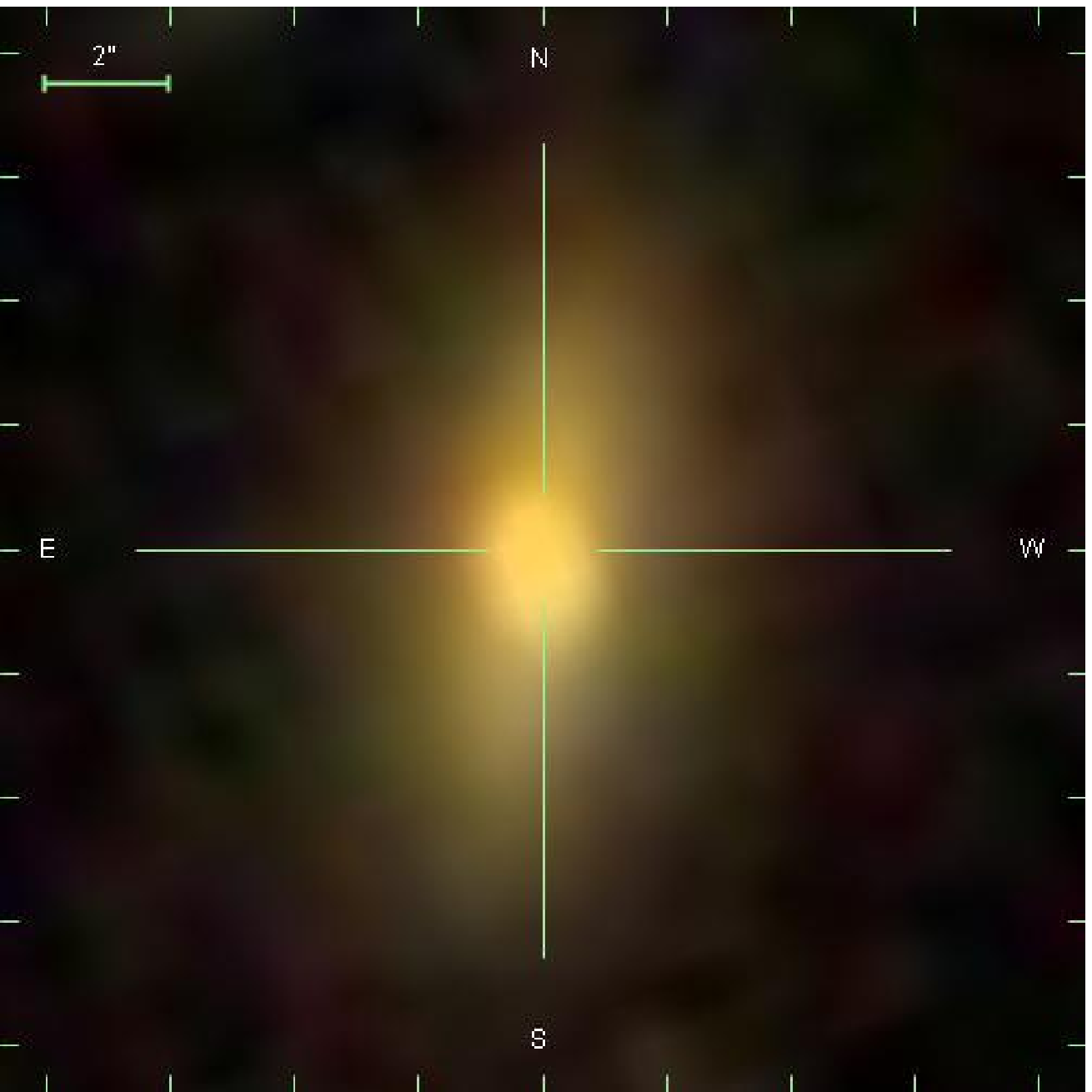}
\includegraphics[width=0.12\textwidth]{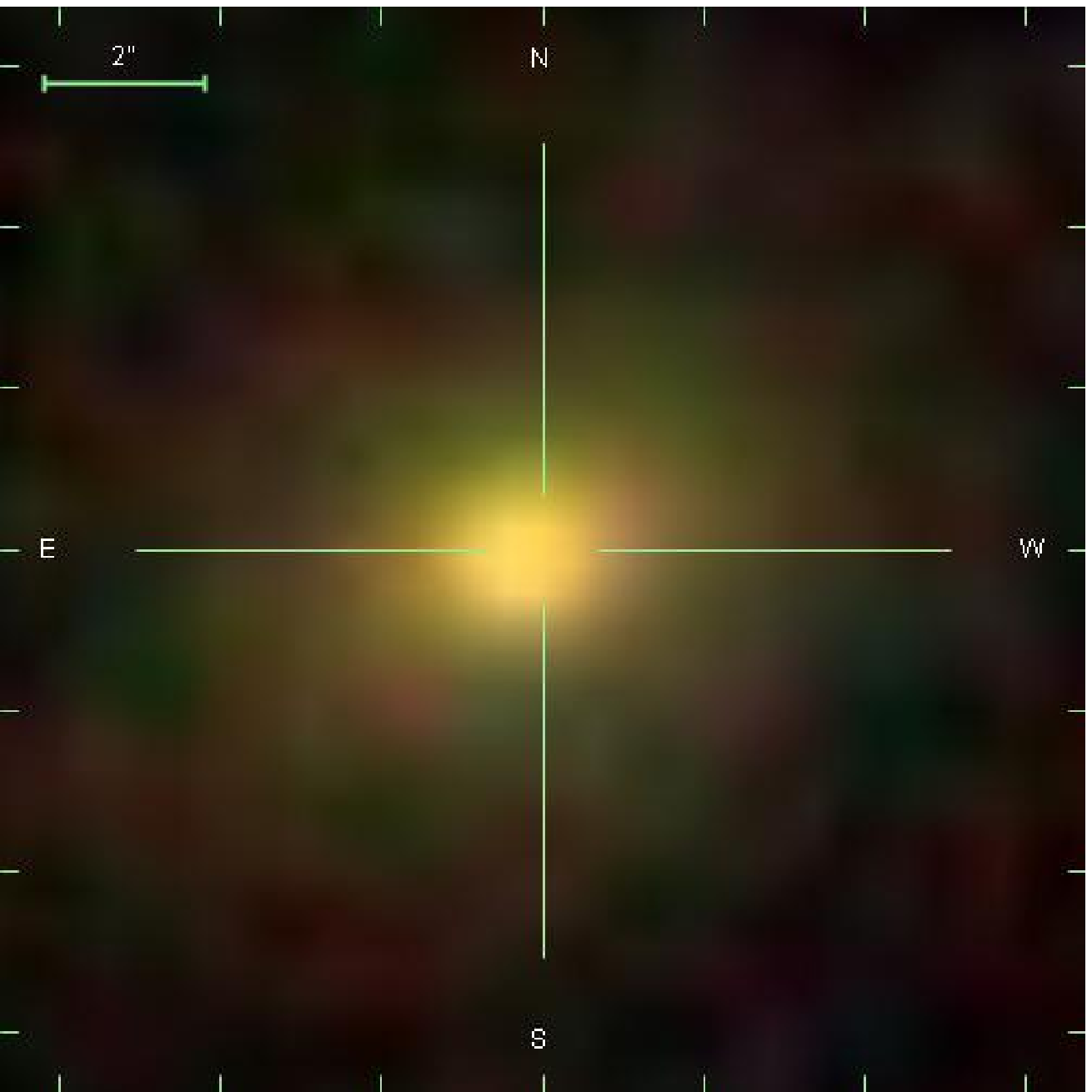}
\includegraphics[width=0.12\textwidth]{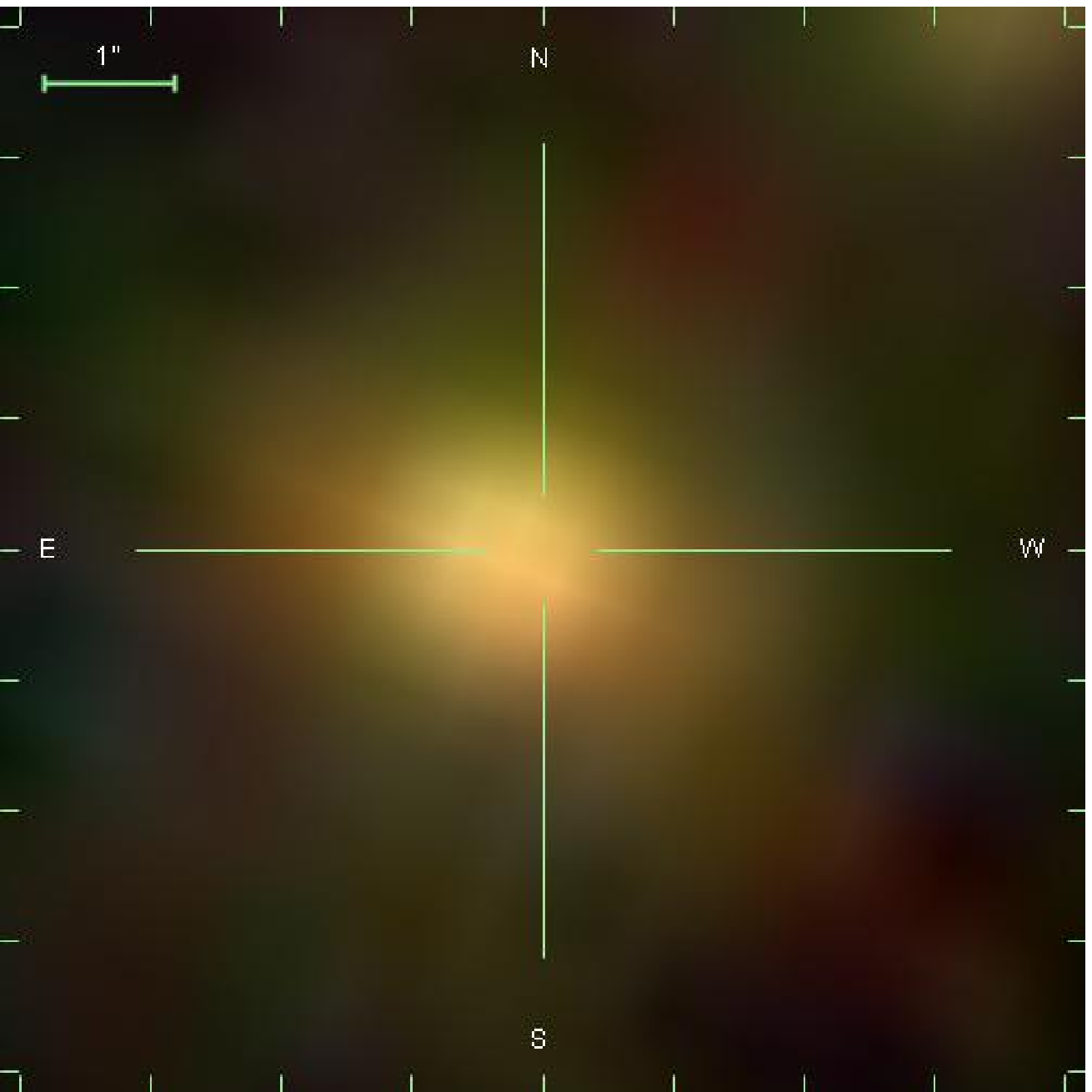}\\
\includegraphics[width=0.12\textwidth]{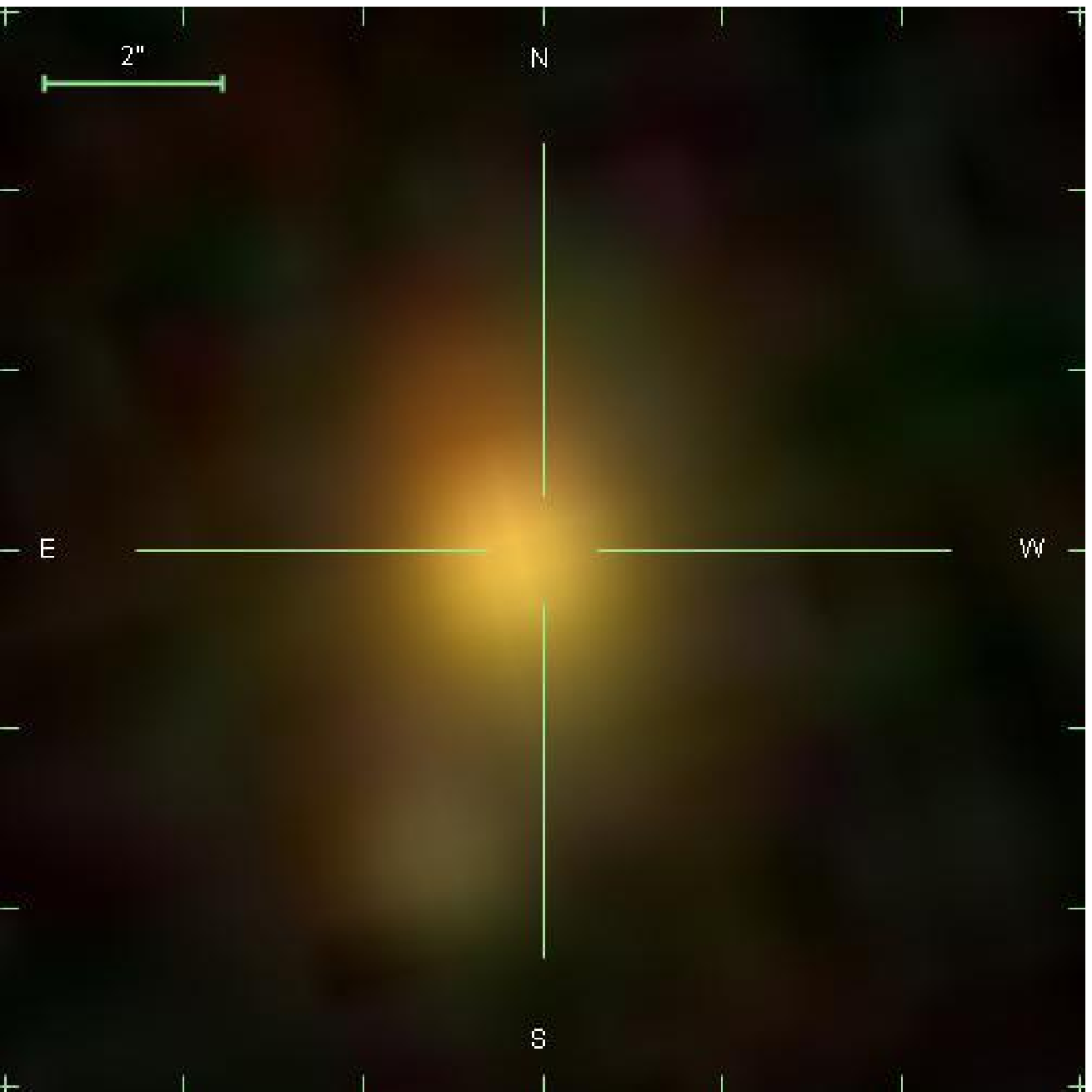}
\includegraphics[width=0.12\textwidth]{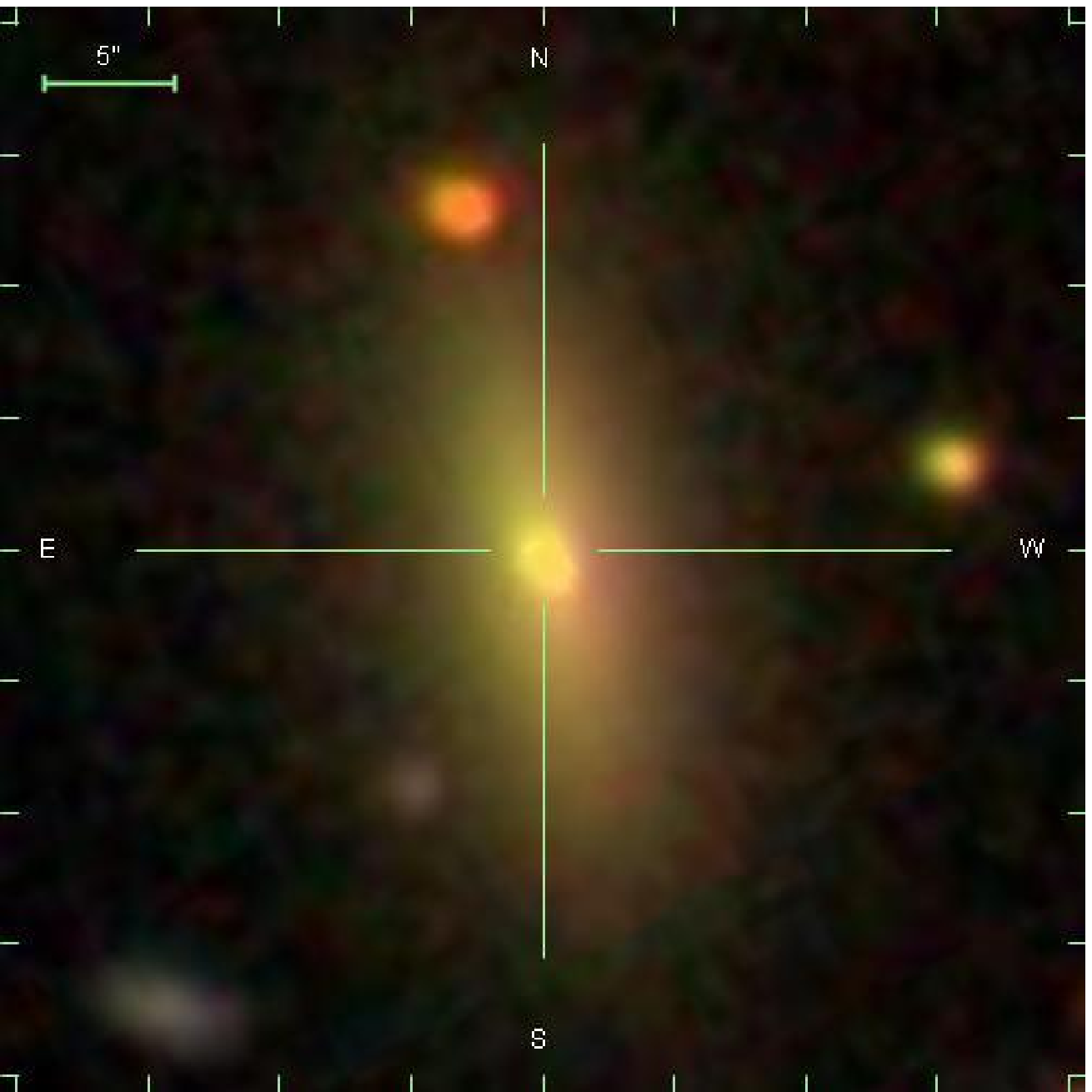}
\includegraphics[width=0.12\textwidth]{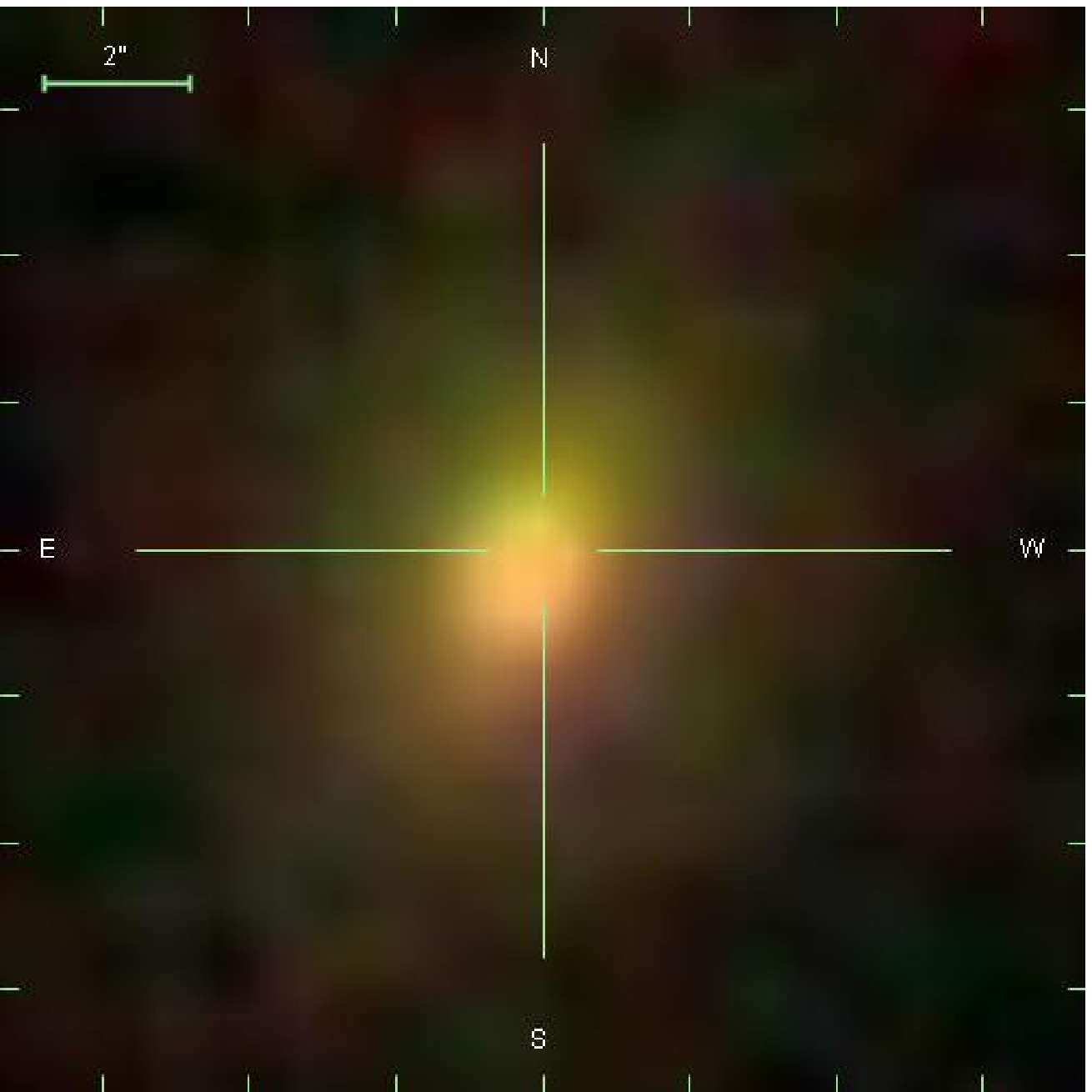}
\includegraphics[width=0.12\textwidth]{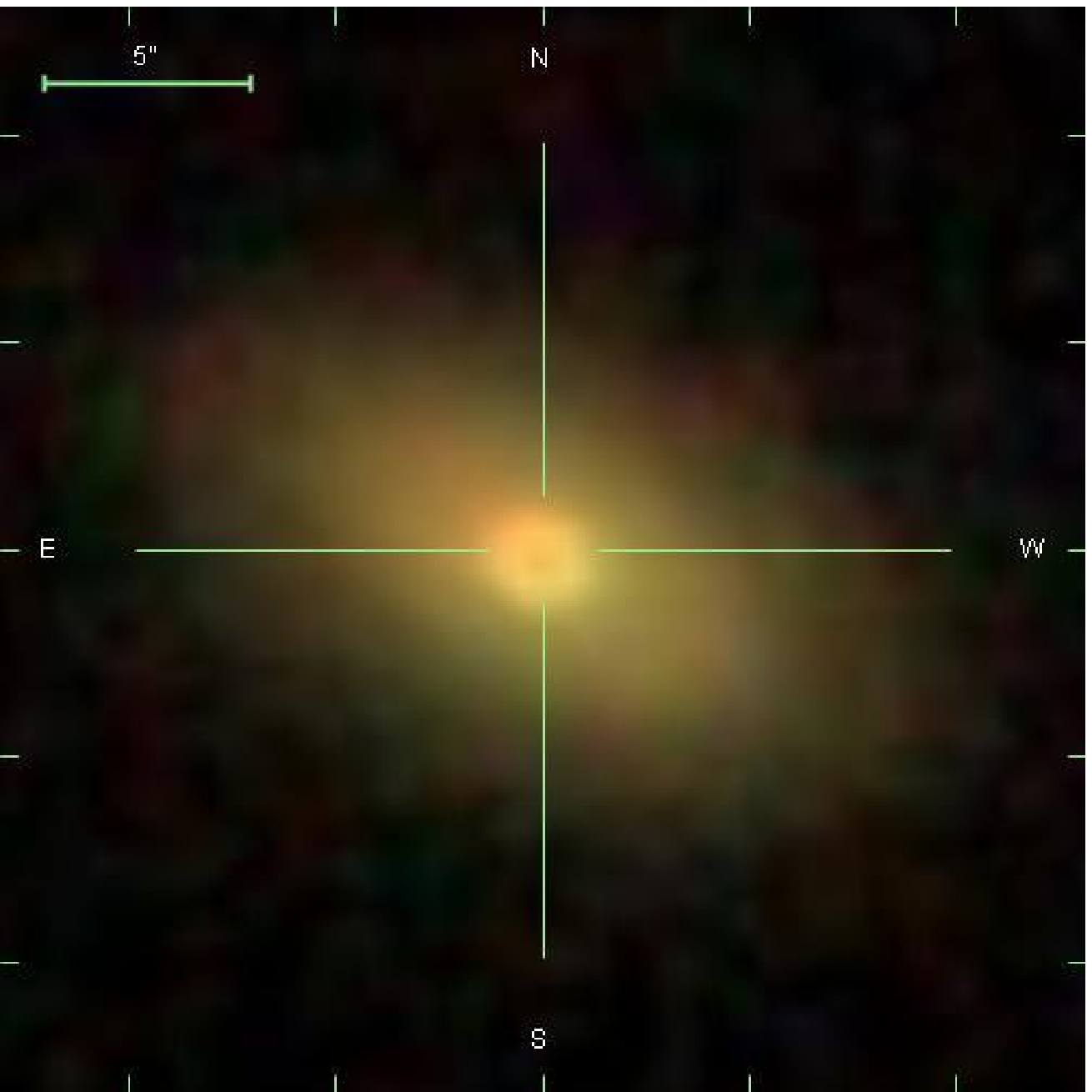}
\includegraphics[width=0.12\textwidth]{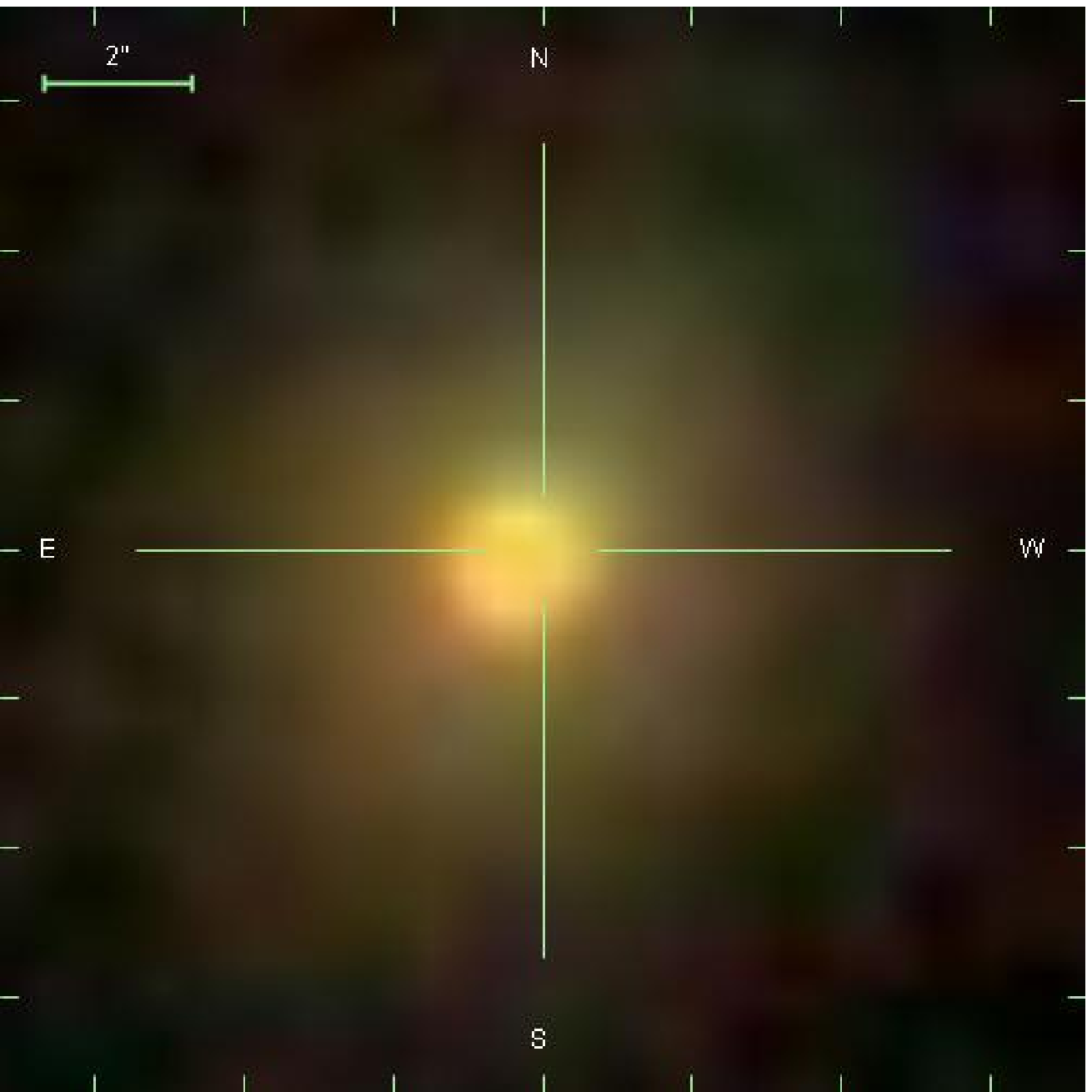}
\includegraphics[width=0.12\textwidth]{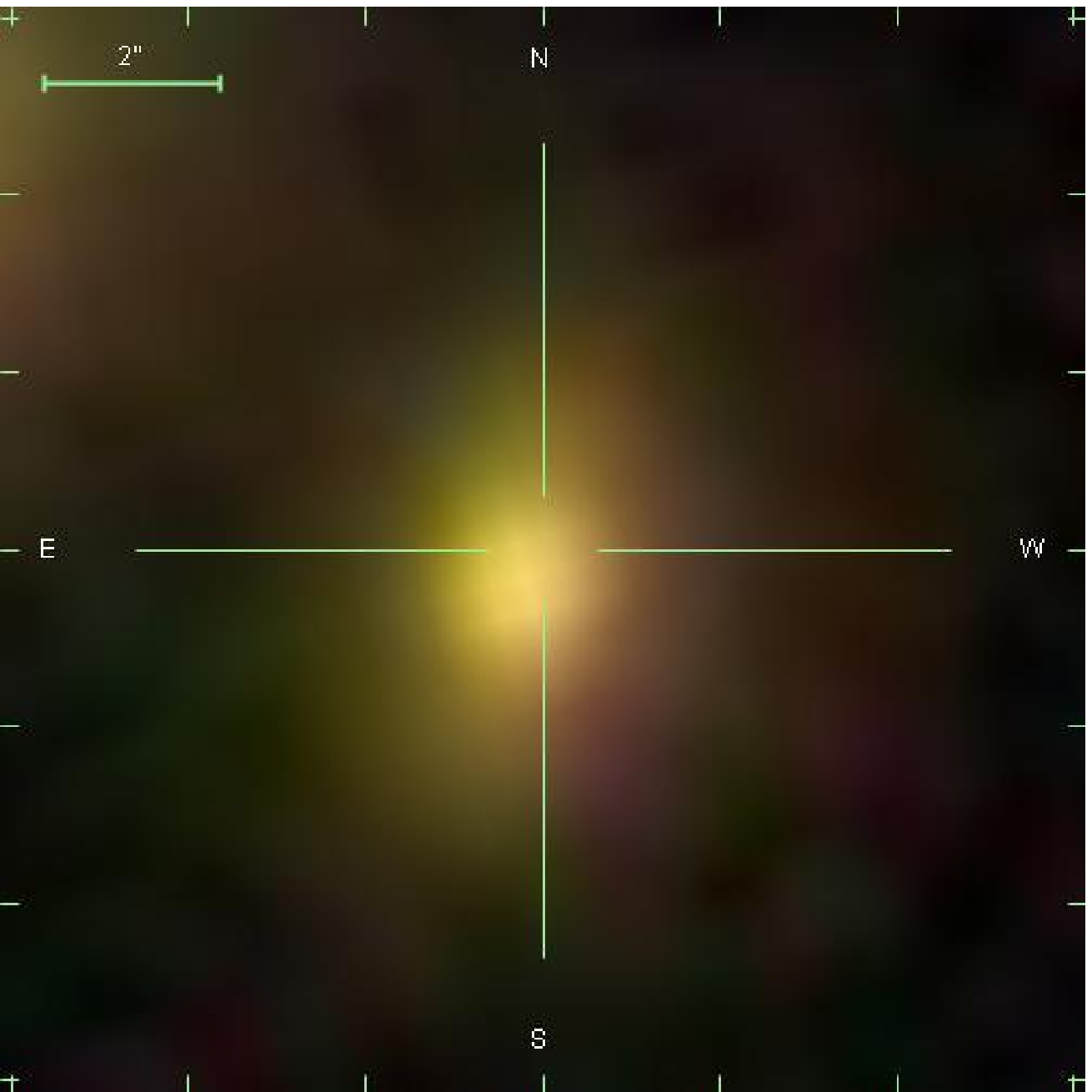}
\includegraphics[width=0.12\textwidth]{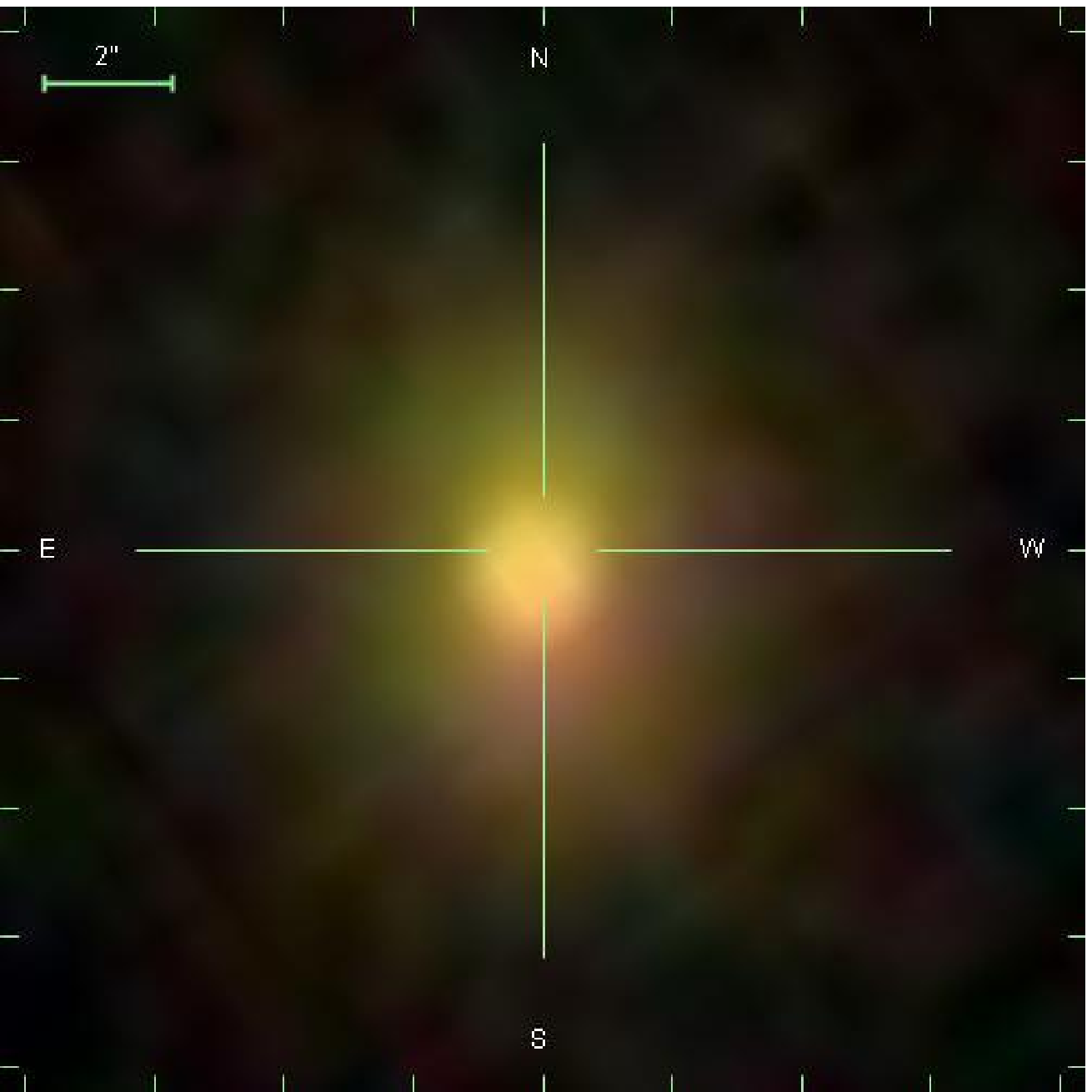}
\includegraphics[width=0.12\textwidth]{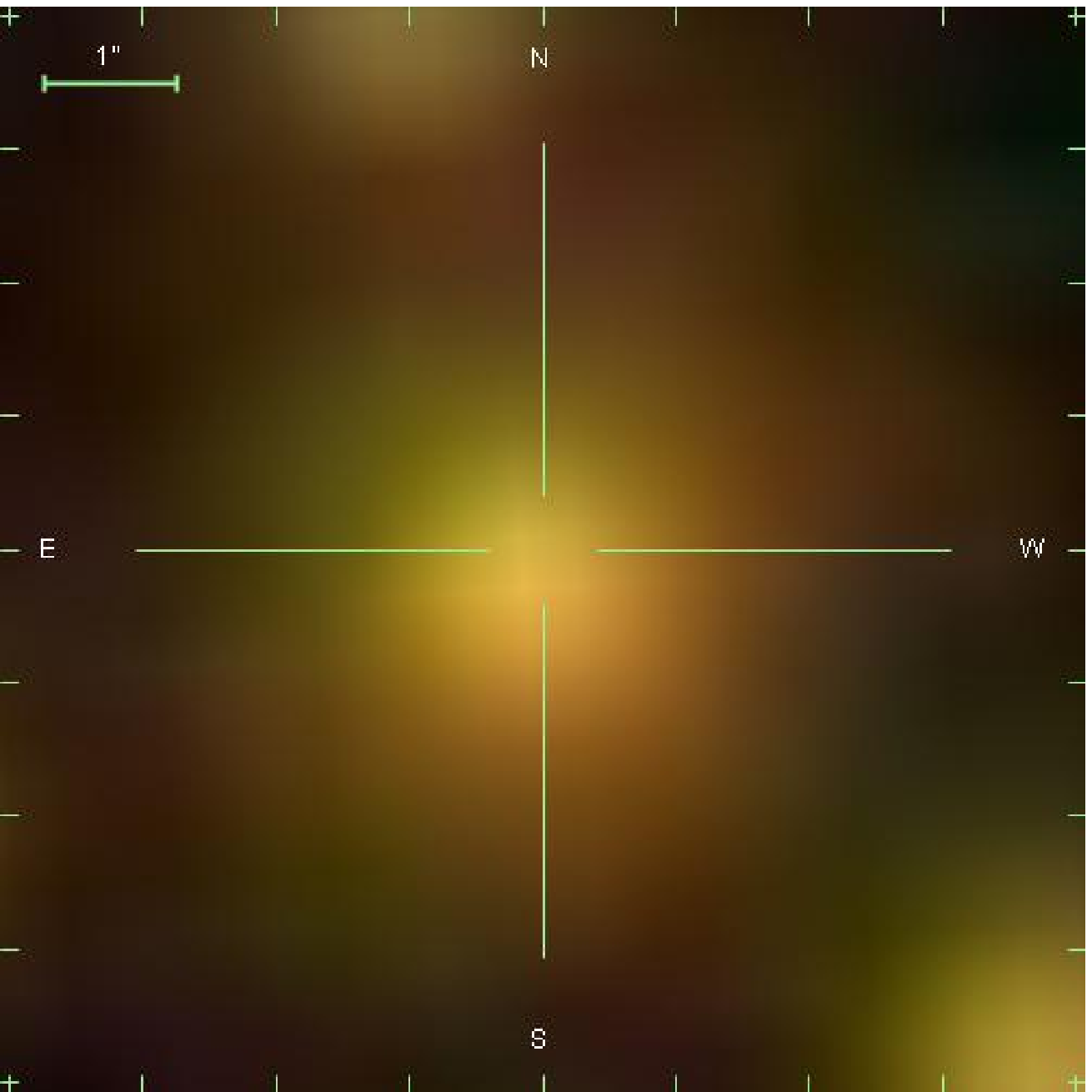}\\
\includegraphics[width=0.12\textwidth]{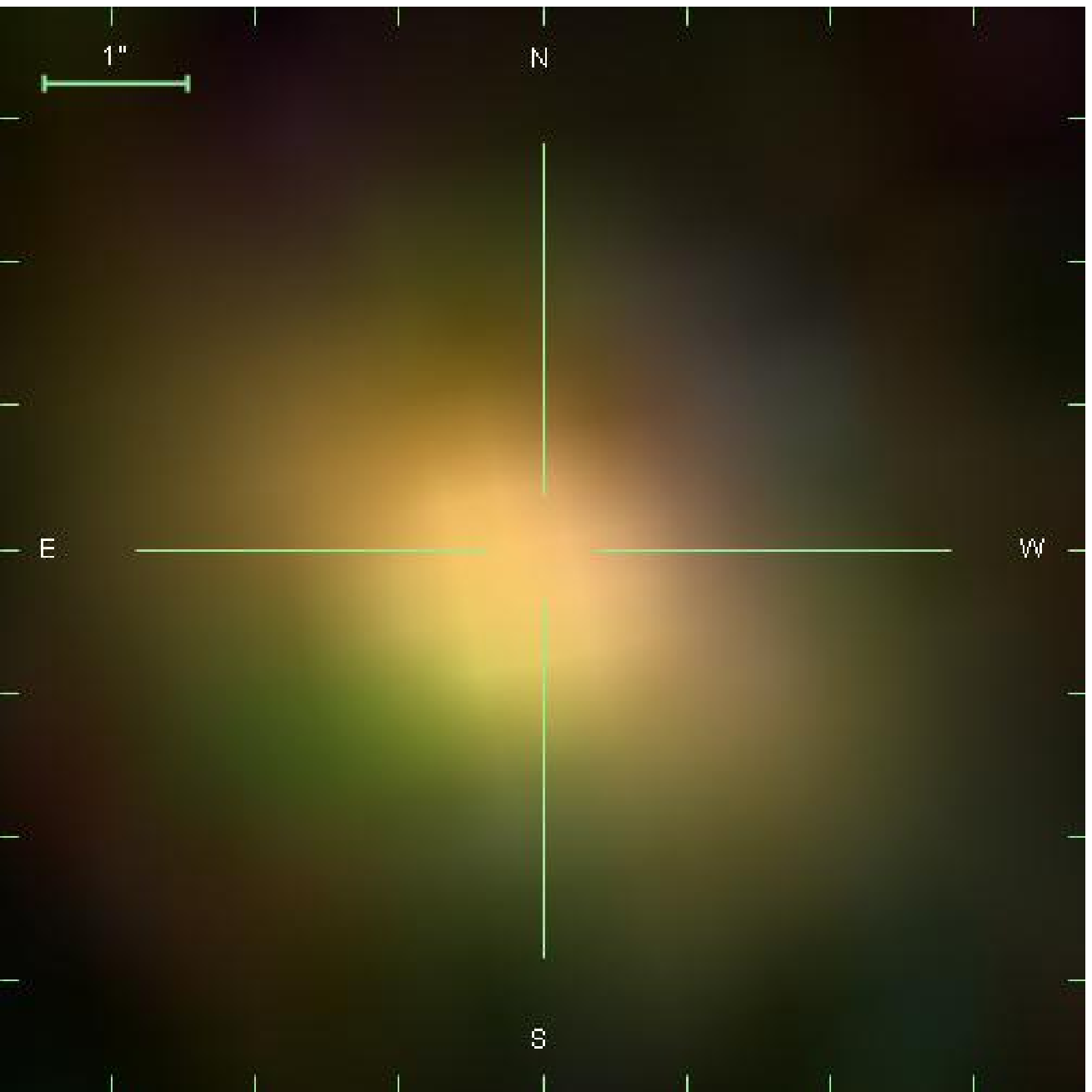}
\includegraphics[width=0.12\textwidth]{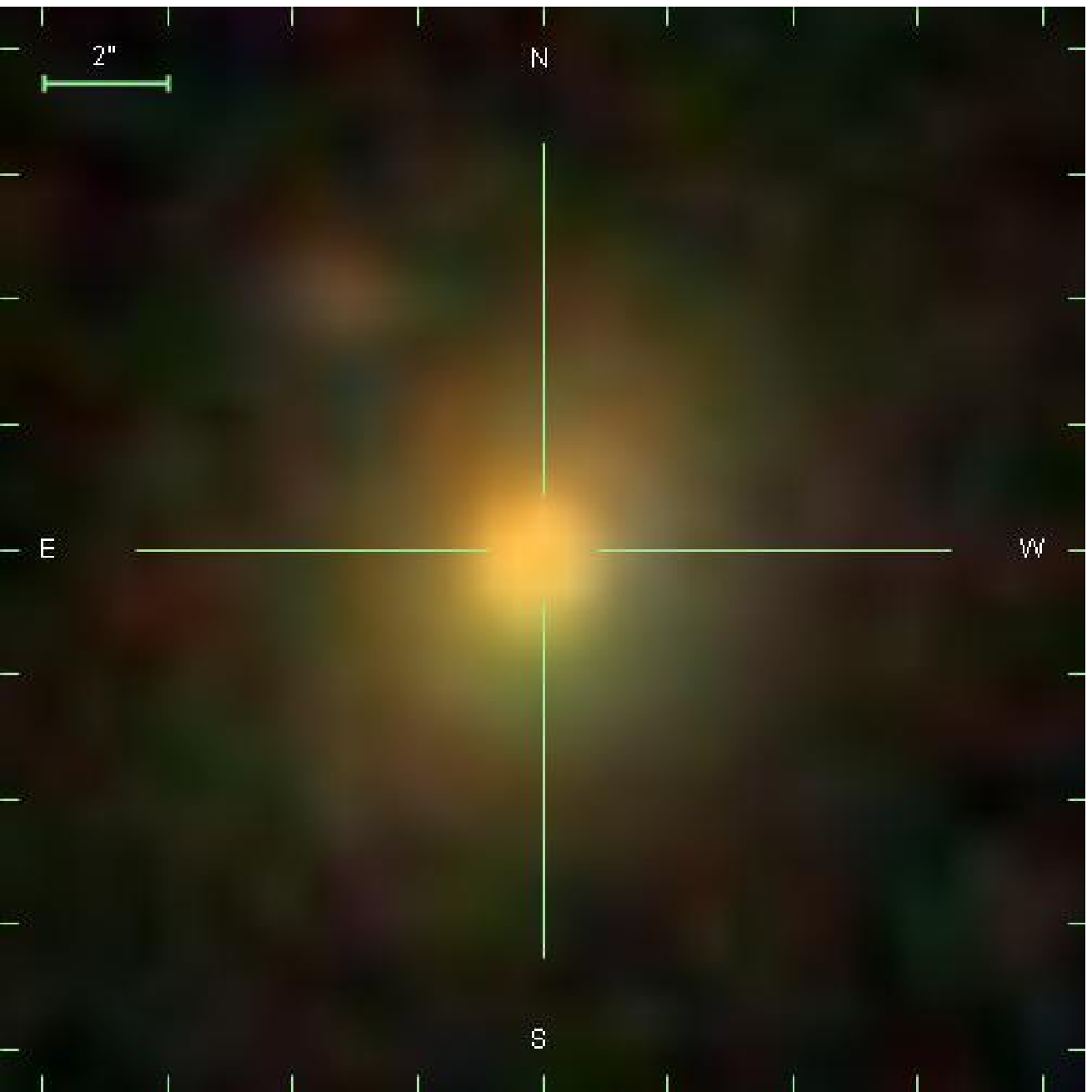}
\includegraphics[width=0.12\textwidth]{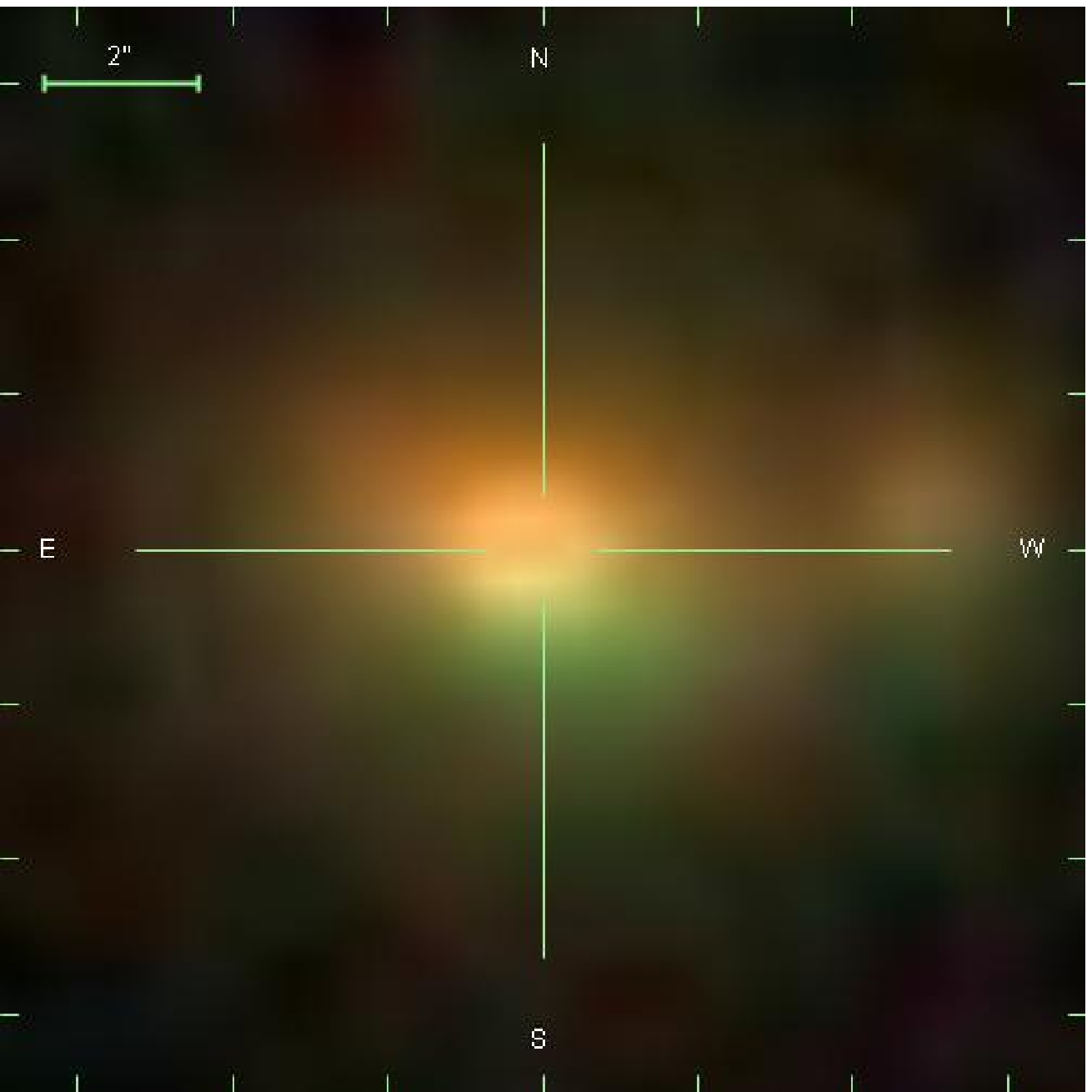}
\includegraphics[width=0.12\textwidth]{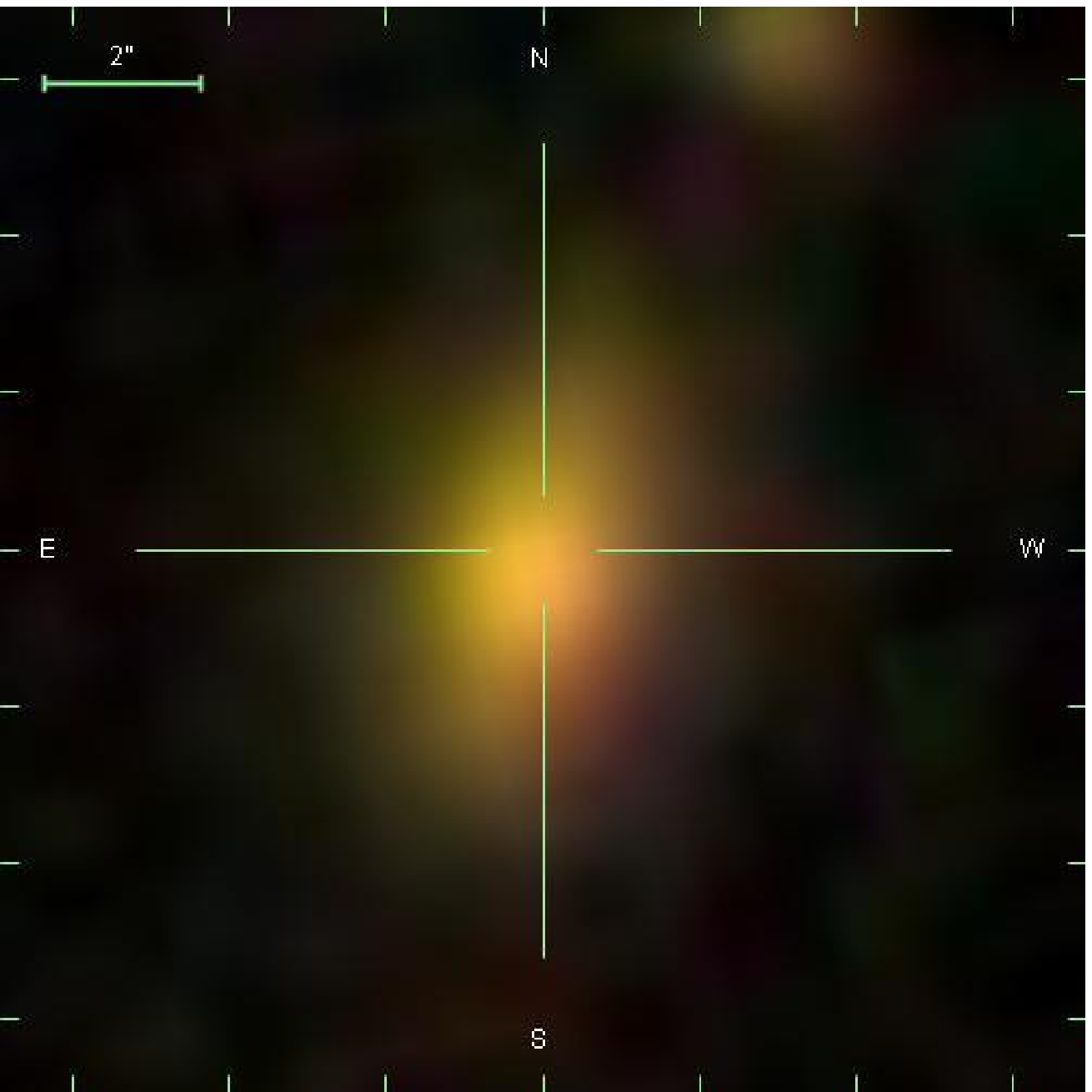}
\includegraphics[width=0.12\textwidth]{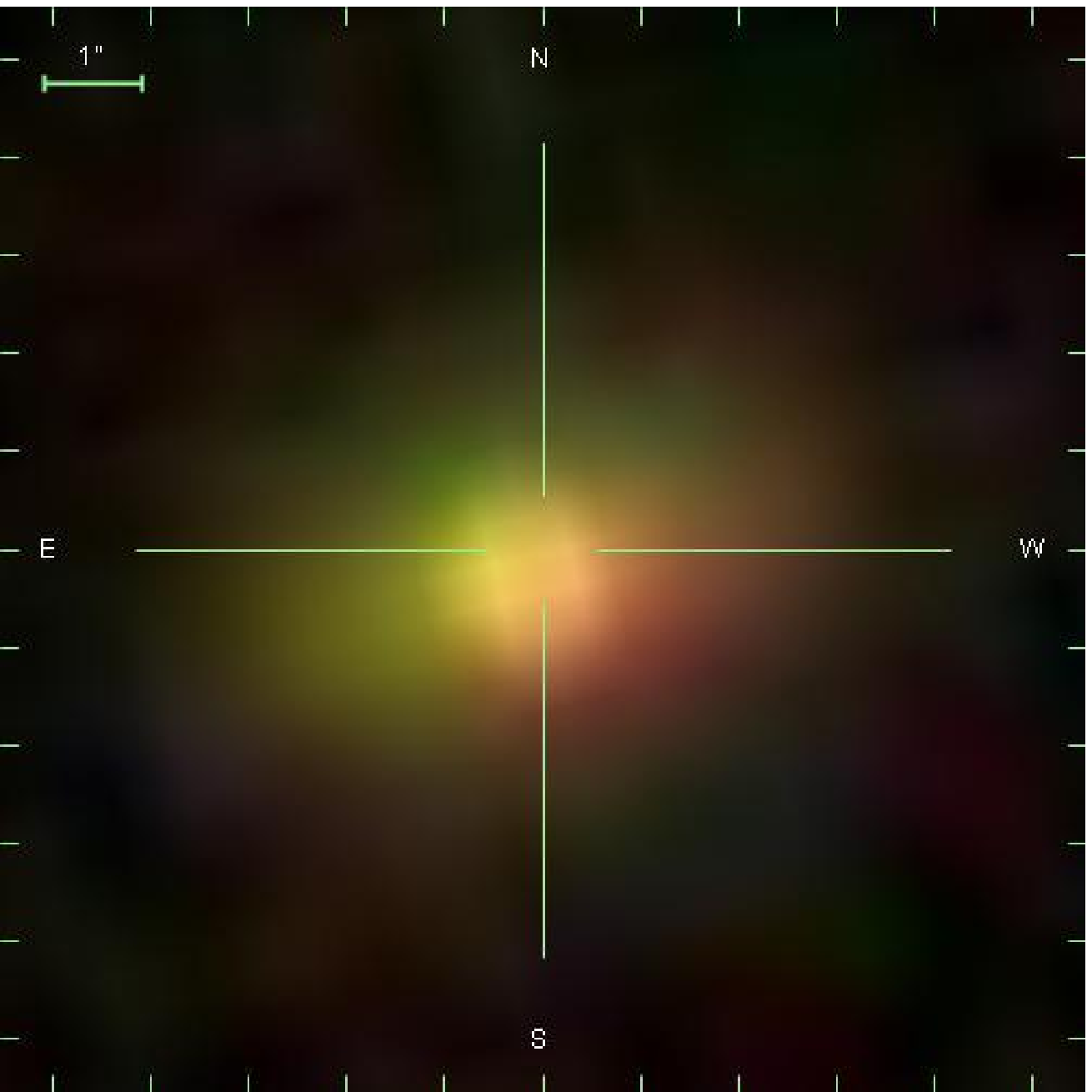}
\includegraphics[width=0.12\textwidth]{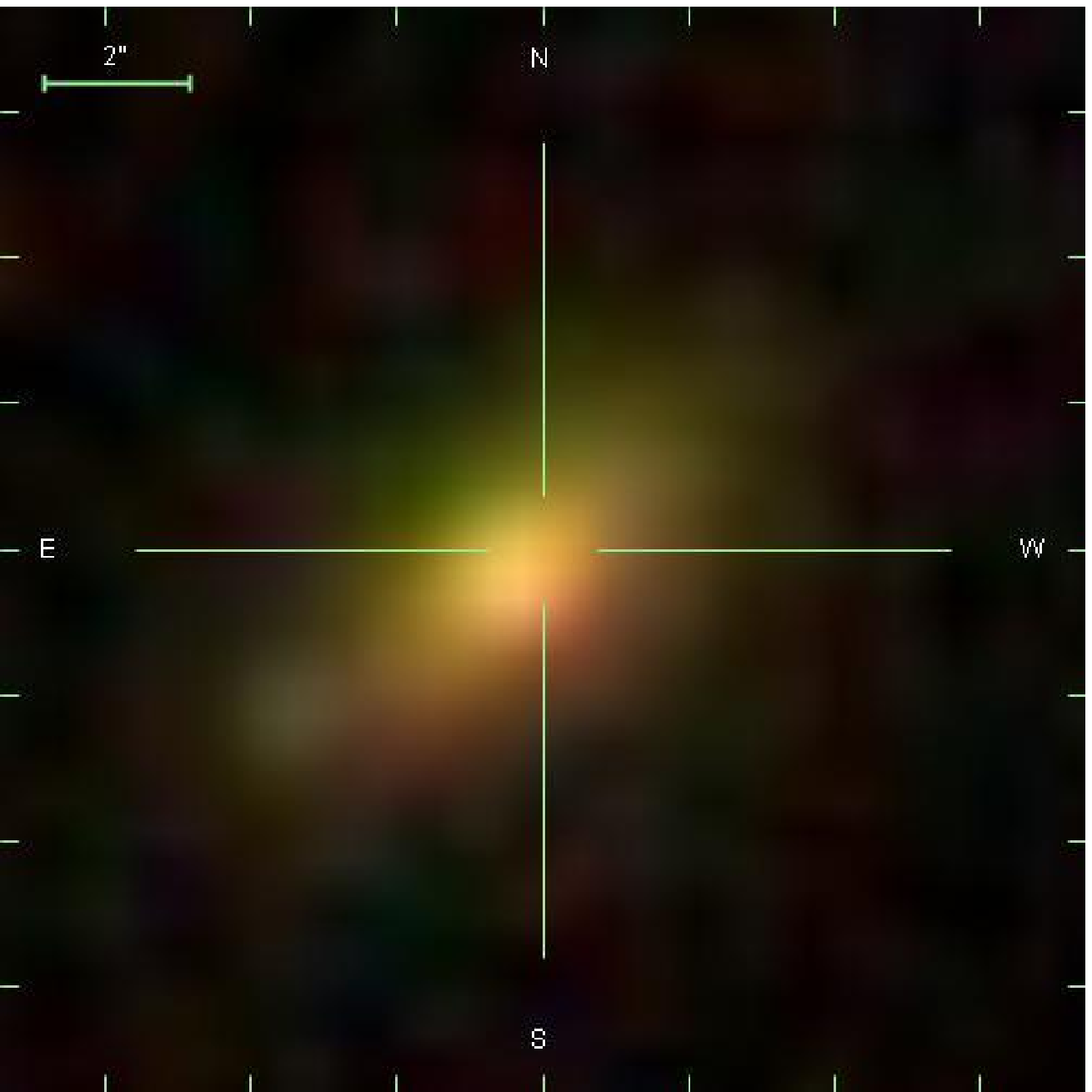}
\includegraphics[width=0.12\textwidth]{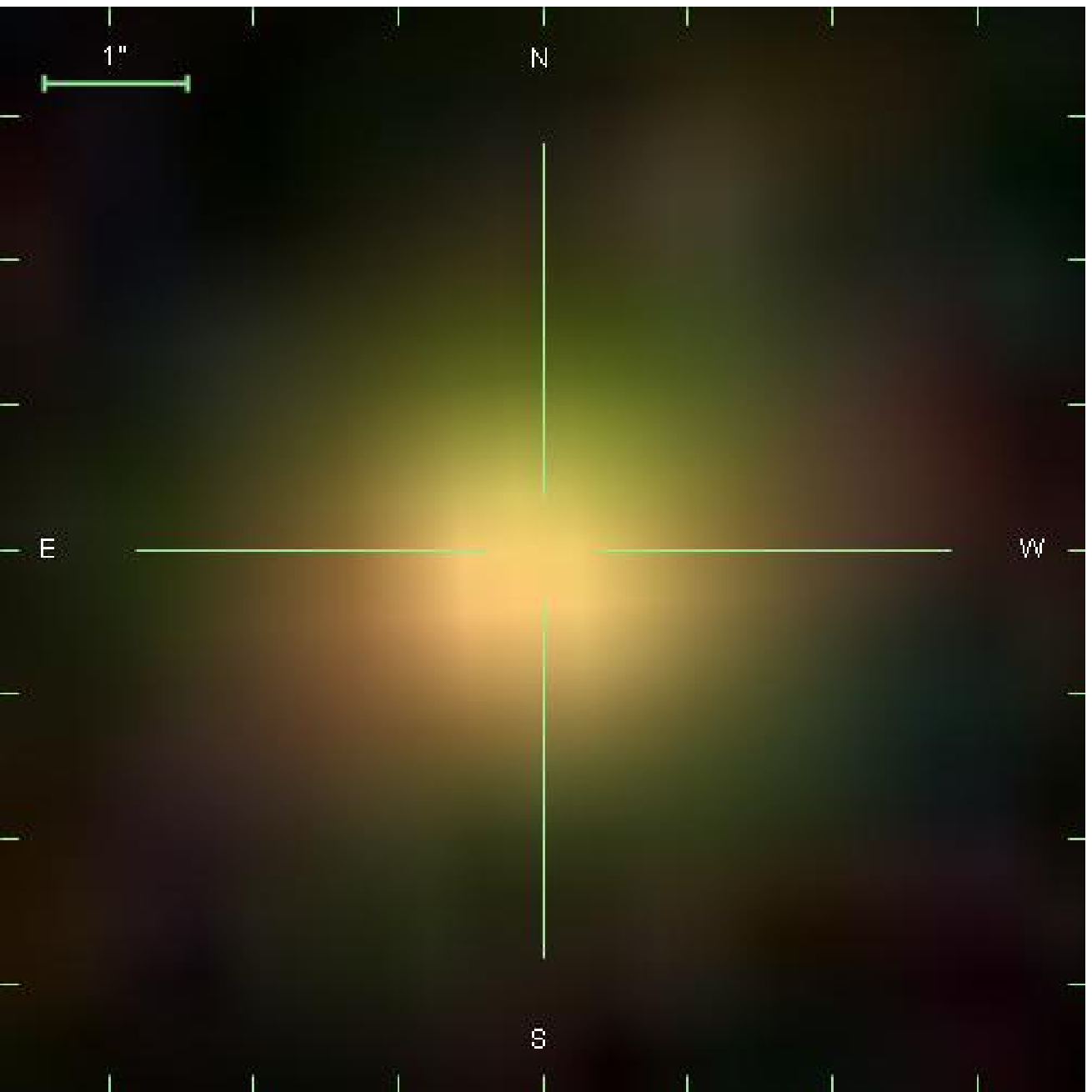}
\includegraphics[width=0.12\textwidth]{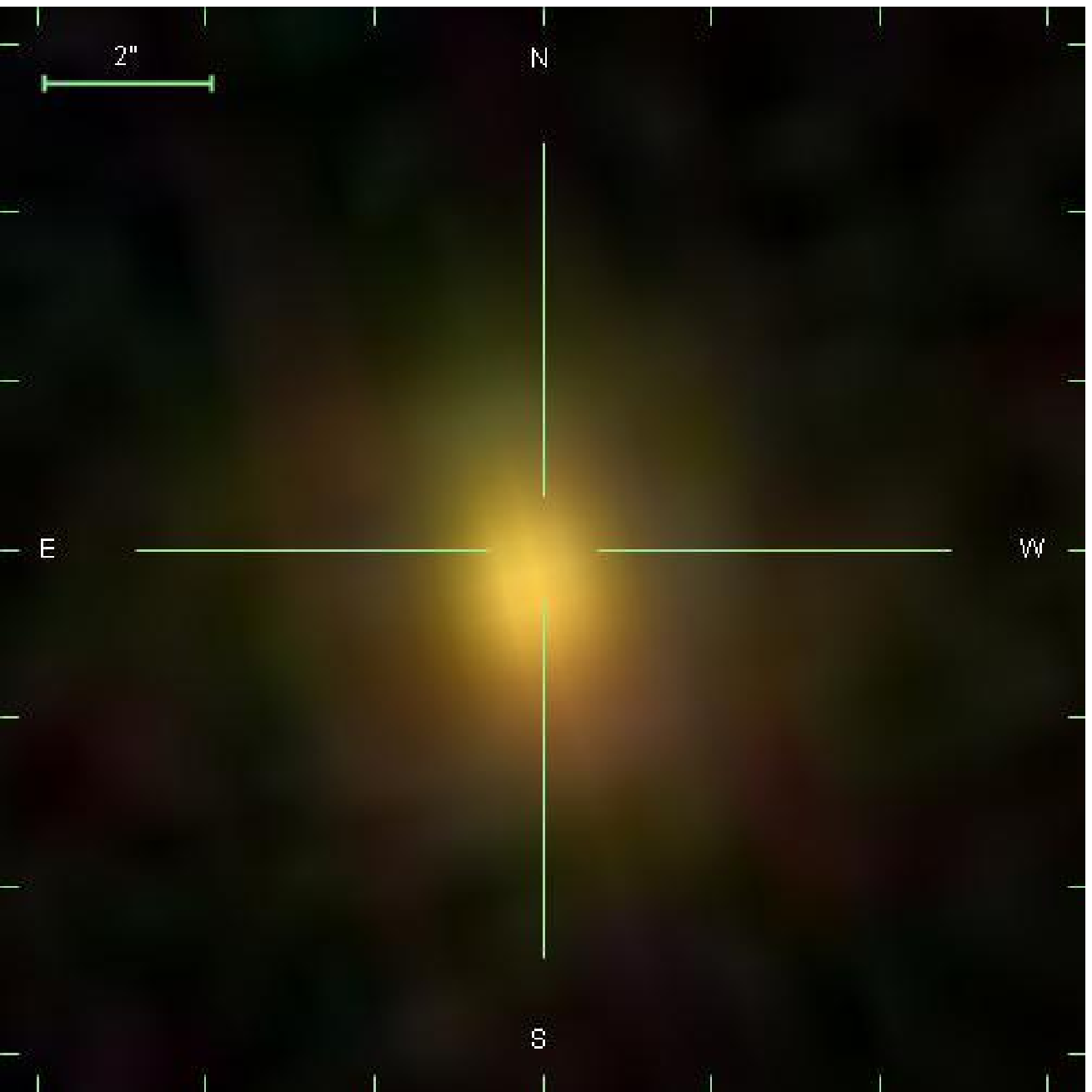}\\
\includegraphics[width=0.12\textwidth]{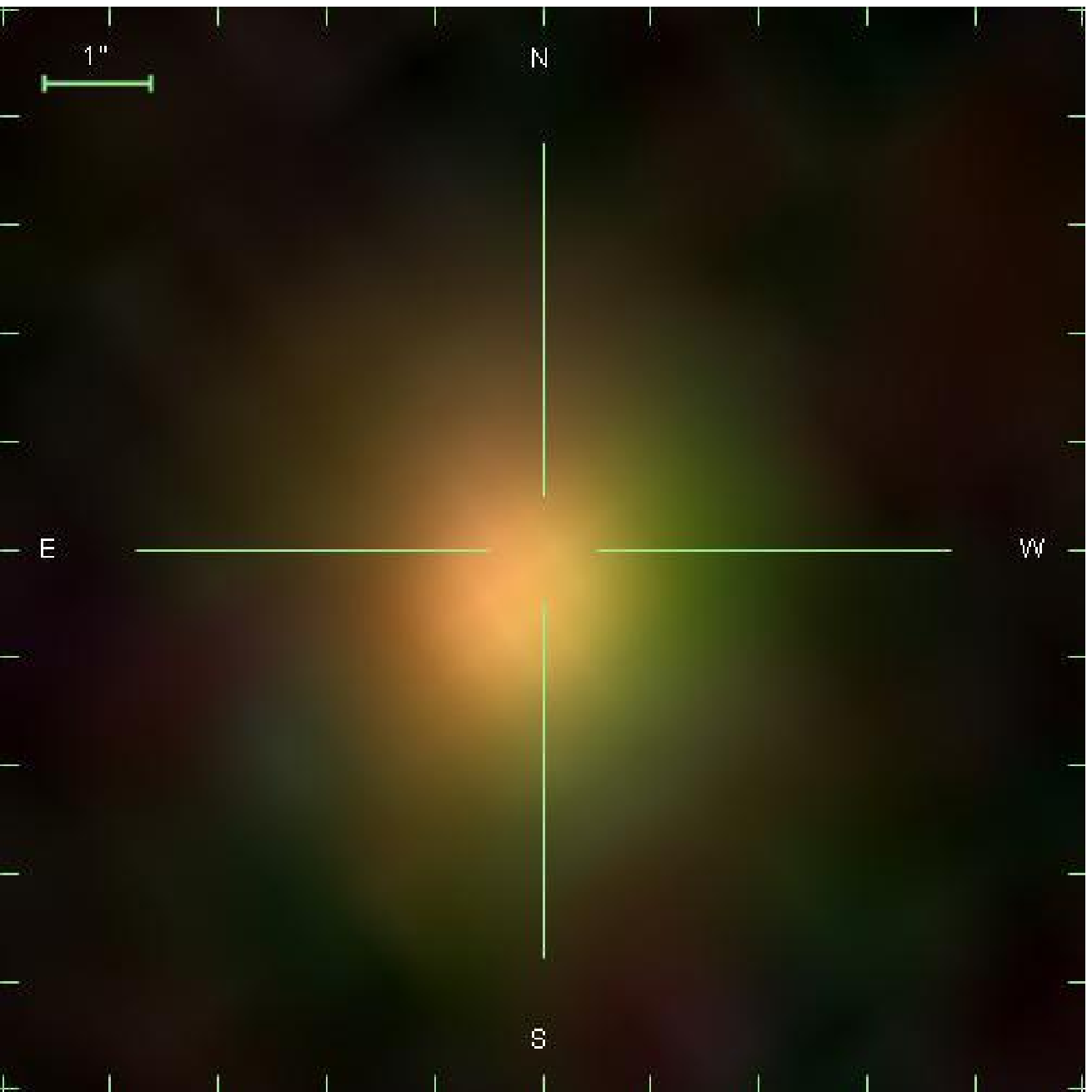}
\includegraphics[width=0.12\textwidth]{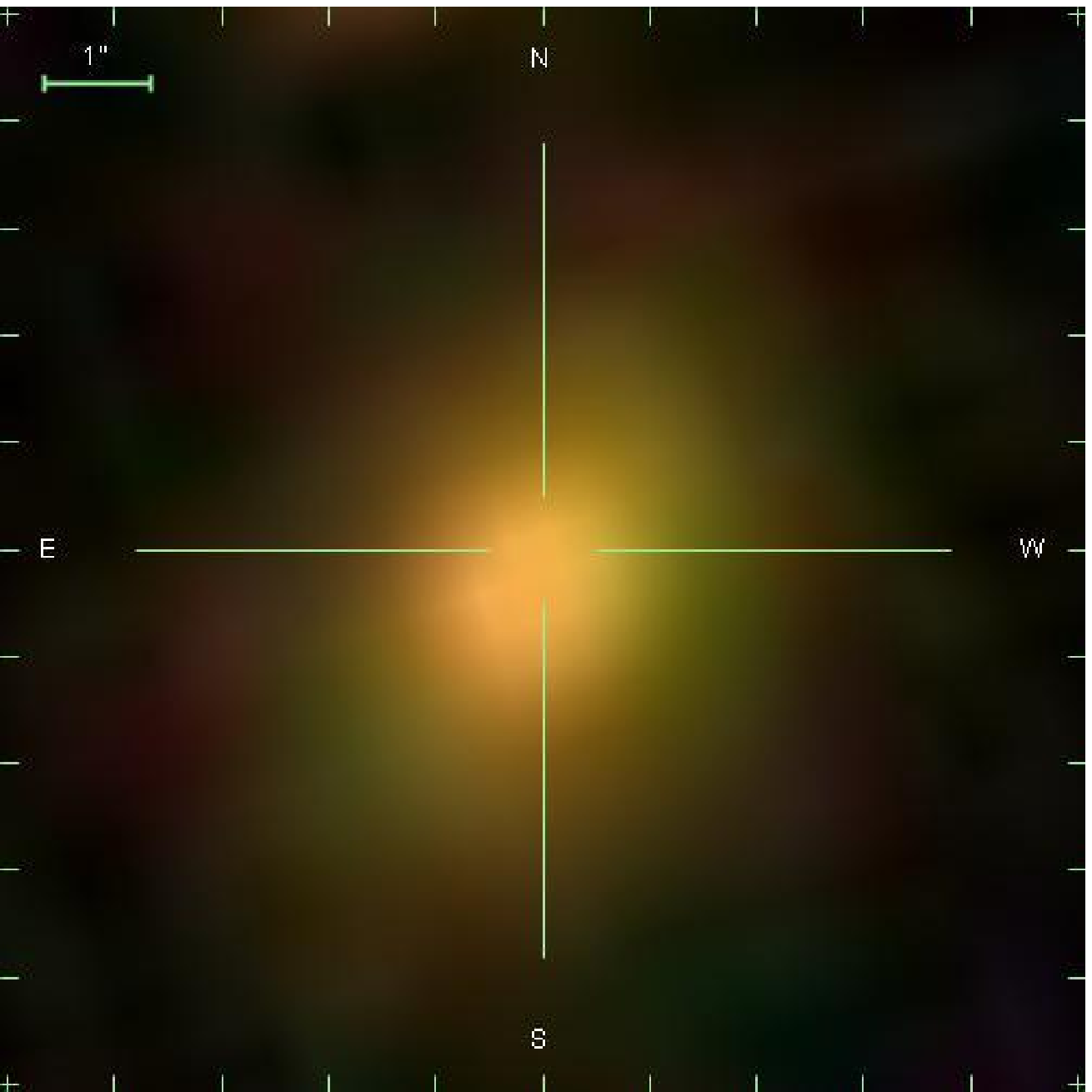}
\includegraphics[width=0.12\textwidth]{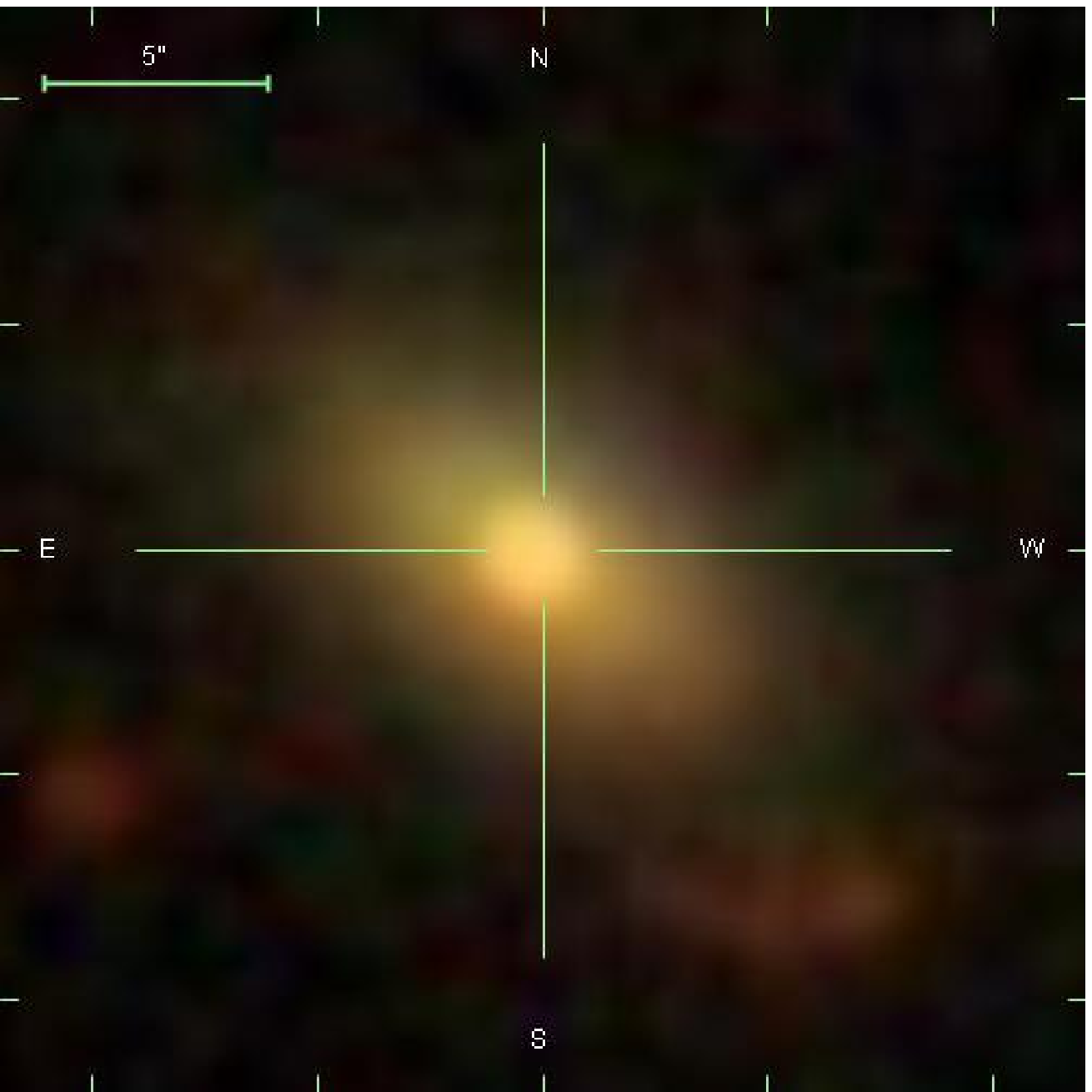}
\includegraphics[width=0.12\textwidth]{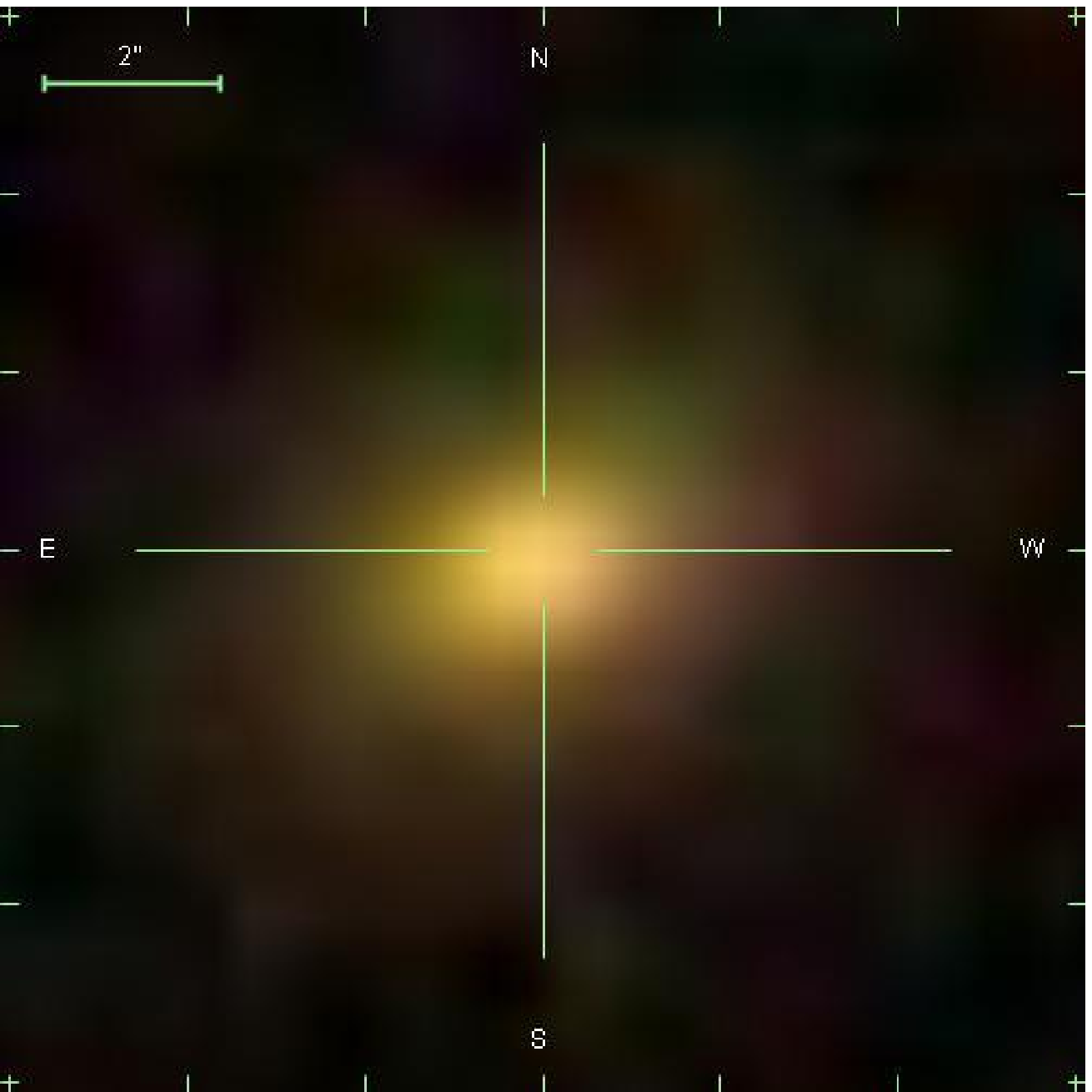}
\includegraphics[width=0.12\textwidth]{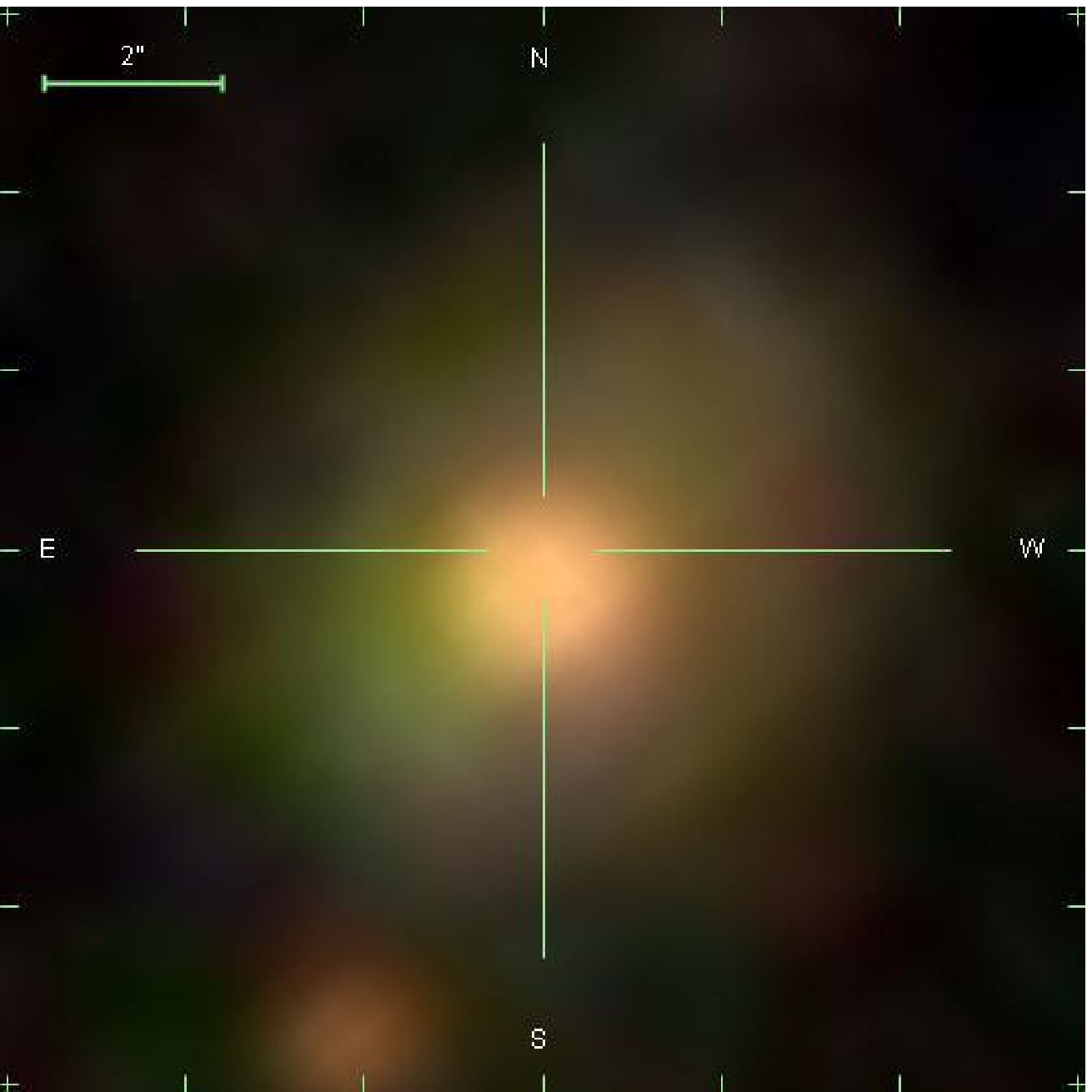}
\includegraphics[width=0.12\textwidth]{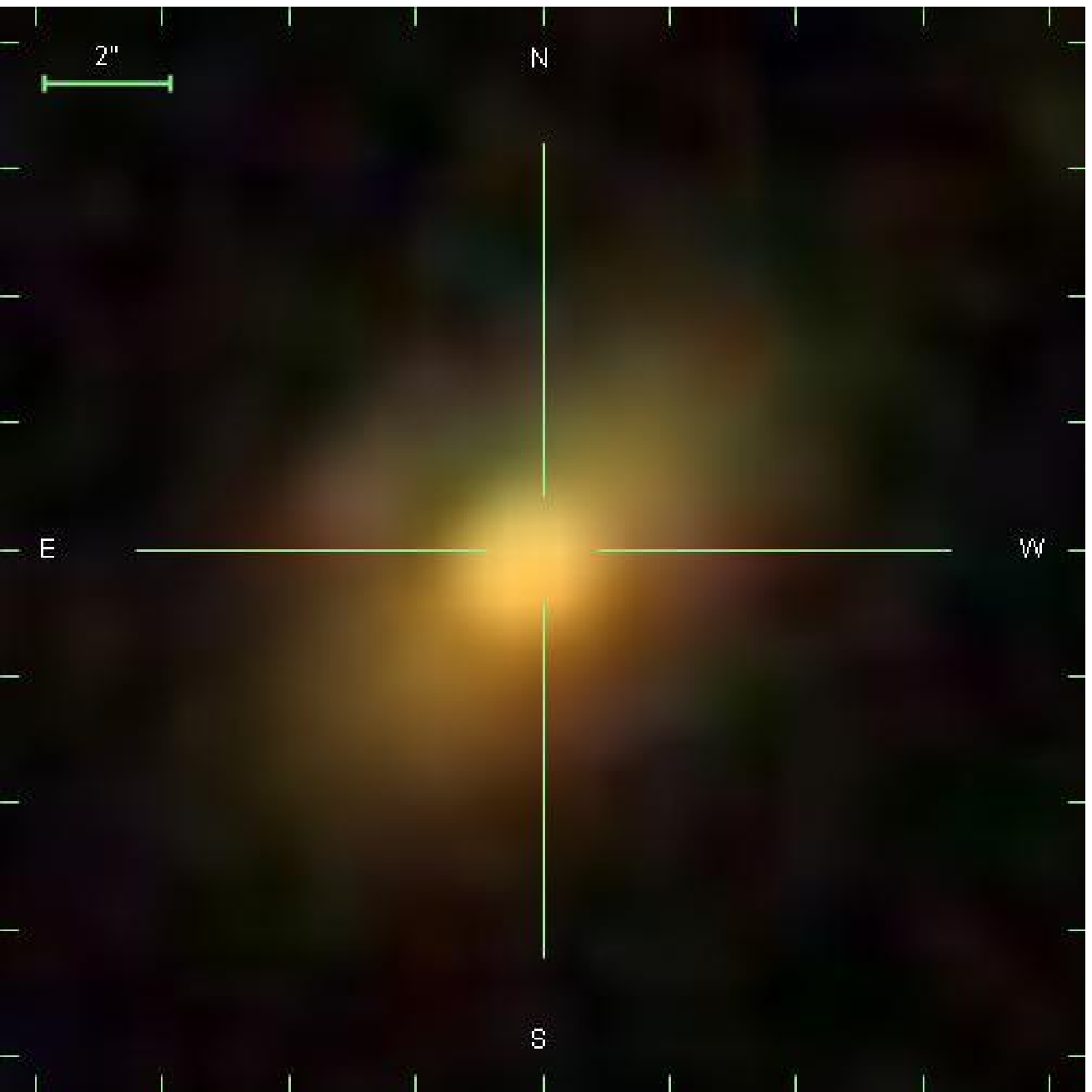}
\includegraphics[width=0.12\textwidth]{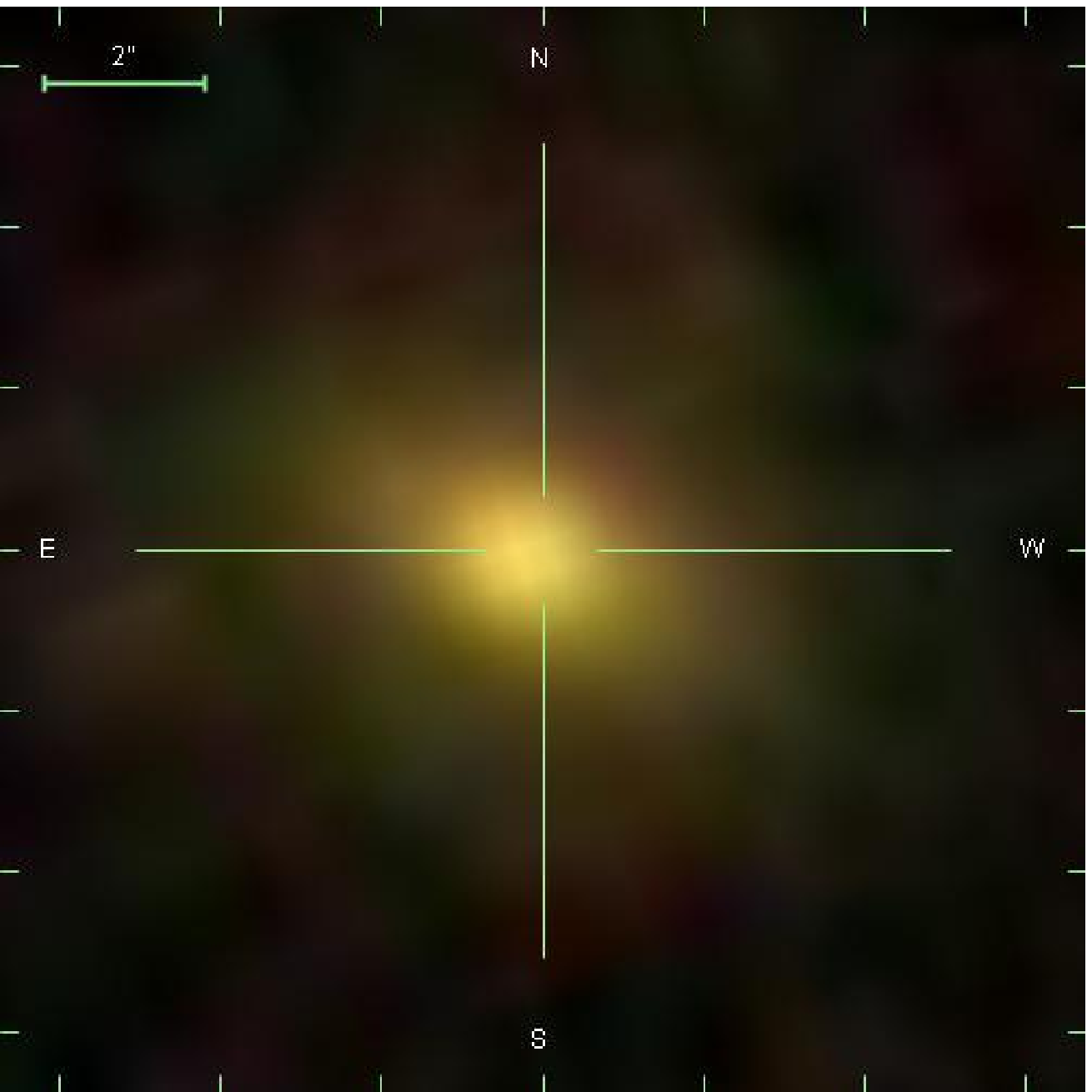}
\includegraphics[width=0.12\textwidth]{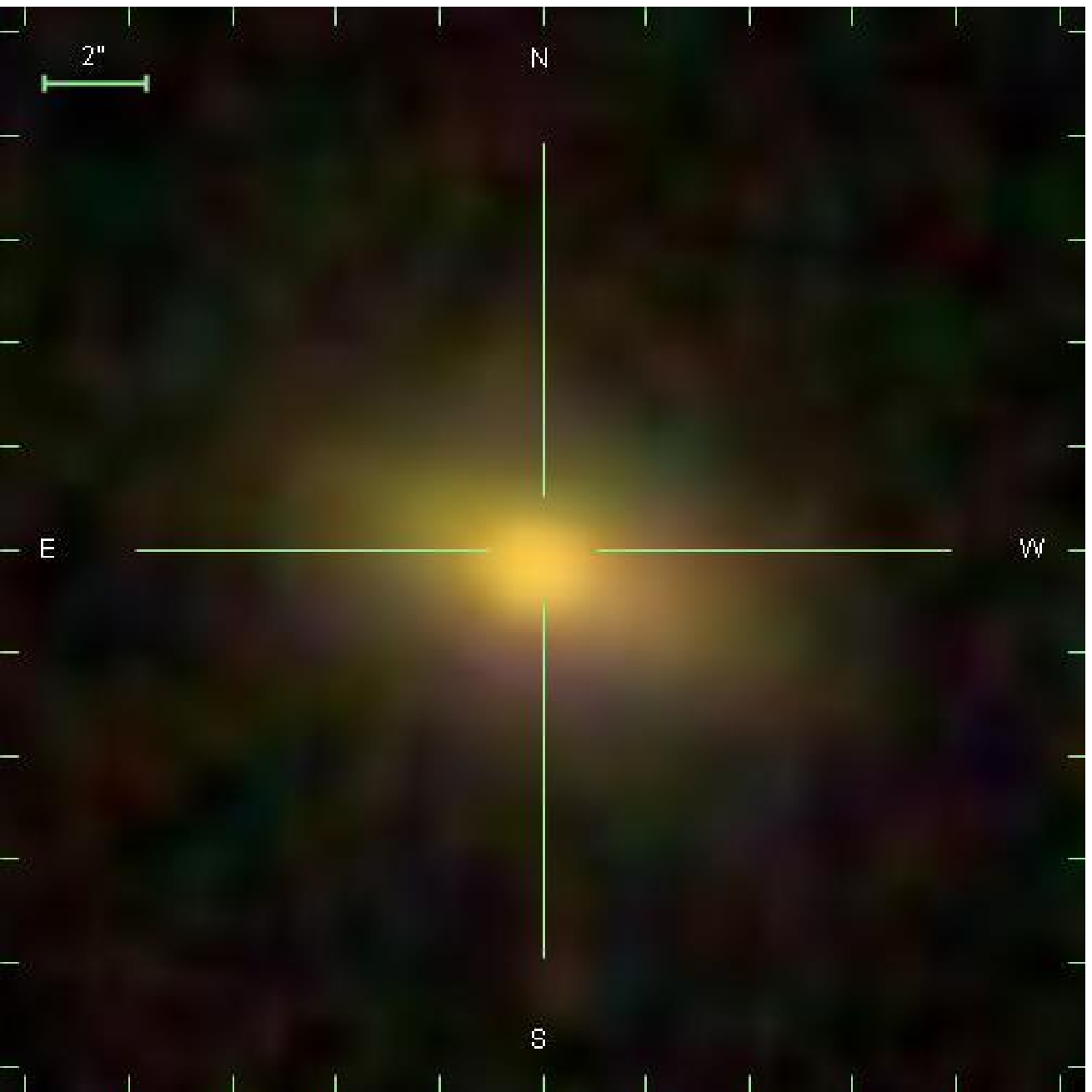}\\
\includegraphics[width=0.12\textwidth]{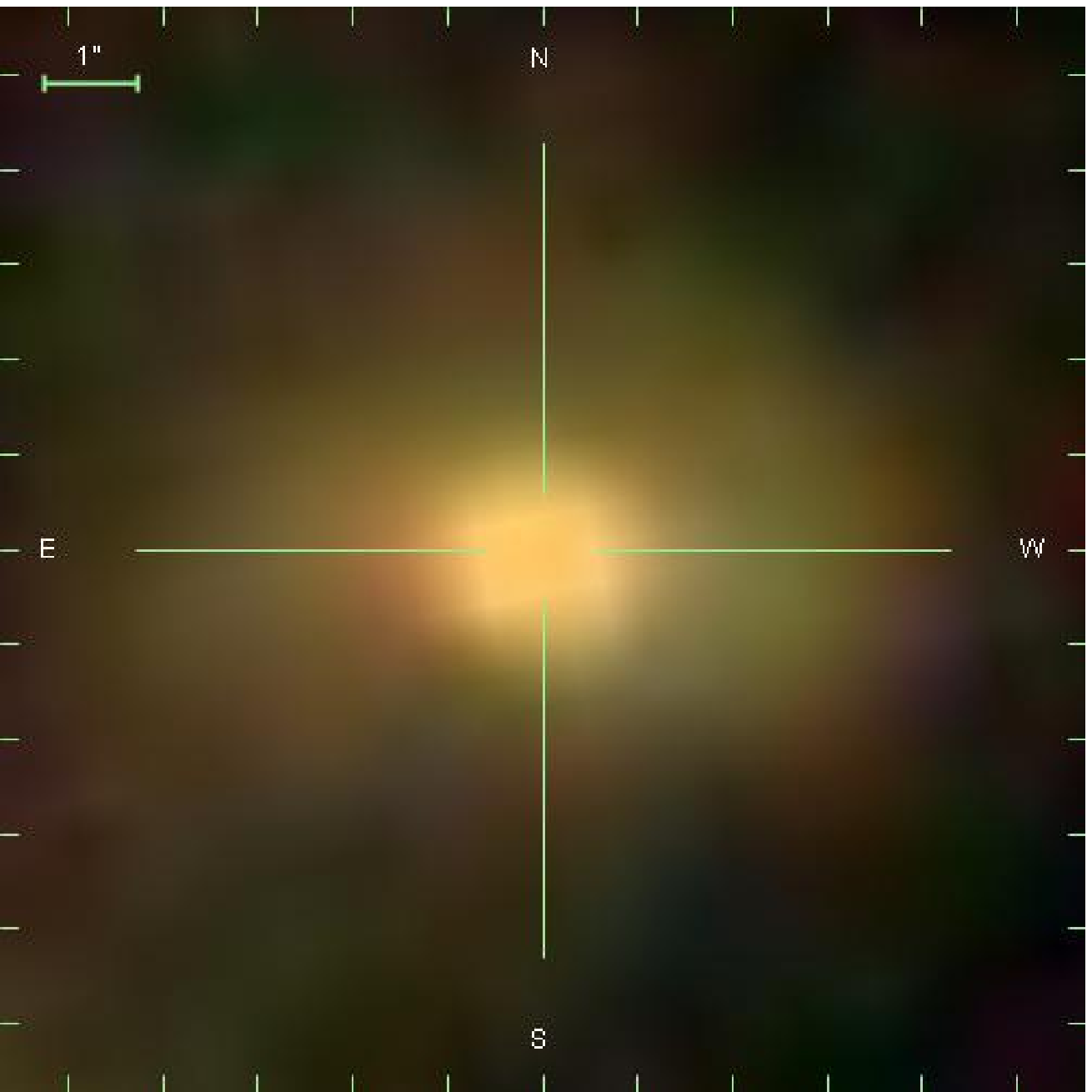}
\includegraphics[width=0.12\textwidth]{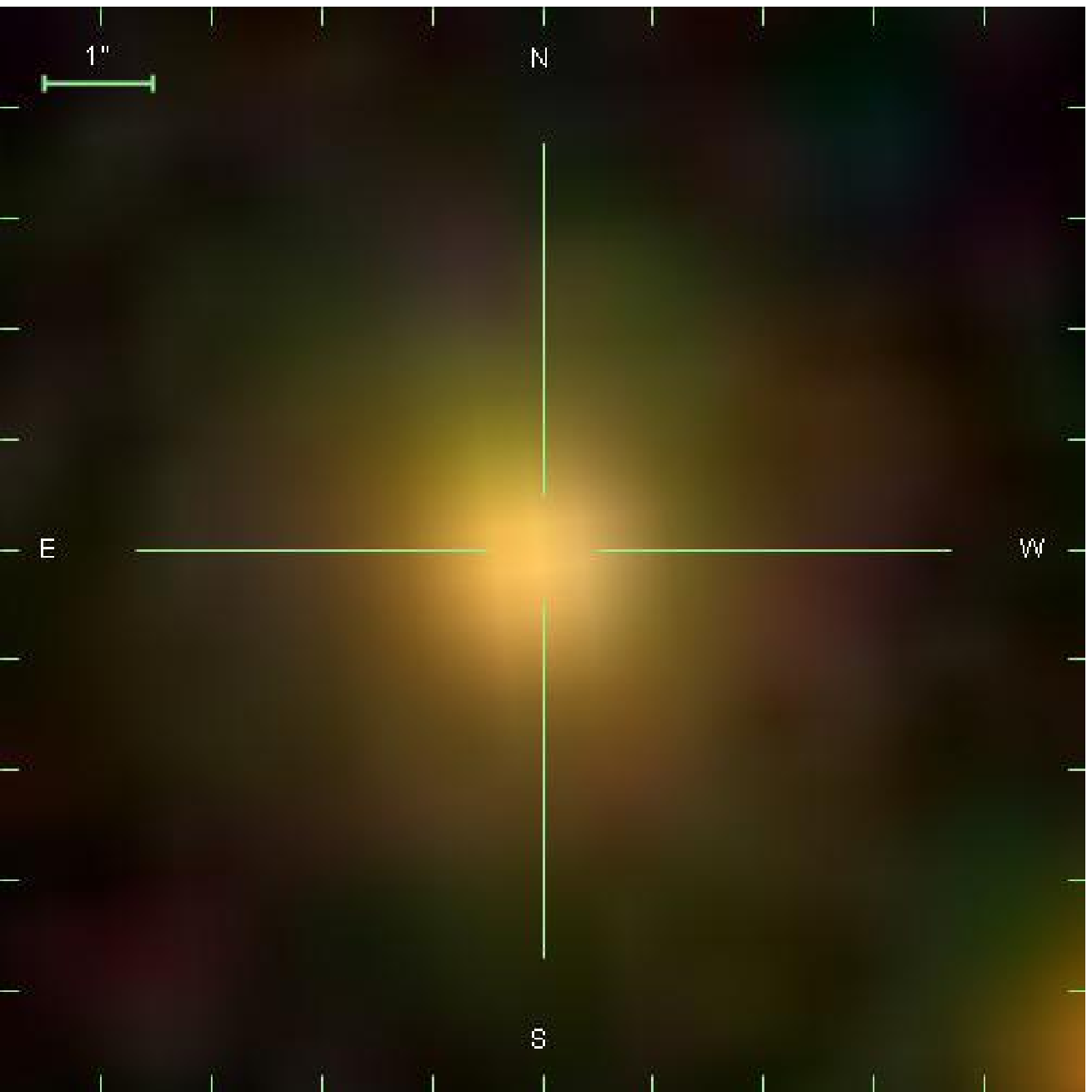}
\includegraphics[width=0.12\textwidth]{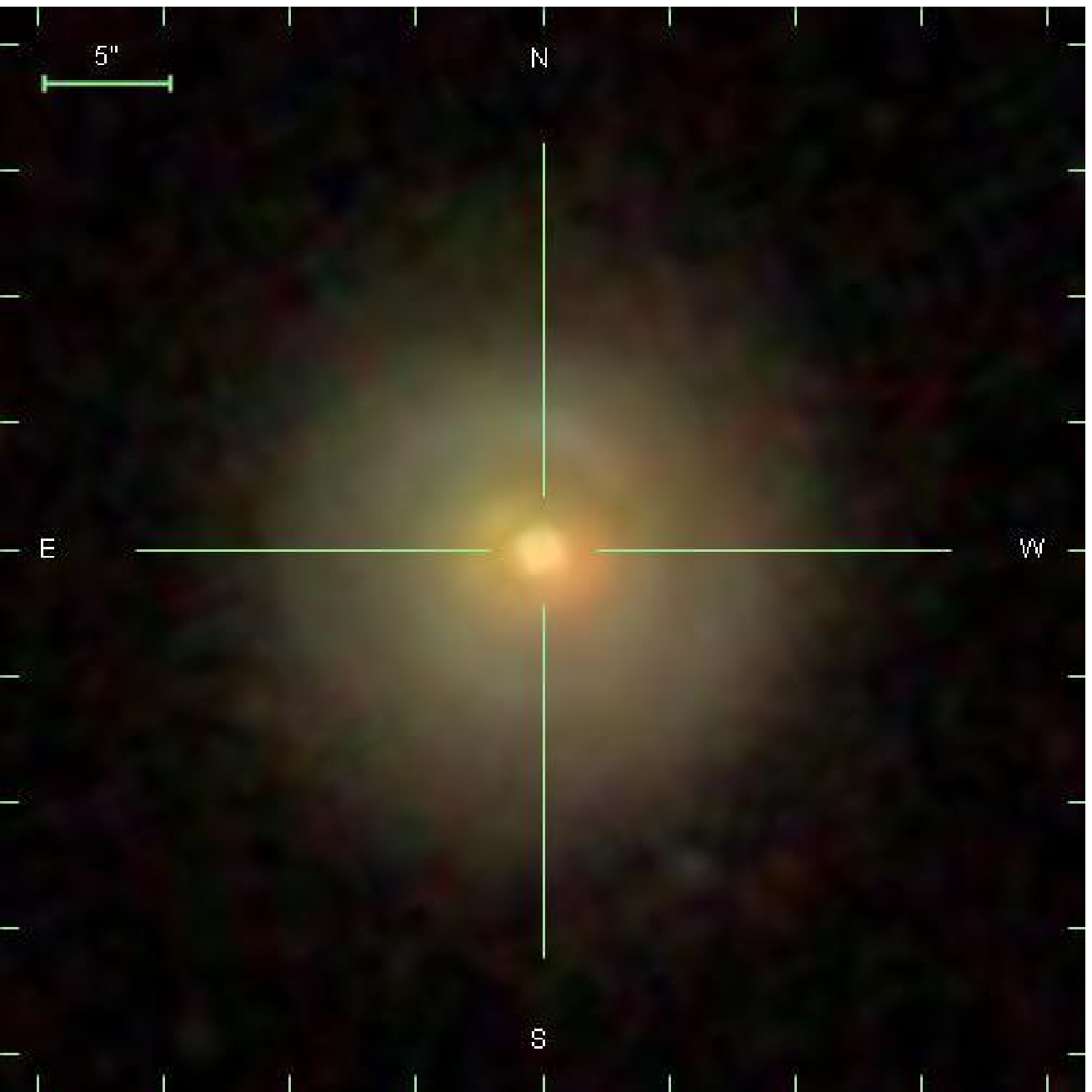}
\includegraphics[width=0.12\textwidth]{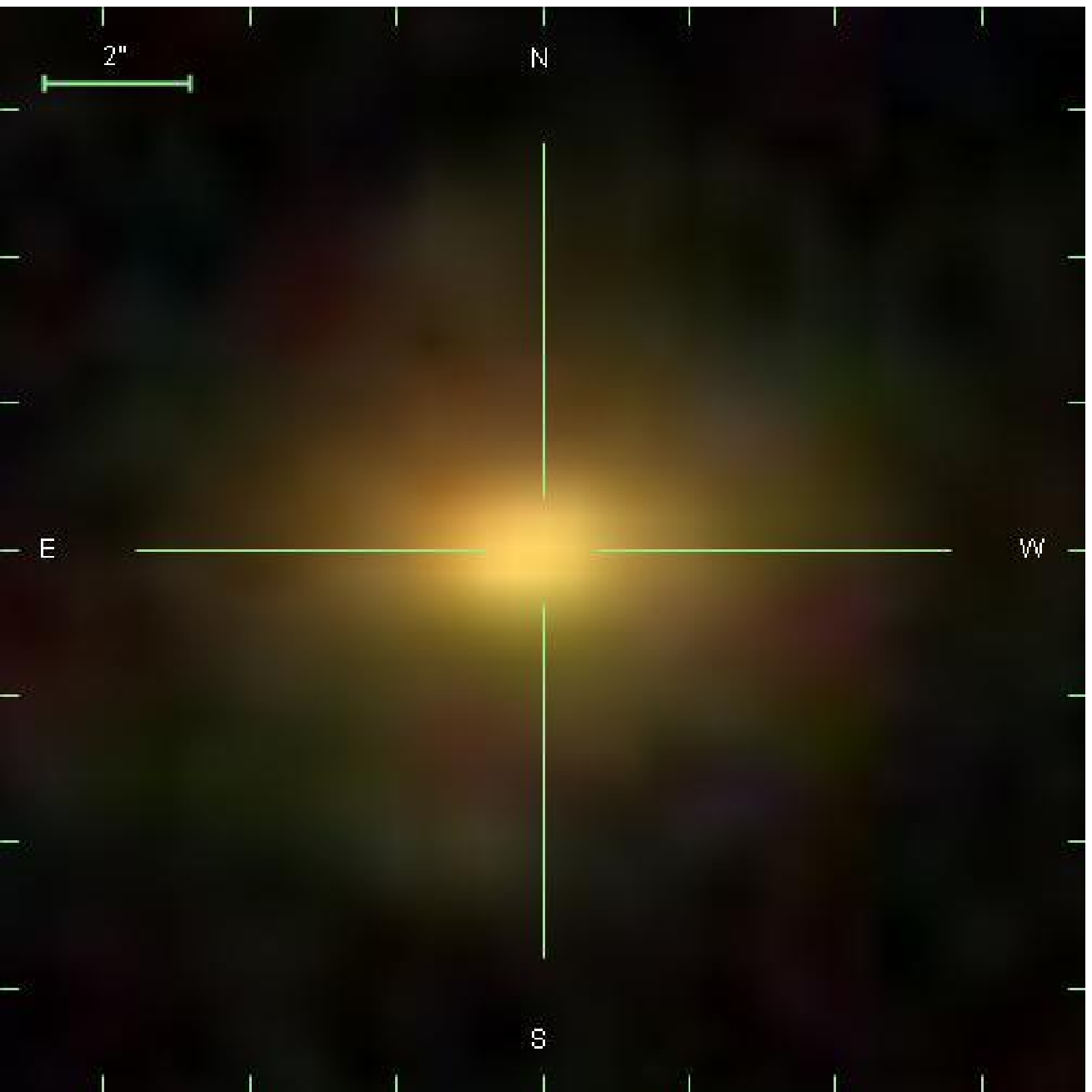}
\caption{SDSS DR10 colour thumbnails for our 76 compact massive galaxy candidates. They are arranged by their internal ID with galaxy 1 in upper left corner and then in ascending order from left to right and top to bottom. B19 is the second galaxy in the top row. The thumbnails show an square area with a side length corresponding to 12 $a_{\textrm{sdss}}$ of the displayed galaxy. There is also a small scale in the top left corner of each image.}
\label{gal_pictures}
\end{center}
\end{figure*}

The main idea behind this paper is to find galaxies with properties comparable to b19 \citep{Lasker:2013} and NGC 1277 \citep{vdBosch:2012} and to investigate whether they are unique objects or not. B19 is characterized by a relatively small scale radius, but a relatively high central velocity dispersion that implies a high dynamical mass for its given radius. 

In the following we define a set of criteria that provides us with galaxies in the same region of the $\textrm{log}_{10}(R_{0})$-$\textrm{log}_{10}(\sigma_{0})$ diagram as b19. The selection criteria have to be restrictive enough that only the most massive and most compact galaxies are included, but still generous enough to include b19. To avoid too much arbitrariness, we used the samples averages and standard deviations as a basis for our definitions. We adopted the following selection criteria: 
\begin{itemize}
\item $\textrm{log}_{10}\left(R_{0}\right)$ < $\overline{\textrm{log}_{10}\left(R_{0}\right)}$ - $\sigma_{\textrm{log}_{10}\left(R_{0}\right)}$
\item $\textrm{log}_{10}\left(\sigma_{0}\right)$ > $\overline{\textrm{log}_{10}\left(\sigma_{0}\right)}$ + 2 $\sigma_{\textrm{log}_{10}\left(\sigma_{0}\right)}$
\item $\textrm{log}_{10}\left(\sigma_{0}\right)$ - $k_{R\sigma} \cdot \textrm{log}_{10}\left(R_{0}\right)$ < $d_{R\sigma}$ + 3 $s_{\epsilon,R\sigma}$.
\end{itemize}

The first criterion means that the logarithm of the physical radius $R_{0}$ has to be smaller than the sample's average $\overline{\textrm{log}_{10}\left(R_{0}\right)}$ by at least one standard deviation $\sigma_{\textrm{log}_{10}\left(R_{0}\right)}$, which provides us with an upper limit for $R_{0}$ of 2.18 kpc for the de Vaucouleurs fit parameters. The lower limit for the central velocity dispersion $\sigma_{0}$ is requiring by demanding it to be at least two standard deviation $\sigma_{\textrm{log}_{10}\left(\sigma_{0}\right)}$ higher than the mean of the logarithm of the central velocity dispersion $\overline{\textrm{log}_{10}\left(\sigma_{0}\right)}$. This yields a lower limit of $\sigma_{0} = $323.2 km s$^{-1}$.  The last criterion ensures that all candidates are more than three root mean square $s_{\epsilon,R\sigma}$ off from the $\textrm{log}_{10}(R_{0})-\textrm{log}_{10}(\sigma_{0})$ relation: $\textrm{log}_{10}(\sigma_{0})=k_{R\sigma} \cdot \textrm{log}_{10}(R_{0})+ d_{R\sigma}$, for which the coefficients $k_{R\sigma}$ and $d_{R\sigma}$ were obtained by a linear fit to the data points of the basic sample. The selection criteria are illustrated in Figure \ref{R0_sigma_V}.

By applying the above selection criteria to the basic sample, one finds 76 galaxies. All candidates are listed with their basic parameters in Table \ref{list_candiates_basics_dV} and their derived parameters in Table \ref{list_candiates_dV}. B19 itself has the internal ID 2. The others are new compact massive galaxies similar to b19, whose global properties will be investigated over the course of this paper. A set of SDSS thumbnail images for all our candidates is provided in Figure \ref{gal_pictures}.

In Appendix \ref{sersic_cand}, we provide an alternative sample of candidates using the Sersic fit parameters from \citet{Simard:2011} instead of the de Vaucouleurs fit directly from SDSS.
\FloatBarrier
\section{Results}

\label{sec_results}
In this section, we discuss the distribution of our candidates along known scaling relations for early-type galaxies. We compare our sample to the work of \citet{Taylor:2010}, who listed 63 compact massive red-sequence galaxies in a similar redshift range. When cross-matching their sample with our data, we find 60 of their galaxies that are in our basic sample. Another sample of possible low-redshift, compact, massive red-sequence galaxies is the sample of \citet{Trujillo:2009}, which contains 29 such galaxies, of which we detect 23 in our basic sample. All samples are based on SDSS. In the following, we compare our compact galaxy sample of 76 galaxies to the 60 galaxies that are in both our basic sample and the \citet{Taylor:2010} sample, as well as to the 23 galaxies, which are in both our basic sample and the \citet{Trujillo:2009} sample. For simplicity, we call the 60 galaxies of \citet{Taylor:2010}, which are in our basic sample, the Ta10 sample from here on, and they are listed with their basic and derived parameters from SDSS in Tables \ref{list_candiates_basics_T} and \ref{list_candiates_T}. The 23 galaxies of the \citet{Trujillo:2009}, which are in our basic sample, are called the Tr09 sample from here, and they are listed with their basic and derived parameters from SDSS in Tables \ref{list_candiates_basics_trujillo} and \ref{list_candiates_trujillo}.

A comparison of the Ta10 sample with our compact galaxy sample reveals that they only have five galaxies in common (see Table \ref{ID_crossmatches}). It is surprising to only find so few galaxies in common with a sample that should be similar to our own. The difference between our candidate sample and the Tr09 is even more striking, since they do not share a single galaxy. Aside from local samples, we compare our candidates also to various samples of intermediate-and high redshift data (see Figs.\ref{M_star_vs_R0_dV} and following). We used the recent intermediate redshift sample of \citet{Zahid:2015}, the classic high redshift sample of \citet{Damjanov:2009}, the new high redshift sample of \citet{Belli:2014}, and the catalogue of \citet{vdSande:2013}, which contains a composition of various high redshift samples, such as \citet{Bezanson:2013}, \citet{vanDokkum:2009}, \citet{Onodera:2012}, \citet{Cappellari:2009}, \citet{Newman:2010}, \citet{vdWel:2008}, \citet{Blakeslee:2006}, \citet{Toft:2012}, and their own work. We cannot perform a comparison with these datasets in every figure, because sometimes some samples do not contain the required parameters.

\subsection{The fundamental plane}
\begin{figure}[ht]
\begin{center}
\includegraphics[width=0.45\textwidth]{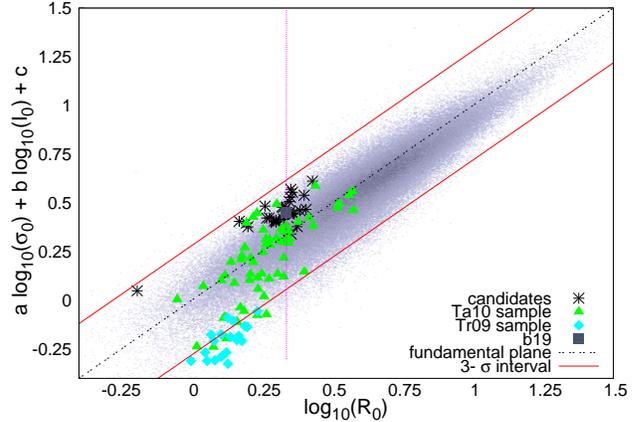}\\ 
\caption{Location of the candidate galaxies on the fundamental plane. The candidates are indicated by black stars. The galaxies belonging to the Ta10 sample are represented using filled green triangles, and the Tr09 sample is denoted by filled cyan diamonds. B19, the starting point of our investigation, is indicated by a filled grey square. The magenta dotted lines show the limiting physical radius used in the sample sample selection. The black dashed lines are the fundamental plane fits from Appendix \ref{newFPparameter} with their corresponding 3-$\sigma$ confidence intervals shown as red solid lines. The fit appears to be slightly offset due to the volume weights used to correct the Malmquist bias in the fitting process.}
\label{fp_dV}
\end{center}
\end{figure}

As illustrated in Figure \ref{fp_dV}, the fundamental plane is a tight relation for early-type galaxies and a good starting point for our investigation. According to \citet{Lasker:2013}, b19 is a clear outlier of the fundamental plane of \citet{Bernardi:2003c}. In contrast to this, we found that b19 is only slightly more than 1-$\sigma$ off the fundamental plane using the new coefficients listed in Appendix \ref{newFPparameter}, which are based on the work of \citet{Saulder:2013}. Furthermore, all candidate galaxies can be found clearly within 3-$\sigma$ of the fundamental plane (see Figure \ref{fp_dV}). Almost all of them are located on the same side above the fundamental plane and are grouped in a similar region. The Ta10 sample is much more distributed over the fundamental plane than our sample. Some galaxies in the Ta10 sample are even beyond the 3-$\sigma$ boundary on the opposite side to the clustering of our candidates. The Tr09 sample forms a relatively tight group around and beyond the 3-$\sigma$ boundary at the opposite side of our candidate sample on the fundamental plane. The Ta10 sample appears to be distributed between the Tr09 sample and our sample, which are opposite extremes of the Ta10 sample distribution.

\subsection{The colour-magnitude diagram}
\begin{figure}[ht]
\begin{center}
\includegraphics[width=0.45\textwidth]{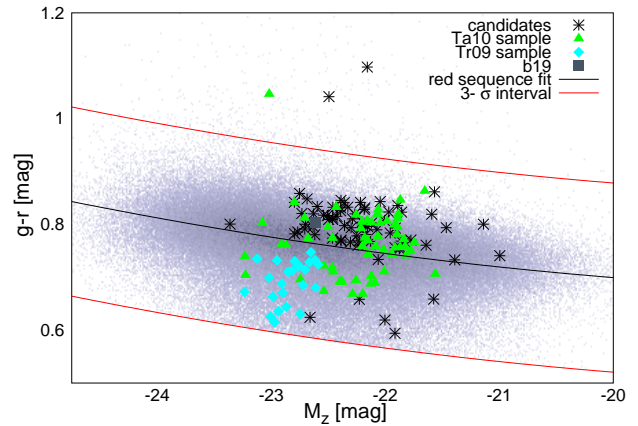}\\ 
\caption{Distribution of the candidate galaxies in a colour-magnitude diagram. The galaxies belonging to the Ta10 sample are represented using filled green triangles, and the Tr09 sample is denoted by filled cyan diamonds. B19 is indicated by a filled grey square. The black dashed line represents the fit on the red sequence performed in \citet{Saulder:2013} with the corresponding 3-$\sigma$ confidence intervals shown as solid red lines.}
\label{CMD_plot}
\end{center}
\end{figure}

In Figure \ref{CMD_plot}, we plot the z band absolute magnitudes vs. the g-r colours. Galaxies in the colour-magnitude plane can generally be divided into two main groups: the red sequence and the blue cloud \citep{Chilingarian:2012}, which are only connected by a relatively sparsely populated `green valley'. While the blue cloud is mainly composed of late-type galaxies, the red sequence mainly consists of early-type galaxies, such as the galaxies discussed in this paper. The selection criteria of our basic sample reduces the galaxies used in this paper to the red sequence. In Figure \ref{CMD_plot} we indicate the red sequence fit from \citet{Saulder:2013} in the g-r colour vs. the absolute z band magnitude plane. 

At a given absolute magnitude M$_z$, the galaxies of our candidate sample are systematically redder than the average red sequence galaxy by about 0.05 mag in g-r. At the same time, except for two outliers, all galaxies are well within 3-$\sigma$ limits of the overall distribution, and there are only a few galaxies, which are blue than the average red sequence galaxy. Most of the galaxies of the Ta10 sample are also redder than the average and are associated to the grouping of galaxies from the candidates (see Figure \ref{CMD_plot}). The systematic offset of our sample towards redder colours is consistent with a higher stellar metallicity than that of the average early-type galaxy at the same luminosity. All galaxies of the Tr09 sample are bluer than the average red sequence galaxy, which contrasts with our candidate sample.

\subsection{The mass-size relations}
\begin{figure*}[ht]
\begin{center}
\includegraphics[width=0.90\textwidth]{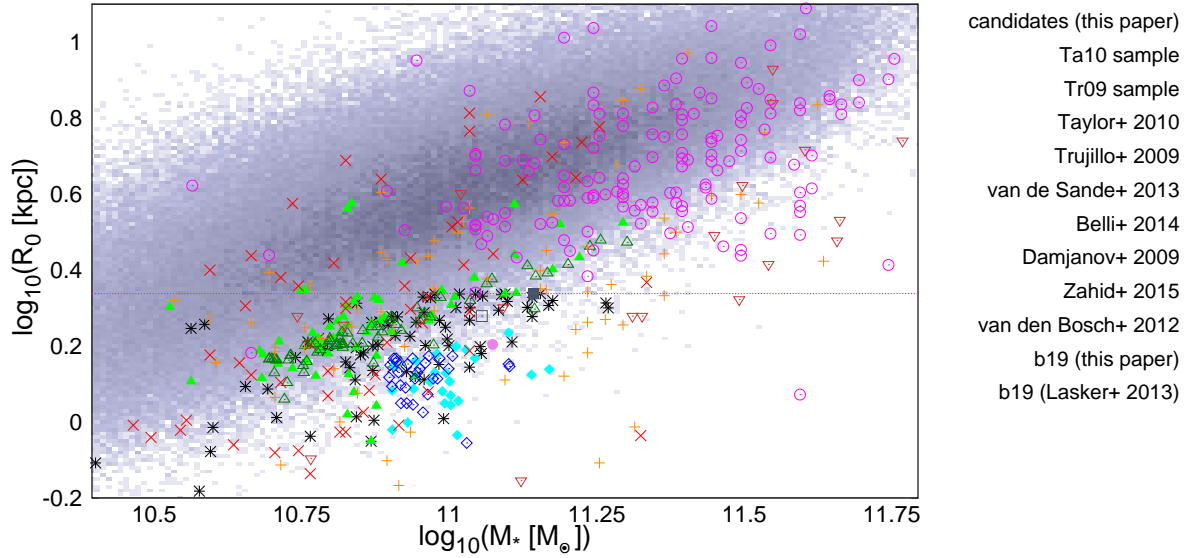}\\ 
\caption{Stellar mass-size relation for our basic sample and several other samples of compact massive early-type galaxies from the literature in comparison to our own data. The blueish cloud indicates the early-type galaxies of our basic sample. The black stars represent the candidates of our sample. The galaxies of the Ta10 sample are shown using filled green triangles and the galaxies of the Tr09 sample are denoted by filled cyan diamonds. The galaxies from \citet{Taylor:2010}, using the values of their paper, are indicated by open dark green triangles. The open blue diamonds represent the galaxies of \citet{Trujillo:2009}. Orange crosses mark the catalogue of various high redshift samples by \citet{vdSande:2013}. The high redshift sample of \citet{Belli:2014} is indicated by red Xs. Open brown nabla symbols mark the high redshift galaxies of \citet{Damjanov:2009}. Open magenta circles indicate the intermediate redshift sample of \citet{Zahid:2015}. NGC 1277 of \citet{vdBosch:2012}, which is the only galaxy of their sample for which we have a stellar mass is represented by an filled violet circle. b19 using our calibration of SDSS data is shown by a filled grey square and b19 using the calibration of \citet{Lasker:2013} is indicated by an open grey square. Because we use the values available in the literature to mark the positions of the galaxies in this plot, one has to consider potential systematics, especially in the effective radius $R_{0}$, which was measured in different filters by different authors. The dashed magenta line marks the limiting scaling radius for our sample selection.}
\label{M_star_vs_R0_dV}
\end{center}
\end{figure*}

\begin{figure*}[ht]
\begin{center}
\includegraphics[width=0.90\textwidth]{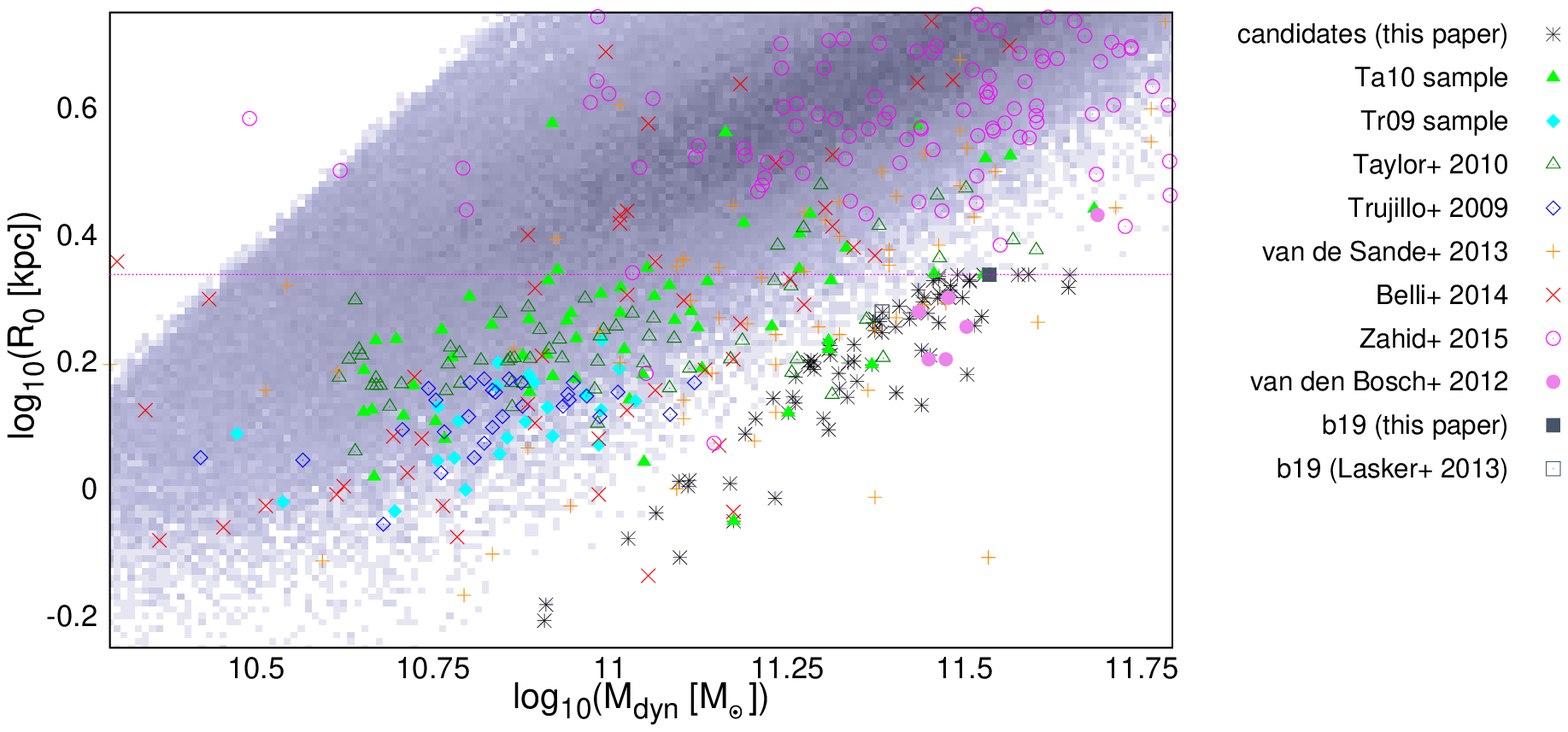}\\ 
\caption{Dynamical mass-size relation for our basic sample and several other samples of compact massive early-type galaxies from the literature in comparison to our own data. The blueish cloud indicates the early-type galaxies of our basic sample. The black stars represent the candidates of our sample. The galaxies of the Ta10 sample are shown using filled green triangles and the galaxies of the Tr09 sample are marked by filled cyan diamonds. The galaxies from \citet{Taylor:2010} using the values of their paper are indicated by open dark green triangles. The open blue diamonds represent the galaxies of \citet{Trujillo:2009}. Orange crosses mark the catalogue of various high redshift samples by \citet{vdSande:2013}. The high redshift sample of \citet{Belli:2014} is indicated by red Xs. We do not have dynamical masses for the high-redshift sample of \citet{Damjanov:2009}. However we can calculate dynamical masses for the intermediate-redshift sample of \citet{Zahid:2015}, which is indicated by open magenta circles. The six galaxies of \citet{vdBosch:2012} are represented by filled violet circles. b19 using our calibration of SDSS data is shown by a filled grey square and b19 using the calibration of \citet{Lasker:2013} is indicated by an open grey square. Because we use the values available in the literature to mark the positions of the galaxies in this plot, one has to consider potential systematics, especially in the effective radius $R_{0}$, which was measured in different filters by different authors. The dashed magenta line denotes the limiting scaling radius for our sample selection.}
\label{M_dyn_vs_R0_dV}
\end{center}
\end{figure*}

In Figures \ref{M_star_vs_R0_dV} and \ref{M_dyn_vs_R0_dV}, we plot the stellar masses and the dynamical masses, respectively, against the physical radii of the galaxies. Relations between the size and the mass of dynamically hot stellar systems are frequently used to distinguish them into different classes, as done, for example, in \citet{Misgeld:2011} for dwarf galaxies vs. star clusters. Also bulges, large elliptical galaxies and similar objects can be found in very distinct areas of a mass-size diagram. The galaxies in which we are interested in this paper are early-type galaxies with small radii and relatively high masses. These galaxies are located on the edge of the so-called zone of exclusion \citep{Burstein:1997,Misgeld:2011,Norris:2014}. This zone is empirically defined by a limit of stellar mass beyond which (most) hot stellar systems cannot grow at fixed sizes.

In Figure \ref{M_star_vs_R0_dV}, one finds, in contrast to the previous figures, that there seems to be rough overall agreement on the distribution of our galaxies and the galaxies from \citet{Taylor:2010} as well as our galaxies and the galaxies from \citet{Trujillo:2009}. We found that the galaxies from the Ta10 sample tend to contain less stellar mass for their sizes than our candidates. In contrast, the galaxies of the Tr09 sample tend to be more compact for their stellar masses than most of our galaxies. Almost all galaxies are at the edge of the distribution, as expected. When plotting the dynamical mass instead of the stellar mass against the scale radius (see Figure \ref{M_dyn_vs_R0_dV}), the Ta10 and the Tr09 samples are detached from our candidates again. Since Figure \ref{M_dyn_vs_R0_dV} is basically a rescaled and tilted version of the selection criteria (see Figure \ref{R0_sigma_V}), because of the definition of the dynamical mass (see Equation \ref{dynmass}), it highlights the differences in the sample selection between this work and \citet{Taylor:2010} as well as \citet{Trujillo:2009}, who used stellar masses, when compared to Figure \ref{M_star_vs_R0_dV}. Over the course of this paper, we found that our selection criteria yield a more cohesive sample than the Ta10 sample or the Tr09 sample. The sample of \citet{Zahid:2015} apparently contains many galaxies with larger radii than the low redshift samples. Most galaxies in the various high redshift samples can be found in areas close to our candidates and the other low redshift sample. They are located close to the edge of the zone of exclusion. A more detailed discussion of the differences between our sample and the samples of various other authors can be found in Sections \ref{othersamples_local} and \ref{othersamples_high}.

\subsection{The mass-to-light ratio}

Since the starting point of our investigation, b19, is said \citep{Lasker:2013} to have a bottom-heavy initial mass function, the mass-to-light ratio $\vernal$ will contain valuable information for us. We investigated both the dynamical mass-to-light ratio $\vernal_{\textrm{dyn}}$ and the stellar mass-to-light ratio $\vernal_{\textrm{*}}$. The dynamical mass-to-light ratio is derived directly from measured SDSS parameters using Equation \ref{dynmass}, while the stellar mass-to-light ratio requires some additional modelling to derive the stellar masses, which was done by \citet{Mendel:2014}. They used a stellar population synthesis to derive the stellar masses from spectral energy distributions based on the SDSS broadband photometry.

\begin{figure}[ht]
\begin{center}
\includegraphics[width=0.45\textwidth]{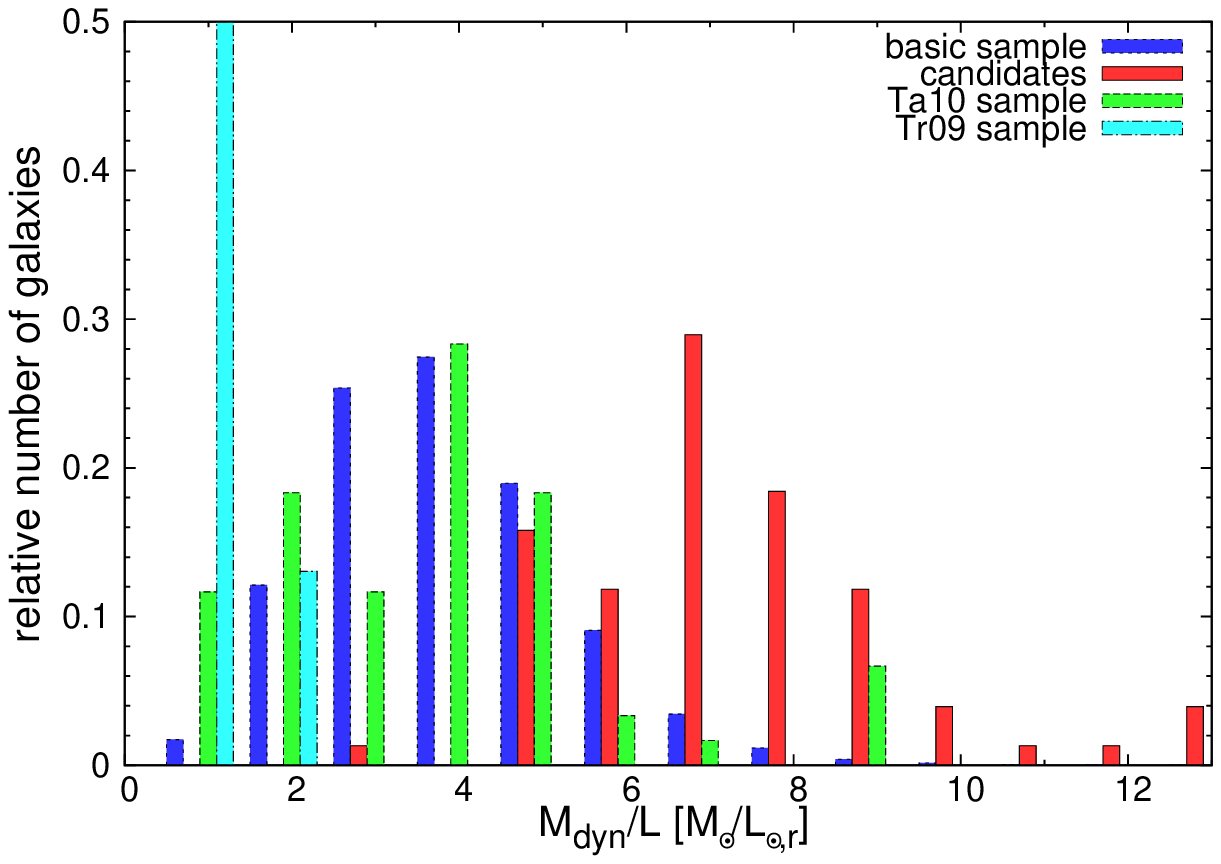}\\ 
\caption{Distribution of the dynamical mass-to-light ratios $\vernal_{\textrm{dyn}}$. The blue histogram corresponds to our basic sample, which consists of early-type galaxies alone.The green histogram represents the Ta10 sample, while the cyan histogram corresponds to Tr09 sample. The red histogram indicates our 76 candidates.}
\label{ml_dyn_ratiosdist_dV}
\end{center}
\end{figure}
\begin{figure}[ht]
\begin{center}
\includegraphics[width=0.45\textwidth]{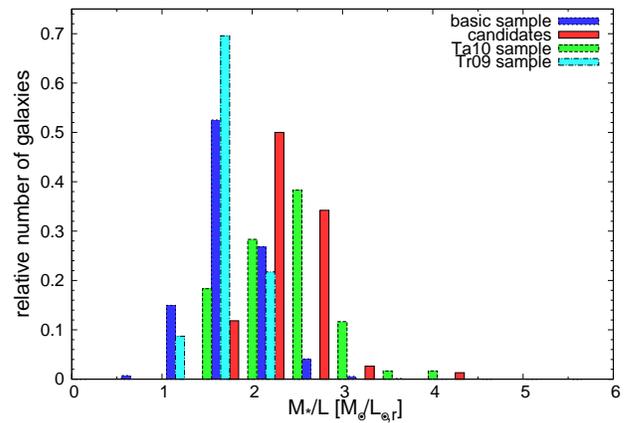}\\ 
\caption{Distribution of the stellar mass-to-light ratios $\vernal_{*}$. The blue histogram corresponds to our basic sample, which consists of early-type galaxies alone. The green histogram represents the Ta10 sample, while the cyan histogram corresponds to Tr09 sample. The red histogram indicates our 76 candidates.}
\label{ml_star_ratiosdist_dV}
\end{center}
\end{figure}

Figures \ref{ml_dyn_ratiosdist_dV} and \ref{ml_star_ratiosdist_dV} illustrate the distribution of the r band dynamical and stellar mass-to-light ratios, respectively, for our candidate sample, the Ta10 sample, the Tr09 sample, and the basic sample. Comparing the mass-to-light ratios of our candidates to the basic sample, we found them clearly elevated. 
The average dynamical mass-to-light of the basic sample is 3.75$\pm$0.46 $M_{\astrosun}/L_{\astrosun ,\textrm{r}}$, and the average stellar mass-to-light of the basic sample is 2.07$\pm$0.20 $M_{\astrosun}/L_{\astrosun ,\textrm{r}}$. The average dynamical mass-to-light ratio of our candidate sample is 7.60$\pm$2.45 $M_{\astrosun}/L_{\astrosun ,\textrm{r}}$, which is about twice the number of the average of the basic sample. Also the average stellar mass-to-light ratio of our candidates is with, 2.66$\pm$0.38 $M_{\astrosun}/L_{\astrosun ,\textrm{r}}$, notably higher than the one of the basic sample. The average mass-to-light ratios of the Ta10 sample are, however, relatively close to the averages of the basic sample with a $\vernal_{\textrm{dyn}}$ of 3.81$\pm$1.98 $M_{\astrosun}/L_{\astrosun ,\textrm{r}}$ and a $\vernal_{\textrm{*}}$ of 2.27$\pm$0.51 $M_{\astrosun}/L_{\astrosun ,\textrm{r}}$. The average mass-to-light ratios of the Tr09 sample are extremely low: $\vernal_{\textrm{dyn}}$ = 1.15$\pm$0.31 $M_{\astrosun}/L_{\astrosun ,\textrm{r}}$ and $\vernal_{\textrm{*}}$ = 1.60$\pm$0.24 $M_{\astrosun}/L_{\astrosun ,\textrm{r}}$.

\begin{figure}[ht]
\begin{center}
\includegraphics[width=0.45\textwidth]{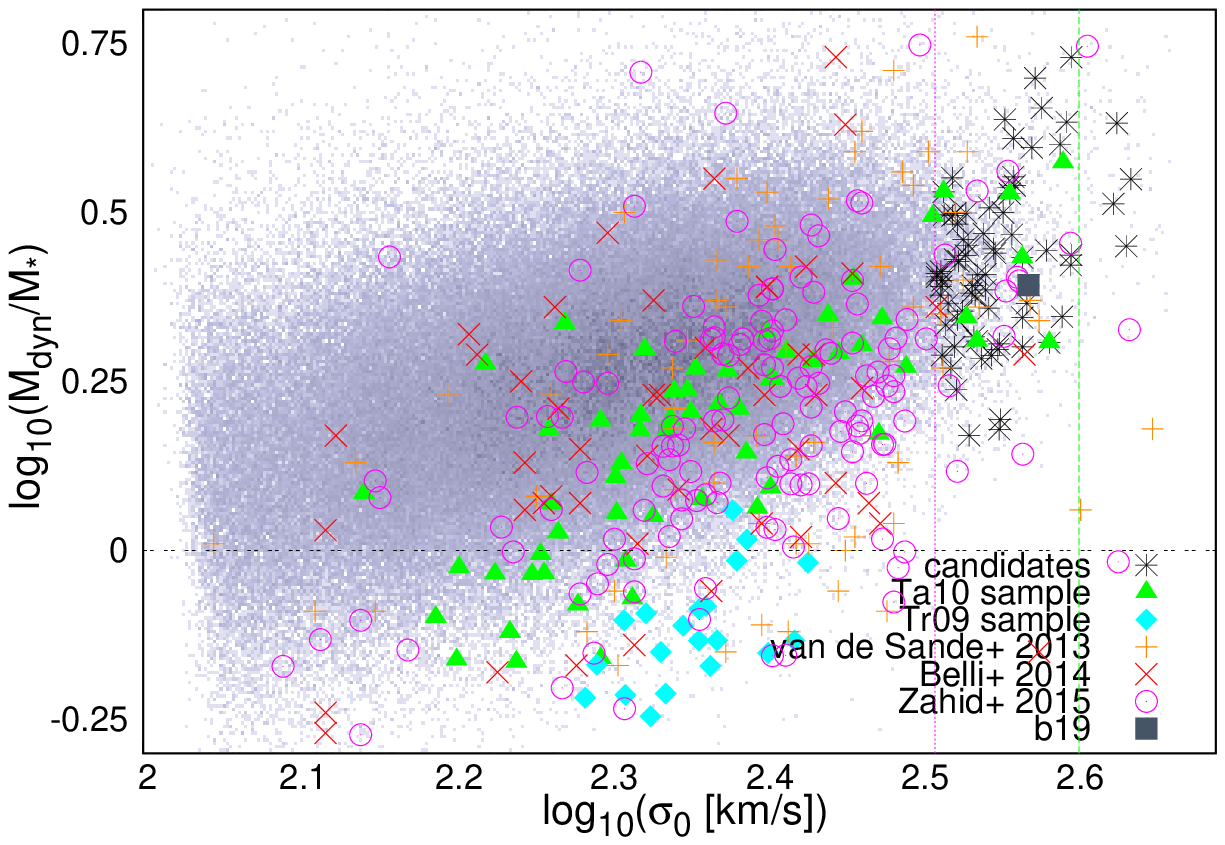}\\ 
\caption{Dependence of the dynamical mass $M_{\textrm{dyn}}$ to stellar mass $M_{*}$ ratio on central velocity dispersion $\sigma_{0}$. The candidates are indicated by black stars. The galaxies belonging to the Ta10 sample are represented using filled green triangles, and the Tr09 sample is indicated by filled cyan diamonds. Orange crosses indicate the catalogue of various high redshift samples by \citet{vdSande:2013}. The high redshift sample of \citet{Belli:2014} is indicated by red Xs. The intermediate redshift sample of \citet{Zahid:2015} is indicated by open magenta circles. B19 is indicated by a filled grey square. The magenta dashed line marks the limiting scaling central velocity dispersion for our sample selection. The green dashed line indicates a central velocity dispersion of 400 km/s. The area below the black dashed line is considered to be unphysical, because $M_{*}$ would exceed $M_{\textrm{dyn}}$.}
\label{sigma_mstar_mdyn_dV}
\end{center}
\end{figure}

\citet{Conroy:2013} show that there is an increasing difference between the dynamical and the stellar mass-to-light ratio for compact galaxies with higher central velocity dispersion using the same data (fits from \citet{Simard:2011} and stellar masses from \citet{Mendel:2014}). They argue that this indicates a systematic variation in the initial mass function. In Figure \ref{sigma_mstar_mdyn_dV}, we plot the ratio of dynamical over stellar mass against velocity dispersion. The increase in this ratio with increasing velocity dispersion is clearly visible. The area in Figure \ref{sigma_mstar_mdyn_dV} below a logarithm of the dynamical-mass-to-stellar-mass ratio of zero is only sparsely populated, and most galaxies in that region are consistent with a log ratio of 0 thanks to measurement uncertainties (0.15 dex for the stellar masses according to \citet{Mendel:2014}). The Ta10 sample is scattered widely over the general distribution with some galaxies even in the forbidden area, while our candidates form the high end in Figure \ref{sigma_mstar_mdyn_dV} owing to our selection criteria. We found that the galaxies of our sample have a $M_{\textrm{dyn}}$ to $M_{*}$ ratio as one might expect for galaxies with such high $\sigma_{0}$, following the general trend of the galaxy distribution in Figure \ref{sigma_mstar_mdyn_dV}. Galaxies of the basic sample with a central velocity dispersion between 323.2 and 400.0 km/s have  $\textrm{log}_{10}(M_{\textrm{dyn}}/M_{*})=$0.432, which is almost the same value as our candidates with $\textrm{log}_{10}(M_{\textrm{dyn}}/M_{*})=$0.441. The vast majority of the Tr09 sample have $M_{\textrm{dyn}}$ to $M_{*}$ ratios below one and are thus located in a zone of exclusion, indicating possible problems in the measurement of the stellar masses of these galaxies. Although the galaxies of the various intermediate and high redshift samples are scattered widely the distribution of our basic sample, there is a tendency toward higher central velocity dispersion, but few of them reach values as high as our candidates.

\begin{figure}[ht]
\begin{center}
\includegraphics[width=0.45\textwidth]{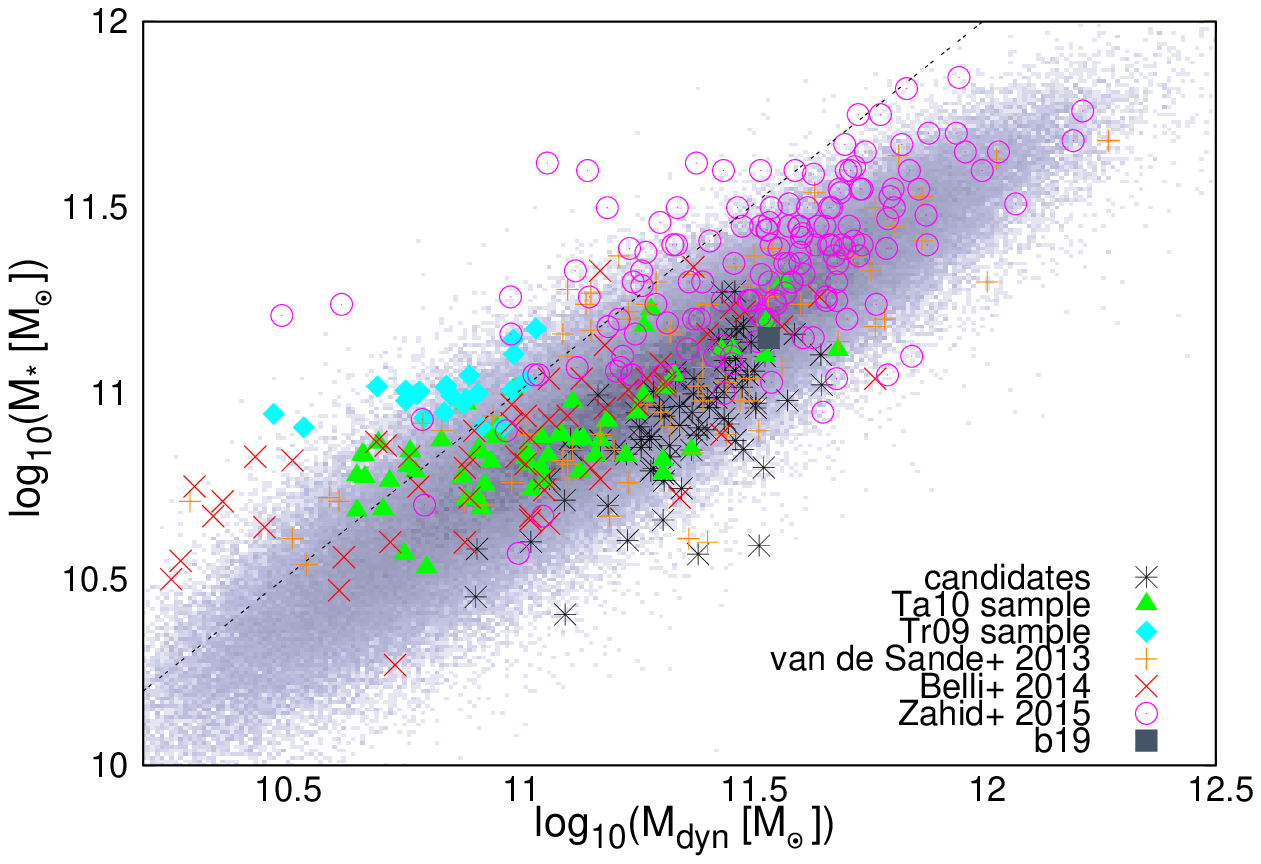}\\ 
\caption{Distribution of the sample's galaxies in the dynamical mass $M_{\textrm{dyn}}$ vs. stellar mass $M_{*}$ plane. The candidates are indicated by black stars. The galaxies belonging to the Ta10 sample are represented using filled green triangles, and the Tr09 sample is indicated by filled cyan diamonds. Orange crosses indicate the catalogue of various high redshift samples by \citet{vdSande:2013}. The high redshift sample of \citet{Belli:2014} is indicated by red Xs. The intermediate redshift sample of \citet{Zahid:2015} is indicated by open magenta circles. B19 is indicated by a filled grey square. The magenta dashed line marks the limiting scaling central velocity dispersion for our sample selection. The black dashed line marks the limit of the $M_{\textrm{dyn}}$ to $M_{*}$ ratio, which is still considered to be physical, because above it $M_{*}$ would exceed $M_{\textrm{dyn}}$.}
\label{masses_dV}
\end{center}
\end{figure}

In Figure \ref{masses_dV}, we plot the distribution of our sample in the dynamical mass vs. stellar mass plane. The difference between our own sample and the Ta10 sample becomes very clear in Figure \ref{masses_dV}. While the Ta10 sample has several objects with lower dynamical to stellar mass-to-light ratios, our galaxies are in general more massive in both dynamical and stellar mass, and they show a tendency for elevated dynamical mass compared to their stellar mass, as already shown in Figure \ref{sigma_mstar_mdyn_dV}. The Tr09 sample is again concentrated in what is normally a zone of exclusion, with formally higher stellar than dynamical masses.

\subsection{Sersic indices}
\begin{figure}[ht]
\begin{center}
\includegraphics[width=0.45\textwidth]{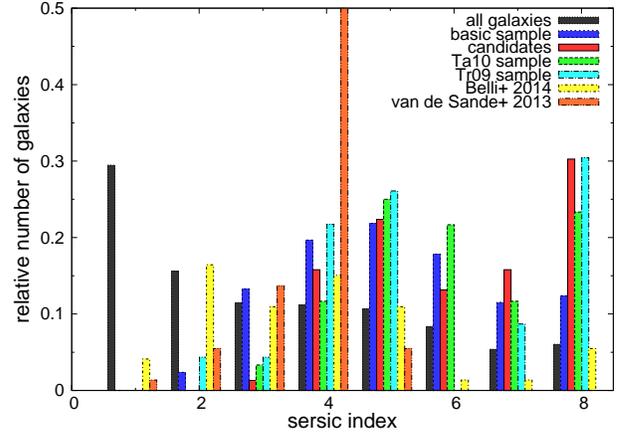}\\ 
\caption{Distribution of the Sersic indices of different samples of galaxies. The black histogram corresponds to all galaxies in SDSS DR7 for which \citet{Simard:2011} did their refits. The blue histogram corresponds to our basic sample, which consists of early-type galaxies alone. The green histogram represents the Ta10 sample, and the cyan histogram the Tr09 sample. The yellow histogram corresponds to sample of \citet{Belli:2014} and the orange histogram indicated the distribution of the sample of \citet{vdSande:2013}. The red histogram shows our 76 candidates.}
\label{sersic_distribution_dV}
\end{center}
\end{figure}

Thanks to the SDSS refits from \citet{Simard:2011}, we have Sersic profiles for almost all SDSS galaxies. In Figure \ref{sersic_distribution_dV}, we compare the Sersic indices of different samples. It should be pointed out that the algorithm used by \citet{Simard:2011} only allows a maximum Sersic index n=8, so we have some clustering around this value for all samples. In Figure \ref{sersic_distribution_dV} there is a clear difference between our basic sample, which only consists of red sequence galaxies because of the colour cut and GalaxyZoo classification \citep{GalaxyZoo_data} used in its selection, and the sample of all SDSS DR7 galaxies, that qualified for the refits done by \citet{Simard:2011}, which thereby consists of a mixed population. The early-type galaxies have clearly higher Sersic indices than the full SDSS DR7 sample. The Ta10 sample and our 76 candidates do not show any significant peculiarities compared to the distribution of the Sersic indices of the basic sample, except for a less smooth distribution due to small number statistics and a weak trend toward higher Sersic indices. The Tr09 sample shows an outstanding peak in its distribution around a Sersic index of 4.5, but is otherwise in agreement with our basic sample. The sample of \citet{Belli:2014} shows a preference for lower Sersic indices in contrast to all other samples.

\subsection{Semi-axis ratios}
\begin{figure}[ht]
\begin{center}
\includegraphics[width=0.45\textwidth]{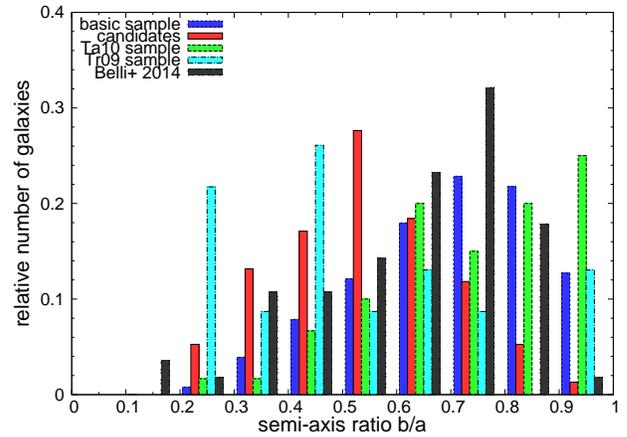}\\ 
\caption{Distribution of the semi-minor to semi-major axis ratios $q_{b/a}$ of different samples of galaxies. The blue histogram corresponds to our basic sample. The green histogram represents the Ta10 sample, and the cyan histogram the Tr09 sample. The black histogram corresponds to sample of \citet{Belli:2014}. The red histogram shows our 76 candidates.}
\label{bin_ba_ratios}
\end{center}
\end{figure}
We investigate the distribution of the semi-minor to semi-major axis ratios $q_{b/a}$ of our sample and the samples that we used for comparison. A ratio of 1 indicates a perfectly round galaxy, and the value decreases to zero for increasing ellipticity of the galaxy. Our candidates are clearly more elongated than the galaxies of our basic sample. The Ta10 sample appears to be rounder than the basic sample, but the Tr09 contains several galaxies with high ellipticity (higher than our sample), but also a larger number of very round galaxies than our sample. The sample of \citet{Belli:2014} seems to contain on average rounder galaxies than our candidate sample.

\section{Discussion}

\label{sec_discussion}
We started our investigation with b19, which is known to be a very compact massive elliptical galaxy in SDSS in the local universe \citep{Lasker:2013}. In this paper, we identify a sample of 76 galaxies (including b19) that have similar global properties, with dispersions of $\sigma_{0} > $323.2 km s$^{-1}$ and sizes smaller than $R_{0} < $2.18 kpc. These selection criteria, which are described in detail in section \ref{sec_candsel}, place these objects at the edge of the $\textrm{log}_{10}(R_{0})$-$\textrm{log}_{10}(\sigma_{0})$.

\subsection{Global properties}
Our candidate sample forms a relatively homogeneous group in most of the scaling relations and parameter spaces we investigated. The observed parameters (see Table \ref{list_candiates_basics_dV}), such as the observed apparent magnitude $m_{\textrm{sdss}}$, the angular semi-major axis $a_{\textrm{sdss}}$, and the central velocity dispersion $\sigma_{\textrm{sdss}}$, are in a range where SDSS measurements are reliable (statistical error of less than 4$\%$). 

We found that all our candidate galaxies are located within the 3-$\sigma$ limits of the fundamental plane using the new coefficients listed in Appendix \ref{newFPparameter}. They occupy  the same corner of the overall distribution of the galaxies on the fundamental plane (see Figure \ref{fp_dV}), indicating that our candidates are more compact than the average galaxy with similar surface brightness and central velocity dispersion. 

Our candidates are also redder than the average early-type galaxies, suggesting metallicity enhancements compared to less compact galaxies at the same luminosity. The vast majority of them can be found above our fit on the red sequence in the CMD diagram (see Figure \ref{CMD_plot}), but still within the 3-$\sigma$ limits (except for two very red outliers).

When analysing the mass-size relations, we found that the galaxies of the candidate sample are located within or close to the zone of exclusion \citep{Burstein:1997,Misgeld:2011,Norris:2014}. This is a direct consequence of the sample's definition, because we were looking specifically for galaxies in this area. We wanted the most massive galaxies for their given small sizes, and Figures \ref{M_star_vs_R0_dV} and \ref{M_dyn_vs_R0_dV} illustrate nicely that we got them. One should keep in mind that the sizes may carry an additional systematic uncertainty, because the intrinsic angular sizes of our galaxies are close to the size of the SDSS PSF.

We found that both the stellar and the dynamical mass-to-light ratios of our candidates are elevated compared to the rest of the sample. Figure \ref{sigma_mstar_mdyn_dV} illustrates that there is an increasing difference between the stellar masses and the dynamical masses with increasing central velocity dispersion $\sigma_{0}$ as reported in \citet{Conroy:2013}. Since we specifically selected for high central velocity dispersion galaxies, the candidates belong to the high $\sigma_{0}$ tip of this correlation. Our galaxies have not only high stellar and dynamical masses for their small sizes, but also high dynamical masses for their given stellar mass as illustrated in Figure \ref{masses_dV}. 

As shown in Figure \ref{sersic_distribution_dV}, our sample does not have any significant difference in their Sersic indices compared to regular early-type galaxy. Figure \ref{bin_ba_ratios} shows that our candidate galaxies have a higher ellipticity than the galaxies of the basic sample. In the context of the SAURON results \citep{Krajnovic:2008,Cappellari:2007,Emsellem:2007}, we interpret these results such that most of our our candidates are lenticular galaxies that host a significant disc component. This result is consistent with the observation that red nuggets are disc-dominated \citep{vdWel:2011,Chevance:2012}.

\subsection{Comparison to the Ta10 sample}
The most important difference between our candidate sample and the sample of \citet{Taylor:2010} is the way in which it was selected. \citet{Taylor:2010} selected their sample using a colour cut demanding that their galaxies be redder than than 2.5 in $^{0.1}$(u-r)\footnote{This denotes the SDSS u-r band colour at a redshift of 0.1}, and they also restricted their sample to have a minimum stellar mass of $10^{10.7}M_{\astrosun}$. Furthermore, their galaxies have to be located between the redshifts of 0.066 and 0.12, while our candidate sample allows redshifts up to 0.4, although we did only detect one galaxy beyond 0.2. When cross-matching their 63 galaxies with our own basic sample, we found 60 galaxies. These 60 galaxies, the so-called Ta10 sample, are listed with their parameters in Tables \ref{list_candiates_basics_T} and \ref{list_candiates_T}. There are only five galaxies that are shared between the Ta10 sample and our candidate sample (see Table \ref{ID_crossmatches}), the galaxies with the internal IDs 6, 13, 39, 52, and 53. We attribute the difference between our candidates and the Ta10 sample to the different selection criteria and want to point out that most of our galaxies tend to contain more stellar mass and definitely more dynamical mass for their sizes than the galaxies of the Ta sample. 

In general, we found that the galaxies of the Ta10 sample occupy different regions in the various considered scaling relations and diagrams than our candidate galaxies. Furthermore, they are a less homogeneous sample than the galaxies presented in this paper. In a stellar mass-size diagram (see Figure \ref{M_star_vs_R0_dV}), they occupy a very similar corner to our candidates, although they are less massive for their size than a large portion of our galaxies. Their distribution in this specific diagram is restricted to a small region, which is a consequence of the selection criteria for this sample. Figure \ref{M_star_vs_R0_dV} also contains the positions in the stellar mass-size plane of all galaxies of \citet{Taylor:2010} using the parameters of their paper. They occupy a similar area in that plot. Comparing Figures \ref{M_star_vs_R0_dV} and \ref{M_dyn_vs_R0_dV} highlights the difference between the Ta10 sample and our candidates owing to their selection criteria. We found that the galaxies of the Ta10 sample largely behave like average early-type galaxies, while our candidates always occupy off-average regions in the parameter space, because they are in the extreme tail of the general distribution of early-type galaxies. We note, though, that there is a small overlap of the parameter range occupied by the Ta10 sample with our sample, as seen in the various plots provided in this paper.

\subsection{Comparison to the Tr09 sample}

While the Ta10 sample still has a small overlap with the parameter range of our candidates, the Tr09 sample behaves totally differently from our candidate sample in most scaling relations. The Tr09 does not have a single galaxy in common with our sample or the Ta10 sample.

The galaxies of the Tr09 are bluer than almost all our candidates (see Figure \ref{CMD_plot}), related to their indeed being bluer than the average red sequence galaxies, and some are close to the green valley. This is certainly connected to the younger stellar ages that \citet{Trujillo:2009} derived for their sample of 29 ``superdense massive galaxies''. 

The Tr09 sample occupies a different region on the fundamental plane (see Figure \ref{fp_dV}) around and beyond the $3-\sigma$ limit, rendering their galaxies outliers on the opposite side to our sample. The galaxies of Tr09 sample have stellar masses comparable to our galaxies (see Figure \ref{M_star_vs_R0_dV}), but lower dynamical masses (see Figure \ref{M_dyn_vs_R0_dV}), which are comparable to those of the Ta10 sample. This and the fact that they have very low mass-to-light ratios (see Figures \ref{ml_dyn_ratiosdist_dV} and \ref{ml_star_ratiosdist_dV}) indicate a potential problem. As illustrated in Figure \ref{sigma_mstar_mdyn_dV} and more clearly in Figure \ref{masses_dV}, the galaxies of the Tr09 sample appear to contain more stellar mass than dynamical mass, which hints at stellar population peculiarities in these objects. The stellar masses that we used have uncertainties of about $40\%$. We know that these galaxies are very young (about 2 Gyr \citep{Trujillo:2009}). It seems reasonable to assume that the contrast between the young Tr09 sample and our candidate galaxies may indicate that our sample contains old objects, which are true survivors of the red nuggets from the early universe.

\subsection{Comparison to other local samples}
\label{othersamples_local}
Aside from the comparison to the Ta10 and the Tr09 samples, we cross-matched our data with other samples of compact massive early-type galaxies as well. A visual comparison of the location of compact massive early-type galaxies from different authors on the stellar mass-size plane is provided in Figure \ref{M_star_vs_R0_dV}. We only have the stellar mass of one galaxy in the sample of \citet{vdBosch:2012}, namely NGC 1277, which is located next to the bulk of our sample and b19 in the plot, and it is even one of the denser objects of our sample. None of our candidates has been covered by the HETMGS \citep{vdBosch:2015}. The other samples using galaxies from the local universe (ours, \citet{Trujillo:2009}, and \citet{Taylor:2010}) occupy a distinct region in the stellar mass-size plane at the edge or within the zone of exclusion \citep{Burstein:1997,Misgeld:2011,Norris:2014}, but with only a small overlap between the individual samples. 

As a visual comparison of the different samples in the dynamical mass-size plane, Figure \ref{M_dyn_vs_R0_dV} shows that the samples of \citet{Taylor:2010} (the Ta10 sample and the one with the parameters from their paper) and \citet{Trujillo:2009} (the Tr09 sample and the one with the parameters from their paper) are distributed in a large area of the plot partially overlapping. Although small, they are clearly less massive than the galaxies of our sample. The six galaxies of \citet{vdBosch:2012} are within the compact distribution of our candidates in the dynamical mass-size plane, which agrees with our intention to find galaxies similar to them and b19. In Figures \ref{M_star_vs_R0_dV} and \ref{M_dyn_vs_R0_dV}, there may be small systematic deviations between the different samples because the effective radii were measured in different filters. While our sample and the Ta10 samples uses the r band, the sample of \citet{Taylor:2010} was measured in the i band, the sample of \citet{Trujillo:2009} was measured in the z band and the Tr09 sample uses the r band, and the sample of \citet{vdBosch:2012} was measured in the K band. Our candidates in the redder SDSS bands are smaller than what we measured in the r band by an average of 0.30 kpc ($\sim 14\%$) in the z band and 0.11 kpc ($\sim 5\%$) in the i band. This suggests that at least the points of the samples of \citet{Taylor:2010} and \citet{Trujillo:2009} are shifted downwards a little bit in relation to the r band measured points of our sample, the Ta10 sample, and the Tr09 sample in Figures \ref{M_star_vs_R0_dV} and \ref{M_dyn_vs_R0_dV}.

Since b19 was first analysed in \citet{Bernardi:2008}, we have cross-matched our candidate sample with their list of 43 massive early-type galaxies. Although they also selected their sample by high central velocity dispersion, the only other galaxy in common with our candidate sample, aside from b19, is b17, which has the internal ID 3. However, they do not impose any restriction on the effective radii of their sample, and only a fraction of our sample can match their minimum $\sigma_{0}$ of 350 km/s. Furthermore, the redshifts of most of their galaxies are higher than of the galaxies in our sample, and small galaxies will not be resolved and detected any more at this distance. The sample of \citet{Bernardi:2008} is also a subset of SDSS DR1 \citep{SDSS_DR1}, which covered a much smaller area of the sky than SDSS DR10, which we used. An interesting result of their paper was that almost half (20 out of 43) of their sample of high-$\sigma_{0}$ galaxies are either superposition of two or more galaxies, which is something to consider in the light of follow-up observations.

\subsection{Comparison to intermediate and high redshift samples}
\label{othersamples_high}

We compared our candidates and the other low redshift samples to various intermediate and high redshift samples. We found quite some difference between them and the local samples, but also between the different high redshift samples themselves. Our candidate sample corresponds best to the subsample of the catalogue of \citet{vdSande:2013} with high dynamical masses.

We used a very recent sample of compact massive galaxies at intermediate redshifts of \citet{Zahid:2015}, which is based on COSMOS results \citep{Damjanov:2015}. Their galaxies are in general more massive, but also less compact than ours. A large number of them are located at the edge of or within the zone of exclusion, as can been seen in comparison to our basic sample in Figure \ref{M_star_vs_R0_dV}. They seem to form an extension of our sample and the other low redshift sample, but restricted to higher masses and larger radii. As illustrated in Figure \ref{M_dyn_vs_R0_dV}, the intermediate redshift sample of \citet{Zahid:2015} contains galaxies of similar (and also higher) dynamical masses to our candidates, but they are  generally less compact than our galaxies or other low redshift samples. In Figure \ref{masses_dV}, the galaxies of \citet{Zahid:2015} tend to be more massive than ours and behave like the more massive galaxies of the basic sample. In Figure \ref{sigma_mstar_mdyn_dV}, they show a tendency to higher central velocity dispersion than does the basic sample, but few of them reach values that are as high as ours.

We used a variety of high redshift samples for comparison, such as the classic high redshift sample of \citet{Damjanov:2009}, the new high redshift sample of \citet{Belli:2014}, and the catalogue of \citet{vdSande:2013}, which contains a composition of various high redshift samples, such as \citet{Bezanson:2013}, \citet{vanDokkum:2009}, \citet{Onodera:2012}, \citet{Cappellari:2009}, \citet{Newman:2010}, \citet{vdWel:2008}, \citet{Blakeslee:2006}, \citet{Toft:2012}, and their own work. In the size-stellar mass plane (see Figure \ref{M_star_vs_R0_dV}), all the high redshift samples are distributed along the edge of the zone of exclusion with some objects deeper in to it than any of the low redshift samples. The sample of \citet{Belli:2014} consists of more compact and lower mass objects to which the low redshift samples match best. Although the \citet{vdSande:2013} sample is distributed over a wider range of masses, a significant fraction of it shares the same areas with our low redshift samples. The sample of \citet{Damjanov:2009} mainly consists of higher mass galaxies, which are on average deeper in the zone of exclusion than the other high redshift samples. Things look differently in the size-dynamical mass plane (see Figure \ref{M_dyn_vs_R0_dV}). The galaxies of \citet{Belli:2014} take less extreme positions than our candidates in this diagram. However, the \citet{Belli:2014} galaxies agree well with the low redshift samples of \citet{Taylor:2010} and \citet{Trujillo:2009}. The sample of \citet{vdSande:2013} contains galaxies with higher dynamical masses and many of its most massive and most compact members agree well with our candidate sample and the sample of \citet{vdBosch:2012}. 

When analysing Figure \ref{masses_dV}, we get a similar result. The sample of  \citet{Belli:2014} agrees well with Ta10 sample, while a large number of galaxies from the \citet{vdSande:2013} catalogue can be found around our candidates. We have to bear in mind that \citet{vdSande:2013} report a decrease in $M_{*}/M_{\textrm{dyn}}$ over time, which would explain the shift of our candidates in relation to most of the high redshift sample. In Figure \ref{sigma_mstar_mdyn_dV}, the high redshift samples are all over the place. There are some tendencies towards higher central velocity dispersions for them, but only a few galaxies (from the \citet{vdSande:2013} sample) posses as high values as our candidates. The distribution of the Sersic indices of the \citet{Belli:2014} sample is different from the low redshift sample, since it favours lower values for the Sersic indices as illustrated in Figure \ref{sersic_distribution_dV}. The Sersic indices of the \citet{vdSande:2013} catalogue strongly peak at four, which is the value for de Vaucouleurs profiles. Since the \citet{vdSande:2013} catalogue is a composite of various high redshift samples, where not all of them performed a Sersic model fit, but only a de Vaucouleurs fit, this result is not surprising. As illustrated in Figure \ref{bin_ba_ratios}, the galaxies of the high redshift sample of \citet{Belli:2014} are also slightly rounder than our candidates or the Tr09 sample, but not as round as the galaxies of the Ta10 sample.

After this comprehensive analysis, we conclude that the various high redshift samples do not form a very uniform group and that there are differences between the various samples, which raises the question of whether this is due to systematic differences between selection and fitting methods applied to the samples (discussed in \citet{vdSande:2013}) or to the red nuggets themselves being a relatively diverse population. Furthermore, none of the low redshift samples agrees in every aspect with the high redshift data. However, one has to keep in mind that ten billion years lie between them, in which the red nuggets may have undergone significant changes.

\begin{figure*}[ht]
\begin{center}
\includegraphics[width=0.90\textwidth]{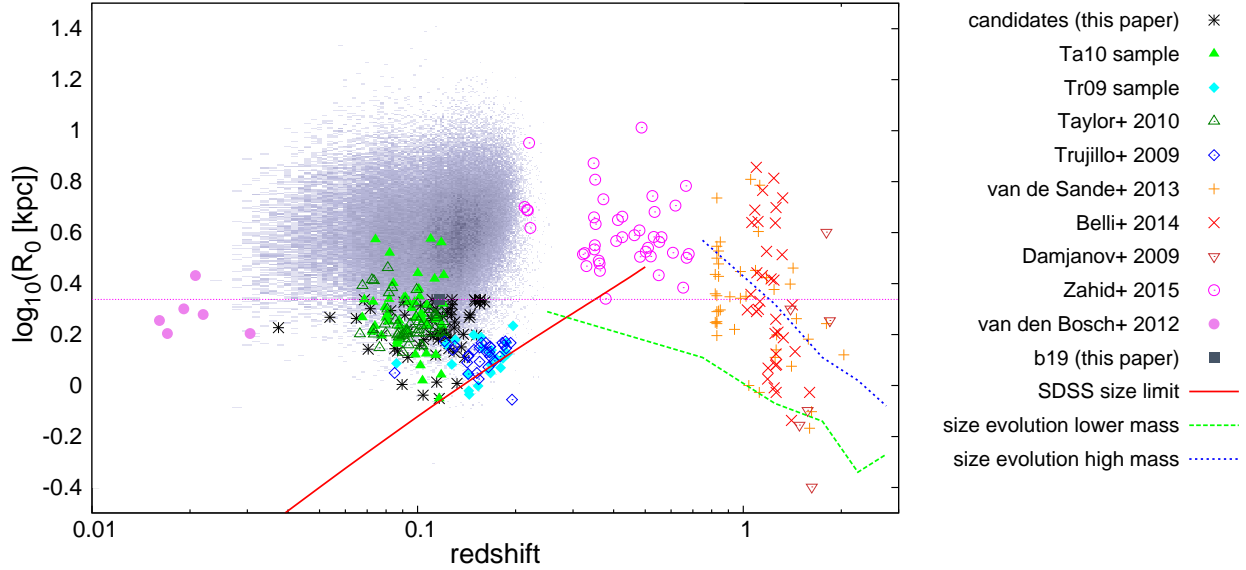}\\ 
\caption{Redshift vs. size distribution for galaxies of different samples within a stellar mass range between  $10^{10.75} M_{\astrosun}$ and $10^{11.25} M_{\astrosun}$. The blueish cloud in the background represents all galaxies of the basic sample with stellar masses greater than $10^{10.75} M_{\astrosun}$ and less than $10^{11.25} M_{\astrosun}$. The black stars represent the candidates of our sample. The galaxies of the Ta10 sample are shown using filled green triangles and the galaxies of the Tr09 sample are indicated by filled cyan diamonds. The galaxies from \citet{Taylor:2010} using the values of their paper are indicated by open dark green triangles. The open blue diamonds represent the galaxies of \citet{Trujillo:2009}. Orange crosses denote the catalogue of various high redshift samples by \citet{vdSande:2013}. The high redshift sample of \citet{Belli:2014} is indicated by red Xs. Open brown nabla symbols indicate the high redshift galaxies of \citet{Damjanov:2009}. Open magenta circles indicate the intermediate redshift sample of \citet{Zahid:2015}. The six galaxies of \citet{vdBosch:2012} are represented by filled violet circles. Using our calibration of SDSS data, b19 is shown by a filled grey square. The magenta dashed line denotes the limiting scaling radius for our sample selection. The red solid line corresponds to the angular resolution limit of SDSS of 0.50 arcsec. The green dashed line denotes the 16-percentile range of the redshift evolution of an early-type galaxy with a stellar masses of $10^{10.75} M_{\astrosun}$ according to Table A1 of \citet{vanderWel:2014a} and the blue dotted line indicates the 16-percentile range of the redshift evolution of an early-type galaxy with a stellar masses of $10^{11.25} M_{\astrosun}$ based on the same work.}
\label{z_vs_size}
\end{center}
\end{figure*}

We explored the connections of our candidate sample to compact galaxies at higher redshifts in Figure \ref{z_vs_size} further, where for reasons of comparability, we plot only galaxies whose stellar masses lie between $10^{10.75} M_{\astrosun}$ and $10^{11.25} M_{\astrosun}$. It is highlighted in the plot that our candidates are indeed amongst the most compact galaxies of the stellar mass range in SDSS and indeed border the resolution limit of SDSS. We adopt a value of 0.50 arcsec for the observed Vaucouleurs radii in the SDSS r band as resolution limit, which is the lower lower 3-$\sigma$-limit of all angular sizes (de Vaucouleurs radii) of early-type galaxies in SDSS. The Ta10 sample and the original values of \citet{Taylor:2010} can both be found in the same region of the plot as our candidates, but at slightly lower redshifts. The \citet{Trujillo:2009} sample (and naturally the Tr09 sample) appears to be around and partly even beyond the resolution limits of SDSS. 

The galaxies of \citet{vdBosch:2012} are much nearer than the galaxies of any other sample and would be beyond the saturation limit of SDSS. The nearest galaxies of our candidate sample can be found close to them. The new intermediate redshift sample of \citet{Zahid:2015} mainly contains larger galaxies than the other samples, but their most compact objects are on the expected evolutionary path of the most compact high redshift objects on the way to our candidates. The high redshift samples of \citet{Damjanov:2009}, \citet{Belli:2014}, and \citet{vdSande:2013} contain many galaxies that are as compact as the most compact galaxies in the local universe, while containing more mass at the same time. We included the redshift evolution of early-type galaxies of the two highest mass bins from \citet{vanderWel:2014a} in our plot. The 16-percentile range of the redshift evolution of an early-type galaxy with a stellar masses of $10^{10.75} M_{\astrosun}$ (only 16$\%$ of all early-type galaxies at this mass range are more compact than indicated by the line) shows a connection between the two of the most compact galaxies of various high redshift samples, the most compact galaxies of the intermediate redshift sample of \citet{Zahid:2015}, and our candidates (if extrapolated to the redshift ranges of our sample). 

The 16-percentile range of the redshift evolution of an early-type galaxy with a stellar masses of $10^{11.25} M_{\astrosun}$ only extends down to a redshift of 0.75 in \citet{vanderWel:2014a}, because for some unknown reason, the value at a redshift of 0.25 for this percentile and mass is missing in their table for circularized radii, but an educated guess based on the other values and other tables would yield a value of $\textrm{log}_{10}(R_{0}) \sim 0.6 $. The stellar masses of almost all of our candidates range between $10^{10.75} M_{\astrosun}$ and $10^{11.25} M_{\astrosun}$ (we only plotted those within that range in Figure \ref{z_vs_size}), which means that when comparing with the extrapolation of the redshift evolution of \citet{vanderWel:2014a}, we found that our candidates are clearly amongst the most compact galaxies of their mass range and that they may be relics of the red nuggets.

\subsection{Space density}
We calculated the space density of candidate sample considering the Malmquist-bias and a resolution limit of 0.50 arcsec (lower 3-$\sigma$-limit). The value that we obtained this way is 4$\cdot 10^{-7}$ galaxies/Mpc${^3}$, which is about 400 times lower than the space density of red nuggets at $z=2$ \citep{Quilis:13}. However, when comparing space densities of different samples, one has to be aware of the selection criteria used to define them and biases affecting them. Our sample only contains one galaxy above a redshift of 0.2, while the rest are clearly below it. This is a consequence of the selection bias from using spectroscopic data from SDSS. Galaxies with very small angular sizes (below the resolution limit mentioned above) are not included in the basic sample, which explains the dearth of compact galaxies at higher redshifts in our sample. Most samples \citep{Trujillo:2009,Taylor:2010,Quilis:13} use stellar masses as a selection criterion. We tried to derive more comparable quantities by considering additional selection criteria for our candidate galaxies, which are similar to the selection criteria of other authors. We restricted our candidate sample to dynamical masses greater than 8 $\cdot 10^{10} M_{\astrosun}$ and physical radii of less than 2 kpc to be better comparable with the predication for old (formed before a redshift of 2) compact massive galaxies by \citet{Damjanov:2014}. Fifty-eight of 76 galaxies in the candidate sample fulfil this condition, and they obtain a space density of $3.6 \cdot 10^{-7}$ galaxies/Mpc${^3}$ for our sample, which is located at redshifts below 0.2. The lowest redshift interval considered in \citet{Damjanov:2014} is 0.2 to 0.3, and they predict a space density of $2.6^{1.2}_{2.4}\cdot 10^{-7}$ galaxies/Mpc${^3}$, which is about seven times higher than ours. 

For our comparison with the space density of \citet{Trujillo:2009}, we applied the same restriction as they do on our candidate sample. The stellar masses have to be higher than 8 $\cdot 10^{10} M_{\astrosun}$, and the z band physical radii smaller than 1.5 kpc. This reduces our candidate sample to merely 16 galaxies, and we obtained a space density for them of $5.9 \cdot 10^{-8}$ galaxies/Mpc${^3}$, which is by more than a factor of two lower than the upper limit of the space density of \citet{Trujillo:2009}, which is $1.3 \cdot 10^{-7}$ galaxies/Mpc${^3}$ \citep{Quilis:13}. Comparing to space densities to the \citet{Taylor:2010} is more difficult due to their definition. We only considered their restriction that the stellar masses have to be higher than $10^{10.7} M_{\astrosun}$. Hence the space density, which we derived using a subsample of 67 galaxies of our candidates, is an upper limit. We obtained a value of $2.7 \cdot 10^{-7}$ galaxies/Mpc${^3}$, which is almost an order of magnitude higher than the value of \citet{Taylor:2010} of $3 \cdot 10^{-8}$ galaxies/Mpc${^3}$ \citep{Quilis:13}. We thus found that the space densities we derived are comparable within an order of magnitude to those of other samples in the local (and intermediate redshift) universe, but we refrain from a strong interpretation of our numbers compared to the high redshift universe given the dominating influence of selection effects on the results.

\subsection{Individual galaxies}
Although the candidate galaxies form a very homogeneous group in all their properties, one can identify particularly peculiar objects by focusing on some individual galaxies. There are seven galaxies in our sample with radii less than a kpc. The smallest one has the internal ID 63 and a physical radius of $R_{0}= (0.62 \pm 0.01)$ kpc. At its redshift of 0.0877, this means that the observed angular semi-major axis $a_{\textrm{sdss}}$ is only $(0.47 \pm 0.01)$ arcsec and therefore at the limit of SDSS resolution. Galaxy 63 also happens to be the second faintest object of our sample, with an r-band absolute magnitude of $(-20.51 \pm 0.01)$, and it has the lowest surface brightness with $(17.12 \pm 0.05)$ mag/arcsec$^{2}$ in the r band. The galaxy also stands out by its mass: it possesses the lowest dynamical mass of our candidates with $M_{\textrm{dyn}} = \left(8.13 \pm 0.38\right) 10^{10} M_{\astrosun}$ and the second lowest stellar mass of our sample with $M_{*} = \left(2.82 \pm 1.16\right) 10^{10} M_{\astrosun}$. By being a peculiarly compact and faint, but also relatively low mass object in our sample, we consider galaxy 63 as one of the most interesting objects for our follow-up observations.

The other extremely small objects in our sample show similar properties to galaxy 63. The faintest object with an absolute r-band magnitude of only $(-20.35 \pm 0.02)$ is, with a physical radius of $R_{0}= (0.78 \pm 0.02)$ kpc, also the third smallest galaxy in the candidate sample. It has an internal ID of 31 and is located at a redshift of 0.0784. Galaxy 31 is with an observed angular semi-major axis $a_{\textrm{sdss}}$ of only $(0.64 \pm 0.02)$ arcsec one of the galaxies close to the resolution limit of SDSS. Although this galaxy has the lowest stellar mass of our candidate sample with $M_{*} = \left(2.57 \pm 1.06\right) 10^{10} M_{\astrosun}$, its dynamical mass $M_{\textrm{dyn}} = \left(1.25 \pm 0.06\right) 10^{11} M_{\astrosun}$ is fairly average for our candidate sample. In combination with the fact that it is the faintest galaxy, this results in a relatively high dynamical mass-to-light ratio of $\vernal_{\textrm{dyn}}=(11.99 \pm 0.82) M_{\astrosun}/L_{\astrosun,\textrm{r}}$. This makes it a promising object for follow-up observations.

At the other extreme of the angular sizes, we have our three best-resolved galaxies with $a_{\textrm{sdss}}$ greater than 3 arcsec. The galaxy with an internal ID of 75 is with an angular semi-major axis of $(3.59 \pm 0.02)$ arcsec not only the apparently biggest object in the sky of our candidate sample, but also the nearest. It is located at a redshift of 0.0260. Galaxy 75 is also outstanding because it has the by far highest dynamical mass-to-light ratio of our sample with $\vernal_{\textrm{dyn}}=(20.56 \pm 2.13) M_{\astrosun}/L_{\astrosun,\textrm{r}}$ and, with 0.51 (which is just over the selection criterion of 0.50), the lowest GalaxyZoo probability $\mathcal{L}_{\textrm{ETG}}$ for a galaxy to be classified as an early type within our sample. A visual inspection of the galaxy (see Figure \ref{gal_pictures}) shows a face-on featureless disc, which differs from the other candidate galaxies. A manual analysis of the SDSS classification of its spectrum confirms our suspicion that this galaxy is no regular early-type galaxy, but rather a starburst galaxy. We therefore consider it a false positive in our sample.

The next largest galaxy in angular size has the internal ID 50, which also happens to be, with a redshift of 0.0374, the second nearest candidate galaxy. It possesses a physical radius of $R_{0}= (1.69 \pm 0.01)$ kpc and an absolute r-band magnitude of $(-21.56 \pm 0.01)$. With a dynamical mass of $M_{\textrm{dyn}} = \left(2.24 \pm 0.05\right) 10^{11} M_{\astrosun}$, a stellar mass of $M_{*} = \left(8.13 \pm 3.35\right) 10^{10} M_{\astrosun}$, and no outstanding mass-to-light ratios, galaxy 50 is an average example of our candidates. Its low redshift allows for easier follow-up observations of this object, so we rank it as one of our priority candidates.

The brightest galaxy in our sample has the internal ID 66 and shines with an absolute magnitude of $(-22.69 \pm 0.01)$ mag in the r band. It is located at a redshift of 0.2018\footnote{It is also the only galaxy beyond a redshift of 0.2 in our candidate sample.} and is therefore the most distant galaxy of our candidates. It has a physical radius of $R_{0}= (2.06 \pm 0.08)$ kpc and is one of the largest candidates. Galaxy 66 possesses a dynamical mass of $M_{\textrm{dyn}} = \left(2.75 \pm 0.20\right) 10^{11} M_{\astrosun}$, and it stands out with the second highest stellar mass of our candidates with $M_{*} = \left(1.86 \pm 0.77\right) 10^{11} M_{\astrosun}$. Another outstanding property of galaxy 66 is its dynamical mass-to-light ratio of $\vernal_{\textrm{dyn}}=(3.03 \pm 0.19) M_{\astrosun}/L_{\astrosun,\textrm{r}}$ is the lowest of our candidate sample, and we found that is also has the fifth lowest stellar mass-to-light ratio with $\vernal_{*}=(2.05 \pm 0.84) M_{\astrosun}/L_{\astrosun,\textrm{r}}$. We consider galaxy 66 a priority candidate for follow-up observations, although its high redshift will make them more difficult, but on the other hand the high luminosity of the galaxy will help a bit.

The galaxy with the highest stellar mass has the internal ID 56. It has a stellar mass of $M_{*} = \left(1.91 \pm 0.79\right) 10^{11} M_{\astrosun}$ and a dynamical mass of $M_{\textrm{dyn}} = \left(2.95 \pm 0.28\right) 10^{11} M_{\astrosun}$. The galaxy with a physical radius of $R_{0}= (2.00 \pm 0.14)$ kpc is also the second most distant candidate with a redshift of 0.1978. It is the second reddest object with a g-r colour of $(1.04 \pm 0.03)$ mag, which makes it one of only two outliers of our candidate sample above the upper $3-\sigma$ limit of the red sequence. Galaxy 56 also stands out for having the highest stellar mass-to-light ratio of our candidate sample with $\vernal_{*}=(4.27 \pm 1.77) M_{\astrosun}/L_{\astrosun,\textrm{r}}$. This and its average dynamical mass-to-light ratio of $\vernal_{\textrm{dyn}}=(6.67 \pm 0.71) M_{\astrosun}/L_{\astrosun,\textrm{r}}$ contrasts it with the previously discussed galaxy 66. Galaxy 56 is a very interesting object and qualifies as priority target for our follow-up observations.

The galaxy with the highest dynamical mass has the internal ID 23. It contains a stellar mass of $M_{*} = \left(1.10 \pm 0.45\right) 10^{11} M_{\astrosun}$ and a dynamical mass of $M_{\textrm{dyn}} = \left(4.57 \pm 0.22\right) 10^{11} M_{\astrosun}$. This is due to its high central velocity dispersion of $\sigma_{0}=(423 \pm 17)$ km s$^{-1}$ and its relatively large physical radius $R_{0}= (2.17 \pm 0.04)$ kpc, which is close to the limit of our sample selection. Galaxy 23 has the fourth highest dynamical mass-to-light ratio with $\vernal_{\textrm{dyn}}=(12.70 \pm 0.64) M_{\astrosun}/L_{\astrosun,\textrm{r}}$ and an average stellar mass-to-light ratio of $\vernal_{*}=(2.96 \pm 1.22) M_{\astrosun}/L_{\astrosun,\textrm{r}}$.

Aside from the troublesome galaxy 75, galaxy 30 has the highest dynamical mass-to-light ratio with $\vernal_{\textrm{dyn}}=(13.21 \pm 0.53) M_{\astrosun}/L_{\astrosun,\textrm{r}}$. In contrast to this, its stellar mass-to-light ratio is one of the lowest with only $\vernal_{*}=(1.99 \pm 0.82) M_{\astrosun}/L_{\astrosun,\textrm{r}}$. This agrees with Galaxy 30, which has the third lowest stellar mass of the candidate sample with $M_{*} = \left(3.72 \pm 1.53\right) 10^{10} M_{\astrosun}$. Galaxy 30 has a dynamical mass of $M_{\textrm{dyn}} = \left(2.45 \pm 0.12\right) 10^{11} M_{\astrosun}$. It is worth pointing out that it has neither an extraordinarily high central velocity dispersion of $\sigma_{0}=(346 \pm 11)$ km s$^{-1}$ nor a large physical radius $R_{0}= (1.76 \pm 0.03)$ kpc for the candidate galaxies.

The galaxy with the highest central velocity dispersion is galaxy 76 with $\sigma_{0}=(432 \pm 18)$ km s$^{-1}$. With an absolute r-band magnitude of $(-22.07 \pm 0.01)$, this galaxy is one of the brighter objects in our candidate sample. It is also the second most massive galaxy in our sample in terms of dynamical mass with $M_{\textrm{dyn}} = \left(4.47 \pm 0.21\right) 10^{11} M_{\astrosun}$.

The reddest galaxy in our candidate sample has the internal ID 12. With a g-r colour of $(-1.10 \pm 0.02)$ mag, it is a clear outlier on the red sequence and significantly redder than all galaxies in our candidate sample (except one) and even than most galaxies in the basic sample (see Figure \ref{CMD_plot}). Surprisingly, galaxy 12 is not outstanding in any other parameters than colour, and it appears to be an average member of the candidate sample.

We visually inspected the images of all our candidates (see Figure \ref{gal_pictures}) and find that several have other (foreground or background) objects that are only less than five angular scale radii $a_{\textrm{sdss}}$ from their centre. The galaxies with the internal IDs 18, 24, 25, 33, 35, and 50 have other prominent objects (galaxies or stars) within their immediate vicinity. We have to be careful when using the parameters measured for these galaxy, since there is a chance that their values might suffer from some contamination.

Last but not least, we return to the starting point of our investigation, b19, which has the internal ID 2. It is a member of our candidate sample with rather average properties compared to the other 75 candidates. It is always a central part of the group of data points formed by our candidates in the diagrams. The only feature of b19 that is a little outstanding compared to the other members of the candidate sample is its Sersic parameter $n_{S}$, which borders the maximum allowed value of eight by the algorithm used by \citet{Simard:2011}. This is close to the Sersic index of 6.9 found by \citet{Lasker:2013} based on HST/ACS/HRC imaging.

\section{Summary and conclusions}
\label{summary}

Our sample of 76 candidates (including b19) for compact massive early-type galaxies forms an ideal basis for future follow-up observations using high-resolution spectroscopy and imaging. We found that our homogeneous sample, which has been defined as extreme outliers in the $\textrm{log}_{10}(R_{0})$-$\textrm{log}_{10}(\sigma_{0})$ plane does not behave as do outliers in other relations for early-type galaxies except for those that are directly related to the selection criteria. The candidates seem to form the compact massive tail of the general distribution of early-type galaxies, and are not a separate population of particularly peculiar objects. We confirmed that there is an increase in the dynamical and stellar mass-to-light ratios at higher central velocity dispersions. 

Furthermore, we saw the same tendency as \citet{Conroy:2013} that the difference between the dynamical mass and the stellar mass derived using simple models \citep{Mendel:2014} increases at higher central velocity dispersions. This is usually considered to be an indication of a systematic variation in the initial mass function of those galaxies. As a result our candidates may have an extremely bottom-heavy initial mass function as proposed for b19 \citep{Lasker:2013}. The high central velocity dispersion, which is crucial for determining of the dynamical mass, could also be due to over-massive central black holes \citep{vdBosch:2012,Mieske:2013,Seth:2014}. However, a robust detection of such a black hole is only possible for the nearest objects with high spatial resolution spectroscopy \citep{Lasker:2013, Emsellem:2013, Yildirim:2014}. It is unlikely that all these objects contain such a large black hole. 

We also tried to draw a connection between the compact massive galaxies in our sample and even more massive and more compact galaxies from high redshifts. Based on the previous observation of quiescent high-redshift galaxies \citep{Kriek:2006,Kriek:2008}, \citet{vanDokkum:2008} found that there are already fully formed early-type galaxies with scale radii that are much smaller and stellar masses that are much higher than any object known in the local universe. For comparison, we used several samples of these galaxies at high redshifts, such as those of \citet{Damjanov:2009}, \cite{vdSande:2013}, and \cite{Belli:2014}, and also at intermediate redshifts as in \citet{Zahid:2015}. Many of these galaxies must have evolved, most likely by mergers, into more regular early-type galaxies. The galaxies in our sample are those that still resemble to a specific subgroup of these red nuggets. Some of these exotic galaxies in the early universe possess high dynamical masses and small sizes. From some local galaxies, such as b19 \citep{Lasker:2013} and NGC 1277, and others \citep{vdBosch:2012}, we know similar features, and they are remnants of red nugget galaxies. It is shown in \citet{vdSande:2015} that although the initial mass function of most red nuggets shows a rather shallow slope, the one for those galaxies with high mass-to-light ratios tend to be bottom-heavy. The galaxy b19 is known for its bottom heavy initial mass function and our candidates have an elevated mass-to-light ratio in comparison to normal early-type galaxies. Therefore, one may suspect that some galaxies of our sample are remnants of these ancient objects or that they are at least somehow related to the subgroup of these objects where the galaxies of \citet{Lasker:2013} and \citet{vdBosch:2012} originated. The advantage of our sample is that it is located in the local universe and is thus easier to study than galaxies at high redshifts. 

By selecting galaxies based on stellar velocity dispersion, we found 76 compact, massive early-type galaxies below a redshift of $z=0.2018$. These are excellent targets for further studies of various scientific questions, such as the variation in the initial mass function, over-massive black holes, and potential remnants of exotic galaxies from the early universe. High-resolution imaging data can confirm their sizes and rule out superpositions. Spectroscopy of these objects will enable us to study their dynamical mass distribution and kinematics, and it will allow for (resolved) stellar population analysis.

\section*{Acknowledgments}

Funding for SDSS-III has been provided by the Alfred P. Sloan Foundation, the Participating Institutions, the National Science Foundation, and the U.S. Department of Energy Office of Science. The SDSS-III web site is \url{http://www.sdss3.org/}.

SDSS-III is managed by the Astrophysical Research Consortium for the Participating Institutions of the SDSS-III Collaboration including the University of Arizona, the Brazilian Participation Group, Brookhaven National Laboratory, University of Cambridge, Carnegie Mellon University, University of Florida, the French Participation Group, the German Participation Group, Harvard University, the Instituto de Astrofisica de Canarias, the Michigan State/Notre Dame/JINA Participation Group, Johns Hopkins University, Lawrence Berkeley National Laboratory, Max Planck Institute for Astrophysics, Max Planck Institute for Extraterrestrial Physics, New Mexico State University, New York University, Ohio State University, Pennsylvania State University, University of Portsmouth, Princeton University, the Spanish Participation Group, University of Tokyo, University of Utah, Vanderbilt University, University of Virginia, University of Washington, and Yale University. 

CS acknowledges the support from an ESO studentship. 

CS acknowledges support from the MPIA for hosting him in Heidelberg for a couple of days.

CS acknowledges helpful advise from Jens-Kristian Krogager and fruitful discussions with Ronald L\"asker and Karina Voggel.
\appendix
\FloatBarrier
\section{Updated fundamental plane coefficients}
\label{newFPparameter}
The fundamental plane, which was first mentioned in \citet{Terlevich:1981} and properly defined and discussed \citet{Dressler:1987} and \citet{Djorgovski:1987}, is an empirical relation between three global parameters of elliptical galaxies: the central velocity dispersion $\sigma_{0}$, the physical effective radius $R_{0}$, and the mean surface brightness $\mu_{0}$ within the effective radius. The last parameter is usually expressed as $I_{0}$, which is a renormalized surface brightness $\mu_{0}$: $\textrm{log}_{10} \left( I_{0} \right) = -\frac{\mu_{0}}{2.5}$. The coefficients $a$, $b$, and $c$ are obtained by fitting
\begin{equation}
\textrm{log}_{10}\left(R_{0}\right)=a \cdot \textrm{log}_{10}\left(\sigma_{0}\right) + b \cdot \textrm{log}_{10}\left(I_{0}\right) + c .
\label{fundamentalplane}
\end{equation}

We provide updated values of the fundamental plane coefficients presented in \citet{Saulder:2013}. The main improvements are that we now use SDSS DR10 \citep{SDSS_DR10} instead of SDSS DR8 \citep{SDSS_DR8} and that we do not use any constraints on or information about the SDSS u band, which we found to be quite problematic. Therefore, we have 133~107 galaxies instead of 100~427 for our basic sample (for definitions see \citet{Saulder:2013}), and after all filtering we end up with 119~085 galaxies instead of the 92~953 that are used for the final fit. This again makes it the largest sample ever used for calibrating the fundamental plane so far. In addition to improved fits, which are based on the de Vaucouleurs fit parameters directly from SDSS, we provide new fits for the g and r bands using the Sersic parameters from \citet{Simard:2011}. To this end, we use 121~443 galaxies selected after some 3-$\sigma$ clipping from the basic sample in this paper. 

\begin{figure}[ht]
\begin{center}
\includegraphics[width=0.45\textwidth]{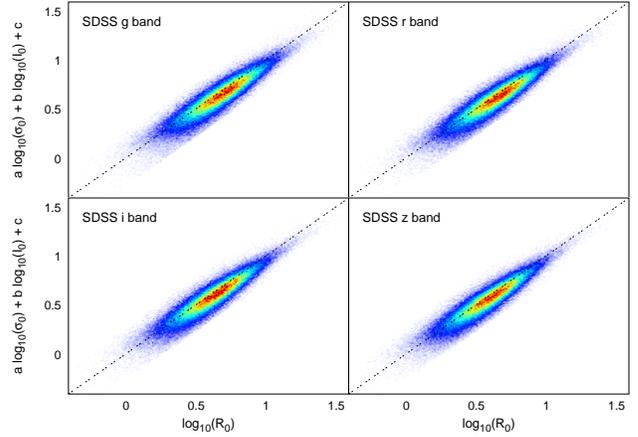}\\ 
\caption{Edge-on projections of the fundamental plane of elliptical galaxies for four different SDSS filters using the de Vaucouleurs fit parameters. The g band is shown in the top left panel and the r band in the top right panels. The bottom left panel displays the i band and the bottom right the z band. The dashed black lines indicate the fundamental plane fits in the corresponding filters.}
\label{fp_all}
\end{center}
\end{figure}
\begin{figure}[ht]
\begin{center}
\includegraphics[width=0.45\textwidth]{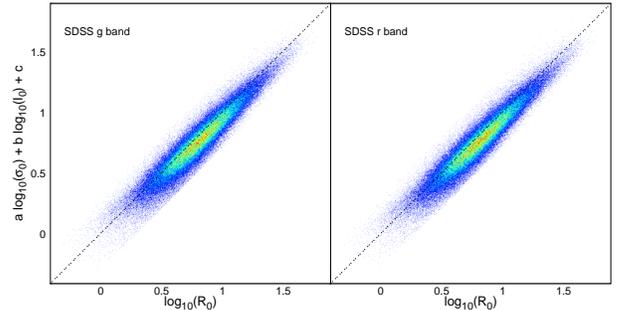}\\ 
\caption{Edge-on projections of the fundamental plane of elliptical galaxies for the SDSS g band (left panel) and r band (right panel) using Sersic fit parameters. The dashed black lines indicate the fundamental plane fits in the corresponding filters. }
\label{fp_all_sersic}
\end{center}
\end{figure}
\begin{table*}[ht]
\begin{center}
\begin{tabular}{c|ccccc}
bands & $a$ & $b$ & $c$ & $s_{\varepsilon}$ & $\bar{\sigma}_{\textrm{dist}}$ [\%] \\ \hline \hline
 g (dV)& $  0.999 \pm   0.026 $  & $ -0.754 \pm  0.011 $  & $  -7.93 \pm   0.10  $ & $  0.0942 $ & $   19.3 $\\
 r (dV)& $   1.070  \pm   0.026 $  & $ -0.770 \pm   0.011 $  & $  -7.98 \pm   0.11   $ & $  0.0935 $ & $   19.0  $\\
 i (dV)& $   1.100   \pm   0.026 $  & $ -0.775 \pm   0.012 $  & $  -7.96\pm   0.11  $ & $  0.0919 $ & $   18.6    $\\
 z (dV)& $   1.145  \pm   0.025 $  & $ -0.781 \pm   0.012 $  & $  -8.02  \pm   0.11  $ & $  0.0920 $ & $   18.5     $\\ \hline
  g (S)& $  0.966 \pm   0.026 $  & $ -0.726 \pm  0.009 $  & $  -7.62 \pm   0.09  $ & $  0.0977 $ & $   20.6 $\\
 r (S)& $   1.029  \pm   0.026 $  & $ -0.729 \pm   0.009 $  & $  -7.56 \pm   0.09   $ & $  0.0972 $ & $   20.4  $\\ 
\end{tabular}
\end{center}
\caption{Results of the best fits for the fundamental plane. The coefficients $a$, $b$, and $c$ for 4 SDSS bands using redshift evolution, volume weights, 3-$\sigma$ clipping, and the radii and magnitudes of de Vaucouleurs (dV) fits are provided in this table. Furthermore, it contains the coefficients for 2 SDSS bands using the same calibration, but the radii and magnitudes from the Sersic (S) fits of \citet{Simard:2011}. The root mean square $s_{\varepsilon}$ of the fits and the relative distance error $\bar{\sigma}_{\textrm{dist}}$ of the fundamental plane are also provided.}
\label{fitparameters}
\end{table*}

Aside from the extended sample, there are a couple of other minor changes and improvements over the old paper \citep{Saulder:2013}. First of all, we corrected a minor mistake in the calculation of the average distance error. This mistake caused the values of the error estimate in the old paper to be systematically lower by a couple of percentage points than they actually are. Even with the slightly larger error, it is still the best fit of the fundamental plane using a large sample at this wave-length range \citep{Bernardi:2003c,Hyde:2009,LaBarbera:2010}. 

Another improvement on the fit is that the volume weights are now considering that the sample only covers a limited redshift from 0.01 to 0.2 (or from 0.05 to 0.4 for the Sersic fits based on the basic sample in this paper). In our previous analysis, the very luminous galaxies were slightly under-represented, because their volume weights assumed a larger volume (the one in which they are theoretically still visible) than the volume of sample (redshift cut at 0.2). We then also subtract the volume corresponding to a redshift of 0.01 from the volume weights, where all galaxies were removed from the sample. The saturation limit of SDSS spectroscopy is also measured and included in the new volume weights by removing the volume associated with it in the same fashion as for the the Malmquest bias limitation. The negligence of these two corrections caused the volume weights of very faint galaxies to be underestimated. Both corrections are relatively tiny, and the new coefficients are only slightly different from the old ones. In particular, the $a$ coefficient is moderately larger, hence closer to the values from the literature (see Table 1 in \citet{Saulder:2013}). The new coefficients are listed in Table \ref{fitparameters}, edge-on projects of the fundamental plane for all four bands used for the de Vaucouleurs fit parameters in the calibration can be found in Figure \ref{fp_all}, and the edge-on projects of the fundamental plane for the g and r SDSS bands using the Sersic fit parameters are displayed in Figure \ref{fp_all_sersic}.
\FloatBarrier

 %\afterpage{\clearpage}
%\FloatBarrier
\section{The Sersic fit sample}

\label{sersic_cand}
\begin{figure}[H]
\begin{center}
\includegraphics[width=0.45\textwidth]{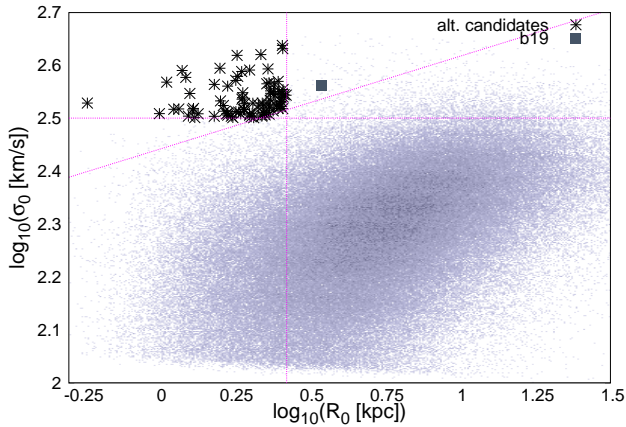}\\ 
\caption{Selection of the alternative candidates in the $R_{0}$-$\sigma_{0}$ plane. The restrictions, which define our alternative candidates, are indicated by the dashed magenta lines. The black stars represent the 85 candidates for galaxies with similar properties in Sersic fit parameters as b19, and b19 itself is represented by a grey filled square in the plot.}
\label{R0_sigma_S}
\end{center}
\end{figure}

In addition to the candidate sample defined using the de Vaucouleurs fit parameters, we provide an alternative sample using the Sersic fit parameters from \citet{Simard:2011}.  

\begin{figure}[htb]
\begin{center}
\includegraphics[width=0.45\textwidth]{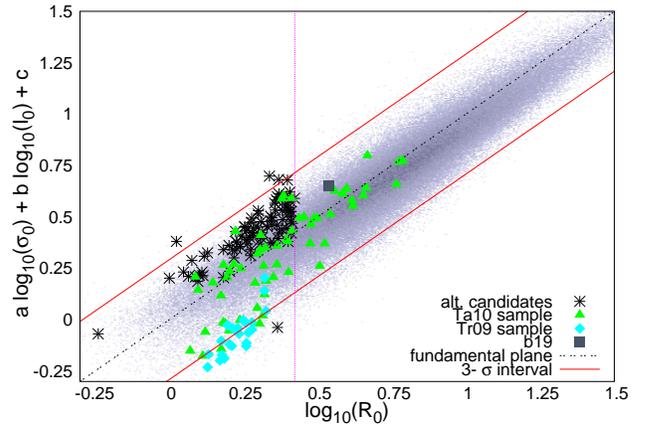}\\ 
\caption{Location of the candidate galaxies on the fundamental plane using Sersic fit parameters. The candidates are indicated by black stars. The galaxies belonging to the Ta10 sample are represented using filled green triangles, and the Tr09 sample is marked by filled cyan diamonds. The starting point of our investigation, b19, is indicated by a filled grey square. The magenta dotted lines show the limiting physical radius used in the sample sample selection. The black dashed lines are the fundamental plane fits from Appendix \ref{newFPparameter} with their corresponding 3-$\sigma$ confidence intervals shown as red solid lines. The fit appears to be slightly offset owing to the volume weights used to correct the Malmquist bias in the fitting process.}
\label{fp_S}
\end{center}
\end{figure}

\begin{figure}[!htb]
\begin{center}
\includegraphics[width=0.45\textwidth]{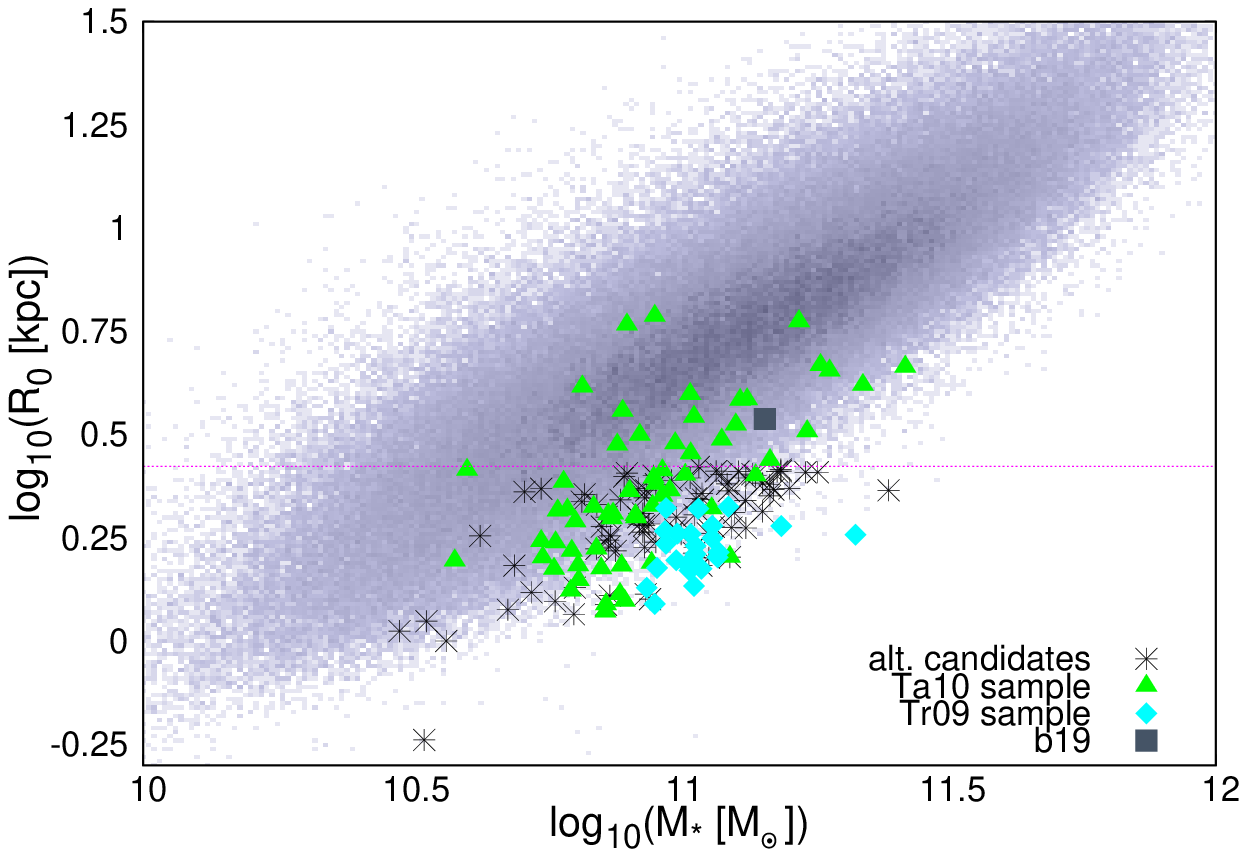}\\ 
\caption{Stellar mass-size relation for our basic sample using the Sersic fit parameters. The alternative candidates are indicated by black stars. The galaxies belonging to the Ta10 sample are represented using filled green triangles, and the Tr09 sample is indicated by filled cyan diamonds. B19 is indicated by a filled grey square. The magenta dashed line denotes the limiting scaling radius for our sample selection.}
\label{M_star_vs_R0_S}
\end{center}
\end{figure}

\begin{figure}[!htb]
\begin{center}
\includegraphics[width=0.45\textwidth]{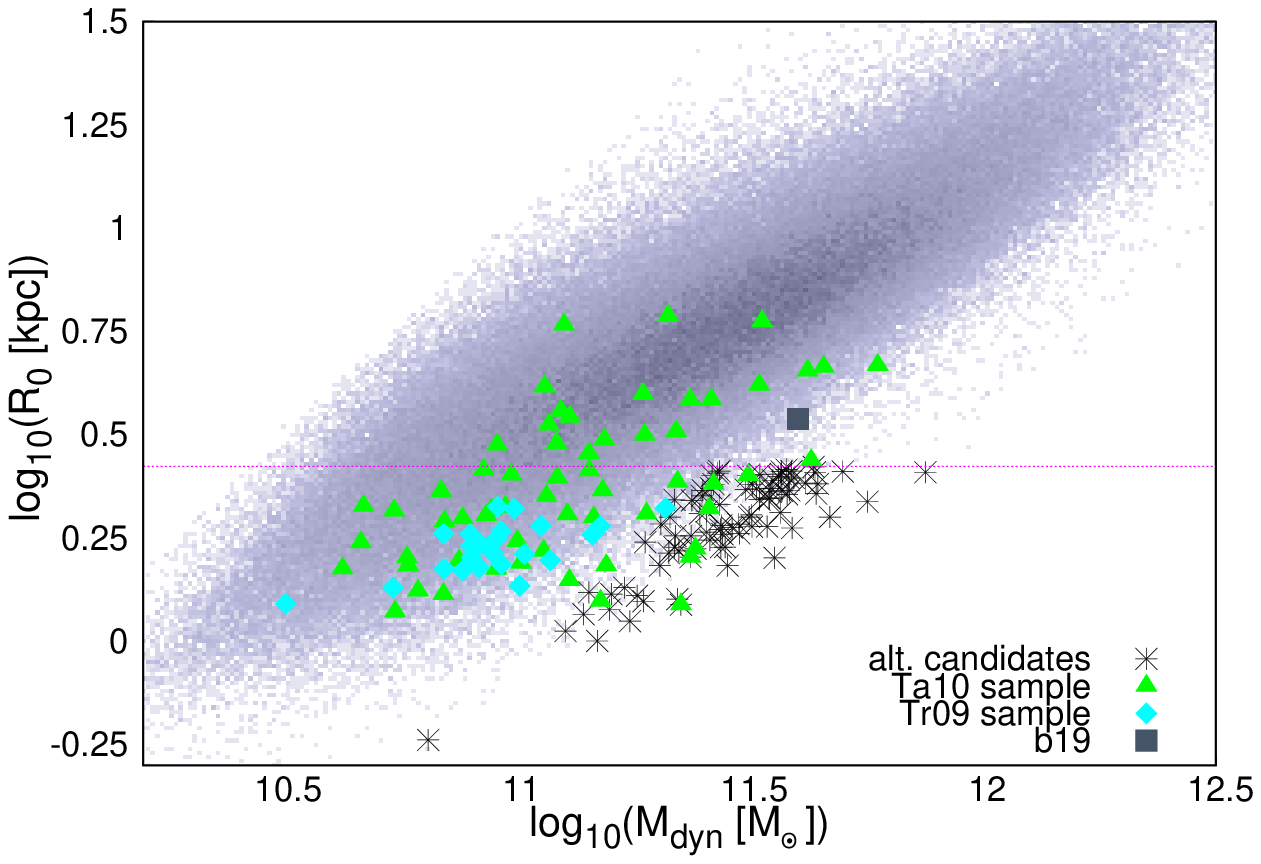}\\ 
\caption{Dynamical mass-size relation for our basic sample using the Sersic fit parameters. The alternative candidates are indicated by black stars. The galaxies belonging to the Ta10 sample are represented using filled green triangles, and the Tr09 sample is indicated by filled cyan diamonds. B19 is indicated by a filled grey square. The magenta dashed line denotes the limiting scaling radius for our sample selection.}
\label{M_dyn_vs_R0_S}
\end{center}
\end{figure}

\begin{figure}[!htb]
\begin{center}
\includegraphics[width=0.45\textwidth]{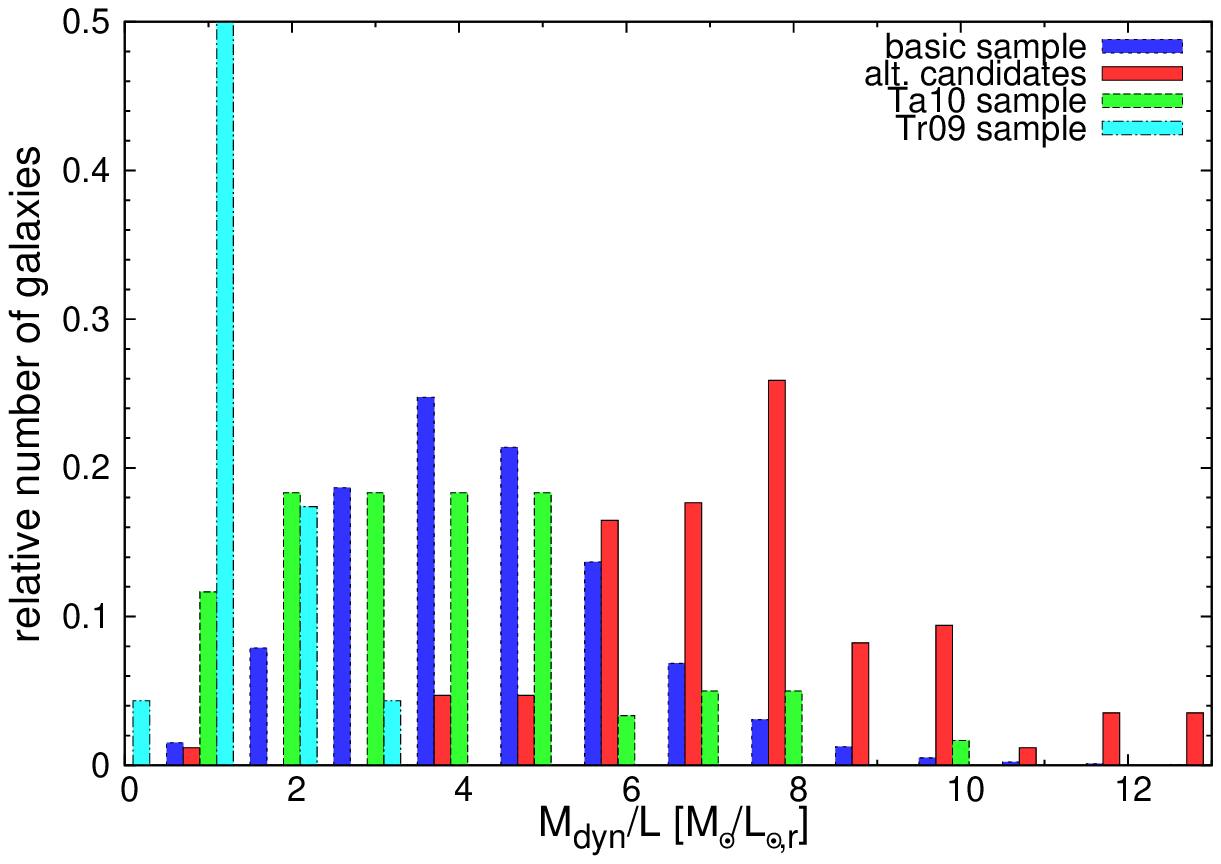}\\ 
\caption{Distribution of the dynamical mass-to-light ratios $\vernal_{\textrm{dyn}}$ using the Sersic fit parameters. The blue histogram corresponds to our basic sample, which only consists of early-type galaxies. The green histogram represents the Ta10 sample, while the cyan histogram corresponds to Tr09 sample. The red histogram denotes our 85 alternative candidates.}
\label{ml_dyn_ratiosdist_S}
\end{center}
\end{figure}

\begin{figure}[!htb]
\begin{center}
\includegraphics[width=0.45\textwidth]{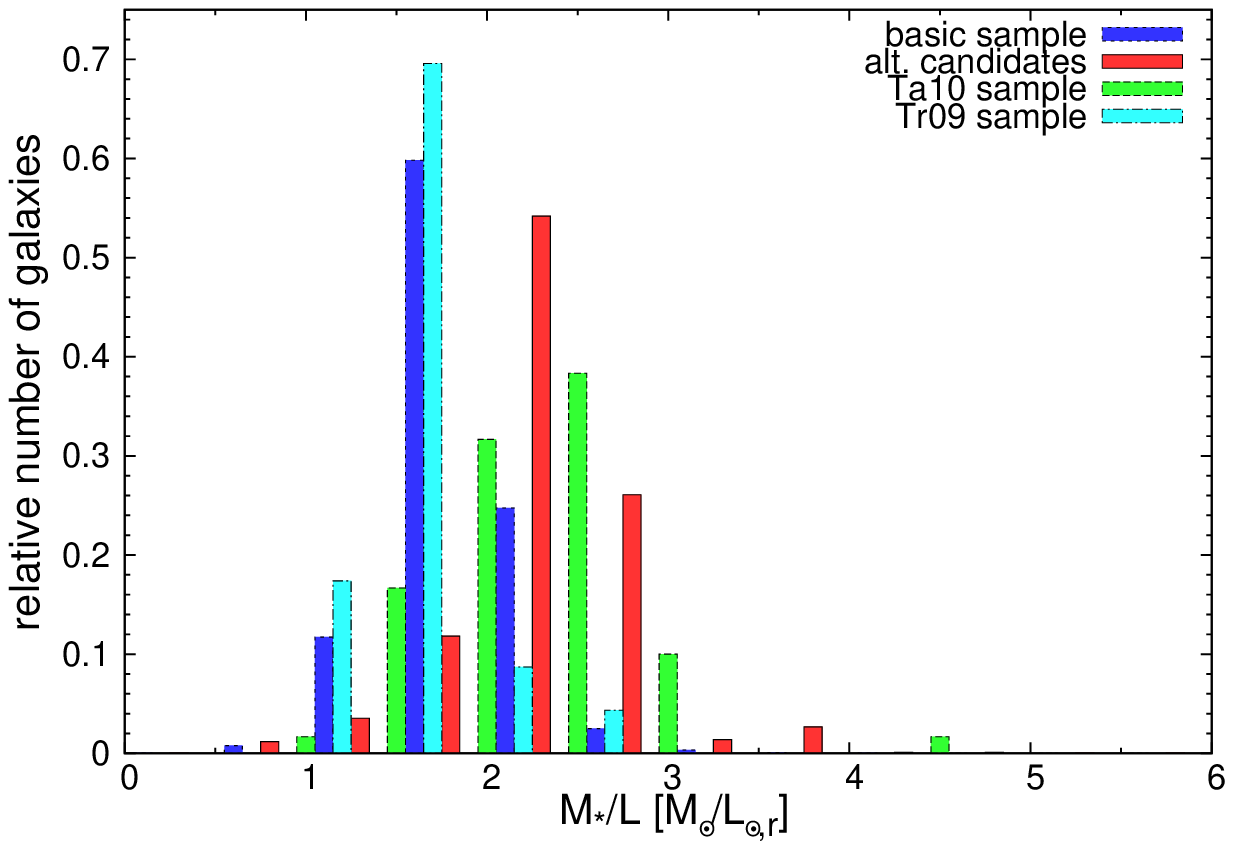}\\ 
\caption{Distribution of the stellar mass-to-light ratios $\vernal_{*}$ using the Sersic fit parameters. The blue histogram corresponds to our basic sample, which only consists of early-type galaxies. The green histogram represents the Ta10 sample, while the cyan histogram corresponds to Tr09 sample. The red histogram denotes our 85 alternative candidates.}
\label{ml_star_ratiosdist_S}
\end{center}
\end{figure}

\begin{figure}[!htb]
\begin{center}
\includegraphics[width=0.45\textwidth]{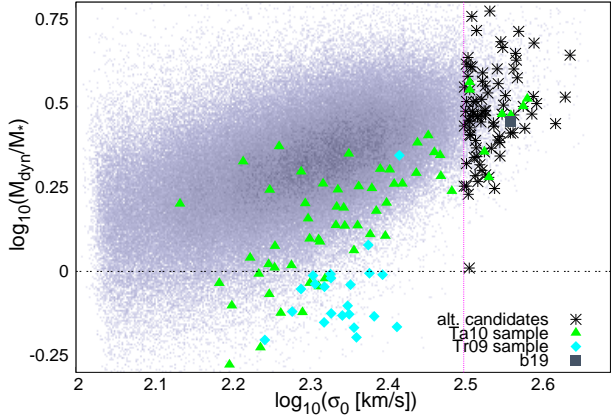}\\ 
\caption{Dependence of the dynamical mass $M_{\textrm{dyn}}$ to stellar mass $M_{*}$ ratio on central velocity dispersion $\sigma_{0}$ using the Sersic fit parameters. The alternative candidates are indicated by black stars. The galaxies belonging to the Ta10 sample are represented using filled green triangles and the Tr09 sample is marked by filled cyan diamonds. B19 is indicated by a filled grey square. The magenta dashed line marks the limiting scaling central velocity dispersion for our sample selection. The area below the black dashed line is considered to be unphysical, because $M_{*}$ would exceed $M_{\textrm{dyn}}$.}
\label{sigma_mstar_mdyn_S}
\end{center}
\end{figure}

\begin{figure}[!htb]
\begin{center}
\includegraphics[width=0.45\textwidth]{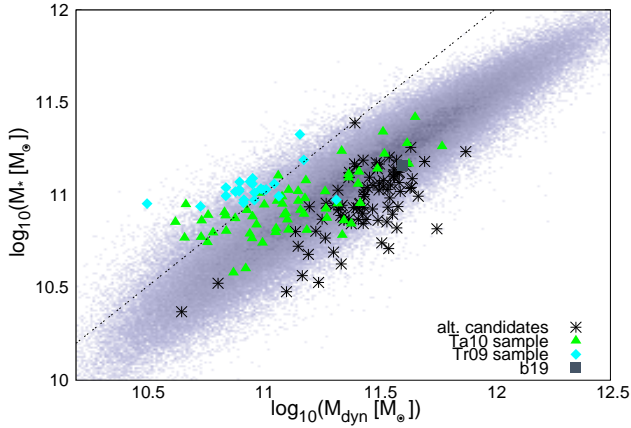}\\ 
\caption{Distribution of the sample's galaxies in the dynamical mass $M_{\textrm{dyn}}$ vs. stellar mass $M_{*}$ plane using the Sersic fit parameters. The alternative candidates are indicated by black stars. The galaxies belonging to the Ta10 sample are represented using filled green triangles, and the Tr09 sample is indicated by filled cyan diamonds. B19 is indicated by a filled grey square.The magenta dashed line marks the limiting scaling central velocity dispersion for our sample selection. The black dashed line denotes the limit of the $M_{\textrm{dyn}}$ to $M_{*}$ ratio, which is still considered to be physical, because $M_{*}$ would exceed $M_{\textrm{dyn}}$ above it.}
\label{masses_S}
\end{center}
\end{figure}

\begin{figure}[!htb]
\begin{center}
\includegraphics[width=0.45\textwidth]{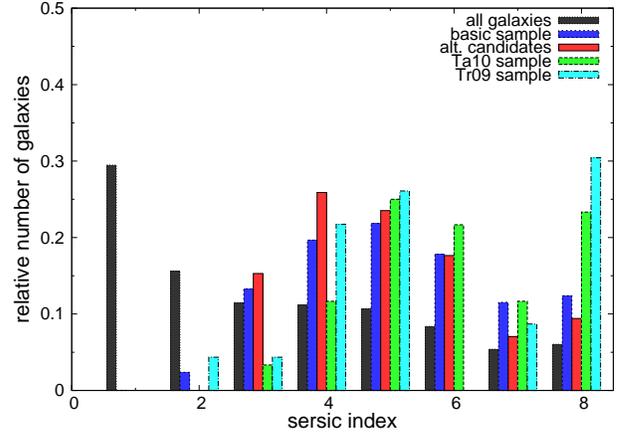}\\ 
\caption{Distribution of the Sersic indices of different samples of galaxies. The black histogram of the Sersic indices stands for all galaxies in SDSS DR7 for which \citet{Simard:2011} did their refits. The blue histogram indicates the distribution of Sersic parameters for our basic sample, which only consists of early-type galaxies. The green histogram represents the Ta10 sample, while the cyan histogram corresponds to Tr09 sample. The red histogram denotes our 85 alternative candidates.}
\label{sersic_distribution_S}
\end{center}
\end{figure}

The sample is defined in the same fashion as the main candidate sample, and we find 85 galaxies fulfilling the requirements (listed in Table \ref{list_candiates_basics_S}, together with b19, which was assigned the Sersic ID 1). The logarithm of the physical radius $R_{0}$ has to be smaller than the sample's average by at least one standard deviation, which provides us with an upper limit for $R_{0}$ of $\sim$2.65 kpc. The lower limit for the central velocity dispersion $\sigma_{0}$ of $\sim$316.6 km s$^{-1}$ is obtained by requiring it to be at least two standard deviations higher than the mean of the logarithm of the central velocity dispersion. The last criterion ensures that all candidates are more than three standard deviations off from the $\textrm{log}_{10}(R_{0})-\textrm{log}_{10}(\sigma_{0})$ relation, which was obtained by a linear fit to the data points. The selection criteria is illustrated in Figure \ref{R0_sigma_S}. We find that b19 fails to fulfil the radial size requirement in the case of the Sersic fit parameters (see Table \ref{list_candiates_S} for numbers), and it is not included in the 85 alternative candidates. However, we keep on providing its position in the plots and tables. As illustrated in Figures \ref{fp_S} to \ref{sersic_distribution_S}, the alternative sample has generally speaking similar properties to the main candidate sample, but it is less cohesive and more scattered. We therefore prefer our main sample to this one. There are 51 galaxies, which the two candidate lists have in common (see Table \ref{ID_crossmatches}). We consider these galaxies as candidates with increased priority for any follow-up observations. 

%\FloatBarrier
   \afterpage{\onecolumn
 \begin{landscape}
 \centering
   \bottomcaption{List of the basic parameters of our alternative candidate galaxies using the Sersic fit parameters of \citet{Simard:2011} instead of the de Vaucouleurs fit parameters used for our main candidate sample. First column: internal Sersic IDs, which are used to identify the galaxies. The numbering is essentially random and only based on the order the galaxies were drawn from the basic sample. The galaxy b19 has the internal ID 3. Second column: object ID used by SDSS DR10. Third and fourth column: equatorial coordinates of the galaxies. Fifth column: redshift $z$, already corrected for our motion relative to the CMB. Sixth, seventh, and eighth columns: observed uncorrected refitted SDSS parameters in the following order: observed apparent magnitude $m_{\textrm{sdss}}$, angular semi-major axis $a_{\textrm{sdss}}$, central velocity dispersion $\sigma_{\textrm{sdss}}$. Ninth column: axis ratio $q_{b/a}$. Tenth column: GalaxyZoo probability $\mathcal{L}_{\textrm{ETG}}$ of the galaxy being classified as an early-type.}
 \tablehead{Sersic ID & SDSS DR10 ID & ra & dec & $z$ & $m_{\textrm{S},r}$ &$a_{\textrm{S}}$ & $\sigma_{\textrm{sdss}}$ & $n_{S}$ & $\mathcal{L}_{\textrm{ETG}}$ \\ 
  &  & [$^{\circ}$]&  [$^{\circ}$]&   & [mag] & [arcsec] & [km/s] & & \\ \hline}
 \begin{supertabular}{cccccccccc}
           1 &1237648703523520846&  229.4240  &   -0.7049  &    0.1166  &   16.71    $\pm$    0.01    &    1.62    $\pm$    0.02    &         336$\pm$          12&    7.99    $\pm$    0.04    &    0.81    \\
           2 &1237651191892607189&  125.5691  &   48.2553  &    0.1276  &   17.56    $\pm$    0.03    &    0.87    $\pm$    0.02    &         351$\pm$          14&    2.92    $\pm$    0.30    &    0.75    \\
           3 &1237651252557513010&  125.0735  &   48.8830  &    0.1338  &   17.44    $\pm$    0.03    &    1.00    $\pm$    0.03    &         318$\pm$          13&    3.22    $\pm$    0.34    &    0.83    \\
           4 &1237652934037536913&  327.3491  &   -8.6752  &    0.1014  &   17.13    $\pm$    0.02    &    1.01    $\pm$    0.03    &         320$\pm$          16&    3.50    $\pm$    0.19    &    0.70    \\
           5 &1237654342254002376&  212.1645  &   61.1317  &    0.1215  &   17.19    $\pm$    0.03    &    1.13    $\pm$    0.02    &         338$\pm$          16&    5.66    $\pm$    0.36    &    0.76    \\
           6 &1237652900773298301&   58.0541  &   -5.8611  &    0.1137  &   17.13    $\pm$    0.03    &    0.77    $\pm$    0.02    &         306$\pm$          14&    4.76    $\pm$    0.37    &    0.66    \\
           7 &1237651252589363420&  247.9117  &   46.2683  &    0.1321  &   17.61    $\pm$    0.02    &    0.53    $\pm$    0.01    &         311$\pm$          14&    3.79    $\pm$    0.21    &    0.76    \\
           8 &1237655502424769160&  256.4241  &   33.4779  &    0.1022  &   17.12    $\pm$    0.02    &    1.32    $\pm$    0.02    &         326$\pm$          16&    5.63    $\pm$    0.17    &    0.77    \\
           9 &1237651539246186637&  167.7205  &   66.7862  &    0.1362  &   17.52    $\pm$    0.02    &    0.65    $\pm$    0.02    &         350$\pm$          14&    3.51    $\pm$    0.13    &    0.59    \\
          10 &1237655742944248167&  223.0119  &    5.2335  &    0.0639  &   16.05    $\pm$    0.01    &    1.37    $\pm$    0.02    &         294$\pm$          10&    2.51    $\pm$    0.10    &    0.77    \\
          11 &1237651714798125236&  248.3287  &   47.1274  &    0.1229  &   17.33    $\pm$    0.01    &    0.81    $\pm$    0.02    &         335$\pm$          12&    7.98    $\pm$    0.08    &    0.66    \\
          12 &1237660615586611373&  175.2231  &   11.0085  &    0.0809  &   16.75    $\pm$    0.02    &    1.00    $\pm$    0.02    &         289$\pm$          14&    4.48    $\pm$    0.23    &    0.82    \\
          13 &1237658206124507259&  193.5474  &   50.8170  &    0.1209  &   17.07    $\pm$    0.02    &    0.83    $\pm$    0.01    &         341$\pm$          16&    5.70    $\pm$    0.33    &    0.80    \\
          14 &1237652944786424004&    1.1323  &   16.0719  &    0.1144  &   17.46    $\pm$    0.02    &    0.62    $\pm$    0.01    &         291$\pm$          15&    4.48    $\pm$    0.29    &    0.55    \\
          15 &1237658423007707334&  138.8689  &    4.6676  &    0.1431  &   17.49    $\pm$    0.03    &    0.91    $\pm$    0.03    &         303$\pm$          14&    3.75    $\pm$    0.38    &    0.71    \\
          16 &1237657242435584230&  146.2765  &   47.8321  &    0.1170  &   16.99    $\pm$    0.02    &    1.24    $\pm$    0.03    &         319$\pm$          19&    4.25    $\pm$    0.27    &    0.62    \\
          17 &1237657856067830007&  161.5842  &   49.4468  &    0.1306  &   17.41    $\pm$    0.03    &    0.88    $\pm$    0.02    &         286$\pm$          14&    5.01    $\pm$    0.49    &    0.71    \\
          18 &1237654952670003535&  253.9937  &   39.4776  &    0.1496  &   17.30    $\pm$    0.02    &    0.97    $\pm$    0.02    &         393$\pm$          18&    2.85    $\pm$    0.20    &    0.78    \\
          19 &1237670956787695816&   23.2042  &   -9.1208  &    0.1336  &   17.00    $\pm$    0.02    &    1.09    $\pm$    0.03    &         318$\pm$           9&    5.36    $\pm$    0.27    &    0.79    \\
          20 &1237652948530102500&   10.3768  &   -9.2352  &    0.0538  &   15.05    $\pm$    0.00    &    2.46    $\pm$    0.02    &         310$\pm$           5&    4.38    $\pm$    0.01    &    0.53    \\
          21 &1237656243317113067&  354.1646  &   15.8222  &    0.1179  &   17.43    $\pm$    0.03    &    0.79    $\pm$    0.02    &         290$\pm$          16&    5.00    $\pm$    0.34    &    0.73    \\
          22 &1237657610723655845&  158.1494  &   53.3763  &    0.1340  &   17.50    $\pm$    0.04    &    0.92    $\pm$    0.02    &         293$\pm$          15&    3.29    $\pm$    0.46    &    0.69    \\
          23 &1237655474503024820&  245.6049  &   44.7856  &    0.0716  &   15.65    $\pm$    0.00    &    1.89    $\pm$    0.02    &         333$\pm$           8&    5.46    $\pm$    0.03    &    0.81    \\
          24 &1237658424616616162&  134.8572  &    5.6269  &    0.1625  &   17.65    $\pm$    0.03    &    0.79    $\pm$    0.03    &         288$\pm$          16&    6.21    $\pm$    0.28    &    0.72    \\
          25 &1237657596224209238&  123.8014  &   38.6793  &    0.1259  &   17.02    $\pm$    0.02    &    1.06    $\pm$    0.02    &         333$\pm$          13&    4.70    $\pm$    0.17    &    0.89    \\
          26 &1237665569297203655&  254.5120  &   41.8378  &    0.0375  &   15.18    $\pm$    0.00    &    1.50    $\pm$    0.01    &         303$\pm$           7&    3.54    $\pm$    0.04    &    0.64    \\
          27 &1237654605857751221&  148.8860  &    4.3722  &    0.0937  &   16.32    $\pm$    0.01    &    1.07    $\pm$    0.01    &         352$\pm$           9&    3.99    $\pm$    0.05    &    0.52    \\
          28 &1237653614796865660&  143.0592  &   56.4013  &    0.1173  &   16.73    $\pm$    0.01    &    1.09    $\pm$    0.01    &         302$\pm$          10&    5.21    $\pm$    0.12    &    0.69    \\
          29 &1237655465916170402&  184.8400  &   63.5358  &    0.1039  &   17.33    $\pm$    0.01    &    0.60    $\pm$    0.01    &         292$\pm$          14&    6.00    $\pm$    0.28    &    0.52    \\
          30 &1237658204493185306&  130.8260  &   34.6824  &    0.0658  &   16.10    $\pm$    0.01    &    1.79    $\pm$    0.03    &         301$\pm$          11&    2.62    $\pm$    0.03    &    0.86    \\
          31 &1237657628456190055&  187.6884  &   51.7060  &    0.1517  &   17.46    $\pm$    0.02    &    0.70    $\pm$    0.01    &         307$\pm$          14&    4.51    $\pm$    0.22    &    0.62    \\
          32 &1237660025032081578&  340.4373  &   -0.8113  &    0.1293  &   17.38    $\pm$    0.02    &    0.78    $\pm$    0.02    &         373$\pm$          22&    7.95    $\pm$    0.16    &    0.77    \\
          33 &1237661064411349290&  138.3286  &    8.1161  &    0.0934  &   16.71    $\pm$    0.01    &    0.95    $\pm$    0.01    &         295$\pm$           9&    4.78    $\pm$    0.14    &    0.61    \\
          34 &1237661849849430137&  156.3195  &   40.3153  &    0.0682  &   16.36    $\pm$    0.03    &    1.79    $\pm$    0.02    &         317$\pm$          10&    5.30    $\pm$    0.28    &    0.58    \\
          35 &1237661069261209757&  180.2716  &   14.5850  &    0.0831  &   16.24    $\pm$    0.01    &    1.29    $\pm$    0.02    &         291$\pm$           9&    2.86    $\pm$    0.06    &    0.57    \\
          36 &1237662663746060502&  221.9296  &   34.6657  &    0.0974  &   17.30    $\pm$    0.02    &    0.72    $\pm$    0.01    &         284$\pm$          14&    6.17    $\pm$    0.38    &    0.59    \\
          37 &1237663277928022281&    0.6027  &    0.5352  &    0.0784  &   17.30    $\pm$    0.03    &    0.71    $\pm$    0.02    &         331$\pm$          17&    7.97    $\pm$    0.11    &    0.77    \\
          38 &1237661383314702588&  160.1959  &   39.9311  &    0.1394  &   17.64    $\pm$    0.02    &    0.68    $\pm$    0.02    &         324$\pm$          15&    5.06    $\pm$    0.25    &    0.69    \\
          39 &1237662697568796852&  226.2857  &   30.1184  &    0.1450  &   16.99    $\pm$    0.02    &    1.00    $\pm$    0.02    &         314$\pm$           9&    7.99    $\pm$    0.04    &    0.71    \\
          40 &1237661812272857187&  180.2528  &   12.2175  &    0.1295  &   17.60    $\pm$    0.03    &    0.87    $\pm$    0.03    &         291$\pm$          17&    5.55    $\pm$    0.46    &    0.81    \\
          41 &1237665532252520624&  223.1388  &   22.5927  &    0.1551  &   17.51    $\pm$    0.02    &    0.94    $\pm$    0.02    &         318$\pm$          16&    4.06    $\pm$    0.28    &    0.54    \\
          42 &1237667255083991162&  170.3135  &   29.9694  &    0.1237  &   17.64    $\pm$    0.02    &    0.97    $\pm$    0.03    &         378$\pm$          24&    3.26    $\pm$    0.18    &    0.65    \\
          43 &1237667324323758158&  166.9049  &   27.1948  &    0.1502  &   17.45    $\pm$    0.04    &    0.83    $\pm$    0.02    &         287$\pm$          14&    6.32    $\pm$    0.69    &    0.63    \\
          44 &1237667735054647478&  206.9225  &   20.9708  &    0.1232  &   17.41    $\pm$    0.03    &    0.78    $\pm$    0.01    &         286$\pm$          14&    6.30    $\pm$    0.49    &    0.76    \\
          45 &1237662193459986552&  206.5336  &   39.4248  &    0.1297  &   17.47    $\pm$    0.03    &    0.80    $\pm$    0.03    &         289$\pm$          15&    3.77    $\pm$    0.35    &    0.68    \\
          46 &1237667736104861820&  149.3027  &   19.2625  &    0.0975  &   17.01    $\pm$    0.03    &    0.97    $\pm$    0.02    &         287$\pm$          10&    3.41    $\pm$    0.26    &    0.84    \\
          47 &1237665549429899544&  223.0734  &   22.4871  &    0.1165  &   17.29    $\pm$    0.02    &    0.58    $\pm$    0.01    &         335$\pm$          13&    4.63    $\pm$    0.19    &    0.62    \\
          48 &1237667209978380503&  149.1117  &   23.9641  &    0.1193  &   17.14    $\pm$    0.02    &    1.06    $\pm$    0.02    &         356$\pm$          25&    6.92    $\pm$    0.26    &    0.68    \\
          49 &1237663278461944053&  353.8668  &    1.0467  &    0.0827  &   16.16    $\pm$    0.02    &    1.62    $\pm$    0.02    &         320$\pm$           9&    5.18    $\pm$    0.24    &    0.80    \\
          50 &1237662340012638220&  239.5694  &   27.2367  &    0.0896  &   16.81    $\pm$    0.02    &    0.77    $\pm$    0.01    &         296$\pm$          12&    5.82    $\pm$    0.28    &    0.75    \\
          51 &1237664667887140986&  128.6548  &   24.3250  &    0.0705  &   16.13    $\pm$    0.01    &    1.63    $\pm$    0.02    &         296$\pm$           9&    7.17    $\pm$    0.08    &    0.77    \\
          52 &1237664093432119636&  121.7265  &   20.7624  &    0.1247  &   17.48    $\pm$    0.04    &    0.72    $\pm$    0.02    &         299$\pm$          14&    5.55    $\pm$    0.53    &    0.66    \\
          53 &1237665533335175692&  243.8410  &   16.3942  &    0.0818  &   15.72    $\pm$    0.01    &    1.63    $\pm$    0.01    &         311$\pm$           7&    6.03    $\pm$    0.13    &    0.81    \\
          54 &1237661850400260193&  199.4989  &   43.6141  &    0.1140  &   17.64    $\pm$    0.04    &    0.65    $\pm$    0.02    &         287$\pm$          16&    5.15    $\pm$    0.48    &    0.62    \\
          55 &1237664852035174654&  219.1545  &   31.3943  &    0.0850  &   16.06    $\pm$    0.01    &    0.95    $\pm$    0.01    &         331$\pm$           9&    3.96    $\pm$    0.04    &    0.77    \\
          56 &1237667321652248694&  199.8561  &   25.5487  &    0.2810  &   17.48    $\pm$    0.03    &    0.52    $\pm$    0.01    &         284$\pm$          12&    6.49    $\pm$    0.44    &    0.89    \\
          57 &1237667730736873763&  134.9926  &   14.7626  &    0.1008  &   17.22    $\pm$    0.02    &    1.07    $\pm$    0.02    &         291$\pm$          14&    2.98    $\pm$    0.09    &    0.57    \\
          58 &1237662664290402490&  239.6933  &   27.2131  &    0.0879  &   16.99    $\pm$    0.01    &    1.01    $\pm$    0.02    &         292$\pm$          15&    3.91    $\pm$    0.11    &    0.74    \\
          59 &1237665535469486145&  243.3042  &   17.8080  &    0.0374  &   14.57    $\pm$    0.00    &    2.89    $\pm$    0.01    &         316$\pm$           7&    5.76    $\pm$    0.05    &    0.68    \\
          60 &1237665016311840908&  163.4479  &   32.8694  &    0.1307  &   17.34    $\pm$    0.03    &    0.97    $\pm$    0.02    &         332$\pm$          14&    3.67    $\pm$    0.44    &    0.64    \\
          61 &1237667212115050932&  124.1550  &   16.1777  &    0.1511  &   17.62    $\pm$    0.02    &    0.77    $\pm$    0.02    &         303$\pm$          13&    2.74    $\pm$    0.19    &    0.64    \\
          62 &1237665564997976239&  218.1283  &   20.4259  &    0.1216  &   17.28    $\pm$    0.02    &    1.11    $\pm$    0.03    &         304$\pm$          17&    4.62    $\pm$    0.23    &    0.83    \\
          63 &1237663478723969457&  338.0784  &   -0.4059  &    0.0865  &   16.74    $\pm$    0.02    &    1.44    $\pm$    0.02    &         327$\pm$          17&    7.96    $\pm$    0.10    &    0.80    \\
          64 &1237667910055100586&  181.7985  &   23.8744  &    0.0775  &   16.27    $\pm$    0.01    &    1.63    $\pm$    0.02    &         328$\pm$          11&    7.99    $\pm$    0.03    &    0.86    \\
          65 &1237667734526492801&  227.3075  &   16.4333  &    0.1159  &   17.24    $\pm$    0.04    &    1.21    $\pm$    0.02    &         310$\pm$          17&    7.62    $\pm$    0.37    &    0.61    \\
          66 &1237670450522816720&  137.8481  &   16.5697  &    0.0900  &   16.59    $\pm$    0.01    &    1.09    $\pm$    0.02    &         295$\pm$           9&    4.42    $\pm$    0.08    &    0.77    \\
          67 &1237663789032669425&  125.7014  &   59.7435  &    0.1344  &   17.34    $\pm$    0.02    &    0.99    $\pm$    0.03    &         296$\pm$          14&    4.86    $\pm$    0.25    &    0.69    \\
          68 &1237665429169242591&  209.7906  &   27.9501  &    0.0811  &   17.07    $\pm$    0.01    &    0.65    $\pm$    0.01    &         287$\pm$          10&    3.67    $\pm$    0.32    &    0.62    \\
          69 &1237668299281662070&  194.2881  &   20.8064  &    0.0868  &   16.39    $\pm$    0.03    &    1.54    $\pm$    0.01    &         307$\pm$           9&    7.40    $\pm$    0.52    &    0.56    \\
          70 &1237668349753950509&  232.0499  &   12.1307  &    0.1225  &   17.56    $\pm$    0.03    &    0.85    $\pm$    0.02    &         311$\pm$          14&    3.77    $\pm$    0.21    &    0.81    \\
          71 &1237667783900135493&  164.2812  &   22.2115  &    0.1206  &   17.11    $\pm$    0.02    &    1.01    $\pm$    0.02    &         303$\pm$          14&    3.80    $\pm$    0.17    &    0.77    \\
          72 &1237668271372501042&  227.9714  &   14.2653  &    0.1221  &   17.43    $\pm$    0.03    &    0.90    $\pm$    0.02    &         291$\pm$          16&    6.61    $\pm$    0.40    &    0.75    \\
          73 &1237648721758978188&  160.3022  &    0.2285  &    0.1300  &   17.45    $\pm$    0.02    &    0.94    $\pm$    0.02    &         305$\pm$          15&    4.40    $\pm$    0.18    &    0.64    \\
          74 &1237668495782117442&  176.4731  &   17.3242  &    0.0928  &   16.88    $\pm$    0.01    &    0.72    $\pm$    0.01    &         285$\pm$           9&    3.40    $\pm$    0.11    &    0.59    \\
          75 &1237664671640715458&  191.2284  &   36.1838  &    0.0877  &   17.44    $\pm$    0.02    &    0.35    $\pm$    0.01    &         293$\pm$          15&    7.03    $\pm$    0.49    &    0.68    \\
          76 &1237662335717015837&  236.8248  &   33.1773  &    0.1265  &   17.53    $\pm$    0.03    &    0.84    $\pm$    0.03    &         296$\pm$          16&    7.01    $\pm$    0.48    &    0.78    \\
          77 &1237667782274187688&  128.9043  &   12.6627  &    0.1054  &   17.07    $\pm$    0.02    &    0.98    $\pm$    0.02    &         297$\pm$          12&    4.29    $\pm$    0.13    &    0.71    \\
          78 &1237661087497126080&  132.4080  &   29.6033  &    0.1059  &   16.55    $\pm$    0.01    &    1.27    $\pm$    0.02    &         302$\pm$           7&    3.75    $\pm$    0.12    &    0.80    \\
          79 &1237661358617067696&  181.3091  &   48.4216  &    0.0648  &   15.80    $\pm$    0.00    &    1.54    $\pm$    0.02    &         311$\pm$           8&    4.09    $\pm$    0.07    &    0.71    \\
          80 &1237653651837026391&    4.8387  &   14.9802  &    0.1277  &   17.31    $\pm$    0.03    &    1.10    $\pm$    0.03    &         303$\pm$          14&    5.41    $\pm$    0.29    &    0.74    \\
          81 &1237668298203070641&  182.4650  &   20.0535  &    0.1116  &   17.43    $\pm$    0.02    &    0.88    $\pm$    0.02    &         293$\pm$          12&    4.96    $\pm$    0.30    &    0.72    \\
          82 &1237654604796985469&  178.6502  &    4.3536  &    0.0761  &   17.50    $\pm$    0.02    &    0.18    $\pm$    0.01    &         277$\pm$          13&    2.66    $\pm$    0.22    &    0.53    \\
          83 &1237667917032980629&  189.9670  &   21.1529  &    0.1085  &   16.58    $\pm$    0.01    &    1.17    $\pm$    0.01    &         321$\pm$           9&    5.19    $\pm$    0.12    &    0.74    \\
          84 &1237661950244945934&  162.5130  &   11.8190  &    0.0812  &   16.42    $\pm$    0.01    &    1.51    $\pm$    0.02    &         340$\pm$          11&    5.14    $\pm$    0.16    &    0.88    \\
          85 &1237668333640810655&  225.5537  &   14.6343  &    0.0697  &   16.37    $\pm$    0.01    &    0.89    $\pm$    0.01    &         351$\pm$          14&    8.00    $\pm$    0.02    &    0.52    \\
          86 &1237667917030555837&  184.0304  &   21.1393  &    0.1278  &   16.79    $\pm$    0.01    &    1.12    $\pm$    0.02    &         389$\pm$          16&    6.35    $\pm$    0.13    &    0.79    \\
 \end{supertabular}
 \label{list_candiates_basics_S}
 % \twocolumn
 %\end{landscape}}

 %  \afterpage{\onecolumn
 %\begin{landscape}
% \centering
   \bottomcaption{List of the derived parameters based on the Sersic fits from \citet{Simard:2011} for our alternative candidate galaxies. First column: internal Sersic IDs of our galaxies. Second column: scale radius $R_{\textrm{r}}$ of the galaxies measured in the SDSS r band (in kpc). Third column: corrected central velocity dispersion $\sigma_{0}$ (in km/s). Fourth column: surface brightness $\mu_{r}$ measured in the SDSS r band (in mag/arcsec$^{2}$). Fifth column: absolute magnitude in r band $M_{\textrm{r}}$. Sixth column: g-r colour $(M_{\textrm{g}}-M_{\textrm{r}})$ (in mag). Seventh column: logarithm of the dynamical mass $M_{\textrm{dyn}}$ (in solar masses). Eighth column: logarithm of the stellar mass $M_{\textrm{*}}$ (in solar masses). Ninth column: dynamical mass-to-light ratio $\vernal_{\textrm{dyn}}$ (in solar units $M_{\astrosun}/L_{\astrosun ,\textrm{r}}$). Tenth column: stellar mass-to-light ratio $\vernal_{*}$ (in solar units $M_{\astrosun}/L_{\astrosun ,\textrm{r}}$).}
 \tablehead{Sersic ID & $R_{\textrm{r}}$ & $\sigma_{0}$ & $\mu_{r}$ & $M_{\textrm{r}}$ & $(M_{\textrm{g}}-M_{\textrm{r}})$ &  log$_{10}$($M_{\textrm{dyn}}$) &  log$_{10}$($M_{*}$) &$\vernal_{\textrm{dyn}}$ & $\vernal_{*}$  \\
  & [kpc] &[km s$^{-1}$] & [$\frac{\textrm{mag}}{\textrm{arcsec}^{2}}$] & [mag] & [mag] & [log$_{10}(M_{\astrosun})$] & [log$_{10}(M_{\astrosun})$] & [$M_{\astrosun}/L_{\astrosun,\textrm{r}}$]& [$M_{\astrosun}/L_{\astrosun,\textrm{r}}$] \\ \hline}
 \begin{supertabular}{cccccccccc}
           1 &    3.46    $\pm$    0.05    &         364$\pm$          13&   19.11    $\pm$    0.01    &  -22.28    $\pm$    0.01    &    0.80    $\pm$    0.02    &   11.60    $\pm$    0.02    &   11.16    $\pm$    0.15    &    6.44    $\pm$    0.30    &    2.31    $\pm$    0.96    \\
           2 &    2.00    $\pm$    0.05    &         390$\pm$          16&   18.59    $\pm$    0.03    &  -21.62    $\pm$    0.03    &    0.84    $\pm$    0.04    &   11.67    $\pm$    0.05    &   10.99    $\pm$    0.15    &   13.84    $\pm$    1.84    &    2.90    $\pm$    1.22    \\
           3 &    2.41    $\pm$    0.06    &         351$\pm$          14&   18.76    $\pm$    0.03    &  -21.86    $\pm$    0.03    &    0.83    $\pm$    0.05    &   11.65    $\pm$    0.05    &   11.09    $\pm$    0.15    &   10.43    $\pm$    1.41    &    2.89    $\pm$    1.21    \\
           4 &    1.90    $\pm$    0.05    &         354$\pm$          18&   18.62    $\pm$    0.02    &  -21.45    $\pm$    0.02    &    0.67    $\pm$    0.04    &   11.54    $\pm$    0.03    &   10.86    $\pm$    0.15    &   11.86    $\pm$    1.05    &    2.46    $\pm$    1.02    \\
           5 &    2.49    $\pm$    0.05    &         371$\pm$          17&   18.92    $\pm$    0.03    &  -21.76    $\pm$    0.03    &    0.86    $\pm$    0.04    &   11.60    $\pm$    0.03    &   11.02    $\pm$    0.15    &   10.16    $\pm$    1.10    &    2.70    $\pm$    1.13    \\
           6 &    1.60    $\pm$    0.04    &         341$\pm$          15&   17.78    $\pm$    0.03    &  -21.92    $\pm$    0.03    &    0.83    $\pm$    0.04    &   11.37    $\pm$    0.04    &   11.09    $\pm$    0.15    &    5.24    $\pm$    0.61    &    2.76    $\pm$    1.15    \\
           7 &    1.27    $\pm$    0.03    &         352$\pm$          16&   17.67    $\pm$    0.02    &  -21.55    $\pm$    0.02    &    0.87    $\pm$    0.04    &   11.35    $\pm$    0.03    &   10.95    $\pm$    0.15    &    6.95    $\pm$    0.60    &    2.76    $\pm$    1.15    \\
           8 &    2.51    $\pm$    0.05    &         356$\pm$          17&   19.24    $\pm$    0.02    &  -21.43    $\pm$    0.02    &    0.87    $\pm$    0.03    &   11.56    $\pm$    0.02    &   10.90    $\pm$    0.15    &   12.79    $\pm$    0.99    &    2.76    $\pm$    1.15    \\
           9 &    1.59    $\pm$    0.05    &         393$\pm$          16&   18.00    $\pm$    0.02    &  -21.73    $\pm$    0.02    &    0.88    $\pm$    0.03    &   11.55    $\pm$    0.02    &   11.05    $\pm$    0.15    &    9.53    $\pm$    0.71    &    3.02    $\pm$    1.26    \\
          10 &    1.69    $\pm$    0.03    &         321$\pm$          11&   18.37    $\pm$    0.01    &  -21.41    $\pm$    0.01    &    0.81    $\pm$    0.02    &   11.45    $\pm$    0.02    &   10.87    $\pm$    0.15    &   10.00    $\pm$    0.61    &    2.63    $\pm$    1.09    \\
          11 &    1.80    $\pm$    0.03    &         373$\pm$          14&   18.38    $\pm$    0.01    &  -21.59    $\pm$    0.01    &    0.59    $\pm$    0.02    &   11.34    $\pm$    0.02    &   10.63    $\pm$    0.15    &    6.62    $\pm$    0.30    &    1.28    $\pm$    0.53    \\
          12 &    1.53    $\pm$    0.03    &         319$\pm$          16&   18.25    $\pm$    0.02    &  -21.32    $\pm$    0.02    &    0.76    $\pm$    0.03    &   11.31    $\pm$    0.03    &   10.69    $\pm$    0.15    &    7.85    $\pm$    0.69    &    1.90    $\pm$    0.79    \\
          13 &    1.83    $\pm$    0.03    &         380$\pm$          18&   18.15    $\pm$    0.02    &  -21.86    $\pm$    0.02    &    0.84    $\pm$    0.04    &   11.48    $\pm$    0.03    &   11.05    $\pm$    0.15    &    7.11    $\pm$    0.64    &    2.67    $\pm$    1.11    \\
          14 &    1.29    $\pm$    0.02    &         328$\pm$          17&   17.83    $\pm$    0.02    &  -21.41    $\pm$    0.02    &    0.81    $\pm$    0.04    &   11.26    $\pm$    0.03    &   10.87    $\pm$    0.15    &    6.47    $\pm$    0.62    &    2.64    $\pm$    1.10    \\
          15 &    2.31    $\pm$    0.07    &         336$\pm$          16&   18.57    $\pm$    0.03    &  -21.97    $\pm$    0.03    &    0.84    $\pm$    0.04    &   11.57    $\pm$    0.05    &   11.09    $\pm$    0.15    &    7.89    $\pm$    1.06    &    2.63    $\pm$    1.10    \\
          16 &    2.64    $\pm$    0.06    &         350$\pm$          21&   18.96    $\pm$    0.02    &  -21.84    $\pm$    0.02    &    0.73    $\pm$    0.04    &   11.64    $\pm$    0.04    &   11.04    $\pm$    0.15    &   10.40    $\pm$    1.05    &    2.60    $\pm$    1.08    \\
          17 &    2.06    $\pm$    0.05    &         317$\pm$          16&   18.54    $\pm$    0.03    &  -21.74    $\pm$    0.03    &    0.96    $\pm$    0.04    &   11.41    $\pm$    0.05    &   11.15    $\pm$    0.15    &    6.73    $\pm$    0.89    &    3.76    $\pm$    1.58    \\
          18 &    2.57    $\pm$    0.05    &         434$\pm$          20&   18.58    $\pm$    0.02    &  -22.19    $\pm$    0.02    &    0.87    $\pm$    0.04    &   11.88    $\pm$    0.04    &   11.23    $\pm$    0.15    &   13.05    $\pm$    1.29    &    2.97    $\pm$    1.24    \\
          19 &    2.60    $\pm$    0.06    &         350$\pm$          10&   18.54    $\pm$    0.02    &  -22.25    $\pm$    0.02    &    0.85    $\pm$    0.03    &   11.58    $\pm$    0.02    &   11.19    $\pm$    0.15    &    6.27    $\pm$    0.47    &    2.55    $\pm$    1.06    \\
          20 &    2.58    $\pm$    0.02    &         330$\pm$           5&   18.67    $\pm$    0.00    &  -22.01    $\pm$    0.01    &    0.86    $\pm$    0.01    &   11.57    $\pm$    0.01    &   11.11    $\pm$    0.15    &    7.62    $\pm$    0.16    &    2.65    $\pm$    1.09    \\
          21 &    1.69    $\pm$    0.04    &         323$\pm$          18&   18.24    $\pm$    0.03    &  -21.59    $\pm$    0.03    &    0.83    $\pm$    0.04    &   11.34    $\pm$    0.04    &   10.94    $\pm$    0.15    &    6.60    $\pm$    0.76    &    2.60    $\pm$    1.09    \\
          22 &    2.22    $\pm$    0.06    &         325$\pm$          17&   18.75    $\pm$    0.04    &  -21.69    $\pm$    0.04    &    0.79    $\pm$    0.06    &   11.54    $\pm$    0.06    &   10.94    $\pm$    0.15    &    9.57    $\pm$    1.71    &    2.36    $\pm$    1.00    \\
          23 &    2.58    $\pm$    0.02    &         358$\pm$           9&   18.71    $\pm$    0.00    &  -21.99    $\pm$    0.01    &    0.85    $\pm$    0.01    &   11.59    $\pm$    0.01    &   11.07    $\pm$    0.15    &    8.10    $\pm$    0.24    &    2.43    $\pm$    1.00    \\
          24 &    2.25    $\pm$    0.07    &         321$\pm$          18&   18.35    $\pm$    0.03    &  -22.15    $\pm$    0.03    &    0.86    $\pm$    0.05    &   11.40    $\pm$    0.03    &   11.17    $\pm$    0.15    &    4.50    $\pm$    0.46    &    2.65    $\pm$    1.11    \\
          25 &    2.42    $\pm$    0.04    &         367$\pm$          14&   18.53    $\pm$    0.02    &  -22.09    $\pm$    0.02    &    0.88    $\pm$    0.03    &   11.62    $\pm$    0.02    &   11.15    $\pm$    0.15    &    7.97    $\pm$    0.58    &    2.73    $\pm$    1.14    \\
          26 &    1.12    $\pm$    0.01    &         329$\pm$           7&   17.83    $\pm$    0.00    &  -21.02    $\pm$    0.01    &    0.83    $\pm$    0.01    &   11.24    $\pm$    0.01    &   10.53    $\pm$    0.15    &    8.94    $\pm$    0.28    &    1.72    $\pm$    0.71    \\
          27 &    1.88    $\pm$    0.02    &         388$\pm$          10&   17.95    $\pm$    0.01    &  -22.09    $\pm$    0.01    &    0.82    $\pm$    0.02    &   11.59    $\pm$    0.01    &   11.12    $\pm$    0.15    &    7.44    $\pm$    0.29    &    2.53    $\pm$    1.05    \\
          28 &    2.33    $\pm$    0.03    &         332$\pm$          11&   18.37    $\pm$    0.01    &  -22.16    $\pm$    0.01    &    0.86    $\pm$    0.02    &   11.49    $\pm$    0.02    &   11.18    $\pm$    0.15    &    5.56    $\pm$    0.27    &    2.68    $\pm$    1.11    \\
          29 &    1.16    $\pm$    0.02    &         329$\pm$          16&   17.75    $\pm$    0.01    &  -21.25    $\pm$    0.01    &    0.84    $\pm$    0.02    &   11.14    $\pm$    0.03    &   10.80    $\pm$    0.15    &    5.74    $\pm$    0.42    &    2.62    $\pm$    1.08    \\
          30 &    2.27    $\pm$    0.04    &         324$\pm$          11&   19.00    $\pm$    0.01    &  -21.42    $\pm$    0.01    &    0.80    $\pm$    0.02    &   11.58    $\pm$    0.02    &   10.82    $\pm$    0.15    &   13.45    $\pm$    0.64    &    2.35    $\pm$    0.97    \\
          31 &    1.88    $\pm$    0.04    &         344$\pm$          16&   18.03    $\pm$    0.02    &  -22.07    $\pm$    0.02    &    0.88    $\pm$    0.04    &   11.46    $\pm$    0.03    &   11.10    $\pm$    0.15    &    5.63    $\pm$    0.47    &    2.42    $\pm$    1.01    \\
          32 &    1.82    $\pm$    0.05    &         416$\pm$          24&   18.16    $\pm$    0.02    &  -21.85    $\pm$    0.02    &    0.83    $\pm$    0.04    &   11.44    $\pm$    0.03    &   11.00    $\pm$    0.15    &    6.60    $\pm$    0.54    &    2.40    $\pm$    1.00    \\
          33 &    1.66    $\pm$    0.02    &         326$\pm$          10&   18.08    $\pm$    0.01    &  -21.69    $\pm$    0.01    &    0.78    $\pm$    0.02    &   11.35    $\pm$    0.02    &   10.88    $\pm$    0.15    &    6.17    $\pm$    0.31    &    2.10    $\pm$    0.87    \\
          34 &    2.34    $\pm$    0.03    &         342$\pm$          11&   19.31    $\pm$    0.03    &  -21.18    $\pm$    0.03    &    0.73    $\pm$    0.04    &   11.52    $\pm$    0.03    &   10.74    $\pm$    0.15    &   14.47    $\pm$    1.38    &    2.44    $\pm$    1.02    \\
          35 &    2.02    $\pm$    0.03    &         318$\pm$          10&   18.34    $\pm$    0.01    &  -21.84    $\pm$    0.01    &    0.85    $\pm$    0.02    &   11.50    $\pm$    0.02    &   11.03    $\pm$    0.15    &    7.64    $\pm$    0.35    &    2.55    $\pm$    1.06    \\
          36 &    1.31    $\pm$    0.02    &         317$\pm$          16&   18.15    $\pm$    0.02    &  -21.12    $\pm$    0.02    &    0.77    $\pm$    0.03    &   11.16    $\pm$    0.03    &   10.72    $\pm$    0.15    &    6.69    $\pm$    0.63    &    2.47    $\pm$    1.03    \\
          37 &    1.06    $\pm$    0.02    &         370$\pm$          19&   18.16    $\pm$    0.03    &  -20.62    $\pm$    0.03    &    0.78    $\pm$    0.04    &   11.11    $\pm$    0.02    &   10.48    $\pm$    0.15    &    9.42    $\pm$    0.90    &    2.22    $\pm$    0.93    \\
          38 &    1.69    $\pm$    0.05    &         364$\pm$          17&   18.17    $\pm$    0.02    &  -21.68    $\pm$    0.02    &    0.85    $\pm$    0.03    &   11.44    $\pm$    0.03    &   11.03    $\pm$    0.15    &    7.62    $\pm$    0.64    &    2.96    $\pm$    1.23    \\
          39 &    2.58    $\pm$    0.06    &         347$\pm$          10&   18.40    $\pm$    0.02    &  -22.39    $\pm$    0.02    &    0.74    $\pm$    0.03    &   11.44    $\pm$    0.01    &   11.19    $\pm$    0.15    &    3.96    $\pm$    0.22    &    2.24    $\pm$    0.93    \\
          40 &    2.04    $\pm$    0.06    &         323$\pm$          19&   18.69    $\pm$    0.03    &  -21.56    $\pm$    0.03    &    0.85    $\pm$    0.04    &   11.39    $\pm$    0.04    &   10.93    $\pm$    0.15    &    7.70    $\pm$    0.97    &    2.66    $\pm$    1.11    \\
          41 &    2.57    $\pm$    0.07    &         353$\pm$          17&   18.65    $\pm$    0.02    &  -22.13    $\pm$    0.02    &    0.89    $\pm$    0.04    &   11.64    $\pm$    0.04    &   11.26    $\pm$    0.15    &    8.00    $\pm$    0.79    &    3.31    $\pm$    1.38    \\
          42 &    2.18    $\pm$    0.07    &         417$\pm$          26&   19.06    $\pm$    0.02    &  -21.33    $\pm$    0.02    &    0.75    $\pm$    0.03    &   11.75    $\pm$    0.03    &   10.82    $\pm$    0.15    &   21.75    $\pm$    2.22    &    2.53    $\pm$    1.05    \\
          43 &    2.19    $\pm$    0.04    &         319$\pm$          16&   18.36    $\pm$    0.04    &  -22.07    $\pm$    0.04    &    0.84    $\pm$    0.06    &   11.38    $\pm$    0.05    &   11.12    $\pm$    0.15    &    4.61    $\pm$    0.71    &    2.57    $\pm$    1.09    \\
          44 &    1.74    $\pm$    0.03    &         319$\pm$          16&   18.30    $\pm$    0.03    &  -21.60    $\pm$    0.03    &    0.80    $\pm$    0.04    &   11.28    $\pm$    0.04    &   10.96    $\pm$    0.15    &    5.66    $\pm$    0.67    &    2.71    $\pm$    1.14    \\
          45 &    1.87    $\pm$    0.06    &         322$\pm$          17&   18.44    $\pm$    0.03    &  -21.63    $\pm$    0.03    &    0.85    $\pm$    0.05    &   11.44    $\pm$    0.04    &   10.99    $\pm$    0.15    &    8.01    $\pm$    1.03    &    2.87    $\pm$    1.20    \\
          46 &    1.76    $\pm$    0.04    &         317$\pm$          11&   18.48    $\pm$    0.03    &  -21.42    $\pm$    0.03    &    0.83    $\pm$    0.04    &   11.41    $\pm$    0.03    &   10.87    $\pm$    0.15    &    9.22    $\pm$    1.03    &    2.61    $\pm$    1.09    \\
          47 &    1.23    $\pm$    0.02    &         378$\pm$          15&   17.53    $\pm$    0.02    &  -21.60    $\pm$    0.02    &    0.82    $\pm$    0.03    &   11.35    $\pm$    0.02    &   10.86    $\pm$    0.15    &    6.74    $\pm$    0.52    &    2.18    $\pm$    0.91    \\
          48 &    2.30    $\pm$    0.05    &         392$\pm$          27&   18.76    $\pm$    0.02    &  -21.74    $\pm$    0.02    &    0.60    $\pm$    0.03    &   11.55    $\pm$    0.03    &   10.71    $\pm$    0.15    &    9.22    $\pm$    0.86    &    1.35    $\pm$    0.56    \\
          49 &    2.54    $\pm$    0.02    &         346$\pm$           9&   18.74    $\pm$    0.02    &  -21.94    $\pm$    0.02    &    0.88    $\pm$    0.03    &   11.57    $\pm$    0.02    &   11.14    $\pm$    0.15    &    8.11    $\pm$    0.59    &    3.00    $\pm$    1.25    \\
          50 &    1.30    $\pm$    0.02    &         330$\pm$          13&   17.76    $\pm$    0.02    &  -21.47    $\pm$    0.02    &    0.89    $\pm$    0.03    &   11.20    $\pm$    0.03    &   10.94    $\pm$    0.15    &    5.38    $\pm$    0.43    &    2.91    $\pm$    1.21    \\
          51 &    2.20    $\pm$    0.02    &         321$\pm$          10&   18.80    $\pm$    0.01    &  -21.56    $\pm$    0.01    &    0.83    $\pm$    0.02    &   11.34    $\pm$    0.01    &   10.89    $\pm$    0.15    &    6.79    $\pm$    0.29    &    2.41    $\pm$    1.00    \\
          52 &    1.63    $\pm$    0.03    &         334$\pm$          15&   18.17    $\pm$    0.04    &  -21.59    $\pm$    0.04    &    0.84    $\pm$    0.06    &   11.33    $\pm$    0.04    &   11.03    $\pm$    0.15    &    6.39    $\pm$    0.92    &    3.20    $\pm$    1.36    \\
          53 &    2.52    $\pm$    0.02    &         337$\pm$           7&   18.30    $\pm$    0.01    &  -22.36    $\pm$    0.01    &    0.78    $\pm$    0.02    &   11.50    $\pm$    0.01    &   11.14    $\pm$    0.15    &    4.67    $\pm$    0.19    &    2.06    $\pm$    0.85    \\
          54 &    1.35    $\pm$    0.04    &         323$\pm$          18&   18.18    $\pm$    0.04    &  -21.16    $\pm$    0.04    &    0.84    $\pm$    0.06    &   11.23    $\pm$    0.04    &   10.80    $\pm$    0.15    &    7.65    $\pm$    1.12    &    2.86    $\pm$    1.22    \\
          55 &    1.53    $\pm$    0.01    &         367$\pm$          10&   17.57    $\pm$    0.01    &  -22.01    $\pm$    0.01    &    0.81    $\pm$    0.02    &   11.45    $\pm$    0.01    &   11.04    $\pm$    0.15    &    5.84    $\pm$    0.24    &    2.26    $\pm$    0.93    \\
          56 &    2.32    $\pm$    0.05    &         322$\pm$          14&   16.94    $\pm$    0.03    &  -23.76    $\pm$    0.03    &    0.71    $\pm$    0.05    &   11.40    $\pm$    0.03    &   11.39    $\pm$    0.15    &    1.03    $\pm$    0.11    &    1.01    $\pm$    0.42    \\
          57 &    2.00    $\pm$    0.04    &         321$\pm$          15&   18.83    $\pm$    0.02    &  -21.35    $\pm$    0.02    &    0.86    $\pm$    0.04    &   11.50    $\pm$    0.02    &   10.86    $\pm$    0.15    &   11.92    $\pm$    0.90    &    2.76    $\pm$    1.15    \\
          58 &    1.67    $\pm$    0.03    &         322$\pm$          17&   18.54    $\pm$    0.01    &  -21.25    $\pm$    0.01    &    0.85    $\pm$    0.03    &   11.38    $\pm$    0.02    &   10.85    $\pm$    0.15    &   10.05    $\pm$    0.67    &    2.90    $\pm$    1.20    \\
          59 &    2.14    $\pm$    0.01    &         335$\pm$           7&   18.56    $\pm$    0.00    &  -21.71    $\pm$    0.01    &    0.84    $\pm$    0.01    &   11.44    $\pm$    0.01    &   10.85    $\pm$    0.15    &    7.41    $\pm$    0.22    &    1.90    $\pm$    0.79    \\
          60 &    2.28    $\pm$    0.05    &         368$\pm$          16&   18.69    $\pm$    0.03    &  -21.80    $\pm$    0.03    &    0.85    $\pm$    0.04    &   11.64    $\pm$    0.05    &   11.04    $\pm$    0.15    &   10.91    $\pm$    1.61    &    2.73    $\pm$    1.14    \\
          61 &    2.05    $\pm$    0.05    &         338$\pm$          14&   18.35    $\pm$    0.02    &  -21.94    $\pm$    0.02    &    0.86    $\pm$    0.03    &   11.57    $\pm$    0.03    &   11.10    $\pm$    0.15    &    8.07    $\pm$    0.76    &    2.77    $\pm$    1.15    \\
          62 &    2.46    $\pm$    0.06    &         335$\pm$          19&   18.92    $\pm$    0.02    &  -21.73    $\pm$    0.02    &    0.80    $\pm$    0.03    &   11.55    $\pm$    0.03    &   10.98    $\pm$    0.15    &    9.42    $\pm$    0.85    &    2.56    $\pm$    1.07    \\
          63 &    2.35    $\pm$    0.03    &         356$\pm$          18&   19.01    $\pm$    0.02    &  -21.50    $\pm$    0.02    &    0.85    $\pm$    0.03    &   11.42    $\pm$    0.02    &   10.92    $\pm$    0.15    &    8.58    $\pm$    0.65    &    2.75    $\pm$    1.15    \\
          64 &    2.40    $\pm$    0.03    &         355$\pm$          12&   18.93    $\pm$    0.01    &  -21.62    $\pm$    0.01    &    0.87    $\pm$    0.03    &   11.42    $\pm$    0.01    &   10.96    $\pm$    0.15    &    7.79    $\pm$    0.35    &    2.66    $\pm$    1.10    \\
          65 &    2.55    $\pm$    0.05    &         340$\pm$          19&   19.11    $\pm$    0.04    &  -21.62    $\pm$    0.04    &    0.81    $\pm$    0.07    &   11.43    $\pm$    0.03    &   10.90    $\pm$    0.15    &    7.97    $\pm$    0.99    &    2.35    $\pm$    1.00    \\
          66 &    1.84    $\pm$    0.04    &         325$\pm$          10&   18.32    $\pm$    0.01    &  -21.67    $\pm$    0.01    &    0.82    $\pm$    0.02    &   11.41    $\pm$    0.02    &   10.93    $\pm$    0.15    &    7.19    $\pm$    0.32    &    2.40    $\pm$    0.99    \\
          67 &    2.39    $\pm$    0.06    &         327$\pm$          15&   18.55    $\pm$    0.02    &  -22.05    $\pm$    0.02    &    0.88    $\pm$    0.04    &   11.51    $\pm$    0.03    &   11.17    $\pm$    0.15    &    6.32    $\pm$    0.54    &    2.90    $\pm$    1.21    \\
          68 &    1.00    $\pm$    0.01    &         323$\pm$          11&   17.76    $\pm$    0.01    &  -20.90    $\pm$    0.01    &    0.79    $\pm$    0.02    &   11.17    $\pm$    0.04    &   10.56    $\pm$    0.15    &    8.48    $\pm$    0.83    &    2.09    $\pm$    0.86    \\
          69 &    2.52    $\pm$    0.02    &         333$\pm$          10&   18.88    $\pm$    0.03    &  -21.78    $\pm$    0.03    &    0.82    $\pm$    0.05    &   11.42    $\pm$    0.03    &   10.95    $\pm$    0.15    &    6.64    $\pm$    0.70    &    2.26    $\pm$    0.95    \\
          70 &    1.89    $\pm$    0.04    &         346$\pm$          16&   18.58    $\pm$    0.03    &  -21.51    $\pm$    0.03    &    0.91    $\pm$    0.05    &   11.51    $\pm$    0.03    &   11.06    $\pm$    0.15    &   10.45    $\pm$    1.09    &    3.78    $\pm$    1.58    \\
          71 &    2.22    $\pm$    0.05    &         334$\pm$          15&   18.59    $\pm$    0.02    &  -21.84    $\pm$    0.02    &    0.84    $\pm$    0.03    &   11.54    $\pm$    0.03    &   11.03    $\pm$    0.15    &    8.37    $\pm$    0.68    &    2.58    $\pm$    1.07    \\
          72 &    2.00    $\pm$    0.04    &         322$\pm$          18&   18.61    $\pm$    0.03    &  -21.59    $\pm$    0.03    &    0.84    $\pm$    0.05    &   11.33    $\pm$    0.03    &   10.92    $\pm$    0.15    &    6.43    $\pm$    0.71    &    2.52    $\pm$    1.06    \\
          73 &    2.20    $\pm$    0.05    &         338$\pm$          16&   18.64    $\pm$    0.02    &  -21.78    $\pm$    0.02    &    0.85    $\pm$    0.03    &   11.52    $\pm$    0.03    &   11.04    $\pm$    0.15    &    8.39    $\pm$    0.67    &    2.79    $\pm$    1.16    \\
          74 &    1.25    $\pm$    0.02    &         319$\pm$          10&   17.72    $\pm$    0.01    &  -21.44    $\pm$    0.01    &    0.78    $\pm$    0.02    &   11.27    $\pm$    0.02    &   10.77    $\pm$    0.15    &    6.53    $\pm$    0.34    &    2.04    $\pm$    0.84    \\
          75 &    0.58    $\pm$    0.01    &         338$\pm$          17&   16.76    $\pm$    0.02    &  -20.71    $\pm$    0.02    &    0.83    $\pm$    0.03    &   10.81    $\pm$    0.04    &   10.52    $\pm$    0.15    &    4.38    $\pm$    0.44    &    2.26    $\pm$    0.94    \\
          76 &    1.92    $\pm$    0.06    &         329$\pm$          18&   18.56    $\pm$    0.03    &  -21.56    $\pm$    0.03    &    0.82    $\pm$    0.04    &   11.31    $\pm$    0.04    &   10.94    $\pm$    0.15    &    6.32    $\pm$    0.73    &    2.71    $\pm$    1.14    \\
          77 &    1.91    $\pm$    0.05    &         328$\pm$          13&   18.48    $\pm$    0.02    &  -21.60    $\pm$    0.02    &    0.81    $\pm$    0.03    &   11.44    $\pm$    0.02    &   10.93    $\pm$    0.15    &    8.24    $\pm$    0.58    &    2.57    $\pm$    1.07    \\
          78 &    2.49    $\pm$    0.03    &         330$\pm$           8&   18.55    $\pm$    0.01    &  -22.11    $\pm$    0.01    &    0.84    $\pm$    0.02    &   11.58    $\pm$    0.02    &   11.12    $\pm$    0.15    &    7.19    $\pm$    0.34    &    2.49    $\pm$    1.03    \\
          79 &    1.92    $\pm$    0.02    &         338$\pm$           9&   18.40    $\pm$    0.00    &  -21.66    $\pm$    0.01    &    0.83    $\pm$    0.01    &   11.48    $\pm$    0.01    &   10.93    $\pm$    0.15    &    8.50    $\pm$    0.30    &    2.39    $\pm$    0.99    \\
          80 &    2.54    $\pm$    0.06    &         333$\pm$          16&   18.84    $\pm$    0.03    &  -21.88    $\pm$    0.03    &    0.86    $\pm$    0.04    &   11.52    $\pm$    0.03    &   11.08    $\pm$    0.15    &    7.67    $\pm$    0.79    &    2.76    $\pm$    1.16    \\
          81 &    1.80    $\pm$    0.05    &         326$\pm$          13&   18.60    $\pm$    0.02    &  -21.36    $\pm$    0.02    &    0.83    $\pm$    0.04    &   11.37    $\pm$    0.03    &   10.87    $\pm$    0.15    &    8.84    $\pm$    0.77    &    2.77    $\pm$    1.15    \\
          82 &    0.26    $\pm$    0.01    &         328$\pm$          15&   15.44    $\pm$    0.02    &  -20.33    $\pm$    0.02    &    0.76    $\pm$    0.03    &   10.65    $\pm$    0.04    &   10.37    $\pm$    0.15    &    4.37    $\pm$    0.47    &    2.27    $\pm$    0.94    \\
          83 &    2.35    $\pm$    0.03    &         353$\pm$          10&   18.37    $\pm$    0.01    &  -22.16    $\pm$    0.01    &    0.89    $\pm$    0.02    &   11.55    $\pm$    0.02    &   11.21    $\pm$    0.15    &    6.29    $\pm$    0.29    &    2.86    $\pm$    1.18    \\
          84 &    2.32    $\pm$    0.03    &         369$\pm$          12&   18.91    $\pm$    0.01    &  -21.58    $\pm$    0.01    &    0.85    $\pm$    0.02    &   11.59    $\pm$    0.02    &   10.94    $\pm$    0.15    &   11.79    $\pm$    0.62    &    2.65    $\pm$    1.10    \\
          85 &    1.19    $\pm$    0.01    &         389$\pm$          15&   17.76    $\pm$    0.01    &  -21.26    $\pm$    0.01    &    0.79    $\pm$    0.02    &   11.20    $\pm$    0.02    &   10.68    $\pm$    0.15    &    6.46    $\pm$    0.32    &    1.95    $\pm$    0.81    \\
          86 &    2.58    $\pm$    0.04    &         428$\pm$          17&   18.43    $\pm$    0.01    &  -22.33    $\pm$    0.01    &    0.83    $\pm$    0.02    &   11.70    $\pm$    0.02    &   11.18    $\pm$    0.15    &    7.67    $\pm$    0.41    &    2.33    $\pm$    0.96    \\
 \end{supertabular}
 \label{list_candiates_S}
 % \twocolumn
% \end{landscape}%}

%\FloatBarrier

\section{The cross-match list and, the Ta10 and Tr09 samples}
\label{taylor_cross}
 % \afterpage{\onecolumn
 %\begin{landscape}
% \centering
   \bottomcaption{Cross-match list of all IDs of all galaxies of used in this investigation. First column: Object IDs used by SDSS DR10. Second column: internal IDs of our candidate sample. Third column: Sersic IDs of the alternative sample based on the parameters from \citet{Simard:2011} provided in Appendix \ref{sersic_cand}. Fourth column: IDs of \citet{Taylor:2010} as listed in their paper. Fifth column: IDNY from \citet{Trujillo:2009}.}
 \tablehead{SDSS DR10 ID & internal ID & Sersic ID & Taylor ID & Trujillo IDNY & SDSS DR10 ID & internal ID& Sersic ID & Taylor ID & Trujillo IDNY \\ \hline}
 \begin{supertabular}{ccccc|ccccc}
 1237648704060129355&-           &-           &          55&-           &1237648721255596242&           1&-           &-           &-           \\
 1237648703523520846&           2&           1&-           &-           &1237651191892607189&           3&           2&-           &-           \\
 1237650760782905596&-           &-           &-           &      155310&1237651538710167661&-           &-           &-           &      225402\\
 1237651252557513010&-           &           3&-           &-           &1237651753466462236&           4&-           &-           &-           \\
 1237652934037536913&           5&           4&-           &-           &1237654880201932994&-           &-           &-           &      460843\\
 1237654342254002376&-           &           5&-           &-           &1237652900773298301&           6&           6&          49&-           \\
 1237652629102067836&           7&-           &-           &-           &1237651252589363420&           8&           7&-           &-           \\
 1237655502424769160&           9&           8&-           &-           &1237656496713892027&-           &-           &-           &      685469\\
 1237657401874710721&-           &-           &          62&-           &1237651539246186637&          10&           9&-           &-           \\
 1237654400765591702&-           &-           &          30&-           &1237651735773708418&          11&-           &-           &-           \\
 1237655742944248167&-           &          10&-           &-           &1237662267538997604&-           &-           &          21&-           \\
 1237662501086691600&-           &-           &          38&-           &1237659329240236080&          12&-           &-           &-           \\
 1237666339727671425&          13&-           &          19&-           &1237651714798125236&          14&          11&-           &-           \\
 1237660615586611373&-           &          12&-           &-           &1237658206124507259&          15&          13&-           &-           \\
 1237652943695184336&-           &-           &-           &      321479&1237658204522807485&-           &-           &-           &      796740\\
 1237652944786424004&          16&          14&-           &-           &1237652629104427133&-           &-           &          22&-           \\
 1237658423007707334&-           &          15&-           &-           &1237660412113912034&-           &-           &-           &      929051\\
 1237652629103968326&-           &-           &          18&-           &1237662267540570526&          17&-           &-           &-           \\
 1237653651308871866&-           &-           &          15&-           &1237654400224592070&-           &-           &-           &      415405\\
 1237650796219662509&-           &-           &          14&-           &1237657242435584230&-           &          16&-           &-           \\
 1237657856067830007&-           &          17&-           &-           &1237654391106896136&-           &-           &-           &      411130\\
 1237654952670003535&-           &          18&-           &-           &1237662524157460585&-           &-           &           1&-           \\
 1237670956787695816&-           &          19&-           &-           &1237658300604809510&-           &-           &-           &      815852\\
 1237652948530102500&          18&          20&-           &-           &1237653665789575334&-           &-           &-           &      417973\\
 1237662524694659165&-           &-           &          42&-           &1237657590319022174&-           &-           &-           &      721837\\
 1237656241159995854&          19&-           &-           &-           &1237656243317113067&          20&          21&-           &-           \\
 1237657610723655845&-           &          22&-           &-           &1237655474503024820&          21&          23&-           &-           \\
 1237661070319091925&-           &-           &           8&-           &1237658423018389671&-           &-           &-           &      824795\\
 1237658424616616162&-           &          24&-           &-           &1237657769628926193&-           &-           &          27&-           \\
 1237657190367297807&-           &-           &          48&-           &1237657596224209238&          22&          25&-           &-           \\
 1237662264318034136&          23&-           &-           &-           &1237671265496006878&-           &-           &          20&-           \\
 1237657401346687209&-           &-           &          36&-           &1237665569297203655&          24&          26&-           &-           \\
 1237674650998341919&-           &-           &          33&-           &1237654605857751221&          25&          27&-           &-           \\
 1237655126084157462&-           &-           &          47&-           &1237653614796865660&-           &          28&-           &-           \\
 1237655463236141124&-           &-           &           9&-           &1237655465916170402&          26&          29&-           &-           \\
 1237658204493185306&-           &          30&-           &-           &1237658206117036087&-           &-           &          26&-           \\
 1237659324945072200&-           &-           &-           &      896687&1237657628456190055&          27&          31&-           &-           \\
 1237657874328715438&-           &-           &          24&-           &1237661356460671120&-           &-           &          12&-           \\
 1237659153685610726&-           &-           &          37&-           &1237660343928750289&-           &-           &          57&-           \\
 1237660962936062177&-           &-           &-           &      986020&1237660025032081578&          28&          32&-           &-           \\
 1237661064411349290&          29&          33&-           &-           &1237661064941929079&-           &-           &          45&-           \\
 1237659161735397586&-           &-           &-           &      890167&1237661068721586383&-           &-           &          61&-           \\
 1237661849849430137&          30&          34&-           &-           &1237661069261209757&-           &          35&-           &-           \\
 1237662663746060502&-           &          36&-           &-           &1237663277928022281&          31&          37&-           &-           \\
 1237661383314702588&          32&          38&-           &-           &1237662619722711187&-           &-           &          50&-           \\
 1237662697568796852&          33&          39&-           &-           &1237665128542044254&-           &-           &          34&-           \\
 1237665531170783414&-           &-           &          56&-           &1237661812272857187&          34&          40&-           &-           \\
 1237665532252520624&          35&          41&-           &-           &1237667255083991162&-           &          42&-           &-           \\
 1237667324323758158&-           &          43&-           &-           &1237667735054647478&-           &          44&-           &-           \\
 1237662193459986552&-           &          45&-           &-           &1237667736104861820&-           &          46&-           &-           \\
 1237662224087974057&          36&-           &-           &-           &1237670965389557929&-           &-           &          39&-           \\
 1237664130483618005&          37&-           &-           &-           &1237664669510074510&          38&-           &-           &-           \\
 1237665549429899544&          39&          47&          35&-           &1237667209978380503&          40&          48&-           &-           \\
 1237663278461944053&          41&          49&-           &-           &1237662340012638220&          42&          50&-           &-           \\
 1237662664292630745&-           &-           &          11&-           &1237662307269804288&-           &-           &          52&-           \\
 1237667781740331285&-           &-           &          41&-           &1237664667887140986&          43&          51&-           &-           \\
 1237664093432119636&          44&          52&-           &-           &1237665533335175692&-           &          53&          54&-           \\
 1237661850400260193&          45&          54&-           &-           &1237664852035174654&          46&          55&-           &-           \\
 1237663543683711270&-           &-           &          43&-           &1237665098466656347&-           &-           &          63&-           \\
 1237667321652248694&-           &          56&-           &-           &1237667429035540562&          47&-           &-           &-           \\
 1237664339328172101&-           &-           &-           &     1780650&1237673808655221213&          48&-           &-           &-           \\
 1237667730736873763&-           &          57&-           &-           &1237665373329096903&-           &-           &          10&-           \\
 1237664853648015625&-           &-           &           3&-           &1237664854715727968&          49&-           &-           &-           \\
 1237662664290402490&-           &          58&          17&-           &1237665535469486145&          50&          59&-           &-           \\
 1237665016311840908&-           &          60&-           &-           &1237667212115050932&-           &          61&-           &-           \\
 1237667442972754078&-           &-           &          53&-           &1237665564997976239&-           &          62&-           &-           \\
 1237663478723969457&          51&          63&-           &-           &1237665440978698364&          52&-           &          59&-           \\
 1237667910055100586&          53&          64&          13&-           &1237667252924842120&-           &-           &-           &     2258945\\
 1237667734526492801&          54&          65&-           &-           &1237667782277071029&-           &-           &          46&-           \\
 1237670450522816720&-           &          66&-           &-           &1237668495245705310&-           &-           &-           &     2402259\\
 1237663789032669425&-           &          67&-           &-           &1237662619725005006&          55&-           &-           &-           \\
 1237664869745230095&          56&-           &-           &-           &1237665429169242591&          57&          68&-           &-           \\
 1237665440975224988&          58&-           &-           &-           &1237668299281662070&          59&          69&-           &-           \\
 1237668349753950509&          60&          70&-           &-           &1237670449986273410&-           &-           &          31&-           \\
 1237661871876669606&-           &-           &           5&-           &1237667783900135493&-           &          71&-           &-           \\
 1237668271372501042&          61&          72&-           &-           &1237648721758978188&          62&          73&-           &-           \\
 1237663547440431315&-           &-           &           2&-           &1237668625165975629&-           &-           &          23&-           \\
 1237668495782117442&-           &          74&-           &-           &1237648720716890184&-           &-           &-           &       54829\\
 1237674365919363403&-           &-           &           7&-           &1237664671640715458&          63&          75&-           &-           \\
 1237661356469387315&-           &-           &          60&-           &1237665351319552146&-           &-           &           6&-           \\
 1237661433237733495&-           &-           &          25&-           &1237661139034046601&-           &-           &-           &     1044397\\
 1237667735062708393&          64&-           &-           &-           &1237662237484646804&-           &-           &           4&-           \\
 1237662335717015837&          65&          76&-           &-           &1237667782274187688&-           &          77&-           &-           \\
 1237661087497126080&-           &          78&-           &-           &1237668585969877156&-           &-           &-           &     2434587\\
 1237668310021440087&          66&-           &-           &-           &1237662195064438832&-           &-           &-           &     1173134\\
 1237661358617067696&          67&          79&-           &-           &1237648721747378408&-           &-           &          28&-           \\
 1237653651837026391&-           &          80&-           &-           &1237668298203070641&          68&          81&-           &-           \\
 1237662336261685637&-           &-           &          29&-           &1237662224621240601&-           &-           &          51&-           \\
 1237654604796985469&-           &          82&-           &-           &1237662236945088747&-           &-           &          58&-           \\
 1237662336794820961&          69&-           &-           &-           &1237667917032980629&          70&          83&-           &-           \\
 1237662224614490342&          71&-           &-           &-           &1237661950244945934&          72&          84&-           &-           \\
 1237668333640810655&          73&          85&-           &-           &1237662236410577091&          74&-           &-           &-           \\
 1237662302971691136&          75&-           &-           &-           &1237661971718799467&-           &-           &          40&-           \\
 1237667917030555837&          76&          86&-           &-           &                   &            &            &            &            \\
 \end{supertabular}
 \label{ID_crossmatches}
%  \twocolumn
% \end{landscape}}

 % \afterpage{\onecolumn
% \begin{landscape}
% \centering
   \bottomcaption{List of the basic parameters of all galaxies in our basic sample that are also parts of the galaxies provided in \citet{Taylor:2010}. First column: IDs used in the table in their paper. Second column: object ID used by SDSS DR10. Third and fourth column: equatorial coordinates of the galaxies. Fifth column: redshift $z$, already corrected for our motion relative to the CMB. Sixth, seventh, and eighth columns: observed uncorrected refitted SDSS parameters in the following order: observed apparent magnitude $m_{\textrm{sdss}}$, angular semi-major axis $a_{\textrm{sdss}}$, central velocity dispersion $\sigma_{\textrm{sdss}}$. Ninth column: axis ratio $q_{b/a}$. Tenth column: GalaxyZoo probability $\mathcal{L}_{\textrm{ETG}}$ of the galaxy being classified as an early-type.}
 \tablehead{Taylor ID & SDSS DR10 ID & ra & dec & $z$ & $m_{\textrm{sdss},r}$ &$a_{\textrm{sdss}}$ & $\sigma_{\textrm{sdss}}$ & $q_{b/a}$ & $\mathcal{L}_{\textrm{ETG}}$\\ 
  &  & [$^{\circ}$]&  [$^{\circ}$]&   & [mag] & [arcsec] & [km/s] & & \\ \hline }
 \begin{supertabular}{cccccccccc}
           55&1237648704060129355&  228.8519  &   -0.3402  &    0.1001  &   16.49    $\pm$    0.00    &    1.62    $\pm$    0.03    &         359$\pm$          11&    0.84    &    0.86    \\
           49&1237652900773298301&   58.0541  &   -5.8611  &    0.1137  &   17.25    $\pm$    0.01    &    1.03    $\pm$    0.02    &         306$\pm$          14&    0.38    &    0.66    \\
           62&1237657401874710721&  121.4479  &   32.8120  &    0.1203  &   16.97    $\pm$    0.01    &    1.38    $\pm$    0.03    &         227$\pm$          10&    0.80    &    0.78    \\
           30&1237654400765591702&  178.7375  &   65.7964  &    0.1071  &   16.74    $\pm$    0.00    &    1.11    $\pm$    0.01    &         192$\pm$           8&    0.62    &    0.75    \\
           21&1237662267538997604&  231.9958  &    5.0639  &    0.0872  &   16.94    $\pm$    0.00    &    1.00    $\pm$    0.02    &         229$\pm$           8&    0.85    &    0.88    \\
           38&1237662501086691600&  253.5743  &   26.9582  &    0.1035  &   17.10    $\pm$    0.00    &    0.81    $\pm$    0.01    &         174$\pm$           8&    0.46    &    0.67    \\
           19&1237666339727671425&   20.8205  &    0.2955  &    0.0928  &   17.10    $\pm$    0.00    &    1.11    $\pm$    0.02    &         296$\pm$          11&    0.73    &    0.88    \\
           22&1237652629104427133&   13.5572  &  -10.6207  &    0.1189  &   17.70    $\pm$    0.01    &    0.83    $\pm$    0.04    &         143$\pm$          14&    0.91    &    0.86    \\
           18&1237652629103968326&   12.4721  &  -10.7547  &    0.0983  &   17.36    $\pm$    0.01    &    1.06    $\pm$    0.04    &         184$\pm$          12&    0.90    &    0.68    \\
           15&1237653651308871866&   25.4214  &   13.6498  &    0.0724  &   16.43    $\pm$    0.00    &    1.25    $\pm$    0.02    &         183$\pm$           6&    0.87    &    0.81    \\
           14&1237650796219662509&  145.3475  &    0.0544  &    0.0913  &   17.08    $\pm$    0.00    &    0.95    $\pm$    0.02    &         163$\pm$           7&    0.98    &    0.84    \\
            1&1237662524157460585&  190.1666  &   13.8156  &    0.0865  &   16.38    $\pm$    0.00    &    1.02    $\pm$    0.02    &         160$\pm$           6&    0.77    &    0.83    \\
           42&1237662524694659165&  190.9405  &   14.1608  &    0.0877  &   16.59    $\pm$    0.00    &    1.43    $\pm$    0.02    &         231$\pm$           8&    0.58    &    0.76    \\
            8&1237661070319091925&  143.0571  &   11.7045  &    0.0821  &   16.49    $\pm$    0.00    &    0.90    $\pm$    0.01    &         166$\pm$           6&    0.86    &    0.70    \\
           27&1237657769628926193&  131.3568  &   41.5528  &    0.1015  &   16.96    $\pm$    0.00    &    1.19    $\pm$    0.03    &         192$\pm$           8&    0.98    &    0.79    \\
           48&1237657190367297807&  357.6717  &   -0.6124  &    0.0794  &   16.07    $\pm$    0.00    &    1.77    $\pm$    0.02    &         187$\pm$           7&    0.58    &    0.68    \\
           20&1237671265496006878&  191.8619  &   -1.5344  &    0.0887  &   16.89    $\pm$    0.00    &    1.16    $\pm$    0.02    &         249$\pm$          10&    0.64    &    0.83    \\
           36&1237657401346687209&  141.6903  &   45.8730  &    0.0799  &   16.53    $\pm$    0.00    &    1.23    $\pm$    0.02    &         221$\pm$           9&    0.98    &    0.79    \\
           33&1237674650998341919&  170.7120  &    0.4215  &    0.1040  &   17.18    $\pm$    0.01    &    1.12    $\pm$    0.02    &         259$\pm$           9&    0.70    &    0.63    \\
           47&1237655126084157462&  187.1464  &    5.5812  &    0.0676  &   15.30    $\pm$    0.00    &    1.52    $\pm$    0.01    &         175$\pm$           4&    0.87    &    0.88    \\
            9&1237655463236141124&  204.6658  &   59.8185  &    0.0707  &   16.43    $\pm$    0.00    &    1.16    $\pm$    0.01    &         235$\pm$           7&    0.78    &    0.82    \\
           26&1237658206117036087&  166.7747  &   49.6303  &    0.1069  &   17.17    $\pm$    0.00    &    1.07    $\pm$    0.03    &         189$\pm$          10&    0.96    &    0.77    \\
           24&1237657874328715438&  139.8645  &   40.1167  &    0.0937  &   16.73    $\pm$    0.00    &    1.29    $\pm$    0.03    &         167$\pm$           7&    0.89    &    0.89    \\
           12&1237661356460671120&  153.4443  &   42.0479  &    0.1060  &   17.03    $\pm$    0.00    &    0.93    $\pm$    0.01    &         143$\pm$           8&    0.71    &    0.81    \\
           37&1237659153685610726&  254.3686  &   26.7014  &    0.1198  &   17.66    $\pm$    0.01    &    1.00    $\pm$    0.04    &         203$\pm$          12&    0.92    &    0.82    \\
           57&1237660343928750289&  127.0233  &   30.0714  &    0.1097  &   17.13    $\pm$    0.01    &    1.90    $\pm$    0.06    &         129$\pm$           9&    0.96    &    0.91    \\
           45&1237661064941929079&  124.0758  &    5.9420  &    0.1032  &   17.00    $\pm$    0.00    &    1.01    $\pm$    0.02    &         221$\pm$          11&    0.98    &    0.81    \\
           61&1237661068721586383&  173.8238  &   13.9530  &    0.0821  &   15.61    $\pm$    0.00    &    2.25    $\pm$    0.02    &         279$\pm$           7&    0.89    &    0.97    \\
           50&1237662619722711187&  235.1621  &   32.1894  &    0.1183  &   17.18    $\pm$    0.01    &    2.25    $\pm$    0.04    &         173$\pm$          15&    0.56    &    0.75    \\
           34&1237665128542044254&  179.5965  &   35.0486  &    0.0807  &   16.69    $\pm$    0.00    &    1.39    $\pm$    0.02    &         199$\pm$           8&    0.79    &    0.87    \\
           56&1237665531170783414&  203.4256  &   25.7488  &    0.0742  &   15.59    $\pm$    0.00    &    2.70    $\pm$    0.02    &         237$\pm$           6&    0.96    &    1.00    \\
           39&1237670965389557929&  149.4396  &   16.5398  &    0.1017  &   16.72    $\pm$    0.00    &    1.25    $\pm$    0.02    &         183$\pm$           7&    0.64    &    0.74    \\
           35&1237665549429899544&  223.0734  &   22.4871  &    0.1165  &   17.39    $\pm$    0.01    &    0.78    $\pm$    0.01    &         335$\pm$          13&    0.29    &    0.62    \\
           11&1237662664292630745&  244.4193  &   24.3831  &    0.0829  &   17.20    $\pm$    0.01    &    1.70    $\pm$    0.04    &         152$\pm$           9&    0.57    &    0.83    \\
           52&1237662307269804288&  232.0260  &   32.5324  &    0.0918  &   16.49    $\pm$    0.00    &    1.56    $\pm$    0.02    &         247$\pm$           8&    0.68    &    0.79    \\
           41&1237667781740331285&  135.7019  &   14.4294  &    0.1141  &   16.97    $\pm$    0.00    &    0.96    $\pm$    0.02    &         163$\pm$           6&    0.82    &    0.87    \\
           54&1237665533335175692&  243.8410  &   16.3942  &    0.0818  &   15.87    $\pm$    0.00    &    1.86    $\pm$    0.01    &         311$\pm$           7&    0.58    &    0.81    \\
           43&1237663543683711270&  331.9419  &    0.3080  &    0.0978  &   17.05    $\pm$    0.01    &    1.20    $\pm$    0.03    &         216$\pm$          12&    0.94    &    0.88    \\
           63&1237665098466656347&  149.9874  &   30.2277  &    0.0833  &   15.61    $\pm$    0.00    &    2.62    $\pm$    0.03    &         289$\pm$           8&    0.66    &    0.90    \\
           10&1237665373329096903&  230.2855  &   24.2198  &    0.0813  &   16.75    $\pm$    0.00    &    1.17    $\pm$    0.02    &         153$\pm$           6&    0.98    &    0.79    \\
            3&1237664853648015625&  225.3171  &   30.5827  &    0.0984  &   17.05    $\pm$    0.00    &    0.92    $\pm$    0.01    &         195$\pm$           6&    0.70    &    0.88    \\
           17&1237662664290402490&  239.6933  &   27.2131  &    0.0879  &   17.04    $\pm$    0.00    &    1.17    $\pm$    0.02    &         292$\pm$          15&    0.79    &    0.74    \\
           53&1237667442972754078&  189.3463  &   27.3214  &    0.1009  &   16.61    $\pm$    0.00    &    1.39    $\pm$    0.02    &         256$\pm$           8&    0.84    &    0.91    \\
           59&1237665440978698364&  194.2722  &   28.9814  &    0.0686  &   15.45    $\pm$    0.00    &    2.19    $\pm$    0.01    &         340$\pm$           8&    0.57    &    0.78    \\
           13&1237667910055100586&  181.7985  &   23.8744  &    0.0775  &   16.55    $\pm$    0.00    &    1.21    $\pm$    0.02    &         328$\pm$          11&    0.77    &    0.86    \\
           46&1237667782277071029&  135.2174  &   14.7181  &    0.0959  &   17.63    $\pm$    0.01    &    1.24    $\pm$    0.04    &         204$\pm$          11&    0.82    &    0.84    \\
           31&1237670449986273410&  138.7474  &   16.3422  &    0.0909  &   16.88    $\pm$    0.00    &    1.61    $\pm$    0.03    &         167$\pm$           8&    0.65    &    0.78    \\
            5&1237661871876669606&  215.4104  &   40.0323  &    0.1003  &   17.51    $\pm$    0.01    &    1.06    $\pm$    0.02    &         176$\pm$          10&    0.42    &    0.74    \\
            2&1237663547440431315&  127.0272  &   55.3799  &    0.0669  &   16.42    $\pm$    0.00    &    1.31    $\pm$    0.03    &         191$\pm$           7&    0.93    &    0.81    \\
           23&1237668625165975629&  199.4207  &   17.6978  &    0.0739  &   15.97    $\pm$    0.00    &    1.50    $\pm$    0.01    &         141$\pm$           5&    0.65    &    0.80    \\
            7&1237674365919363403&  118.8170  &   33.2286  &    0.0985  &   17.06    $\pm$    0.00    &    0.92    $\pm$    0.01    &         154$\pm$           7&    0.62    &    0.67    \\
           60&1237661356469387315&  180.7761  &   46.6946  &    0.0730  &   15.16    $\pm$    0.00    &    1.91    $\pm$    0.01    &         267$\pm$           6&    0.64    &    0.93    \\
            6&1237665351319552146&  222.1299  &   26.4879  &    0.1063  &   16.80    $\pm$    0.00    &    0.71    $\pm$    0.02    &         155$\pm$           6&    0.92    &    0.68    \\
           25&1237661433237733495&  195.3300  &   46.1813  &    0.0914  &   16.77    $\pm$    0.00    &    1.22    $\pm$    0.02    &         212$\pm$           9&    0.63    &    0.80    \\
            4&1237662237484646804&  227.0853  &    7.2533  &    0.0770  &   16.79    $\pm$    0.00    &    1.03    $\pm$    0.02    &         199$\pm$           8&    1.00    &    0.76    \\
           28&1237648721747378408&  133.7980  &    0.2189  &    0.1020  &   16.60    $\pm$    0.00    &    0.95    $\pm$    0.01    &         183$\pm$           6&    0.44    &    0.72    \\
           29&1237662336261685637&  252.5918  &   22.1319  &    0.1182  &   17.62    $\pm$    0.01    &    0.78    $\pm$    0.01    &         262$\pm$          14&    0.43    &    0.73    \\
           51&1237662224621240601&  230.9765  &   29.9078  &    0.1128  &   17.09    $\pm$    0.01    &    1.42    $\pm$    0.03    &         207$\pm$          11&    0.80    &    0.88    \\
           58&1237662236945088747&  220.8522  &    7.6574  &    0.0842  &   15.48    $\pm$    0.00    &    1.64    $\pm$    0.02    &         234$\pm$           5&    0.93    &    0.87    \\
           40&1237661971718799467&  176.6058  &    7.6119  &    0.0867  &   16.17    $\pm$    0.00    &    1.17    $\pm$    0.01    &         206$\pm$           5&    0.61    &    0.74    \\
 \end{supertabular}
 \label{list_candiates_basics_T}
 % \twocolumn
 %\end{landscape}}

 % \afterpage{\onecolumn
% \begin{landscape}
 %\centering
   \bottomcaption{List of the derived parameters of all galaxies in our basic sample that are also parts of the galaxies provided in \citet{Taylor:2010}. First column: IDs used in the table in their paper. Second column: scale radius $R_{\textrm{r}}$ of the galaxies measured in the SDSS r band (in kpc). Third column: corrected central velocity dispersion $\sigma_{0}$ (in km/s). Fourth column: surface brightness $\mu_{r}$ measured in the SDSS r band (in mag/arcsec$^{2}$). Fifth column: absolute magnitude in r band $M_{\textrm{r}}$. Sixth column: g-r colour $(M_{\textrm{g}}-M_{\textrm{r}})$ (in mag). Seventh column: logarithm of the dynamical mass $M_{\textrm{dyn}}$ (in solar masses). Eighth column: logarithm of the stellar mass $M_{\textrm{*}}$ (in solar masses). Ninth column: dynamical mass-to-light ratio $\vernal_{\textrm{dyn}}$ (in solar units $M_{\astrosun}/L_{\astrosun ,\textrm{r}}$). Tenth column: stellar mass-to-light ratio $\vernal_{*}$ (in solar units $M_{\astrosun}/L_{\astrosun ,\textrm{r}}$).}
 \tablehead{Taylor ID & $R_{\textrm{r}}$ & $\sigma_{0}$ & $\mu_{r}$ & $M_{\textrm{r}}$ & $(M_{\textrm{g}}-M_{\textrm{r}})$ & $M_{\textrm{z}}$ & log$_{10}$($M_{\textrm{dyn}}$) &  log$_{10}$($M_{*}$) & $\vernal_{\textrm{dyn}}$ & $\vernal_{*}$ \\
  & [kpc] & [km s$^{-1}$] & [$\frac{\textrm{mag}}{\textrm{arcsec}^{2}}$]  & [mag] & [mag] & [mag] & [log$_{10}(M_{\astrosun})$] & [log$_{10}(M_{\astrosun})$] & [$M_{\astrosun}/L_{\astrosun,\textrm{r}}$]& [$M_{\astrosun}/L_{\astrosun,\textrm{r}}$] \\ \hline}
 \begin{supertabular}{ccccccccccc}
           55&    2.76    $\pm$    0.05    &         391$\pm$          12&   18.80    $\pm$    0.04    &  -22.08    $\pm$    0.01    &    0.77    $\pm$    0.01    &  -22.68    $\pm$    0.01    &   11.69    $\pm$    0.02    &   11.12    $\pm$    0.15    &    9.39    $\pm$    0.40    &    2.51    $\pm$    1.04    \\
           49&    1.32    $\pm$    0.04    &         344$\pm$          15&   17.50    $\pm$    0.06    &  -21.79    $\pm$    0.01    &    0.77    $\pm$    0.01    &  -22.46    $\pm$    0.01    &   11.26    $\pm$    0.02    &   10.95    $\pm$    0.15    &    4.54    $\pm$    0.26    &    2.23    $\pm$    0.92    \\
           62&    2.71    $\pm$    0.06    &         248$\pm$          11&   18.81    $\pm$    0.05    &  -22.05    $\pm$    0.01    &    0.76    $\pm$    0.01    &  -22.91    $\pm$    0.01    &   11.29    $\pm$    0.02    &   11.23    $\pm$    0.15    &    3.84    $\pm$    0.21    &    3.32    $\pm$    1.37    \\
           30&    1.72    $\pm$    0.02    &         213$\pm$           9&   17.99    $\pm$    0.03    &  -21.87    $\pm$    0.01    &    0.69    $\pm$    0.01    &  -22.40    $\pm$    0.01    &   10.96    $\pm$    0.02    &   10.91    $\pm$    0.15    &    2.12    $\pm$    0.10    &    1.89    $\pm$    0.78    \\
           21&    1.51    $\pm$    0.03    &         253$\pm$           8&   18.29    $\pm$    0.04    &  -21.27    $\pm$    0.01    &    0.81    $\pm$    0.01    &  -21.98    $\pm$    0.01    &   11.05    $\pm$    0.02    &   10.80    $\pm$    0.15    &    4.57    $\pm$    0.19    &    2.55    $\pm$    1.05    \\
           38&    1.04    $\pm$    0.03    &         197$\pm$          10&   17.15    $\pm$    0.05    &  -21.62    $\pm$    0.01    &    0.76    $\pm$    0.01    &  -22.20    $\pm$    0.01    &   10.67    $\pm$    0.02    &   10.83    $\pm$    0.15    &    1.38    $\pm$    0.08    &    2.00    $\pm$    0.83    \\
           19&    1.66    $\pm$    0.03    &         327$\pm$          13&   18.54    $\pm$    0.04    &  -21.22    $\pm$    0.01    &    0.76    $\pm$    0.01    &  -21.87    $\pm$    0.01    &   11.31    $\pm$    0.02    &   10.78    $\pm$    0.15    &    8.76    $\pm$    0.41    &    2.58    $\pm$    1.07    \\
           22&    1.72    $\pm$    0.08    &         159$\pm$          16&   18.62    $\pm$    0.10    &  -21.25    $\pm$    0.01    &    0.85    $\pm$    0.02    &  -21.87    $\pm$    0.02    &   10.70    $\pm$    0.05    &   10.87    $\pm$    0.15    &    2.09    $\pm$    0.24    &    3.03    $\pm$    1.25    \\
           18&    1.84    $\pm$    0.07    &         203$\pm$          14&   18.90    $\pm$    0.09    &  -21.10    $\pm$    0.01    &    0.76    $\pm$    0.02    &  -21.83    $\pm$    0.01    &   10.94    $\pm$    0.03    &   10.82    $\pm$    0.15    &    4.18    $\pm$    0.34    &    3.11    $\pm$    1.28    \\
           15&    1.62    $\pm$    0.03    &         201$\pm$           7&   18.33    $\pm$    0.04    &  -21.37    $\pm$    0.01    &    0.75    $\pm$    0.01    &  -22.02    $\pm$    0.01    &   10.88    $\pm$    0.02    &   10.78    $\pm$    0.15    &    2.84    $\pm$    0.12    &    2.21    $\pm$    0.91    \\
           14&    1.61    $\pm$    0.04    &         180$\pm$           8&   18.39    $\pm$    0.05    &  -21.31    $\pm$    0.01    &    0.81    $\pm$    0.01    &  -21.89    $\pm$    0.01    &   10.78    $\pm$    0.02    &   10.79    $\pm$    0.15    &    2.38    $\pm$    0.13    &    2.41    $\pm$    1.00    \\
            1&    1.45    $\pm$    0.02    &         178$\pm$           6&   17.71    $\pm$    0.04    &  -21.76    $\pm$    0.01    &    0.69    $\pm$    0.01    &  -22.26    $\pm$    0.01    &   10.73    $\pm$    0.02    &   10.76    $\pm$    0.15    &    1.38    $\pm$    0.06    &    1.50    $\pm$    0.62    \\
           42&    1.79    $\pm$    0.03    &         254$\pm$           9&   18.34    $\pm$    0.03    &  -21.59    $\pm$    0.01    &    0.79    $\pm$    0.01    &  -22.21    $\pm$    0.01    &   11.13    $\pm$    0.02    &   10.88    $\pm$    0.15    &    4.07    $\pm$    0.17    &    2.27    $\pm$    0.94    \\
            8&    1.30    $\pm$    0.01    &         185$\pm$           7&   17.72    $\pm$    0.03    &  -21.51    $\pm$    0.01    &    0.69    $\pm$    0.01    &  -22.09    $\pm$    0.01    &   10.71    $\pm$    0.02    &   10.69    $\pm$    0.15    &    1.69    $\pm$    0.07    &    1.59    $\pm$    0.66    \\
           27&    2.23    $\pm$    0.05    &         210$\pm$           9&   18.85    $\pm$    0.05    &  -21.56    $\pm$    0.01    &    0.74    $\pm$    0.01    &  -22.15    $\pm$    0.01    &   11.06    $\pm$    0.02    &   10.76    $\pm$    0.15    &    3.54    $\pm$    0.18    &    1.78    $\pm$    0.74    \\
           48&    2.03    $\pm$    0.02    &         204$\pm$           8&   18.31    $\pm$    0.03    &  -21.88    $\pm$    0.01    &    0.71    $\pm$    0.01    &  -22.46    $\pm$    0.01    &   10.99    $\pm$    0.02    &   11.01    $\pm$    0.15    &    2.28    $\pm$    0.10    &    2.36    $\pm$    0.98    \\
           20&    1.55    $\pm$    0.03    &         276$\pm$          11&   18.30    $\pm$    0.04    &  -21.31    $\pm$    0.01    &    0.80    $\pm$    0.01    &  -21.91    $\pm$    0.01    &   11.14    $\pm$    0.02    &   10.79    $\pm$    0.15    &    5.32    $\pm$    0.26    &    2.39    $\pm$    0.99    \\
           36&    1.84    $\pm$    0.03    &         242$\pm$           9&   18.56    $\pm$    0.04    &  -21.42    $\pm$    0.01    &    0.80    $\pm$    0.01    &  -22.12    $\pm$    0.01    &   11.10    $\pm$    0.02    &   10.89    $\pm$    0.15    &    4.40    $\pm$    0.20    &    2.72    $\pm$    1.12    \\
           33&    1.80    $\pm$    0.03    &         286$\pm$          10&   18.53    $\pm$    0.04    &  -21.43    $\pm$    0.01    &    0.80    $\pm$    0.01    &  -22.04    $\pm$    0.01    &   11.23    $\pm$    0.02    &   10.83    $\pm$    0.15    &    6.02    $\pm$    0.26    &    2.39    $\pm$    0.99    \\
           47&    1.85    $\pm$    0.02    &         190$\pm$           5&   17.74    $\pm$    0.02    &  -22.24    $\pm$    0.00    &    0.70    $\pm$    0.01    &  -22.74    $\pm$    0.01    &   10.89    $\pm$    0.01    &   10.97    $\pm$    0.15    &    1.30    $\pm$    0.04    &    1.56    $\pm$    0.64    \\
            9&    1.38    $\pm$    0.02    &         259$\pm$           7&   18.15    $\pm$    0.03    &  -21.20    $\pm$    0.01    &    0.77    $\pm$    0.01    &  -21.84    $\pm$    0.01    &   11.03    $\pm$    0.01    &   10.74    $\pm$    0.15    &    4.67    $\pm$    0.16    &    2.38    $\pm$    0.98    \\
           26&    2.07    $\pm$    0.06    &         209$\pm$          11&   18.81    $\pm$    0.06    &  -21.46    $\pm$    0.01    &    0.76    $\pm$    0.01    &  -22.08    $\pm$    0.01    &   11.02    $\pm$    0.02    &   10.84    $\pm$    0.15    &    3.59    $\pm$    0.22    &    2.38    $\pm$    0.98    \\
           24&    2.13    $\pm$    0.05    &         183$\pm$           8&   18.74    $\pm$    0.05    &  -21.57    $\pm$    0.01    &    0.74    $\pm$    0.01    &  -22.15    $\pm$    0.01    &   10.92    $\pm$    0.02    &   10.85    $\pm$    0.15    &    2.56    $\pm$    0.13    &    2.18    $\pm$    0.90    \\
           12&    1.54    $\pm$    0.03    &         160$\pm$           9&   18.06    $\pm$    0.04    &  -21.55    $\pm$    0.01    &    0.69    $\pm$    0.01    &  -22.13    $\pm$    0.01    &   10.66    $\pm$    0.03    &   10.68    $\pm$    0.15    &    1.42    $\pm$    0.09    &    1.51    $\pm$    0.62    \\
           37&    2.09    $\pm$    0.08    &         225$\pm$          13&   18.91    $\pm$    0.08    &  -21.38    $\pm$    0.01    &    0.82    $\pm$    0.02    &  -22.07    $\pm$    0.02    &   11.09    $\pm$    0.03    &   10.89    $\pm$    0.15    &    4.50    $\pm$    0.32    &    2.81    $\pm$    1.16    \\
           57&    3.76    $\pm$    0.12    &         139$\pm$          10&   19.96    $\pm$    0.07    &  -21.61    $\pm$    0.01    &    0.67    $\pm$    0.01    &  -22.20    $\pm$    0.01    &   10.92    $\pm$    0.03    &   10.84    $\pm$    0.15    &    2.50    $\pm$    0.20    &    2.06    $\pm$    0.85    \\
           45&    1.90    $\pm$    0.04    &         244$\pm$          12&   18.52    $\pm$    0.05    &  -21.56    $\pm$    0.01    &    0.81    $\pm$    0.01    &  -22.16    $\pm$    0.01    &   11.12    $\pm$    0.02    &   10.98    $\pm$    0.15    &    4.11    $\pm$    0.24    &    2.95    $\pm$    1.22    \\
           61&    3.31    $\pm$    0.03    &         299$\pm$           7&   18.83    $\pm$    0.02    &  -22.42    $\pm$    0.01    &    1.05    $\pm$    0.01    &  -23.02    $\pm$    0.01    &   11.54    $\pm$    0.01    &   11.19    $\pm$    0.15    &    4.83    $\pm$    0.14    &    2.19    $\pm$    0.90    \\
           50&    3.64    $\pm$    0.09    &         187$\pm$          16&   19.75    $\pm$    0.05    &  -21.74    $\pm$    0.01    &    0.77    $\pm$    0.01    &  -22.18    $\pm$    0.01    &   11.17    $\pm$    0.04    &   10.83    $\pm$    0.15    &    3.86    $\pm$    0.35    &    1.78    $\pm$    0.74    \\
           34&    1.89    $\pm$    0.03    &         218$\pm$           9&   18.78    $\pm$    0.03    &  -21.26    $\pm$    0.01    &    0.79    $\pm$    0.01    &  -21.89    $\pm$    0.01    &   11.02    $\pm$    0.02    &   10.83    $\pm$    0.15    &    4.29    $\pm$    0.20    &    2.74    $\pm$    1.13    \\
           56&    3.74    $\pm$    0.03    &         252$\pm$           7&   19.37    $\pm$    0.02    &  -22.14    $\pm$    0.01    &    0.81    $\pm$    0.01    &  -22.71    $\pm$    0.01    &   11.44    $\pm$    0.01    &   11.12    $\pm$    0.15    &    5.00    $\pm$    0.15    &    2.37    $\pm$    0.98    \\
           39&    1.89    $\pm$    0.04    &         202$\pm$           8&   18.24    $\pm$    0.04    &  -21.81    $\pm$    0.01    &    0.69    $\pm$    0.01    &  -22.38    $\pm$    0.01    &   10.95    $\pm$    0.02    &   10.90    $\pm$    0.15    &    2.19    $\pm$    0.10    &    1.93    $\pm$    0.80    \\
           35&    0.89    $\pm$    0.02    &         383$\pm$          15&   16.94    $\pm$    0.05    &  -21.50    $\pm$    0.01    &    0.80    $\pm$    0.01    &  -22.20    $\pm$    0.01    &   11.18    $\pm$    0.02    &   10.87    $\pm$    0.15    &    4.96    $\pm$    0.25    &    2.45    $\pm$    1.01    \\
           11&    2.00    $\pm$    0.06    &         166$\pm$           9&   19.22    $\pm$    0.06    &  -20.95    $\pm$    0.01    &    0.71    $\pm$    0.01    &  -21.56    $\pm$    0.02    &   10.81    $\pm$    0.03    &   10.53    $\pm$    0.15    &    3.51    $\pm$    0.23    &    1.86    $\pm$    0.77    \\
           52&    2.22    $\pm$    0.03    &         270$\pm$           9&   18.60    $\pm$    0.03    &  -21.80    $\pm$    0.01    &    0.83    $\pm$    0.01    &  -22.43    $\pm$    0.01    &   11.27    $\pm$    0.02    &   10.99    $\pm$    0.15    &    4.67    $\pm$    0.19    &    2.45    $\pm$    1.01    \\
           41&    1.81    $\pm$    0.03    &         181$\pm$           6&   18.11    $\pm$    0.04    &  -21.87    $\pm$    0.01    &    0.72    $\pm$    0.01    &  -22.47    $\pm$    0.01    &   10.84    $\pm$    0.02    &   10.87    $\pm$    0.15    &    1.62    $\pm$    0.07    &    1.75    $\pm$    0.72    \\
           54&    2.18    $\pm$    0.02    &         339$\pm$           7&   18.14    $\pm$    0.02    &  -22.21    $\pm$    0.01    &    0.76    $\pm$    0.01    &  -22.87    $\pm$    0.01    &   11.46    $\pm$    0.01    &   11.12    $\pm$    0.15    &    4.97    $\pm$    0.14    &    2.25    $\pm$    0.93    \\
           43&    2.12    $\pm$    0.05    &         238$\pm$          13&   18.78    $\pm$    0.05    &  -21.52    $\pm$    0.01    &    0.77    $\pm$    0.01    &  -22.15    $\pm$    0.01    &   11.14    $\pm$    0.02    &   10.88    $\pm$    0.15    &    4.47    $\pm$    0.27    &    2.42    $\pm$    1.00    \\
           63&    3.34    $\pm$    0.05    &         310$\pm$           9&   18.85    $\pm$    0.03    &  -22.42    $\pm$    0.01    &    0.80    $\pm$    0.01    &  -23.08    $\pm$    0.01    &   11.57    $\pm$    0.01    &   11.30    $\pm$    0.15    &    5.22    $\pm$    0.18    &    2.80    $\pm$    1.15    \\
           10&    1.78    $\pm$    0.03    &         168$\pm$           6&   18.60    $\pm$    0.03    &  -21.31    $\pm$    0.01    &    0.78    $\pm$    0.01    &  -21.90    $\pm$    0.01    &   10.77    $\pm$    0.02    &   10.80    $\pm$    0.15    &    2.30    $\pm$    0.10    &    2.49    $\pm$    1.03    \\
            3&    1.42    $\pm$    0.02    &         218$\pm$           7&   18.06    $\pm$    0.04    &  -21.37    $\pm$    0.01    &    0.74    $\pm$    0.01    &  -21.94    $\pm$    0.01    &   10.89    $\pm$    0.02    &   10.71    $\pm$    0.15    &    2.88    $\pm$    0.11    &    1.92    $\pm$    0.79    \\
           17&    1.71    $\pm$    0.03    &         322$\pm$          17&   18.64    $\pm$    0.04    &  -21.19    $\pm$    0.01    &    0.82    $\pm$    0.01    &  -21.89    $\pm$    0.01    &   11.31    $\pm$    0.02    &   10.82    $\pm$    0.15    &    9.00    $\pm$    0.54    &    2.88    $\pm$    1.19    \\
           53&    2.39    $\pm$    0.04    &         280$\pm$           9&   18.70    $\pm$    0.03    &  -21.87    $\pm$    0.01    &    0.79    $\pm$    0.01    &  -22.50    $\pm$    0.01    &   11.34    $\pm$    0.02    &   11.05    $\pm$    0.15    &    5.08    $\pm$    0.20    &    2.60    $\pm$    1.07    \\
           59&    2.17    $\pm$    0.02    &         368$\pm$           9&   18.22    $\pm$    0.02    &  -22.10    $\pm$    0.01    &    0.84    $\pm$    0.01    &  -22.80    $\pm$    0.01    &   11.53    $\pm$    0.01    &   11.10    $\pm$    0.15    &    6.43    $\pm$    0.19    &    2.37    $\pm$    0.98    \\
           13&    1.57    $\pm$    0.03    &         361$\pm$          12&   18.30    $\pm$    0.03    &  -21.33    $\pm$    0.01    &    0.76    $\pm$    0.01    &  -21.91    $\pm$    0.01    &   11.38    $\pm$    0.02    &   10.85    $\pm$    0.15    &    9.11    $\pm$    0.39    &    2.70    $\pm$    1.12    \\
           46&    2.01    $\pm$    0.07    &         224$\pm$          12&   19.33    $\pm$    0.07    &  -20.85    $\pm$    0.01    &    0.86    $\pm$    0.02    &  -21.65    $\pm$    0.01    &   11.07    $\pm$    0.03    &   10.83    $\pm$    0.15    &    6.97    $\pm$    0.47    &    4.04    $\pm$    1.67    \\
           31&    2.21    $\pm$    0.05    &         182$\pm$           8&   19.00    $\pm$    0.05    &  -21.39    $\pm$    0.01    &    0.71    $\pm$    0.01    &  -22.01    $\pm$    0.01    &   10.93    $\pm$    0.02    &   10.75    $\pm$    0.15    &    3.10    $\pm$    0.17    &    2.05    $\pm$    0.85    \\
            5&    1.28    $\pm$    0.04    &         197$\pm$          12&   18.26    $\pm$    0.06    &  -20.95    $\pm$    0.01    &    0.75    $\pm$    0.01    &  -21.78    $\pm$    0.01    &   10.76    $\pm$    0.03    &   10.57    $\pm$    0.15    &    3.13    $\pm$    0.22    &    2.01    $\pm$    0.83    \\
            2&    1.63    $\pm$    0.03    &         209$\pm$           7&   18.54    $\pm$    0.04    &  -21.16    $\pm$    0.01    &    0.74    $\pm$    0.01    &  -21.82    $\pm$    0.01    &   10.92    $\pm$    0.02    &   10.72    $\pm$    0.15    &    3.72    $\pm$    0.16    &    2.35    $\pm$    0.97    \\
           23&    1.71    $\pm$    0.02    &         154$\pm$           6&   18.04    $\pm$    0.02    &  -21.77    $\pm$    0.00    &    0.67    $\pm$    0.01    &  -22.29    $\pm$    0.01    &   10.68    $\pm$    0.02    &   10.77    $\pm$    0.15    &    1.21    $\pm$    0.05    &    1.52    $\pm$    0.63    \\
            7&    1.32    $\pm$    0.02    &         172$\pm$           8&   17.84    $\pm$    0.04    &  -21.44    $\pm$    0.01    &    0.70    $\pm$    0.01    &  -22.12    $\pm$    0.01    &   10.66    $\pm$    0.02    &   10.78    $\pm$    0.15    &    1.58    $\pm$    0.08    &    2.09    $\pm$    0.86    \\
           60&    2.13    $\pm$    0.02    &         290$\pm$           6&   17.73    $\pm$    0.02    &  -22.55    $\pm$    0.00    &    0.74    $\pm$    0.01    &  -23.23    $\pm$    0.01    &   11.32    $\pm$    0.01    &   11.02    $\pm$    0.15    &    2.59    $\pm$    0.07    &    1.29    $\pm$    0.53    \\
            6&    1.33    $\pm$    0.03    &         174$\pm$           7&   17.48    $\pm$    0.05    &  -21.83    $\pm$    0.01    &    0.69    $\pm$    0.01    &  -22.38    $\pm$    0.01    &   10.67    $\pm$    0.02    &   10.84    $\pm$    0.15    &    1.13    $\pm$    0.05    &    1.66    $\pm$    0.68    \\
           25&    1.66    $\pm$    0.03    &         235$\pm$          10&   18.29    $\pm$    0.04    &  -21.47    $\pm$    0.01    &    0.78    $\pm$    0.01    &  -22.07    $\pm$    0.01    &   11.03    $\pm$    0.02    &   10.81    $\pm$    0.15    &    3.58    $\pm$    0.17    &    2.17    $\pm$    0.90    \\
            4&    1.50    $\pm$    0.04    &         220$\pm$           9&   18.44    $\pm$    0.05    &  -21.09    $\pm$    0.01    &    0.77    $\pm$    0.01    &  -21.74    $\pm$    0.01    &   10.93    $\pm$    0.02    &   10.69    $\pm$    0.15    &    4.02    $\pm$    0.20    &    2.34    $\pm$    0.97    \\
           28&    1.20    $\pm$    0.02    &         206$\pm$           6&   17.08    $\pm$    0.03    &  -21.99    $\pm$    0.01    &    0.67    $\pm$    0.01    &  -22.54    $\pm$    0.01    &   10.77    $\pm$    0.01    &   10.84    $\pm$    0.15    &    1.24    $\pm$    0.04    &    1.46    $\pm$    0.60    \\
           29&    1.10    $\pm$    0.03    &         297$\pm$          16&   17.54    $\pm$    0.06    &  -21.37    $\pm$    0.01    &    0.83    $\pm$    0.02    &  -22.06    $\pm$    0.02    &   11.05    $\pm$    0.02    &   10.88    $\pm$    0.15    &    4.20    $\pm$    0.26    &    2.83    $\pm$    1.17    \\
           51&    2.62    $\pm$    0.06    &         227$\pm$          12&   19.08    $\pm$    0.05    &  -21.70    $\pm$    0.01    &    0.82    $\pm$    0.01    &  -22.26    $\pm$    0.01    &   11.19    $\pm$    0.02    &   10.93    $\pm$    0.15    &    4.29    $\pm$    0.26    &    2.31    $\pm$    0.95    \\
           58&    2.52    $\pm$    0.03    &         253$\pm$           5&   18.02    $\pm$    0.03    &  -22.64    $\pm$    0.00    &    0.70    $\pm$    0.01    &  -23.22    $\pm$    0.01    &   11.27    $\pm$    0.01    &   11.18    $\pm$    0.15    &    2.15    $\pm$    0.06    &    1.74    $\pm$    0.72    \\
           40&    1.49    $\pm$    0.02    &         229$\pm$           6&   17.58    $\pm$    0.03    &  -21.95    $\pm$    0.00    &    0.72    $\pm$    0.01    &  -22.54    $\pm$    0.01    &   10.96    $\pm$    0.01    &   10.88    $\pm$    0.15    &    1.97    $\pm$    0.06    &    1.65    $\pm$    0.68    \\
 \end{supertabular}
 \label{list_candiates_T}
%  \twocolumn
 %\end{landscape}}

%  \afterpage{\onecolumn
% \begin{landscape}
% \centering
   \bottomcaption{List of the basic parameters of all galaxies in our basic sample that are also parts of the galaxies provided in \citet{Trujillo:2009}. First column: IDNYs used in the table in their paper. Second column: object ID used by SDSS DR10. Third and fourth column: equatorial coordinates of the galaxies. Fifth column: redshift $z$, already corrected for our motion relative to the CMB. Sixth, seventh, and eighth columns: observed uncorrected refitted SDSS parameters in the following order: observed apparent magnitude $m_{\textrm{sdss}}$, angular semi-major axis $a_{\textrm{sdss}}$, central velocity dispersion $\sigma_{\textrm{sdss}}$. Ninth column: axis ratio $q_{b/a}$. Tenth column: GalaxyZoo probability $\mathcal{L}_{\textrm{ETG}}$ of the galaxy being classified as an early-type.}
 \tablehead{Trujillo IDNY & SDSS DR10 ID & ra & dec & $z$ & $m_{\textrm{sdss},r}$ &$a_{\textrm{sdss}}$ & $\sigma_{\textrm{sdss}}$ & $q_{b/a}$ & $\mathcal{L}_{\textrm{ETG}}$\\ 
  &  & [$^{\circ}$]&  [$^{\circ}$]&   & [mag] & [arcsec] & [km/s] & & \\ \hline }
 \begin{supertabular}{cccccccccc}
       155310&1237650760782905596&  186.7713  &   -3.2216  &    0.1665  &   17.42    $\pm$    0.01    &    0.59    $\pm$    0.01    &         202$\pm$           8&    0.44    &    0.78    \\
       225402&1237651538710167661&  172.5828  &   66.8247  &    0.1441  &   17.09    $\pm$    0.00    &    0.76    $\pm$    0.01    &         188$\pm$           5&    0.22    &    0.55    \\
       460843&1237654880201932994&  219.0524  &    4.0700  &    0.1534  &   17.50    $\pm$    0.01    &    0.49    $\pm$    0.01    &         204$\pm$           9&    0.56    &    0.53    \\
       685469&1237656496713892027&  335.4180  &   13.9873  &    0.1486  &   17.33    $\pm$    0.01    &    0.90    $\pm$    0.01    &         174$\pm$           9&    0.45    &    0.69    \\
       321479&1237652943695184336&  320.2198  &   11.1203  &    0.1274  &   16.64    $\pm$    0.00    &    0.76    $\pm$    0.01    &         216$\pm$           8&    0.48    &    0.82    \\
       796740&1237658204522807485&  221.9016  &   43.4960  &    0.1828  &   17.59    $\pm$    0.01    &    0.70    $\pm$    0.01    &         178$\pm$           9&    0.34    &    0.69    \\
       929051&1237660412113912034&  139.8602  &    6.8893  &    0.1856  &   17.50    $\pm$    0.01    &    0.48    $\pm$    0.01    &         168$\pm$          12&    0.78    &    0.56    \\
       415405&1237654400224592070&  157.7106  &   62.9833  &    0.1675  &   17.65    $\pm$    0.01    &    0.71    $\pm$    0.01    &         189$\pm$          10&    0.29    &    0.53    \\
       411130&1237654391106896136&  127.3659  &   46.2254  &    0.1683  &   17.66    $\pm$    0.01    &    0.80    $\pm$    0.01    &         199$\pm$           9&    0.26    &    0.59    \\
       815852&1237658300604809510&  151.6223  &    7.2351  &    0.1222  &   16.64    $\pm$    0.00    &    0.83    $\pm$    0.01    &         181$\pm$           7&    0.62    &    0.79    \\
       417973&1237653665789575334&  135.8508  &    2.4459  &    0.1890  &   17.38    $\pm$    0.01    &    0.82    $\pm$    0.01    &         221$\pm$          10&    0.25    &    0.55    \\
       721837&1237657590319022174&  167.9007  &   53.6700  &    0.1427  &   16.80    $\pm$    0.00    &    0.51    $\pm$    0.01    &         186$\pm$           5&    0.74    &    0.78    \\
       824795&1237658423018389671&  163.3506  &    6.4059  &    0.1873  &   17.65    $\pm$    0.01    &    0.58    $\pm$    0.01    &         189$\pm$          10&    0.63    &    0.74    \\
       896687&1237659324945072200&  218.9466  &   54.5913  &    0.1305  &   16.59    $\pm$    0.00    &    0.66    $\pm$    0.01    &         187$\pm$           6&    0.96    &    0.80    \\
       986020&1237660962936062177&  129.8227  &   30.6294  &    0.1798  &   17.76    $\pm$    0.01    &    0.58    $\pm$    0.01    &         233$\pm$          11&    0.43    &    0.81    \\
       890167&1237659161735397586&  234.8920  &   44.2979  &    0.1436  &   17.36    $\pm$    0.01    &    0.53    $\pm$    0.01    &         155$\pm$           9&    0.49    &    0.54    \\
      1780650&1237664339328172101&  180.7131  &   38.2790  &    0.1579  &   17.36    $\pm$    0.01    &    0.67    $\pm$    0.01    &         211$\pm$          10&    0.57    &    0.60    \\
      2258945&1237667252924842120&  141.8472  &   21.9347  &    0.1686  &   17.22    $\pm$    0.01    &    0.87    $\pm$    0.01    &         231$\pm$           9&    0.29    &    0.54    \\
      2402259&1237668495245705310&  177.6347  &   17.0510  &    0.1566  &   17.07    $\pm$    0.00    &    0.56    $\pm$    0.02    &         213$\pm$          10&    1.00    &    0.76    \\
        54829&1237648720716890184&  232.5811  &   -0.4885  &    0.0861  &   16.09    $\pm$    0.00    &    0.78    $\pm$    0.01    &         130$\pm$           4&    0.92    &    0.75    \\
      1044397&1237661139034046601&  154.1551  &   39.0343  &    0.1965  &   17.77    $\pm$    0.01    &    0.62    $\pm$    0.02    &         196$\pm$          10&    0.68    &    0.80    \\
      2434587&1237668585969877156&  169.2473  &   17.1548  &    0.1739  &   17.67    $\pm$    0.01    &    0.61    $\pm$    0.01    &         199$\pm$          10&    0.48    &    0.76    \\
      1173134&1237662195064438832&  188.1617  &   42.8557  &    0.1668  &   17.76    $\pm$    0.01    &    0.84    $\pm$    0.02    &         202$\pm$          12&    0.30    &    0.79    \\
 \end{supertabular}
 \label{list_candiates_basics_trujillo}
 % \twocolumn
 %\end{landscape}}

 % \afterpage{\onecolumn
% \begin{landscape}
 %\centering
   \bottomcaption{List of the derived parameters of all galaxies in our basic sample that are also parts of the galaxies provided in \citet{Trujillo:2009}. First column: IDNYs used in the table in their paper. Second column: scale radius $R_{\textrm{r}}$ of the galaxies measured in the SDSS r band (in kpc). Third column: corrected central velocity dispersion $\sigma_{0}$ (in km/s). Fourth column: surface brightness $\mu_{r}$ measured in the SDSS r band (in mag/arcsec$^{2}$). Fifth column: absolute magnitude in r band $M_{\textrm{r}}$. Sixth column: g-r colour $(M_{\textrm{g}}-M_{\textrm{r}})$ (in mag). Seventh column: logarithm of the dynamical mass $M_{\textrm{dyn}}$ (in solar masses). Eighth column: logarithm of the stellar mass $M_{\textrm{*}}$ (in solar masses). Ninth column: dynamical mass-to-light ratio $\vernal_{\textrm{dyn}}$ (in solar units $M_{\astrosun}/L_{\astrosun ,\textrm{r}}$). Tenth column: stellar mass-to-light ratio $\vernal_{*}$ (in solar units $M_{\astrosun}/L_{\astrosun ,\textrm{r}}$).}
 \tablehead{Trujillo IDNY & $R_{\textrm{r}}$ & $\sigma_{0}$ & $\mu_{r}$ & $M_{\textrm{r}}$ & $(M_{\textrm{g}}-M_{\textrm{r}})$ & $M_{\textrm{z}}$ & log$_{10}$($M_{\textrm{dyn}}$) &  log$_{10}$($M_{*}$) & $\vernal_{\textrm{dyn}}$ & $\vernal_{*}$ \\
  & [kpc] & [km s$^{-1}$] & [$\frac{\textrm{mag}}{\textrm{arcsec}^{2}}$]  & [mag] & [mag] & [mag] & [log$_{10}(M_{\astrosun})$] & [log$_{10}(M_{\astrosun})$] & [$M_{\astrosun}/L_{\astrosun,\textrm{r}}$]& [$M_{\astrosun}/L_{\astrosun,\textrm{r}}$] \\ \hline}
 \begin{supertabular}{ccccccccccc}
       155310&    1.14    $\pm$    0.02    &         232$\pm$           9&   16.67    $\pm$    0.05    &  -22.35    $\pm$    0.01    &    0.69    $\pm$    0.01    &  -22.91    $\pm$    0.01    &   10.85    $\pm$    0.02    &   11.02    $\pm$    0.15    &    1.06    $\pm$    0.05    &    1.57    $\pm$    0.65    \\
       225402&    0.92    $\pm$    0.02    &         217$\pm$           5&   16.31    $\pm$    0.04    &  -22.23    $\pm$    0.01    &    0.71    $\pm$    0.01    &  -22.81    $\pm$    0.01    &   10.70    $\pm$    0.01    &   11.02    $\pm$    0.15    &    0.84    $\pm$    0.03    &    1.74    $\pm$    0.72    \\
       460843&    1.00    $\pm$    0.02    &         234$\pm$          10&   16.66    $\pm$    0.05    &  -22.07    $\pm$    0.01    &    0.72    $\pm$    0.01    &  -22.59    $\pm$    0.01    &   10.80    $\pm$    0.02    &   10.94    $\pm$    0.15    &    1.23    $\pm$    0.06    &    1.67    $\pm$    0.69    \\
       685469&    1.58    $\pm$    0.03    &         196$\pm$          10&   17.49    $\pm$    0.04    &  -22.23    $\pm$    0.01    &    0.72    $\pm$    0.01    &  -22.76    $\pm$    0.01    &   10.85    $\pm$    0.02    &   11.02    $\pm$    0.15    &    1.18    $\pm$    0.07    &    1.74    $\pm$    0.72    \\
       321479&    1.21    $\pm$    0.02    &         245$\pm$           9&   16.60    $\pm$    0.03    &  -22.52    $\pm$    0.01    &    0.63    $\pm$    0.01    &  -23.01    $\pm$    0.01    &   10.93    $\pm$    0.02    &   10.91    $\pm$    0.15    &    1.08    $\pm$    0.05    &    1.04    $\pm$    0.43    \\
       796740&    1.28    $\pm$    0.04    &         204$\pm$          10&   16.93    $\pm$    0.07    &  -22.36    $\pm$    0.01    &    0.71    $\pm$    0.01    &  -22.85    $\pm$    0.02    &   10.79    $\pm$    0.03    &   11.01    $\pm$    0.15    &    0.92    $\pm$    0.06    &    1.50    $\pm$    0.62    \\
       929051&    1.35    $\pm$    0.04    &         192$\pm$          14&   16.90    $\pm$    0.06    &  -22.51    $\pm$    0.01    &    0.66    $\pm$    0.01    &  -22.98    $\pm$    0.02    &   10.76    $\pm$    0.03    &   10.98    $\pm$    0.15    &    0.75    $\pm$    0.06    &    1.24    $\pm$    0.51    \\
       415405&    1.12    $\pm$    0.02    &         217$\pm$          12&   16.91    $\pm$    0.04    &  -22.08    $\pm$    0.01    &    0.75    $\pm$    0.01    &  -22.65    $\pm$    0.01    &   10.79    $\pm$    0.02    &   11.00    $\pm$    0.15    &    1.18    $\pm$    0.07    &    1.92    $\pm$    0.79    \\
       411130&    1.20    $\pm$    0.03    &         228$\pm$          10&   17.02    $\pm$    0.06    &  -22.13    $\pm$    0.01    &    0.73    $\pm$    0.01    &  -22.66    $\pm$    0.01    &   10.86    $\pm$    0.02    &   10.99    $\pm$    0.15    &    1.34    $\pm$    0.07    &    1.82    $\pm$    0.75    \\
       815852&    1.45    $\pm$    0.03    &         204$\pm$           8&   17.21    $\pm$    0.04    &  -22.30    $\pm$    0.01    &    0.67    $\pm$    0.01    &  -22.90    $\pm$    0.01    &   10.85    $\pm$    0.02    &   10.95    $\pm$    0.15    &    1.10    $\pm$    0.05    &    1.40    $\pm$    0.58    \\
       417973&    1.33    $\pm$    0.04    &         252$\pm$          12&   16.68    $\pm$    0.06    &  -22.71    $\pm$    0.01    &    0.67    $\pm$    0.01    &  -23.23    $\pm$    0.01    &   10.99    $\pm$    0.02    &   11.14    $\pm$    0.15    &    1.06    $\pm$    0.06    &    1.51    $\pm$    0.62    \\
       721837&    1.11    $\pm$    0.02    &         212$\pm$           6&   16.48    $\pm$    0.04    &  -22.46    $\pm$    0.00    &    0.64    $\pm$    0.01    &  -22.94    $\pm$    0.01    &   10.76    $\pm$    0.01    &   11.01    $\pm$    0.15    &    0.78    $\pm$    0.03    &    1.38    $\pm$    0.57    \\
       824795&    1.47    $\pm$    0.04    &         215$\pm$          12&   17.20    $\pm$    0.06    &  -22.41    $\pm$    0.01    &    0.70    $\pm$    0.01    &  -23.02    $\pm$    0.01    &   10.90    $\pm$    0.03    &   11.05    $\pm$    0.15    &    1.13    $\pm$    0.07    &    1.59    $\pm$    0.66    \\
       896687&    1.51    $\pm$    0.03    &         211$\pm$           7&   17.15    $\pm$    0.04    &  -22.46    $\pm$    0.00    &    0.64    $\pm$    0.01    &  -22.87    $\pm$    0.01    &   10.89    $\pm$    0.02    &   10.98    $\pm$    0.15    &    1.06    $\pm$    0.04    &    1.31    $\pm$    0.54    \\
       986020&    1.17    $\pm$    0.03    &         268$\pm$          12&   16.89    $\pm$    0.06    &  -22.22    $\pm$    0.01    &    0.73    $\pm$    0.01    &  -22.79    $\pm$    0.01    &   10.99    $\pm$    0.02    &   11.01    $\pm$    0.15    &    1.65    $\pm$    0.09    &    1.72    $\pm$    0.71    \\
       890167&    0.95    $\pm$    0.02    &         178$\pm$          10&   16.66    $\pm$    0.05    &  -21.96    $\pm$    0.01    &    0.68    $\pm$    0.01    &  -22.61    $\pm$    0.01    &   10.54    $\pm$    0.03    &   10.91    $\pm$    0.15    &    0.76    $\pm$    0.05    &    1.75    $\pm$    0.72    \\
      1780650&    1.41    $\pm$    0.04    &         239$\pm$          11&   17.29    $\pm$    0.06    &  -22.19    $\pm$    0.01    &    0.69    $\pm$    0.01    &  -22.72    $\pm$    0.01    &   10.97    $\pm$    0.02    &   10.91    $\pm$    0.15    &    1.63    $\pm$    0.09    &    1.42    $\pm$    0.59    \\
      2258945&    1.38    $\pm$    0.04    &         262$\pm$          10&   16.88    $\pm$    0.06    &  -22.56    $\pm$    0.01    &    0.74    $\pm$    0.01    &  -23.12    $\pm$    0.01    &   11.04    $\pm$    0.02    &   11.18    $\pm$    0.15    &    1.36    $\pm$    0.07    &    1.85    $\pm$    0.76    \\
      2402259&    1.55    $\pm$    0.05    &         241$\pm$          11&   17.19    $\pm$    0.06    &  -22.49    $\pm$    0.01    &    0.61    $\pm$    0.01    &  -22.97    $\pm$    0.01    &   11.02    $\pm$    0.02    &   11.03    $\pm$    0.15    &    1.37    $\pm$    0.08    &    1.42    $\pm$    0.59    \\
        54829&    1.22    $\pm$    0.01    &         146$\pm$           5&   16.86    $\pm$    0.03    &  -22.23    $\pm$    0.00    &    0.63    $\pm$    0.01    &  -22.75    $\pm$    0.01    &   10.48    $\pm$    0.01    &   10.95    $\pm$    0.15    &    0.50    $\pm$    0.02    &    1.47    $\pm$    0.61    \\
      1044397&    1.72    $\pm$    0.06    &         223$\pm$          12&   17.56    $\pm$    0.08    &  -22.39    $\pm$    0.01    &    0.73    $\pm$    0.01    &  -22.95    $\pm$    0.01    &   10.99    $\pm$    0.03    &   11.11    $\pm$    0.15    &    1.43    $\pm$    0.09    &    1.84    $\pm$    0.76    \\
      2434587&    1.28    $\pm$    0.04    &         228$\pm$          12&   17.14    $\pm$    0.06    &  -22.14    $\pm$    0.01    &    0.71    $\pm$    0.01    &  -22.70    $\pm$    0.01    &   10.89    $\pm$    0.02    &   10.97    $\pm$    0.15    &    1.40    $\pm$    0.09    &    1.70    $\pm$    0.70    \\
      1173134&    1.34    $\pm$    0.05    &         230$\pm$          13&   17.40    $\pm$    0.08    &  -21.99    $\pm$    0.01    &    0.73    $\pm$    0.01    &  -22.62    $\pm$    0.02    &   10.92    $\pm$    0.03    &   11.00    $\pm$    0.15    &    1.73    $\pm$    0.12    &    2.09    $\pm$    0.86    \\
 \end{supertabular}
 \label{list_candiates_trujillo}
  \twocolumn
 \end{landscape}}

\FloatBarrier

\addcontentsline{toc}{section}{References}
\bibliography{paper}\label{bib}

\end{document}